\documentclass{emulateapj}

\tabletypesize{\scriptsize}
\newcommand{\NUMBERHERERMSLIMIT}{15}
\newcommand{\WASPTwelveEclipseKsBandOneAperturePixels}{16}
\newcommand{\WASPTwelveEclipseKsBandOneEnsembleNStars}{4}
\newcommand{\WASPTwelveEclipseKsBandTwoAperturePixels}{15}
\newcommand{\WASPTwelveEclipseKsBandTwoEnsembleNStars}{6}
\newcommand{\WASPTwelveEclipseKsBandThreeAperturePixels}{21}
\newcommand{\WASPTwelveEclipseKsBandThreeEnsembleNStars}{5}
\newcommand{\WASPTwelveEclipseYBandAperturePixels}{16}
\newcommand{\WASPTwelveEclipseYBandEnsembleNStars}{5}
\newcommand{\WASPTwelveEclipseKContBandAperturePixels}{10}
\newcommand{\WASPTwelveEclipseKContBandEnsembleNStars}{10}
\newcommand{\WASPThreeEclipseKsBandOneAperturePixels}{14}
\newcommand{\WASPThreeEclipseKsBandOneEnsembleNStars}{2}
\newcommand{\WASPThreeEclipseKsBandTwoAperturePixels}{16}
\newcommand{\WASPThreeEclipseKsBandTwoEnsembleNStars}{1}
\newcommand{\QatarOneEclipseKsBandAperturePixels}{14}
\newcommand{\QatarOneEclipseKsBandEnsembleNStars}{5}
\newcommand{\KELTOneEclipseKsBandAperturePixels}{19}
\newcommand{\KELTOneEclipseKsBandEnsembleNStars}{4}
\newcommand{\FpOverFStarPercentAbstractUndilutedRealWASPTwelveKsBand}{0.311}
\newcommand{\FpOverFStarPercentAbstractUndilutedRealPlusWASPTwelveKsBand}{0.021}
\newcommand{\FpOverFStarPercentAbstractUndilutedRealMinusWASPTwelveKsBand}{0.021}
\newcommand{\FpOverFStarPercentAbstractUndilutedRealWASPTwelveKsBandTwentyEleven}{0.319}
\newcommand{\FpOverFStarPercentAbstractUndilutedRealPlusWASPTwelveKsBandTwentyEleven}{0.021}
\newcommand{\FpOverFStarPercentAbstractUndilutedRealMinusWASPTwelveKsBandTwentyEleven}{0.021}
\newcommand{\FpOverFStarPercentAbstractUndilutedRealWASPTwelveKsBandDecTwentyEleven}{0.280}
\newcommand{\FpOverFStarPercentAbstractUndilutedRealPlusWASPTwelveKsBandDecTwentyEleven}{0.049}
\newcommand{\FpOverFStarPercentAbstractUndilutedRealMinusWASPTwelveKsBandDecTwentyEleven}{0.049}
\newcommand{\FpOverFStarPercentAbstractUndilutedRealWASPTwelveYBandTwentyEleven}{0.109}
\newcommand{\FpOverFStarPercentAbstractUndilutedRealPlusWASPTwelveYBandTwentyEleven}{0.014}
\newcommand{\FpOverFStarPercentAbstractUndilutedRealMinusWASPTwelveYBandTwentyEleven}{0.014}
\newcommand{\FpOverFStarPercentAbstractUndilutedRealWASPTwelveKContBandJanTwentyTwelve}{0.301}
\newcommand{\FpOverFStarPercentAbstractUndilutedRealPlusWASPTwelveKContBandJanTwentyTwelve}{0.046}
\newcommand{\FpOverFStarPercentAbstractUndilutedRealMinusWASPTwelveKContBandJanTwentyTwelve}{0.046}

\newcommand{\FpOverFStarPercentAbstractWeatherPlotWASPThreeKsBandCombined}{0.193}
\newcommand{\FpOverFStarPercentAbstractPlusMinusWeatherPlotWASPThreeKsBandCombined}{0.014}
\newcommand{\FpOverFStarPercentDiffWeatherPlotWASPThreeKsBandCombined}{0.041}

\newcommand{\TBrightWeatherPlotWASPThreeKsBandCombined}{2576}
\newcommand{\TBrightPlusWeatherPlotWASPThreeKsBandCombined}{64}
\newcommand{\TBrightMinusWeatherPlotWASPThreeKsBandCombined}{66}
\newcommand{\TBrightDiffHeatherFuhrWeatherPlotWASPThreeKsBandCombined}{187}
\newcommand{\ReducedChiSquaredWeatherPlotWASPThreeKsBandCombined}{1.4}
\newcommand{\FpOverFStarPercentAbstractWeatherPlotWASPTwelveKsBandCombined}{0.296}
\newcommand{\FpOverFStarPercentAbstractPlusMinusWeatherPlotWASPTwelveKsBandCombined}{0.014}
\newcommand{\FpOverFStarPercentDiffWeatherPlotWASPTwelveKsBandCombined}{0.031}

\newcommand{\TBrightWeatherPlotWASPTwelveKsBandCombined}{3050}
\newcommand{\TBrightPlusWeatherPlotWASPTwelveKsBandCombined}{56}
\newcommand{\TBrightMinusWeatherPlotWASPTwelveKsBandCombined}{57}
\newcommand{\TBrightDiffHeatherFuhrWeatherPlotWASPTwelveKsBandCombined}{126}
\newcommand{\ReducedChiSquaredWeatherPlotWASPTwelveKsBandCombined}{0.3}

\newcommand{\FpOverFStarPercentAbstractCorrKELTOneKsBandAll}{0.160}
\newcommand{\FpOverFStarPercentAbstractCorrPlusKELTOneKsBandAll}{0.018}
\newcommand{\FpOverFStarPercentAbstractCorrMinusKELTOneKsBandAll}{0.020}
\newcommand{\TOffsetCorrKELTOneKsBandAll}{-6.5}

\newcommand{\TOffsetCorrPlusKELTOneKsBandAll}{2.2}
\newcommand{\TOffsetCorrMinusKELTOneKsBandAll}{2.0}

\newcommand{\PhaseAbstractCorrKELTOneKsBandAll}{0.4963}
\newcommand{\PhaseAbstractCorrMinusKELTOneKsBandAll}{0.0011}
\newcommand{\PhaseAbstractCorrPlusKELTOneKsBandAll}{0.0012}

\newcommand{\ECosOmegaCorrKELTOneKsBandAll}{-0.0058}
\newcommand{\ECosOmegaCorrPlusKELTOneKsBandAll}{0.0019}
\newcommand{\ECosOmegaCorrMinusKELTOneKsBandAll}{0.0019}
\newcommand{\JDOffsetCorrKELTOneKsBandAll}{16211.8405}
\newcommand{\JDOffsetCorrPlusKELTOneKsBandAll}{0.0015}
\newcommand{\JDOffsetCorrMinusKELTOneKsBandAll}{0.0014}

\newcommand{\FpOverFStarPercentAbstractCorrQatarOneKsBandAll}{0.136}
\newcommand{\FpOverFStarPercentAbstractCorrPlusQatarOneKsBandAll}{0.034}
\newcommand{\FpOverFStarPercentAbstractCorrMinusQatarOneKsBandAll}{0.034}
\newcommand{\TOffsetCorrQatarOneKsBandAll}{-3.4}

\newcommand{\TOffsetCorrPlusQatarOneKsBandAll}{2.6}
\newcommand{\TOffsetCorrMinusQatarOneKsBandAll}{2.6}

\newcommand{\PhaseAbstractCorrQatarOneKsBandAll}{0.4983}
\newcommand{\PhaseAbstractCorrMinusQatarOneKsBandAll}{0.0013}
\newcommand{\PhaseAbstractCorrPlusQatarOneKsBandAll}{0.0013}

\newcommand{\ECosOmegaCorrQatarOneKsBandAll}{-0.0026}
\newcommand{\ECosOmegaCorrPlusQatarOneKsBandAll}{0.0020}
\newcommand{\ECosOmegaCorrMinusQatarOneKsBandAll}{0.0020}
\newcommand{\JDOffsetCorrQatarOneKsBandAll}{16136.8322}
\newcommand{\JDOffsetCorrPlusQatarOneKsBandAll}{0.0018}
\newcommand{\JDOffsetCorrMinusQatarOneKsBandAll}{0.0018}

\newcommand{\FpOverFStarPercentAbstractCorrWASPThreeKsBandEclipseOneAll}{0.234}
\newcommand{\FpOverFStarPercentAbstractCorrPlusWASPThreeKsBandEclipseOneAll}{0.029}
\newcommand{\FpOverFStarPercentAbstractCorrMinusWASPThreeKsBandEclipseOneAll}{0.030}
\newcommand{\TOffsetCorrWASPThreeKsBandEclipseOneAll}{0.0}

\newcommand{\PhaseAbstractCorrWASPThreeKsBandEclipseOneAll}{0.5000}

\newcommand{\ECosOmegaCorrWASPThreeKsBandEclipseOneAll}{0.0000}

\newcommand{\JDOffsetCorrWASPThreeKsBandEclipseOneAll}{14986.9315}

\newcommand{\FpOverFStarPercentAbstractCorrWASPThreeKsBandEclipseTwoAll}{0.159}
\newcommand{\FpOverFStarPercentAbstractCorrPlusWASPThreeKsBandEclipseTwoAll}{0.019}
\newcommand{\FpOverFStarPercentAbstractCorrMinusWASPThreeKsBandEclipseTwoAll}{0.018}
\newcommand{\TOffsetCorrWASPThreeKsBandEclipseTwoAll}{4.8}

\newcommand{\TOffsetCorrPlusWASPThreeKsBandEclipseTwoAll}{3.0}
\newcommand{\TOffsetCorrMinusWASPThreeKsBandEclipseTwoAll}{3.2}

\newcommand{\PhaseAbstractCorrWASPThreeKsBandEclipseTwoAll}{0.5018}
\newcommand{\PhaseAbstractCorrMinusWASPThreeKsBandEclipseTwoAll}{0.0012}
\newcommand{\PhaseAbstractCorrPlusWASPThreeKsBandEclipseTwoAll}{0.0011}

\newcommand{\ECosOmegaCorrWASPThreeKsBandEclipseTwoAll}{0.0028}
\newcommand{\ECosOmegaCorrPlusWASPThreeKsBandEclipseTwoAll}{0.0018}
\newcommand{\ECosOmegaCorrMinusWASPThreeKsBandEclipseTwoAll}{0.0018}
\newcommand{\JDOffsetCorrWASPThreeKsBandEclipseTwoAll}{14998.0158}
\newcommand{\JDOffsetCorrPlusWASPThreeKsBandEclipseTwoAll}{0.0021}
\newcommand{\JDOffsetCorrMinusWASPThreeKsBandEclipseTwoAll}{0.0022}

\newcommand{\FpOverFStarPercentAbstractCorrWASPTwelveKsBand}{0.284}
\newcommand{\FpOverFStarPercentAbstractCorrPlusWASPTwelveKsBand}{0.019}
\newcommand{\FpOverFStarPercentAbstractCorrMinusWASPTwelveKsBand}{0.020}
\newcommand{\TOffsetCorrWASPTwelveKsBand}{-0.8}

\newcommand{\TOffsetCorrPlusWASPTwelveKsBand}{1.4}
\newcommand{\TOffsetCorrMinusWASPTwelveKsBand}{1.3}

\newcommand{\PhaseAbstractCorrWASPTwelveKsBand}{0.4995}
\newcommand{\PhaseAbstractCorrMinusWASPTwelveKsBand}{0.0008}
\newcommand{\PhaseAbstractCorrPlusWASPTwelveKsBand}{0.0009}

\newcommand{\ECosOmegaCorrWASPTwelveKsBand}{-0.0008}
\newcommand{\ECosOmegaCorrPlusWASPTwelveKsBand}{0.0014}
\newcommand{\ECosOmegaCorrMinusWASPTwelveKsBand}{0.0014}
\newcommand{\JDOffsetCorrWASPTwelveKsBand}{15194.9344}
\newcommand{\JDOffsetCorrPlusWASPTwelveKsBand}{0.0010}
\newcommand{\JDOffsetCorrMinusWASPTwelveKsBand}{0.0009}
\newcommand{\FpOverFStarPercentAbstractCorrWASPTwelveKsBandTwentyEleven}{0.289}
\newcommand{\FpOverFStarPercentAbstractCorrPlusWASPTwelveKsBandTwentyEleven}{0.018}
\newcommand{\FpOverFStarPercentAbstractCorrMinusWASPTwelveKsBandTwentyEleven}{0.018}
\newcommand{\TOffsetCorrWASPTwelveKsBandTwentyEleven}{-1.0}

\newcommand{\TOffsetCorrPlusWASPTwelveKsBandTwentyEleven}{1.3}
\newcommand{\TOffsetCorrMinusWASPTwelveKsBandTwentyEleven}{1.3}

\newcommand{\PhaseAbstractCorrWASPTwelveKsBandTwentyEleven}{0.4993}
\newcommand{\PhaseAbstractCorrMinusWASPTwelveKsBandTwentyEleven}{0.0008}
\newcommand{\PhaseAbstractCorrPlusWASPTwelveKsBandTwentyEleven}{0.0009}

\newcommand{\ECosOmegaCorrWASPTwelveKsBandTwentyEleven}{-0.0010}
\newcommand{\ECosOmegaCorrPlusWASPTwelveKsBandTwentyEleven}{0.0013}
\newcommand{\ECosOmegaCorrMinusWASPTwelveKsBandTwentyEleven}{0.0013}
\newcommand{\JDOffsetCorrWASPTwelveKsBandTwentyEleven}{15576.9320}
\newcommand{\JDOffsetCorrPlusWASPTwelveKsBandTwentyEleven}{0.0009}
\newcommand{\JDOffsetCorrMinusWASPTwelveKsBandTwentyEleven}{0.0009}
\newcommand{\FpOverFStarPercentAbstractCorrWASPTwelveKsBandDecTwentyEleven}{0.259}
\newcommand{\FpOverFStarPercentAbstractCorrPlusWASPTwelveKsBandDecTwentyEleven}{0.042}
\newcommand{\FpOverFStarPercentAbstractCorrMinusWASPTwelveKsBandDecTwentyEleven}{0.042}
\newcommand{\TOffsetCorrWASPTwelveKsBandDecTwentyEleven}{-0.4}

\newcommand{\TOffsetCorrPlusWASPTwelveKsBandDecTwentyEleven}{3.0}
\newcommand{\TOffsetCorrMinusWASPTwelveKsBandDecTwentyEleven}{3.0}

\newcommand{\PhaseAbstractCorrWASPTwelveKsBandDecTwentyEleven}{0.4997}
\newcommand{\PhaseAbstractCorrMinusWASPTwelveKsBandDecTwentyEleven}{0.0019}
\newcommand{\PhaseAbstractCorrPlusWASPTwelveKsBandDecTwentyEleven}{0.0019}

\newcommand{\ECosOmegaCorrWASPTwelveKsBandDecTwentyEleven}{-0.0004}
\newcommand{\ECosOmegaCorrPlusWASPTwelveKsBandDecTwentyEleven}{0.0030}
\newcommand{\ECosOmegaCorrMinusWASPTwelveKsBandDecTwentyEleven}{0.0030}
\newcommand{\JDOffsetCorrWASPTwelveKsBandDecTwentyEleven}{15924.0047}
\newcommand{\JDOffsetCorrPlusWASPTwelveKsBandDecTwentyEleven}{0.0021}
\newcommand{\JDOffsetCorrMinusWASPTwelveKsBandDecTwentyEleven}{0.0021}
\newcommand{\FpOverFStarPercentAbstractCorrWASPTwelveYBandTwentyEleven}{0.106}
\newcommand{\FpOverFStarPercentAbstractCorrPlusWASPTwelveYBandTwentyEleven}{0.014}
\newcommand{\FpOverFStarPercentAbstractCorrMinusWASPTwelveYBandTwentyEleven}{0.016}
\newcommand{\TOffsetCorrWASPTwelveYBandTwentyEleven}{0.6}

\newcommand{\TOffsetCorrPlusWASPTwelveYBandTwentyEleven}{2.6}
\newcommand{\TOffsetCorrMinusWASPTwelveYBandTwentyEleven}{2.3}

\newcommand{\PhaseAbstractCorrWASPTwelveYBandTwentyEleven}{0.5004}
\newcommand{\PhaseAbstractCorrMinusWASPTwelveYBandTwentyEleven}{0.0015}
\newcommand{\PhaseAbstractCorrPlusWASPTwelveYBandTwentyEleven}{0.0017}

\newcommand{\ECosOmegaCorrWASPTwelveYBandTwentyEleven}{0.0006}
\newcommand{\ECosOmegaCorrPlusWASPTwelveYBandTwentyEleven}{0.0026}
\newcommand{\ECosOmegaCorrMinusWASPTwelveYBandTwentyEleven}{0.0026}
\newcommand{\JDOffsetCorrWASPTwelveYBandTwentyEleven}{15587.8473}
\newcommand{\JDOffsetCorrPlusWASPTwelveYBandTwentyEleven}{0.0018}
\newcommand{\JDOffsetCorrMinusWASPTwelveYBandTwentyEleven}{0.0016}
\newcommand{\FpOverFStarPercentAbstractCorrWASPTwelveKContBandJanTwentyTwelve}{0.264}
\newcommand{\FpOverFStarPercentAbstractCorrPlusWASPTwelveKContBandJanTwentyTwelve}{0.045}
\newcommand{\FpOverFStarPercentAbstractCorrMinusWASPTwelveKContBandJanTwentyTwelve}{0.055}
\newcommand{\TOffsetCorrWASPTwelveKContBandJanTwentyTwelve}{-2.8}

\newcommand{\TOffsetCorrPlusWASPTwelveKContBandJanTwentyTwelve}{2.1}
\newcommand{\TOffsetCorrMinusWASPTwelveKContBandJanTwentyTwelve}{2.6}

\newcommand{\PhaseAbstractCorrWASPTwelveKContBandJanTwentyTwelve}{0.4982}
\newcommand{\PhaseAbstractCorrMinusWASPTwelveKContBandJanTwentyTwelve}{0.0017}
\newcommand{\PhaseAbstractCorrPlusWASPTwelveKContBandJanTwentyTwelve}{0.0013}

\newcommand{\ECosOmegaCorrWASPTwelveKContBandJanTwentyTwelve}{-0.0028}
\newcommand{\ECosOmegaCorrPlusWASPTwelveKContBandJanTwentyTwelve}{0.0021}
\newcommand{\ECosOmegaCorrMinusWASPTwelveKContBandJanTwentyTwelve}{0.0021}
\newcommand{\JDOffsetCorrWASPTwelveKContBandJanTwentyTwelve}{15946.9229}
\newcommand{\JDOffsetCorrPlusWASPTwelveKContBandJanTwentyTwelve}{0.0015}
\newcommand{\JDOffsetCorrMinusWASPTwelveKContBandJanTwentyTwelve}{0.0018}

\newcommand{\FpOverFStarPercentAbstractWASPTwelveKsBand}{0.277}
\newcommand{\FpOverFStarPercentAbstractMinusWASPTwelveKsBand}{0.020}
\newcommand{\FpOverFStarPercentAbstractPlusWASPTwelveKsBand}{0.017}

\newcommand{\cOneWASPTwelveKsBand}{0.00063}
\newcommand{\cOnePlusWASPTwelveKsBand}{0.00015}
\newcommand{\cOneMinusWASPTwelveKsBand}{0.00015}
\newcommand{\cTwoWASPTwelveKsBand}{0.004}
\newcommand{\cTwoPlusWASPTwelveKsBand}{0.005}
\newcommand{\cTwoMinusWASPTwelveKsBand}{0.004}
\newcommand{\cThreeWASPTwelveKsBand}{0.005}
\newcommand{\cThreePlusWASPTwelveKsBand}{0.018}
\newcommand{\cThreeMinusWASPTwelveKsBand}{0.018}
\newcommand{\TOffsetWASPTwelveKsBand}{-1.5}

\newcommand{\TOffsetPlusWASPTwelveKsBand}{1.3}
\newcommand{\TOffsetMinusWASPTwelveKsBand}{1.4}

\newcommand{\ChiWASPTwelveKsBand}{0.913}
\newcommand{\ChiPlusWASPTwelveKsBand}{0.002}
\newcommand{\ChiMinusWASPTwelveKsBand}{0.002}

\newcommand{\RMSExposureWASPTwelveKsBand}{1.83}
\newcommand{\FpOverFStarPercentAbstractWASPTwelveKsBandTwentyEleven}{0.286}
\newcommand{\FpOverFStarPercentAbstractMinusWASPTwelveKsBandTwentyEleven}{0.023}
\newcommand{\FpOverFStarPercentAbstractPlusWASPTwelveKsBandTwentyEleven}{0.020}

\newcommand{\cOneWASPTwelveKsBandTwentyEleven}{0.00100}
\newcommand{\cOnePlusWASPTwelveKsBandTwentyEleven}{0.00023}
\newcommand{\cOneMinusWASPTwelveKsBandTwentyEleven}{0.00027}
\newcommand{\cTwoWASPTwelveKsBandTwentyEleven}{0.003}
\newcommand{\cTwoPlusWASPTwelveKsBandTwentyEleven}{0.003}
\newcommand{\cTwoMinusWASPTwelveKsBandTwentyEleven}{0.004}
\newcommand{\cThreeWASPTwelveKsBandTwentyEleven}{-0.016}
\newcommand{\cThreePlusWASPTwelveKsBandTwentyEleven}{0.011}
\newcommand{\cThreeMinusWASPTwelveKsBandTwentyEleven}{0.010}
\newcommand{\TOffsetWASPTwelveKsBandTwentyEleven}{-2.2}

\newcommand{\TOffsetPlusWASPTwelveKsBandTwentyEleven}{1.5}
\newcommand{\TOffsetMinusWASPTwelveKsBandTwentyEleven}{1.6}

\newcommand{\ChiWASPTwelveKsBandTwentyEleven}{1.000}
\newcommand{\ChiPlusWASPTwelveKsBandTwentyEleven}{0.001}
\newcommand{\ChiMinusWASPTwelveKsBandTwentyEleven}{0.001}

\newcommand{\RMSExposureWASPTwelveKsBandTwentyEleven}{2.86}
\newcommand{\FpOverFStarPercentAbstractWASPTwelveKsBandDecTwentyEleven}{0.252}
\newcommand{\FpOverFStarPercentAbstractMinusWASPTwelveKsBandDecTwentyEleven}{0.047}
\newcommand{\FpOverFStarPercentAbstractPlusWASPTwelveKsBandDecTwentyEleven}{0.047}

\newcommand{\cOneWASPTwelveKsBandDecTwentyEleven}{-0.00047}
\newcommand{\cOnePlusWASPTwelveKsBandDecTwentyEleven}{0.00057}
\newcommand{\cOneMinusWASPTwelveKsBandDecTwentyEleven}{0.00044}
\newcommand{\cTwoWASPTwelveKsBandDecTwentyEleven}{0.014}
\newcommand{\cTwoPlusWASPTwelveKsBandDecTwentyEleven}{0.014}
\newcommand{\cTwoMinusWASPTwelveKsBandDecTwentyEleven}{0.015}
\newcommand{\cThreeWASPTwelveKsBandDecTwentyEleven}{-0.010}
\newcommand{\cThreePlusWASPTwelveKsBandDecTwentyEleven}{0.062}
\newcommand{\cThreeMinusWASPTwelveKsBandDecTwentyEleven}{0.041}
\newcommand{\TOffsetWASPTwelveKsBandDecTwentyEleven}{-1.2}

\newcommand{\TOffsetPlusWASPTwelveKsBandDecTwentyEleven}{3.0}
\newcommand{\TOffsetMinusWASPTwelveKsBandDecTwentyEleven}{2.6}

\newcommand{\ChiWASPTwelveKsBandDecTwentyEleven}{0.739}
\newcommand{\ChiPlusWASPTwelveKsBandDecTwentyEleven}{0.002}
\newcommand{\ChiMinusWASPTwelveKsBandDecTwentyEleven}{0.002}

\newcommand{\RMSExposureWASPTwelveKsBandDecTwentyEleven}{3.86}
\newcommand{\FpOverFStarPercentAbstractWASPTwelveYBandTwentyEleven}{0.100}
\newcommand{\FpOverFStarPercentAbstractMinusWASPTwelveYBandTwentyEleven}{0.012}
\newcommand{\FpOverFStarPercentAbstractPlusWASPTwelveYBandTwentyEleven}{0.012}

\newcommand{\cOneWASPTwelveYBandTwentyEleven}{0.00096}
\newcommand{\cOnePlusWASPTwelveYBandTwentyEleven}{0.00013}
\newcommand{\cOneMinusWASPTwelveYBandTwentyEleven}{0.00014}
\newcommand{\cTwoWASPTwelveYBandTwentyEleven}{-0.002}
\newcommand{\cTwoPlusWASPTwelveYBandTwentyEleven}{0.003}
\newcommand{\cTwoMinusWASPTwelveYBandTwentyEleven}{0.003}
\newcommand{\cThreeWASPTwelveYBandTwentyEleven}{-0.010}
\newcommand{\cThreePlusWASPTwelveYBandTwentyEleven}{0.009}
\newcommand{\cThreeMinusWASPTwelveYBandTwentyEleven}{0.009}
\newcommand{\TOffsetWASPTwelveYBandTwentyEleven}{-0.1}

\newcommand{\TOffsetPlusWASPTwelveYBandTwentyEleven}{2.6}
\newcommand{\TOffsetMinusWASPTwelveYBandTwentyEleven}{2.4}

\newcommand{\ChiWASPTwelveYBandTwentyEleven}{0.926}
\newcommand{\ChiPlusWASPTwelveYBandTwentyEleven}{0.001}
\newcommand{\ChiMinusWASPTwelveYBandTwentyEleven}{0.002}

\newcommand{\RMSExposureWASPTwelveYBandTwentyEleven}{1.59}
\newcommand{\FpOverFStarPercentAbstractWASPTwelveKContBandJanTwentyTwelve}{0.274}
\newcommand{\FpOverFStarPercentAbstractMinusWASPTwelveKContBandJanTwentyTwelve}{0.036}
\newcommand{\FpOverFStarPercentAbstractPlusWASPTwelveKContBandJanTwentyTwelve}{0.041}

\newcommand{\cOneWASPTwelveKContBandJanTwentyTwelve}{-0.00007}
\newcommand{\cOnePlusWASPTwelveKContBandJanTwentyTwelve}{0.00028}
\newcommand{\cOneMinusWASPTwelveKContBandJanTwentyTwelve}{0.00030}
\newcommand{\cTwoWASPTwelveKContBandJanTwentyTwelve}{0.029}
\newcommand{\cTwoPlusWASPTwelveKContBandJanTwentyTwelve}{0.012}
\newcommand{\cTwoMinusWASPTwelveKContBandJanTwentyTwelve}{0.011}
\newcommand{\cThreeWASPTwelveKContBandJanTwentyTwelve}{-0.091}
\newcommand{\cThreePlusWASPTwelveKContBandJanTwentyTwelve}{0.045}
\newcommand{\cThreeMinusWASPTwelveKContBandJanTwentyTwelve}{0.050}
\newcommand{\TOffsetWASPTwelveKContBandJanTwentyTwelve}{-3.0}

\newcommand{\TOffsetPlusWASPTwelveKContBandJanTwentyTwelve}{2.1}
\newcommand{\TOffsetMinusWASPTwelveKContBandJanTwentyTwelve}{1.9}

\newcommand{\ChiWASPTwelveKContBandJanTwentyTwelve}{1.159}
\newcommand{\ChiPlusWASPTwelveKContBandJanTwentyTwelve}{0.007}
\newcommand{\ChiMinusWASPTwelveKContBandJanTwentyTwelve}{0.008}

\newcommand{\RMSExposureWASPTwelveKContBandJanTwentyTwelve}{1.78}

\newcommand{\FpOverFStarPercentAbstractQatarOneKsBand}{0.121}
\newcommand{\FpOverFStarPercentAbstractMinusQatarOneKsBand}{0.025}
\newcommand{\FpOverFStarPercentAbstractPlusQatarOneKsBand}{0.026}

\newcommand{\cOneQatarOneKsBand}{-0.00005}
\newcommand{\cOnePlusQatarOneKsBand}{0.00024}
\newcommand{\cOneMinusQatarOneKsBand}{0.00022}
\newcommand{\cTwoQatarOneKsBand}{0.000}
\newcommand{\cTwoPlusQatarOneKsBand}{0.008}
\newcommand{\cTwoMinusQatarOneKsBand}{0.007}
\newcommand{\cThreeQatarOneKsBand}{0.044}
\newcommand{\cThreePlusQatarOneKsBand}{0.046}
\newcommand{\cThreeMinusQatarOneKsBand}{0.063}
\newcommand{\TOffsetQatarOneKsBand}{-3.0}

\newcommand{\TOffsetPlusQatarOneKsBand}{2.8}
\newcommand{\TOffsetMinusQatarOneKsBand}{2.8}

\newcommand{\ChiQatarOneKsBand}{0.979}
\newcommand{\ChiPlusQatarOneKsBand}{0.002}
\newcommand{\ChiMinusQatarOneKsBand}{0.003}

\newcommand{\FpOverFStarPercentAbstractWASPThreeKsBandEclipseOne}{0.208}
\newcommand{\FpOverFStarPercentAbstractMinusWASPThreeKsBandEclipseOne}{0.023}
\newcommand{\FpOverFStarPercentAbstractPlusWASPThreeKsBandEclipseOne}{0.026}

\newcommand{\cOneWASPThreeKsBandEclipseOne}{0.00227}
\newcommand{\cOnePlusWASPThreeKsBandEclipseOne}{0.00030}
\newcommand{\cOneMinusWASPThreeKsBandEclipseOne}{0.00030}
\newcommand{\cTwoWASPThreeKsBandEclipseOne}{-0.012}
\newcommand{\cTwoPlusWASPThreeKsBandEclipseOne}{0.006}
\newcommand{\cTwoMinusWASPThreeKsBandEclipseOne}{0.006}
\newcommand{\cThreeWASPThreeKsBandEclipseOne}{-0.004}
\newcommand{\cThreePlusWASPThreeKsBandEclipseOne}{0.030}
\newcommand{\cThreeMinusWASPThreeKsBandEclipseOne}{0.036}
\newcommand{\TOffsetWASPThreeKsBandEclipseOne}{0.0}

\newcommand{\ChiWASPThreeKsBandEclipseOne}{0.700}
\newcommand{\ChiPlusWASPThreeKsBandEclipseOne}{0.002}
\newcommand{\ChiMinusWASPThreeKsBandEclipseOne}{0.002}

\newcommand{\RMSExposureWASPThreeKsBandEclipseOne}{2.07}

\newcommand{\FpOverFStarPercentAbstractWASPThreeKsBandEclipseTwo}{0.164}
\newcommand{\FpOverFStarPercentAbstractMinusWASPThreeKsBandEclipseTwo}{0.027}
\newcommand{\FpOverFStarPercentAbstractPlusWASPThreeKsBandEclipseTwo}{0.022}

\newcommand{\cOneWASPThreeKsBandEclipseTwo}{0.00264}
\newcommand{\cOnePlusWASPThreeKsBandEclipseTwo}{0.00024}
\newcommand{\cOneMinusWASPThreeKsBandEclipseTwo}{0.00031}
\newcommand{\cTwoWASPThreeKsBandEclipseTwo}{-0.027}
\newcommand{\cTwoPlusWASPThreeKsBandEclipseTwo}{0.007}
\newcommand{\cTwoMinusWASPThreeKsBandEclipseTwo}{0.006}
\newcommand{\cThreeWASPThreeKsBandEclipseTwo}{0.065}
\newcommand{\cThreePlusWASPThreeKsBandEclipseTwo}{0.022}
\newcommand{\cThreeMinusWASPThreeKsBandEclipseTwo}{0.027}
\newcommand{\TOffsetWASPThreeKsBandEclipseTwo}{2.5}

\newcommand{\TOffsetPlusWASPThreeKsBandEclipseTwo}{3.0}
\newcommand{\TOffsetMinusWASPThreeKsBandEclipseTwo}{3.2}

\newcommand{\ChiWASPThreeKsBandEclipseTwo}{0.906}
\newcommand{\ChiPlusWASPThreeKsBandEclipseTwo}{0.002}
\newcommand{\ChiMinusWASPThreeKsBandEclipseTwo}{0.003}

\newcommand{\RMSExposureWASPThreeKsBandEclipseTwo}{1.96}
\newcommand{\FpOverFStarPercentAbstractKELTOneKsBand}{0.150}
\newcommand{\FpOverFStarPercentAbstractMinusKELTOneKsBand}{0.015}
\newcommand{\FpOverFStarPercentAbstractPlusKELTOneKsBand}{0.018}

\newcommand{\cOneKELTOneKsBand}{0.00071}
\newcommand{\cOnePlusKELTOneKsBand}{0.00017}
\newcommand{\cOneMinusKELTOneKsBand}{0.00018}
\newcommand{\cTwoKELTOneKsBand}{-0.001}
\newcommand{\cTwoPlusKELTOneKsBand}{0.005}
\newcommand{\cTwoMinusKELTOneKsBand}{0.004}
\newcommand{\cThreeKELTOneKsBand}{-0.001}
\newcommand{\cThreePlusKELTOneKsBand}{0.014}
\newcommand{\cThreeMinusKELTOneKsBand}{0.013}
\newcommand{\TOffsetKELTOneKsBand}{-7.0}

\newcommand{\TOffsetPlusKELTOneKsBand}{2.2}
\newcommand{\TOffsetMinusKELTOneKsBand}{2.0}

\newcommand{\ChiKELTOneKsBand}{0.741}
\newcommand{\ChiPlusKELTOneKsBand}{0.002}
\newcommand{\ChiMinusKELTOneKsBand}{0.002}

\newcommand{\RMSExposureKELTOneKsBand}{1.50}

\shorttitle{Systematics in Ground-based Near-infrared Photometry} 
\shortauthors{Croll et al.}

\begin{document}

\title{Near-infrared Thermal Emission Detections of a number of hot Jupiters and the Systematics of Ground-based Near-infrared Photometry\altaffilmark{*}}
\author{Bryce Croll\altaffilmark{1} \altaffilmark{2},
Loic Albert\altaffilmark{3},
Ray Jayawardhana\altaffilmark{4},
Michael Cushing\altaffilmark{5},
Claire Moutou\altaffilmark{6},
David Lafreniere\altaffilmark{3},
John Asher Johnson\altaffilmark{7},
Aldo S. Bonomo\altaffilmark{8},
Magali Deleuil\altaffilmark{9},
Jonathan Fortney\altaffilmark{10}
}

\altaffiltext{1}{5525 Olund Road, Abbotsford, B.C. Canada}

\altaffiltext{2}{Kavli Institute for Astrophysics and Space Research, Massachusetts Institute
of Technology, Cambridge, MA 02139, USA; croll@space.mit.edu}

\altaffiltext{3}{D\'epartement de physique, Universit\'e de Montr\'eal, C.P.
6128 Succ. Centre-Ville, Montr\'eal, QC, H3C 3J7, Canada}


\altaffiltext{4}{Department of Physics and Astronomy, York University, Toronto, ON L3T 3R1, Canada}

\altaffiltext{5}{Department of Physics and Astronomy, The University of Toledo, 2801 West Bancroft Street, Toledo, OH 43606, USA}

\altaffiltext{6}{Canada-France-Hawaii Telescope Corporation, 65-1238 Mamalahoa Highway, Kamuela, HI 96743, USA}

\altaffiltext{7}{Harvard-Smithsonian Center for Astrophysics; Institute for Theory and Computation, 60 Garden St, MS-51, Cambridge, MA 02138}

\altaffiltext{8}{INAF - Osservatorio Astrofisico di Torino, via Osservatorio 20, 10025 Pino Torinese, Italy}

\altaffiltext{9}{Aix Marseille University, CNRS, LAM (Laboratoire d'Astrophysique de Marseille), UMR 7326, 13388 Marseille cedex 13, France}

\altaffiltext{10}{Department of Astronomy and Astrophysics, University of California, Santa Cruz, CA, 95064}

\altaffiltext{*}{Based on observations obtained with WIRCam, a joint project of CFHT, Taiwan, Korea, Canada, France, at the Canada-France-Hawaii Telescope (CFHT) which is operated by the National Research Council (NRC) of Canada, the Institute National des Sciences de l'Univers of the Centre National de la Recherche Scientifique of France, and the University of Hawaii.}

\begin{abstract}

We present detections of the near-infrared thermal emission of three hot Jupiters and one brown-dwarf
using the Wide-field Infrared Camera (WIRCam) on the Canada-France-Hawaii Telescope (CFHT).
These include Ks-band 									
secondary eclipse detections of the hot Jupiters WASP-3b and Qatar-1b and the
brown dwarf KELT-1b.
We also report Y-band, $K_{CONT}$-band,	
and two new and one reanalyzed Ks-band detections of the thermal
emission of the hot Jupiter WASP-12b.
We present a new reduction pipeline for CFHT/WIRCam data, which is optimized for high precision photometry. 
We also describe
novel techniques for constraining systematic errors in ground-based near-infrared photometry, so
as to return reliable secondary eclipse depths and uncertainties.
We discuss the noise properties of our 
ground-based photometry for wavelengths spanning the near-infrared (the YJHK-bands),
for faint and bright-stars, and for the same object on several occasions.
For the hot Jupiters WASP-3b and WASP-12b we demonstrate the repeatability
of our eclipse depth measurements in the Ks-band; we
therefore place stringent limits
on the systematics of ground-based, near-infrared photometry, and also
rule out violent weather changes in the deep, high pressure
atmospheres of these two hot Jupiters at the epochs of our observations.

\end{abstract}



\section{Introduction}

 The robustness and repeatability of transit and eclipse spectroscopy of exoplanets --
across a wide wavelength range, from the optical to the infrared, and
with a variety of instruments and telescopes -- is an issue of utmost importance
to ensure that we can trust the understanding imparted from 
detections, or lack thereof, of exoplanet atmospheric features.
As a result, the topic of the repeatability of transit and eclipse depth
detections has 
attracted growing interest in recent years (e.g. \citealt{Agol10}; \citealt{Knutson11}; \citealt{Hansen14}; \citealt{Morello14}).
For instance, {\it Spitzer Space Telescope} \citep{Werner04}
measurements of the mid-infrared thermal emission of 
a number of hot Jupiters, largely with the IRAC instrument \citep{Fazio04},
have already undergone a thorough number of analyses, reanalyses and intriguing revisions
(e.g. \citealt{Harrington07}; \citealt{Knutson09}; \citealt{Knutson07};
\citealt{Charbonneau08}; \citealt{Stevenson10}; \citealt{Knutson11}).
{\it Hubble Space Telescope (HST)}/NICMOS \citep{Thompson98} detections of transmission 
features in the atmospheres of exoplanets are in the midst of an ongoing, raging debate as to the
fidelity of these claimed detections (e.g. \citealt{Swain08}; \citealt{Swain09}; \citealt{Swain09b}; \citealt{Gibson11};
\citealt{Mandell11}; \citealt{Gibson12}; \citealt{Waldmann13}; \citealt{Swain14}).


 To-date broadband near-infrared thermal emission measurements of hot Jupiters from the ground
have generally been too few and far between to warrant thorough reanalyses, or attempts at demonstrating
the repeatability of eclipse detections. 
There have been a handful of confirmations of the depths of ground-based, near-infrared
secondary eclipse detections (\citealt{CrollWASPTwelve}; \citealt{Zhao12}; \citealt{CrollThesis}; \citealt{Zhao12WASPThree}). 
However, troublingly, the early indications of the reliability
of these detections has not been overwhelmingly positive.
Reobservations of the thermal emission of TrES-3b in the Ks-band disagreed by 2$\sigma$
\citep{deMooijSnellen09,CrollTrESThree}, while a ground-based H-band upper-limit appears to disagree
with a space-based {\it HST}/Wide Field Camera 3 (WFC3) detection (\citealt{CrollTrESThree,Ranjan14}). 
A suggested, possible detection of a near-infrared, 
feature in the transmission spectrum of GJ 1214b \citep{CrollGJ},
was refuted by other researchers \citep{Bean11}. Two observations of the thermal emission of the hot
Jupiter WASP-19b obtained with the same instrument/telescope configuration, analyzed by the same observer
were only found to agree
at the 2.9$\sigma$ level \citep{Bean13}, and were arguably marginally discrepant from other measurements of the 
ground-based, near-infrared,
thermal emission of this planet (\citealt{Anderson10}; \citealt{Gibson10}; \citealt{Burton12}; \citealt{Lendl13}).
Finally, the obvious systematics in ground-based, near-infrared photometry have caused others
to encourage caution when interpreting the eclipse depths returned via ground-based detections \citep{deMooij11}.

 Complicating this picture, is the question of whether we would actually expect thermal emission depths to be identical
from epoch to epoch in the first place.
To-date multiwavelength constraints have often been obtained at different epochs, and thus the effective comparison
of these measurements relies on the assumption that the thermal emission of these planets is consistent
from epoch to epoch.
One reason this would not be the case is if these planets have
variable weather and intense storms, such as those that have been theoretically
predicted by dynamical atmospheric models of hot Jupiters with simplified radiative transfer. 
Examples include the two-dimensional shallow water models of \citet{LangtonLaughlin08} and 
the three-dimensional Intermediate General Circulation model simulations of \citet{MenouRauscher09}; the latter predict 
brightness temperature variations of $\sim$100$K$ for a hot Jupiter similar to HD 209458.

	One approach to constrain the 
temporal variability of a hot Jupiter is to detect
its thermal emission in a single band at multiple epochs.
This has already been performed with Spitzer in the mid-infrared; the \citet{Agol10} study
placed a stringent 1$\sigma$ upper limit 
on the temporal variability of the thermal emission of the hot Jupiter HD 189733
of 2.7\% 
from seven eclipses in the 8.0 $\mu m$ Spitzer/IRAC channel
(corresponding to a limit on brightness temperature variations of $\pm$ 22 $K$). 
\citet{Knutson11} place a 1$\sigma$ upper limit of 17\% on the temporal variability
of the thermal emission of the hot Neptune GJ 436
in the same 8.0 $\mu m$ Spitzer/IRAC channel 
(corresponding to a limit on brightness temperature variations of $\pm$ 54 $K$). 
Mid-infrared limits are important, but it 
arguably makes more sense to search for temporal variability in the near-infrared.
If we assume the planet radiates like a blackbody, then for a given temperature difference on the planet,
the difference in the Planck function
will be much greater at shorter wavelengths, than at longer wavelengths \citep{Rauscher08}.
For example, if we assume that mid- and near-infrared observations 
probe the same atmospheric layer, then a $\sim$100$K$
temperature difference observed in the \citet{MenouRauscher09} model from eclipse to eclipse
for a canonical HD
209458-like atmosphere will result in eclipse to eclipse variations on the order of $\sim$33\%
of the eclipse depth in Ks-band, compared to $\sim$11\% of the eclipse depth at 8.0 $\mu m$.
Alternatively, as the atmospheres of hot Jupiters are believed to be highly vertically stratified 
(\citealt{MenouRauscher09}, and references therein), and the YJHK-bands are
water opacity windows and therefore should
stare much deeper into the atmospheres of hot Jupiters
than mid-infrared observations \citep{Seager05,Fortney08,Burrows08},
it is possible that near-infrared observations may probe higher pressure
regions that are variable, even if higher altitude layers are not.
A useful analogue could be that of the atmospheres of brown dwarfs:
the phase lags and lack of similarity in the phase curves
observed at different wavelengths for variable brown dwarfs near the L/T transition \citep{Biller13} --
supposedly due to different wavelength of observations penetrating to different depths in the atmosphere --
could be analogous to highly irradiated hot Jupiters.


 In this paper we present a through investigation of how to return
robust detections and errors of secondary eclipses of exoplanets from ground-based, near-infrared
photometry, despite the inherent systematics.
Once accounting for these systematics, we are able to present robust new detections and 
one reanalyzed detection of the near-infrared thermal emission of several hot Jupiters (Qatar-1b,
WASP-3b, and WASP-12b) and one brown dwarf (KELT-1b). The layout of the paper is as follows.
We present our new CFHT/WIRCam observations of the secondary eclipses of hot Jupiters and a brown dwarf
in Section \ref{SecObservations}.
In Section \ref{SecPipelineAll}, 
we describe our new pipeline to reduce
observations obtained with WIRCam
on CFHT in order to optimize the data for high-precision photometry. 
We analyze our secondary eclipse detections in Section \ref{SecAnalysis}, using the techniques
discussed in Section \ref{SecTechniques}; these techniques include how to 
constrain systematic errors, determine the optimal aperture size
and reference star ensemble, and to return honest eclipse depths and uncertainties,
despite the systematics inherent in near-infrared, photometry from the ground.
In Section \ref{SecNoise} we discuss the noise properties of our ground-based,
near-infrared photometry, and how our precision varies for faint and bright-stars, 
for the same object in the same-band on several occasions,
and for the same object in different near-infrared bands (the YJHK-bands).
Lastly, in Section \ref{SecDiscussion} 
we demonstrate the repeatability
of our eclipse depth measurements in the Ks-band for two hot Jupiters; we are therefore able to place a limit
on both the systematics inherent in ground-based, near-infrared photometry,
and on the presence of weather, or large-scale temperature fluctuations, in the deep, high pressure
atmospheres of the hot Jupiters WASP-3b and WASP-12b.

\section{Observations}
\label{SecObservations}

In this paper we analyze an assortment of new and previously presented ground-based, 
near-infrared 
observations of exoplanets obtained with the Wide-field Infrared Camera (WIRCam; \citealt{Puget04})
on the Canada-France-Hawaii Telescope (CFHT).
We discuss the new observations that we present in this paper in Section \ref{SecNewObs}; these new observations
include Ks-band eclipse detections of Qatar-1b, and the brown-dwarf KELT-1b,
two Ks-band eclipse detections of WASP-3b, 
and a eclipse detections of WASP-12b in the Y-band, $K_{CONT}$-band and two detections in the Ks-band.
Other observations in this paper that have been previously discussed
in other papers include our JHKs-band secondary eclipse detections of WASP-12b \citep{CrollWASPTwelve},
and our 2012 September 1 Ks-band observations of the transit of KIC 12557548b \citep{CrollKIC}.
We also reanalyze our Ks-band detection of the thermal emission 
of WASP-12b on 2009 December 28 (discussed previously in \citealt{CrollWASPTwelve}).

\subsection{Our new CFHT/WIRCam observations of hot Jupiters and a brown dwarf}
\label{SecNewObs}

\begin{deluxetable*}{ccccc}
\tablecaption{Observing Log}
\tablehead{
\colhead{Eclipse} 			& \colhead{Date (Hawaiian} 	& \colhead{Duration}	& \colhead{Defocus} 				& \colhead{Exposure} 	\\
\colhead{Light curve}			& \colhead{Standard Time)}	& \colhead{(hours)}	& \colhead{(mm/arc-seconds\tablenotemark{a})}	& \colhead{Time (sec)}	\\
}
\startdata
First WASP-12 Ks-band\tablenotemark{b}	& 2009 December 28		& $\sim$6.2		& 2.0/2.8			& 5.0			\\	
Second WASP-12 Ks-band			& 2011 January 14		& $\sim$8.5 		& 1.5/2.2			& 5.0			\\	
Third WASP-12 Ks-band			& 2011 December 28		& $\sim$6.2 		& 2.0/2.7			& 5.0			\\
WASP-12 Y-band				& 2011 January 25		& $\sim$7.4 		& 1.3/1.9			& 5.0			\\
WASP-12 $K_{CONT}$-band			& 2012 January 19		& $\sim$4.6		& 1.0/1.4			& 30.0			\\
First WASP-3 Ks-band			& 2009 June 3 			& $\sim$4.7 		& 1.0/2.3			& 5.0/4.0		\\
Second WASP-3 Ks-band			& 2009 June 14			& $\sim$5.9		& 1.5/2.4			& 3.5			\\	
Qatar-1 Ks-band				& 2012 July 27			& $\sim$4.0		& 1.5/2.2			& 4.5			\\
KELT-1 Ks-band				& 2012 October 10		& $\sim$6.8		& 2.0/2.9			& 8.0			\\
\enddata
\tablenotetext{a}{The defocus value quoted in arc-seconds is the approximate radius of the maximum flux values of the defocused ``donut'' PSF of the target star, and is not necessarily directly proportional to the defocus value in $mm$.}
\tablenotetext{b}{We note that this data-set is reanalyzed here, and was presented previously in \citet{CrollWASPTwelve}.}
\label{TableDefocus}
\end{deluxetable*}


 We present Ks-band ($\sim$2.15 $\mu m$) observations of the
brown-dwarf hosting star KELT-1 ($K$$\sim$9.44),
and the hot Jupiter hosting stars Qatar-1 ($K$$\sim$10.41), and
WASP-3 ($K$$\sim$9.36); we also present Ks-band, Y-band ($\sim$1.04 $\mu m$)
and $K_{CONT}$-band ($\sim$2.22 $\mu m$) observations of the hot Jupiter hosting star WASP-12 ($K$$\sim$10.19).
We observed these targets using ``Staring Mode'' \citep{CrollTrESTwo,Devost10}, where the target star is observed
continuously for several hours without dithering.
The observing dates and the duration of our observations are listed in Table \ref{TableDefocus}.
We defocused the telescope for each observation,
and list the defocus value and the exposure time in Table \ref{TableDefocus}.
Our exposures were read-out with correlated double sampling; for each exposure the overhead was $\sim$7.8 seconds.


The conditions were photometric for each of our observations.
Our 2011 December 28 observations
of WASP-12b in the Ks-band suffered from poor seeing that
varied from 0.8\arcsec \ at the start of our observations to 2.1\arcsec \ at
the end.
Our observations of WASP-12 in the $K_{CONT}$-band on 2011 January 25 were aborted 
early due to a glycol leak at the telescope; as a result
there is very little out-of-eclipse baseline 
after the secondary eclipse for that data-set.
We note that for our first WASP-3 observation (2009 June 3) we used 5-second exposures for the first five
minutes of the observations;
for these 5-second exposures, 
we found that the brightest pixels in our target aperture were close to saturation, so
we switched to 4-second exposures for the remainder of those observations.

\section{Reduction of the ``Staring Mode'' WIRCam data}
\label{SecPipelineAll}

We have created a pipeline to reduce WIRCam ``Staring Mode'' \citep{Devost10} data, independent of the traditional
WIRCam I'iwi pipeline\footnote{\url{http://www.cfht.hawaii.edu/Instruments/Imaging/WIRCam/  IiwiVersion2Doc.html}}.
The I'iwi pipeline was originally developed to reduce all WIRCam data.
Our new pipeline was developed to optimize the WIRCam data for high-cadence,
photometric accuracy; each step
of the I'iwi version 1.9 pipeline was investigated to determine its effect on the accuracy of the ``Staring Mode'' data.
In Section \ref{SecPipeline} we discuss the details of our pipeline, and in Section \ref{SecPipelineLessons} we summarize the lessons
learned from the development of our pipeline that may be applicable to reduction 
pipelines for other near-infrared arrays, in order to 
optimize these pipelines for high-precision photometry.

\subsection{Our Reduction Pipeline Optimized for high-precision photometry}
\label{SecPipeline}

The original I'iwi pipeline, which was applied to previous iterations of 
our data \citep{CrollTrESTwo,CrollTrESThree,CrollWASPTwelve},
consisted of the following steps: saturated pixel flagging, a non-linearity correction,
reference pixel subtraction, dark subtraction, dome flat-fielding, bad pixel masking, and sky subtraction.
Here we summarize the steps of our pipeline, and note the differences between it and the I'iwi version 1.9 pipeline:
\begin{itemize}
\item {\it Saturated pixel flagging:} Identically to the I'iwi version 1.9 pipeline 
we flag all pixels with CDS values of 36,000 Analog-to-digital unit ($ADU$) as saturated; in a step that is not included
in I'iwi, for these saturated pixels we interpolate their flux from adjacent pixels as discussed below.  

\item {\it The reference pixel subtraction:} In contrast to the I'iwi 1.9 pipeline, we
do not perform the reference pixel subtraction. In I'iwi a reference pixel subtraction was performed
to account for slow bias drifts;
in this step the median of the ``blind'' bottom and top rows of the WIRCam array were subtracted from the WIRCam array for every exposure.
We observed that the reference pixel subtraction did not improve the precision of our resulting light curves;
in most cases it resulted in a small increase in the root mean square (RMS) of the resulting light curves. For this reason we do not
apply a reference pixel subtraction. 
 

\item {\it Bad-pixel masking:}
In the I'iwi pipeline a bad-pixel map is constructed that flags bad-pixels, hot-pixels and those
that experience significant persistence.
This method flagged a greater number of pixels as problematic
than was optimal for our photometry; we discovered that a great number of the pixels
flagged as ``bad'' were suitable for our photometry, and it was better to not flag these pixels as
``bad'', than attempt to recover their flux from interpolation from nearby pixels.
Therefore, we construct a bad pixel map that accepts a larger fraction of bad/poor
pixels than in the normal I'iwi interpretation. Our bad pixel map is constructed from our median sky-flat (discussed below),
and pixels
are flagged as ``bad'' that deviate by more than 2\% from the median of the array.  

\item {\it Dark Subtraction:} Similarly to the I'iwi pipeline, we subtract the dark current using traditional methods.
We use the standard dark frames, produced for each WIRCam run.
The dark current of the WIRCam HAWAII-2RG's \citep{Beletic08} is small ($\sim$0.05 e-/sec/pixel),
therefore this step has negligible impact on our light curves.

\item {\it Non-linearity correction}: We apply a non-linearity correction to our data.
We utilize the WIRCam non-linearity correction using data obtained
from 2007 July 16 to 2008 March 1.
Extensive tests of the non-linearity correction show that for nearly all of our ``Staring Mode'' data-sets --
those featuring relatively low sky backgrounds and modest levels of illumination -- 
the resulting secondary eclipse depths are relatively insensitive to the non-linearity correction.
However, for our Ks-band observations that feature high levels of illumination and a large
sky background (such as our KELT-1 Ks-band photometry),
the resulting eclipse depth depends on the non-linearity correction.

\item {\it Cross-talk correction:} We do not employ a cross-talk correction.
The WIRCam Hawaii-2RG chips suffer from obvious 
artefacts due to either cross-talk, or arguably more noticeably, effects related to the 32-amplifiers on 
each WIRCam chip.
One technique for removing these cross-talk and amplifier artefacts, is 
to take the median of the 32-amplifiers on each WIRCam chip for each exposure,
and to subtract this median amplifier from each of the 32-amplifiers.
Although this appears to remove the effects of these amplifier artefacts, 
it does not improve the precision (the RMS) of the resulting light curves. Therefore, we do not employ
median amplifier subtraction techniques, or any other form of cross-talk correction.

\item {\it Sky frame subtraction:} We do not employ sky frame subtraction; we instead
subtract our sky using an annulus during our aperture photometry.
As our observations are continuous throughout our observations,
we are unable to construct sky frames contemporaneously with our observations. 
Attempts to include sky frame 
subtraction, using the techniques discussed in \citet{CrollTrESTwo}, did not improve
the RMS of the resulting photometry.

\item {\it Division by a sky flat:} We divide our observations by our sky-flat.
We construct a sky-flat for each ``Staring Mode'' sequence (i.e. several hours of observing of a transit or eclipse, for
instance)  by taking the median of a stack of
dithered in-focus images observed before and after our target observations.
Usually this consists of 15 dithered in-focus 
observations before and after our observations.
Due to the high sky and thermal background in the Ks-band, our sky-flats 
in that band typically have $\sim$2000 $ADU$/$pixel$ or more, per exposure.

\item {\it Interpolation over bad/saturated pixels:} In the I'iwi 1.9 pipeline bad or saturated pixels were flagged with ``Not-a-Number'' values
and therefore did not contribute to the flux in our apertures. During times of imperfect guiding, these flagged pixels would move in comparison to the centroids
of our point-spread-functions (PSFs) and would occasionally fall within the aperture we use to determine
the flux of our target or reference stars. When these poor pixels would fall within the aperture of our targets they would result in obvious correlated noise; 
when such flagged pixels fell in our reference star apertures, the effect was more subtle. 
To correct these discrepancies, we interpolate the flux
of all bad/saturated pixels 
from adjoining pixels (the pixels directly above, below and to each side of the affected pixel, as long as they 
themselves are not bad or saturated pixels).
Even still, we have noticed that our attempts at interpolating the flux of the saturated/bad pixel does not precisely capture the true flux 
of the pixel to the level
required for precise measurements of the eclipse or transit depths of exoplanets.
This is likely due in part to the fact that our defocused PSFs are not radially 
symmetric\footnote{This lack of radial symmetry is likely due to either astigmatism, or the effects of the secondary struts projected
onto the image plane due to suboptimal telescope collimation.} and display significant pixel-to-pixel
flux variations, and therefore interpolation cannot precisely account for the flux of a neighbouring pixel.
Due to the precision required for the science goals that we discuss here, we have found that it is preferable
to throw away exposures with a saturated pixel in the aperture of our target star; the interpolation technique that we discuss here 
is useful for our reference stars, so in general
we keep exposures with a saturated pixel in our reference star apertures (as we use the median of our reference star ensemble,
in general a single discrepant point negligibly affects our resulting light cure).

\end{itemize}

\subsection{Lessons for other Near-infrared Pipelines}
\label{SecPipelineLessons}

 In general, the lesson from the reduction pipeline that we describe here, optimized for precise photometry of
``Staring Mode'' observations \citep{CrollTrESTwo,Devost10}, compared
to the original I'iwi pipeline, is to avoid division or subtraction by quantities that vary
from exposure to exposure.		
The significant systematics
observed in near-infrared
photometry from the ground, and the large variations in flux
of our target stars from one exposure to the next, mean that the accuracy we achieve is entirely dependent on our
differential photometry, and the assumption that the flux of our 
target star closely correlates with the flux of our reference stars in each exposure.
Steps, such as the reference pixel subtraction, the cross-talk correction, or the sky subtraction step, that feature subtraction or 
division by values that change
from exposure to exposure, appear to worsen rather than improve the resulting RMS of the light curves; the likely reason is that they cause
small variations, from exposure to exposure,
in the ratio of the flux from the target to the reference stars.
Such variations appear to lead to correlated noise in the resulting light curve.

We emphasize that generic pipelines optimized for other science goals,
may therefore introduce steps that ultimately degrade the quality of the resulting photometry.
One such example is the aforementioned cross-talk correction.
The cross-talk step described above is crucial to remove amplifier effects that are usually visible in the processed WIRCam images;
therefore this cross-talk effect is necessary for individuals performing,
for instance, extragalactic work to detect low-surface brightness features around galaxies.
This same cross-talk correction provides no improvement and occasionally degrades 
the RMS of the resulting ``Staring Mode'' photometry.

\section{Analysis}
\label{SecAnalysis}

\begin{deluxetable*}{cccccc}
\tablecaption{Aperture Sizes and Number of Stars in the Ensemble}
\tablehead{
\colhead{Eclipse} 		& \colhead{Aperture Radius} 			& \colhead{Inner Annuli} 	& \colhead{Outer Annuli}	& \colhead{\# of Stars in the}			& \colhead{RMS of the residuals}	\\
\colhead{Light curve}		& \colhead{(pixels)}				& \colhead{Radius (pixels)}	& \colhead{Radius (pixels)}	& \colhead{reference star ensemble}		& \colhead{per exposure ($10^{-3}$)}	\\
}
\startdata
First WASP-12 Ks-band		& \WASPTwelveEclipseKsBandOneAperturePixels	& 22				& 34				& \WASPTwelveEclipseKsBandOneEnsembleNStars 	& \RMSExposureWASPTwelveKsBand \\
Second WASP-12 Ks-band		& \WASPTwelveEclipseKsBandTwoAperturePixels	& 21				& 29				& \WASPTwelveEclipseKsBandTwoEnsembleNStars 	& \RMSExposureWASPTwelveKsBandTwentyEleven \\
Third WASP-12 Ks-band		& \WASPTwelveEclipseKsBandThreeAperturePixels	& 23				& 31				& \WASPTwelveEclipseKsBandThreeEnsembleNStars 	& \RMSExposureWASPTwelveKsBandDecTwentyEleven \\
WASP-12 Y-band			& \WASPTwelveEclipseYBandAperturePixels		& 21				& 29				& \WASPTwelveEclipseYBandEnsembleNStars 	& \RMSExposureWASPTwelveYBandTwentyEleven \\
WASP-12 $K_{CONT}$-band		& \WASPTwelveEclipseKContBandAperturePixels	& 16				& 24				& \WASPTwelveEclipseKContBandEnsembleNStars 	& \RMSExposureWASPTwelveKContBandJanTwentyTwelve \\
Qatar-1 Ks-band			& \QatarOneEclipseKsBandAperturePixels		& 19				& 28				& \QatarOneEclipseKsBandEnsembleNStars 		& \RMSExposureKELTOneKsBand \\
KELT-1 Ks-band			& \KELTOneEclipseKsBandAperturePixels		& 22				& 30				& \KELTOneEclipseKsBandEnsembleNStars 		& \RMSExposureKELTOneKsBand \\
First WASP-3 Ks-band Eclipse	& \WASPThreeEclipseKsBandOneAperturePixels	& 20				& 27				& \WASPThreeEclipseKsBandOneEnsembleNStars    	& \RMSExposureWASPThreeKsBandEclipseOne \\
Second WASP-3 Ks-band Eclipse	& \WASPThreeEclipseKsBandTwoAperturePixels	& 20				& 29				& \WASPThreeEclipseKsBandTwoEnsembleNStars 	& \RMSExposureWASPThreeKsBandEclipseTwo \\
\enddata
\label{TableApertureNStar}
\end{deluxetable*}

\begin{figure*}
\includegraphics[scale=0.44, angle = 270]{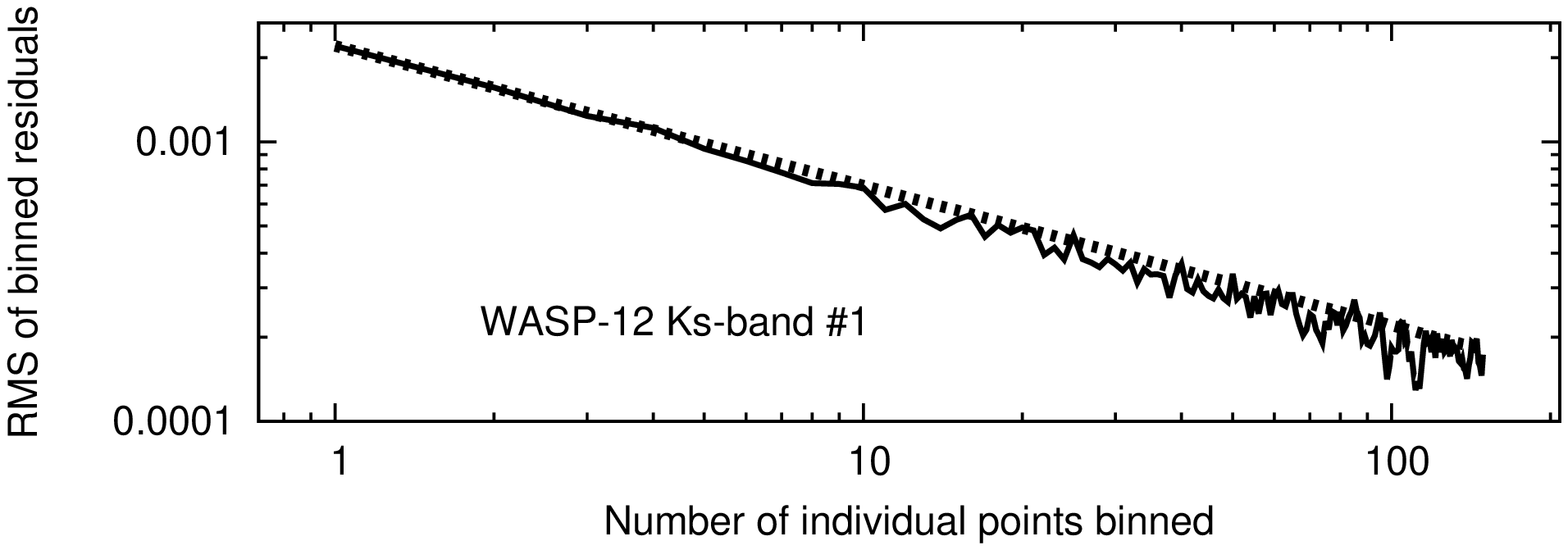}
\includegraphics[scale=0.44, angle = 270]{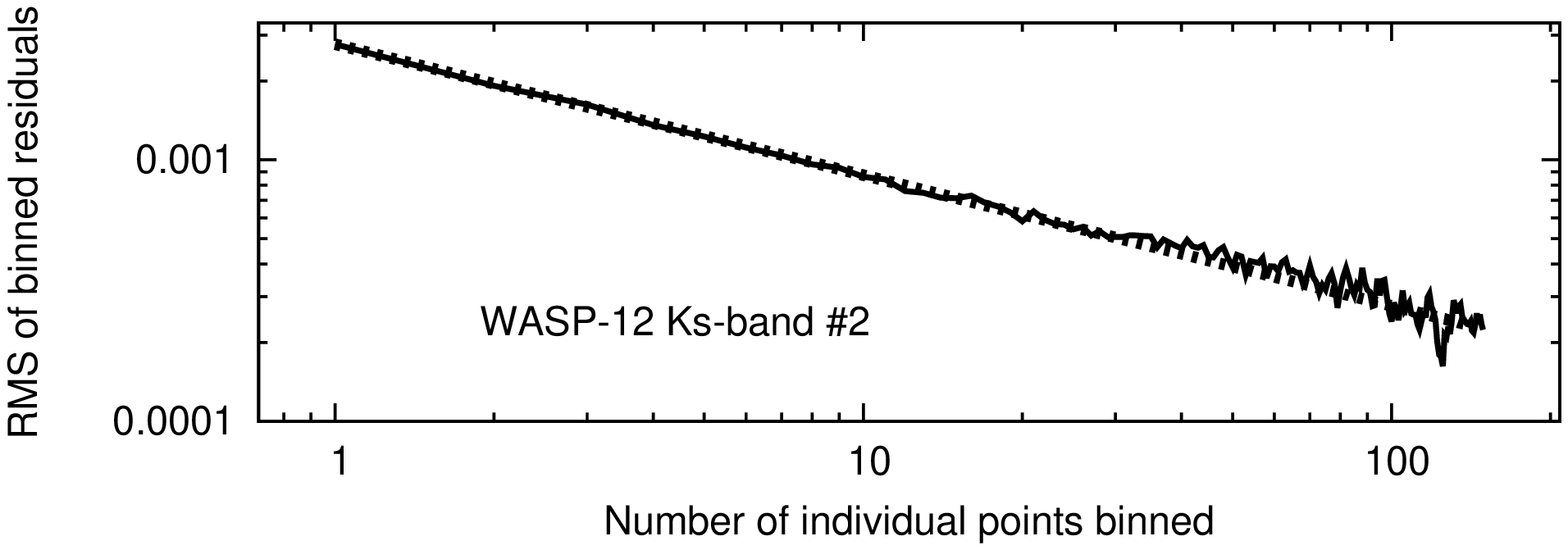}
\includegraphics[scale=0.44, angle = 270]{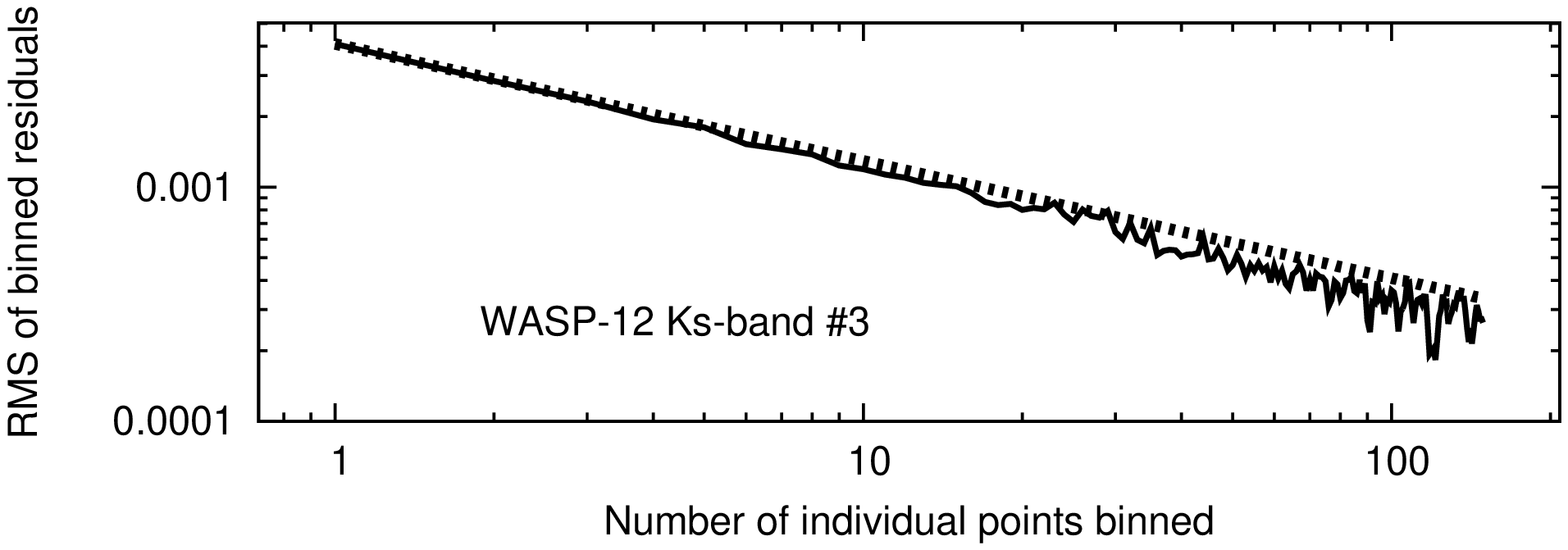}
\includegraphics[scale=0.44, angle = 270]{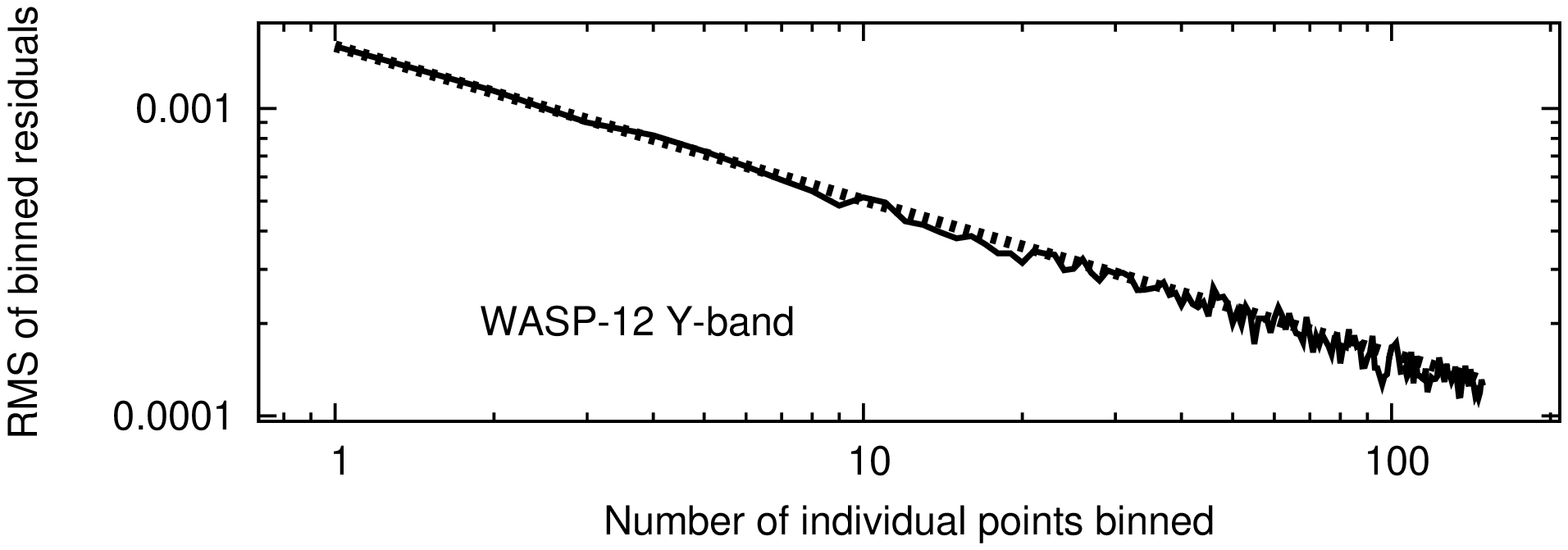}
\includegraphics[scale=0.44, angle = 270]{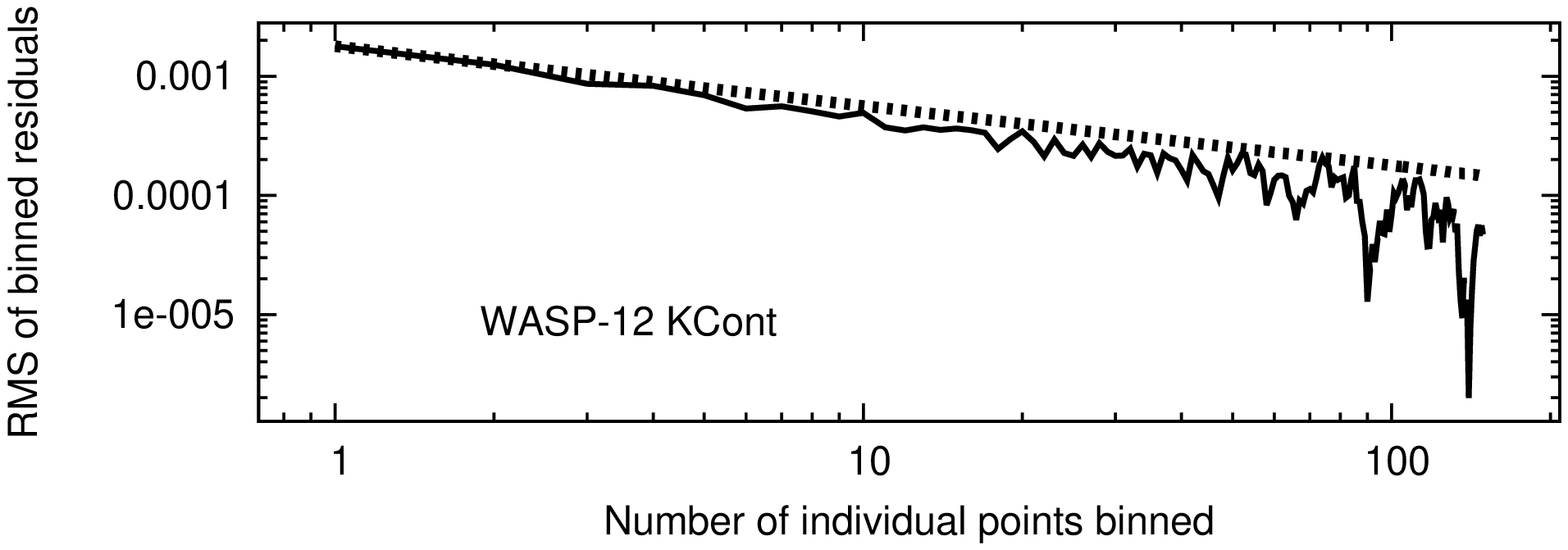}
\includegraphics[scale=0.44, angle = 270]{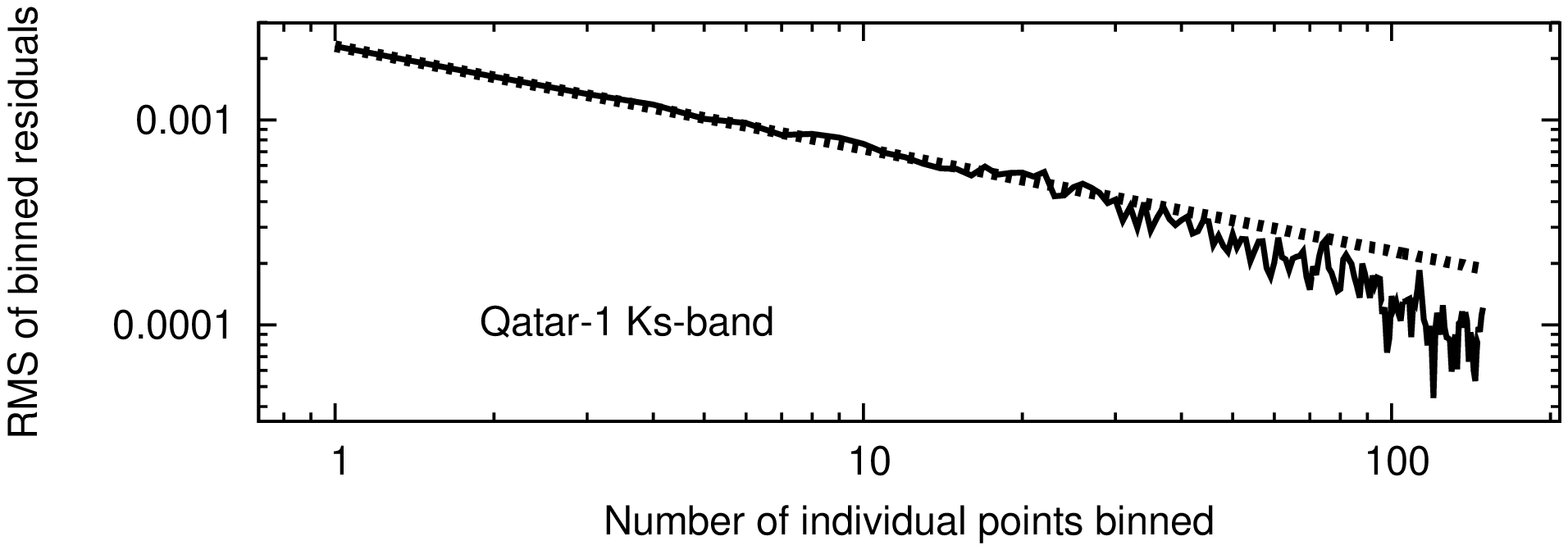}
\includegraphics[scale=0.44, angle = 270]{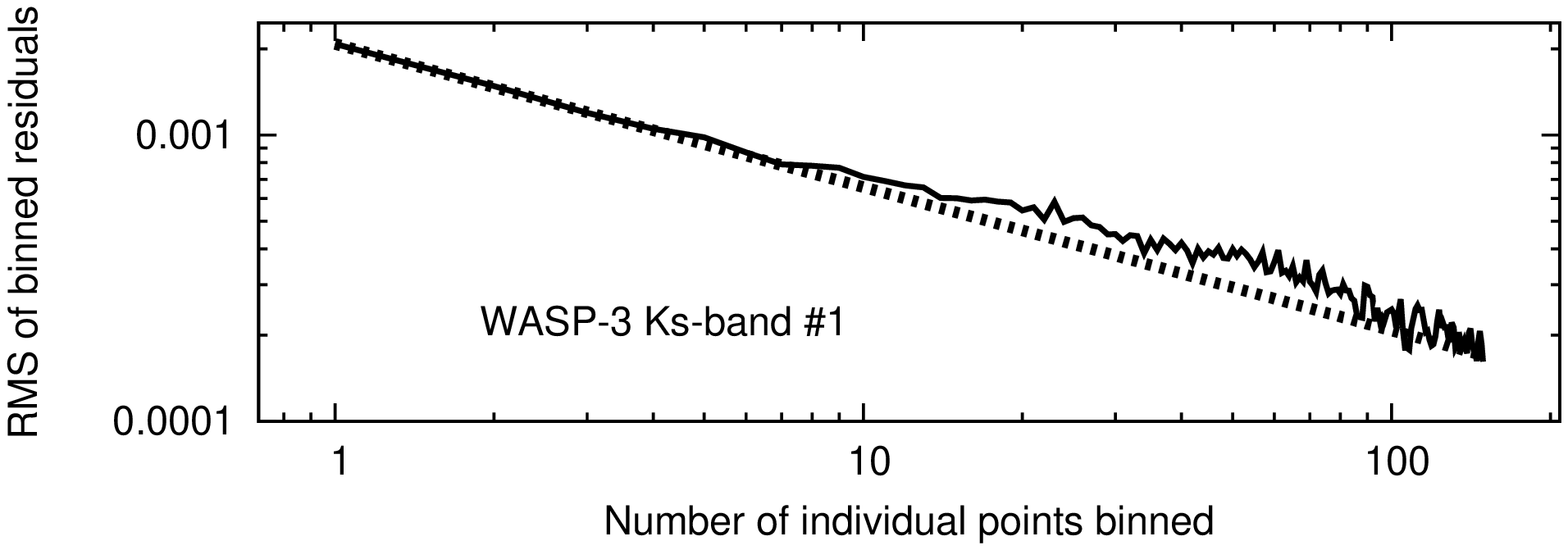}
\includegraphics[scale=0.44, angle = 270]{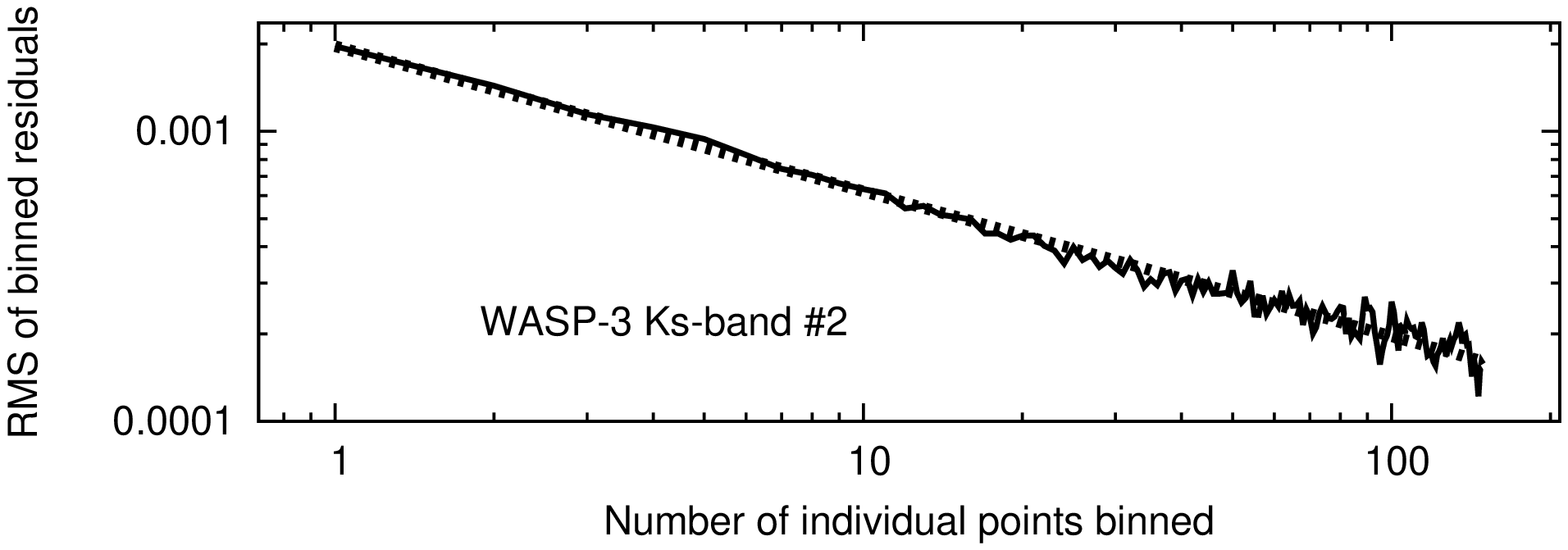}
\includegraphics[scale=0.44, angle = 270]{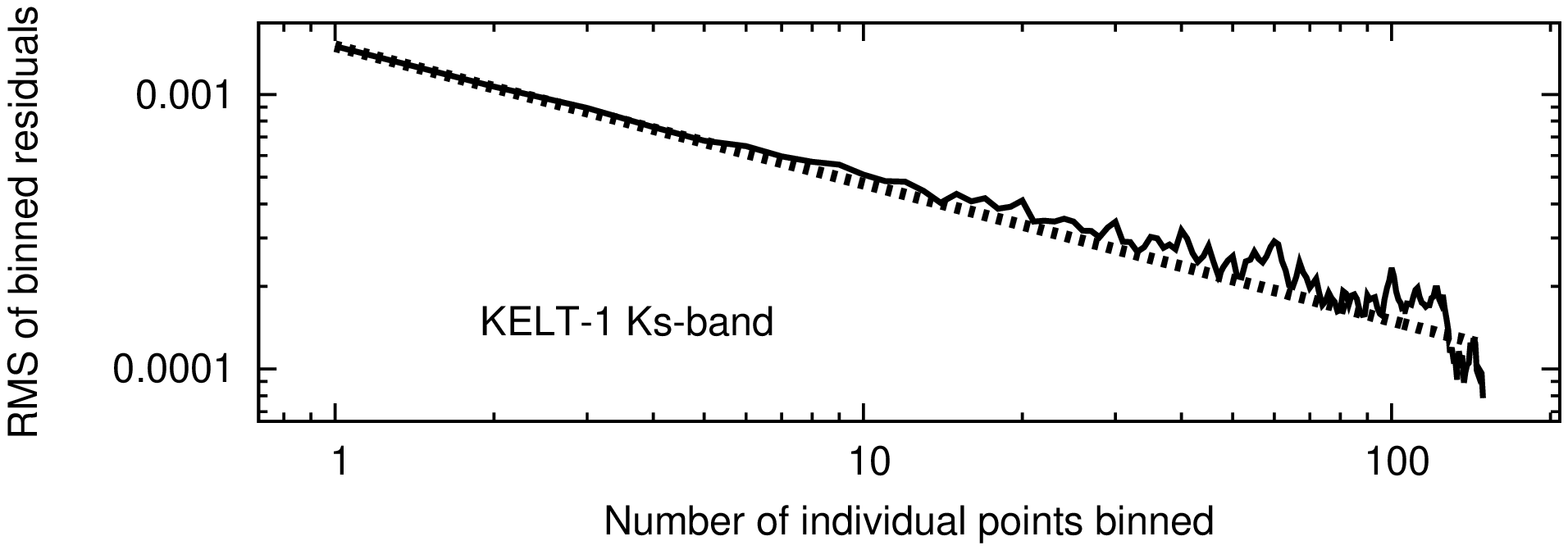}

\caption[RMSEclipses]
	{	Root mean square of our residuals to our best-fit model (solid line)
		for our various data-sets.
		The dashed line in each panel displays the one over the square-root of the
		bin-size expectation for Gaussian noise.
		Most of our data-sets are relatively free of correlated noise.
		Small number statistics are likely responsible for the cases where the data appears to bin
		down below the Gaussian noise expectation.
	}
\label{FigRMSBin}
\end{figure*}

We perform aperture photometry on our target and reference stars; we restrict our choice of reference star to those
on the same WIRCam chip as the 
target\footnote{There are two main reasons that we restrict our choice of reference stars to the same WIRCam chip as the
target; these are: (i) in previous ``Staring Mode'' analyses \citep{CrollTrESTwo,CrollTrESThree,CrollWASPTwelve} we
observed that the reference stars chosen were often on the same chip as the target star (it was unclear whether this
was due to instrumental effects on the chip, or due to telluric affects caused by the spatial separation on the sky),
and (ii) due to the extra computational time involved in running the pipeline discussed in Section \ref{SecPipeline}, 
and the analysis discussed here on all four WIRCam chips, rather than just one.}.
We center our apertures using flux-weighted centroids.
Table \ref{TableApertureNStar} gives the inner and outer radii of the annuli
used to subtract the background for our aperture photometry.
We correct the photometry of our target star with the the best ensemble of reference stars for each
data-set (as discussed in Section \ref{SecOptimalReferenceStars}).
The reference star light curve is produced by taking the median of the light curve of all
the normalized reference star light curves; each normalized reference light curve is produced by normalizing
its light curve to the median flux level of that reference star.
The corrected target light curve is then produced by dividing through by our normalized reference
star ensemble light curve.
To remove obvious outliers from our corrected target light curve,
we apply a 7$\sigma$ cut\footnote{This outlier cut affects only our third WASP-12b Ks-band eclipse (2 points removed),
our KELT-1 Ks-band eclipse (1 point removed), and our first WASP-3 Ks-band eclipse (1 point removed).}.

 We fit our eclipse light curves with a secondary eclipse model calculated
using the routines of \citet{MandelAgol02}, and with a quadratic function with time to fit for what we refer to
as a background trend, $B_f$, defined as:
\begin{equation}
B_f = 1 + c_1 + c_2 dt + c_3 dt^2
\end{equation}
where $dt$ is the interval from the beginning of the observations
and $c_1$, $c_2$ and $c_3$ are fit parameters.
These background trends are likely instrumental, or telluric,
in origin, and have been observed in several of our previous near-infrared
eclipse light curves \citep{CrollTrESTwo,CrollTrESThree,CrollWASPTwelve}.
The secondary eclipse parameters that we fit 
for 
are the 
offset from the mid-point of the eclipse, $t_{offset}$
and the apparent depth of the secondary eclipse, $F_{Ap}/F_{*}$.
For WASP-12b and KELT-1b the apparent depth of the secondary eclipse is not exactly equal to the 
true depth of the secondary eclipse as both stars have nearby companions that dilute their depths.
For our first WASP-3 secondary eclipse we do not fit for the mid-point of the eclipse, $t_{offset}$,
due to the relative lack of pre-eclipse baseline, and instead assume a circular orbit ($t_{offset}$=0).
We employ Markov Chain Monte Carlo (MCMC) techniques as described for our purposes in \citet{CrollMCMC};
we use flat a priori constraints for all parameters.
We obtain our planetary and stellar parameters for WASP-3b from \citet{Pollacco08} and \citet{Southworth10}, 
for Qatar-1 from \citet{Alsubai11},
for WASP-12b from \citet{Hebb09} and \citet{Chan11}, 
and for KELT-1 from \citet{Siverd12}. 

To evaluate the presence of red noise in each of our photometric data-sets, we bin down 
the residuals following the subtraction
of the best-fit model for each data-set, and compare the resulting RMS to the Gaussian
noise expectation of one over the square-root of the bin size (Figure \ref{FigRMSBin}),
for the specific aperture size and reference star ensemble 
given in Table \ref{TableApertureNStar} for each data-set.
To quantify the amount of correlated noise in our data-sets we often use $\beta$, a parameterization
defined by \citet{Winn08} that denotes the factor by which the residuals scale above the Gaussian noise expectation
(see Figure \ref{FigRMSBin}). To determine $\beta$ we take the average of bin sizes between 10 and
80 binned 
points\footnote{In the odd cases where $\beta$ is less than one (and the data therefore scales down below the Gaussian noise limit), we set $\beta$=1.}. 
We note that most of our data-sets are relatively free of time-correlated red-noise;
our second WASP-12 Ks-band eclipse, our KELT-1 Ks-band eclipse, and our first WASP-3 Ks-band eclipse, are the minor exceptions. 

\begin{figure*}
\centering
\includegraphics[scale=0.52, angle = 270]{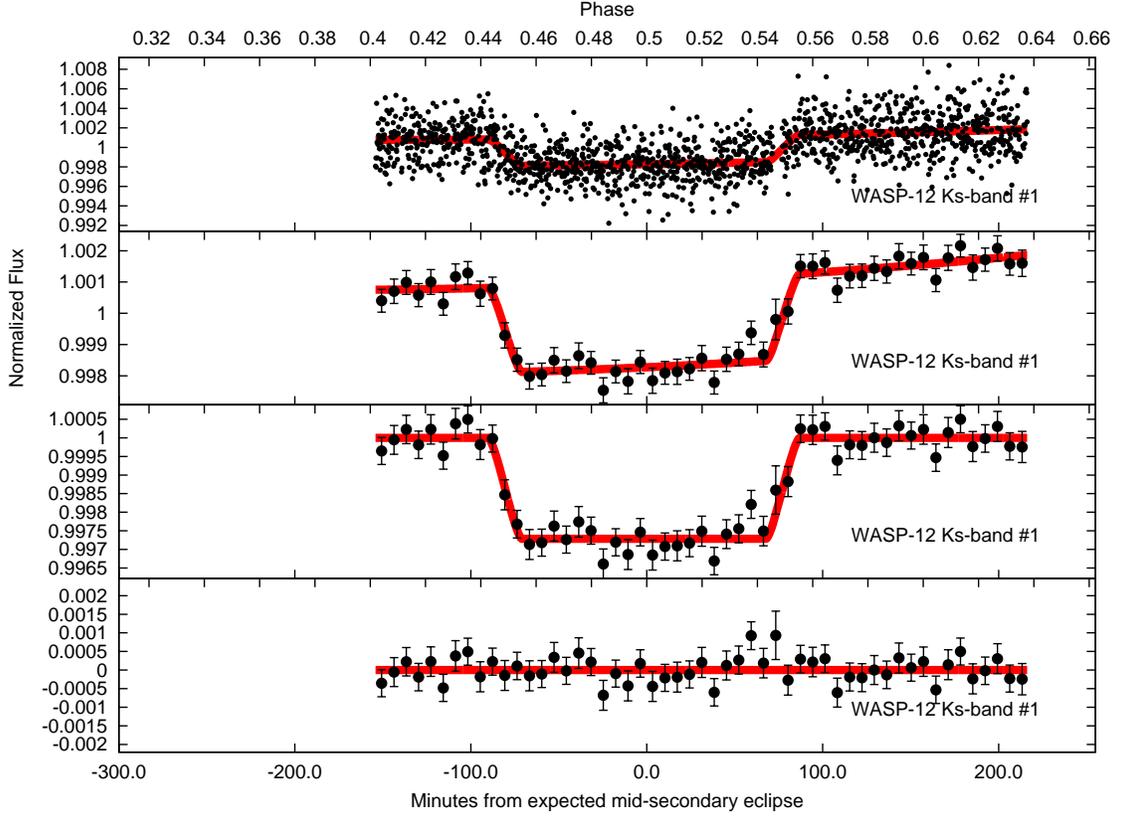}
\caption[WASP-12 Eclipse One]
	{
		CFHT/WIRCam photometry of the secondary eclipse of WASP-12b observed in the Ks-band on 
		2009 December 28. 
		The top panel shows the unbinned 
		light curve with the best-fit secondary eclipse and background from our MCMC analysis (red line). The second
		panel shows the light curve with the data binned every $\sim$7.0 minutes and again our best-fit eclipse and background.
		The third panel shows the binned data after the subtraction of the best-fit background, $B_f$,
		along with the best-fit eclipse model. The bottom panel shows the binned residuals from the best-fit model.
	}
\label{FigWASPTwelveEclipseKsOne}
\end{figure*}

\begin{figure*}
\centering
\includegraphics[scale=0.52, angle = 270]{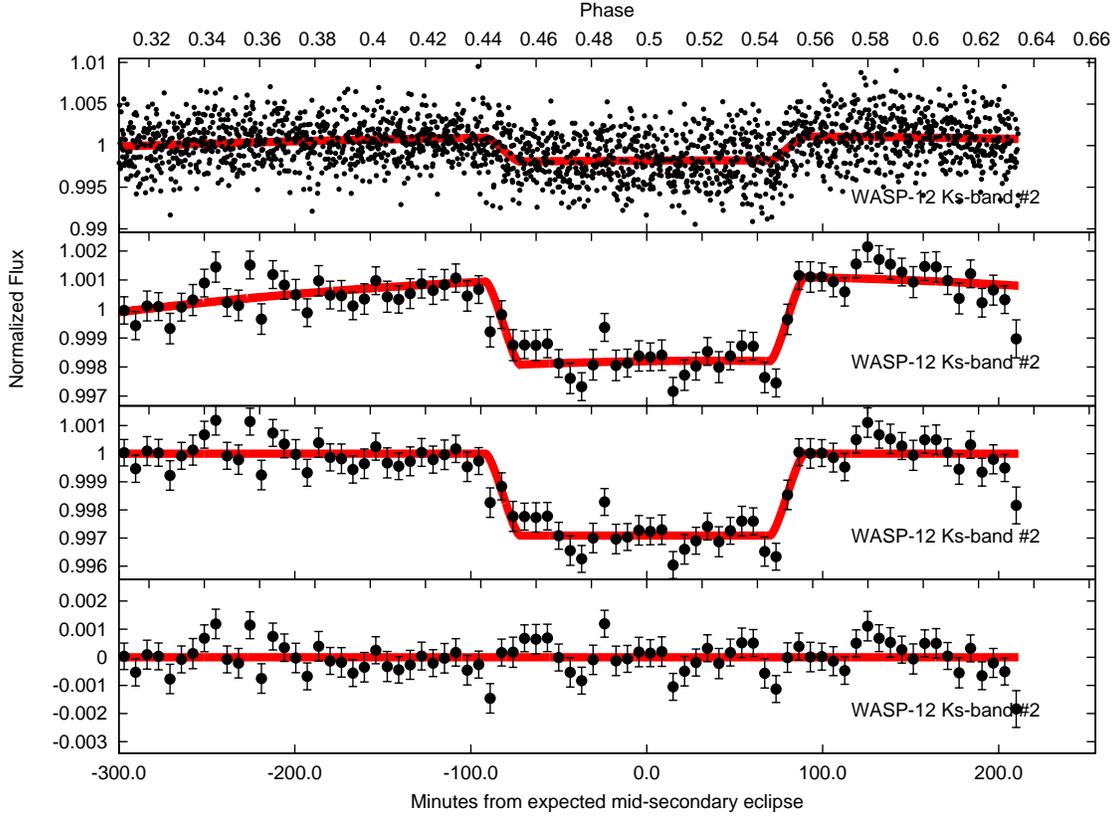}
\caption[WASP-12 Eclipse Two]
	{
		CFHT/WIRCam photometry of the secondary eclipse of WASP-12b observed in the Ks-band on 
		2011 January 14. The format of the Figure is otherwise identical to Figure \ref{FigWASPTwelveEclipseKsOne},
		except the data is binned every $\sim$6.5 minutes in the bottom three panels.
	}
\label{FigWASPTwelveEclipseKsTwo}
\end{figure*}

\begin{figure*}
\centering
\includegraphics[scale=0.52, angle = 270]{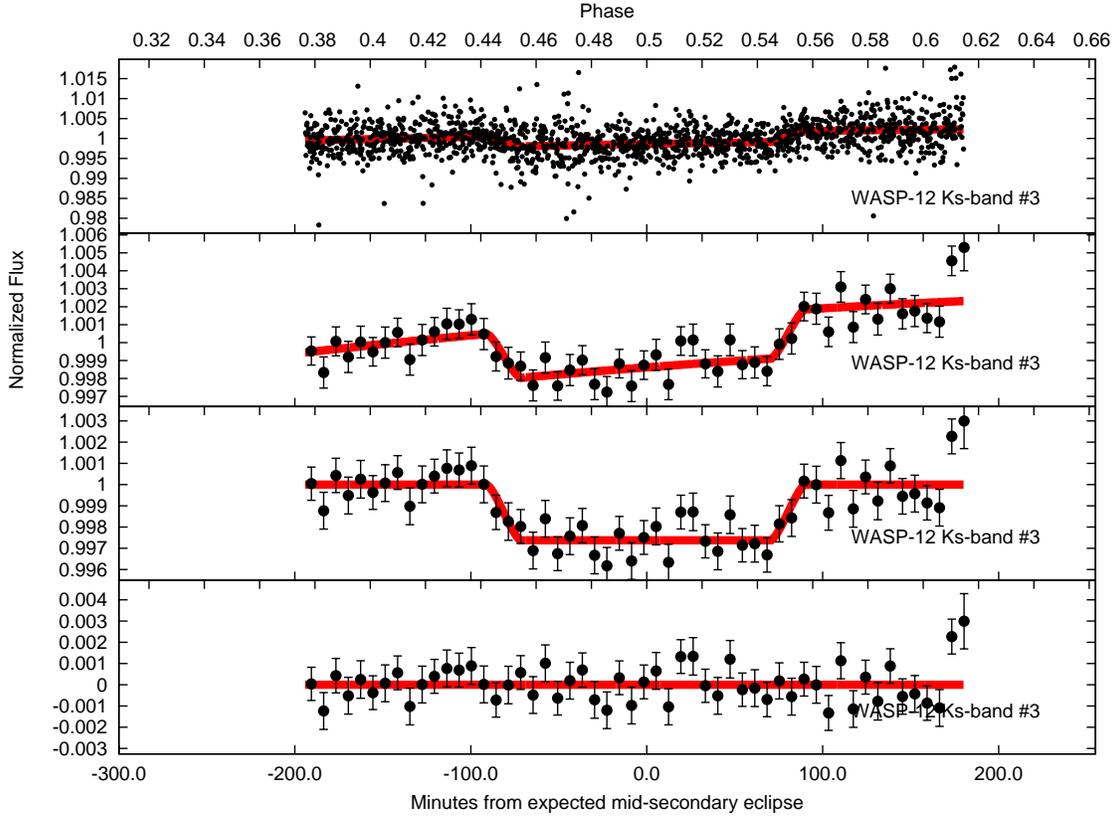}
\caption[WASP-12 Eclipse Three]
	{
		CFHT/WIRCam photometry of the secondary eclipse of WASP-12b observed in the Ks-band on 
		2011 December 28. The format of the Figure is otherwise identical to Figure \ref{FigWASPTwelveEclipseKsOne}.
	}
\label{FigWASPTwelveEclipseKsThree}
\end{figure*}

\begin{figure*}
\centering
\includegraphics[scale=0.52, angle = 270]{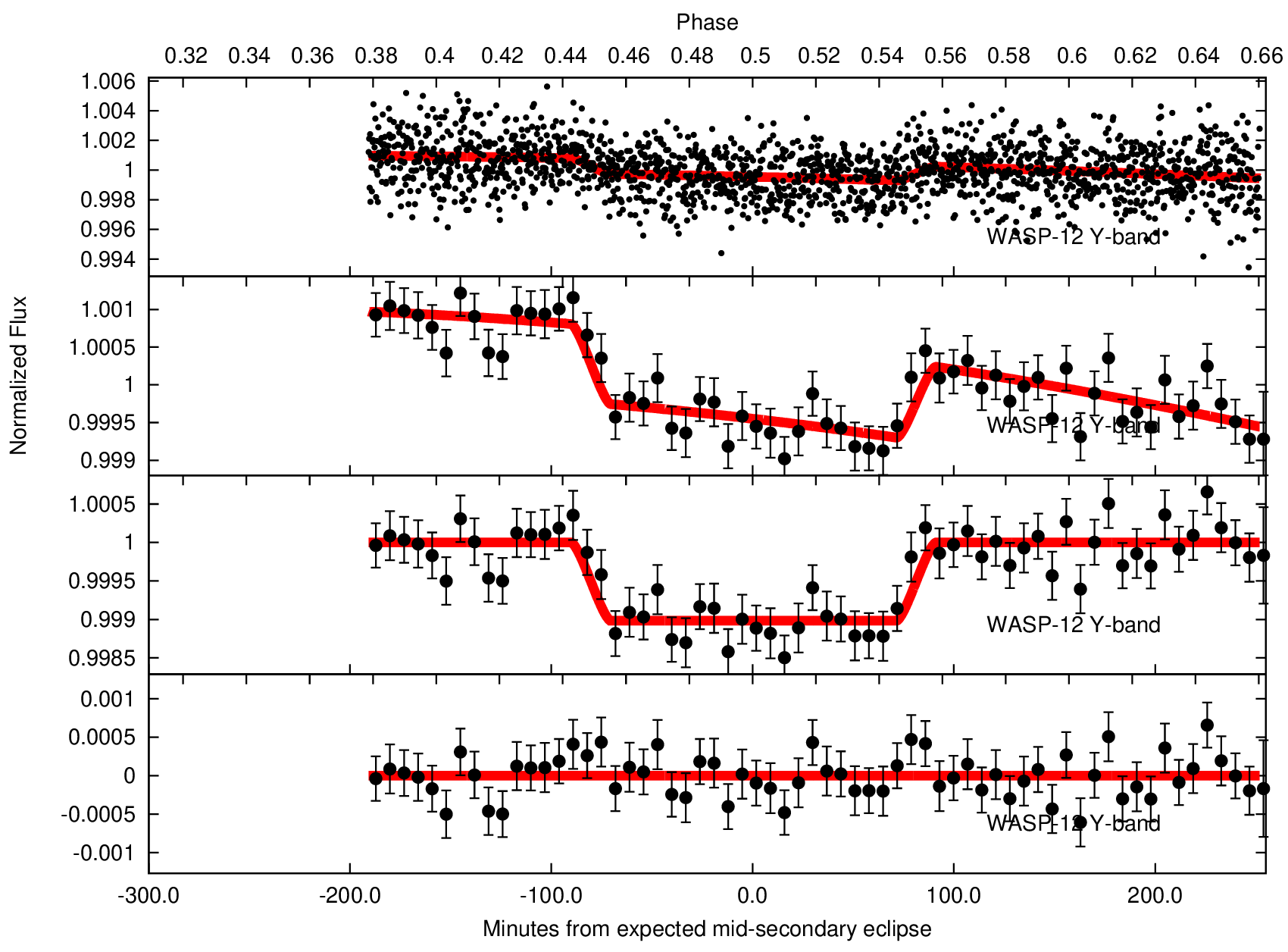}
\caption[WASP-12 Eclipse Y]
	{
		CFHT/WIRCam photometry of the secondary eclipse of WASP-12b observed in the Y-band on 
		2011 January 25. The format of the Figure is otherwise identical to Figure \ref{FigWASPTwelveEclipseKsOne}.
	}
\label{FigWASPTwelveEclipseY}
\end{figure*}

\begin{figure*}
\centering
\includegraphics[scale=0.52, angle = 270]{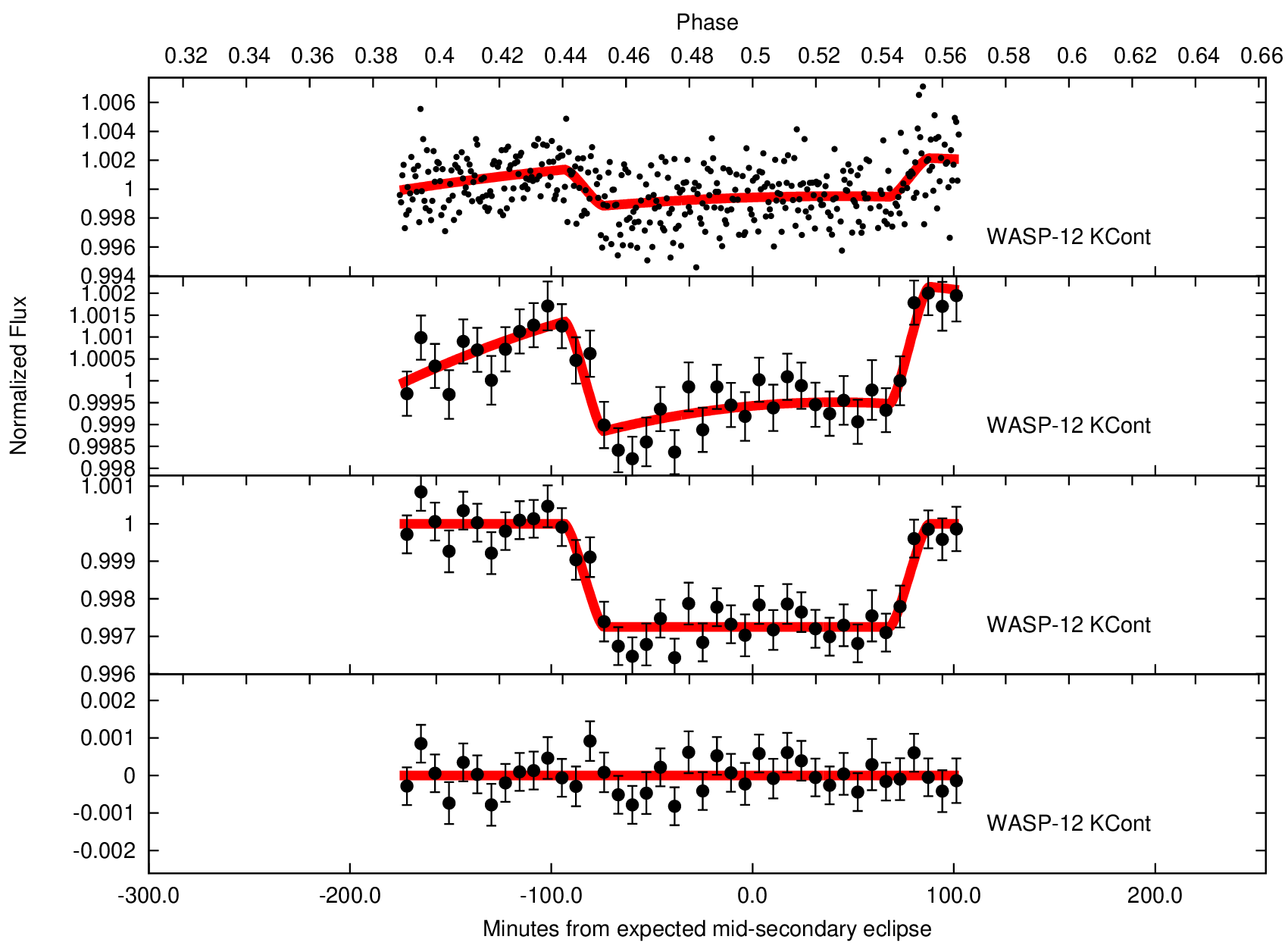}
\caption[WASP-12 Eclipse KCont]
	{
		CFHT/WIRCam photometry of the secondary eclipse of WASP-12b observed in the $K_{CONT}$-band on 
		2012 January 19. The format of the Figure is otherwise identical to Figure \ref{FigWASPTwelveEclipseKsOne}.
	}
\label{FigWASPTwelveEclipseKCont}
\end{figure*}

\begin{figure*}
\centering
\includegraphics[scale=0.52, angle = 270]{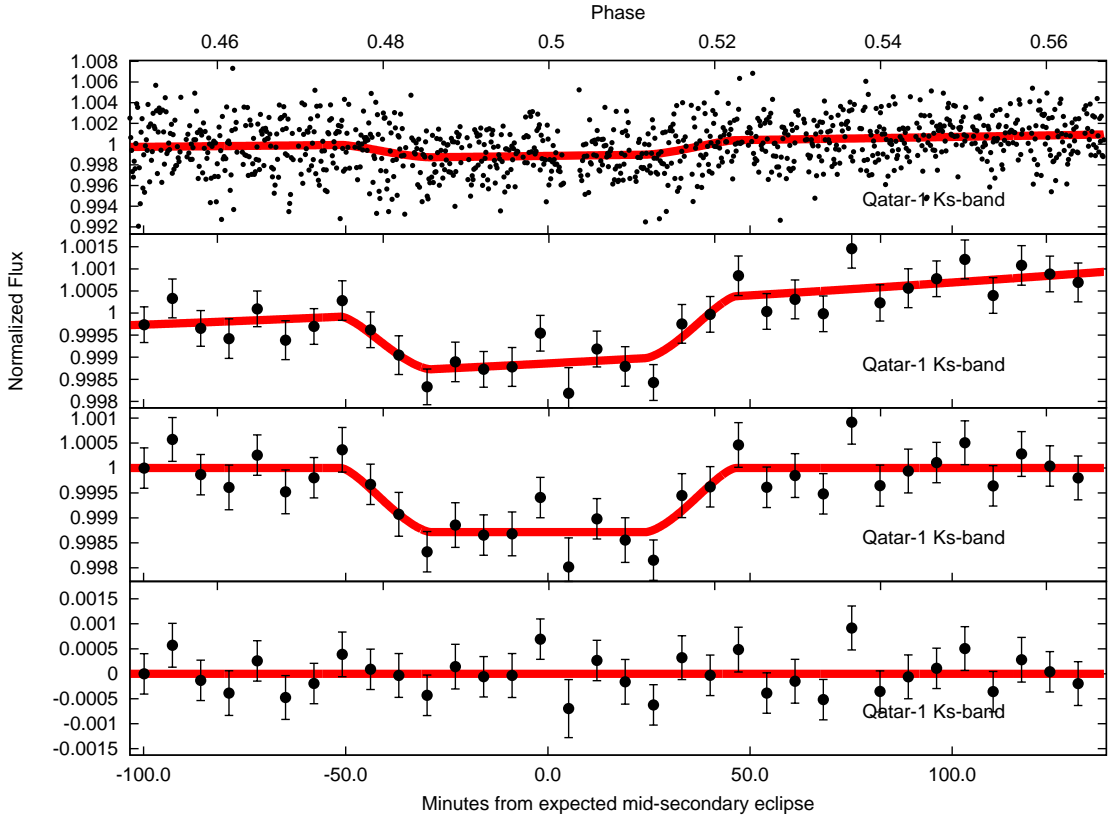}
\caption[Qatar-1 Eclipse]
	{
		CFHT/WIRCam photometry of the secondary eclipse of Qatar-1b observed in the Ks-band on 
		2012 July 28. The format of the Figure is otherwise identical to Figure \ref{FigWASPTwelveEclipseKsOne}.
	}
\label{FigQatarOne}
\end{figure*}

\begin{figure*}
\centering
\includegraphics[scale=0.52, angle = 270]{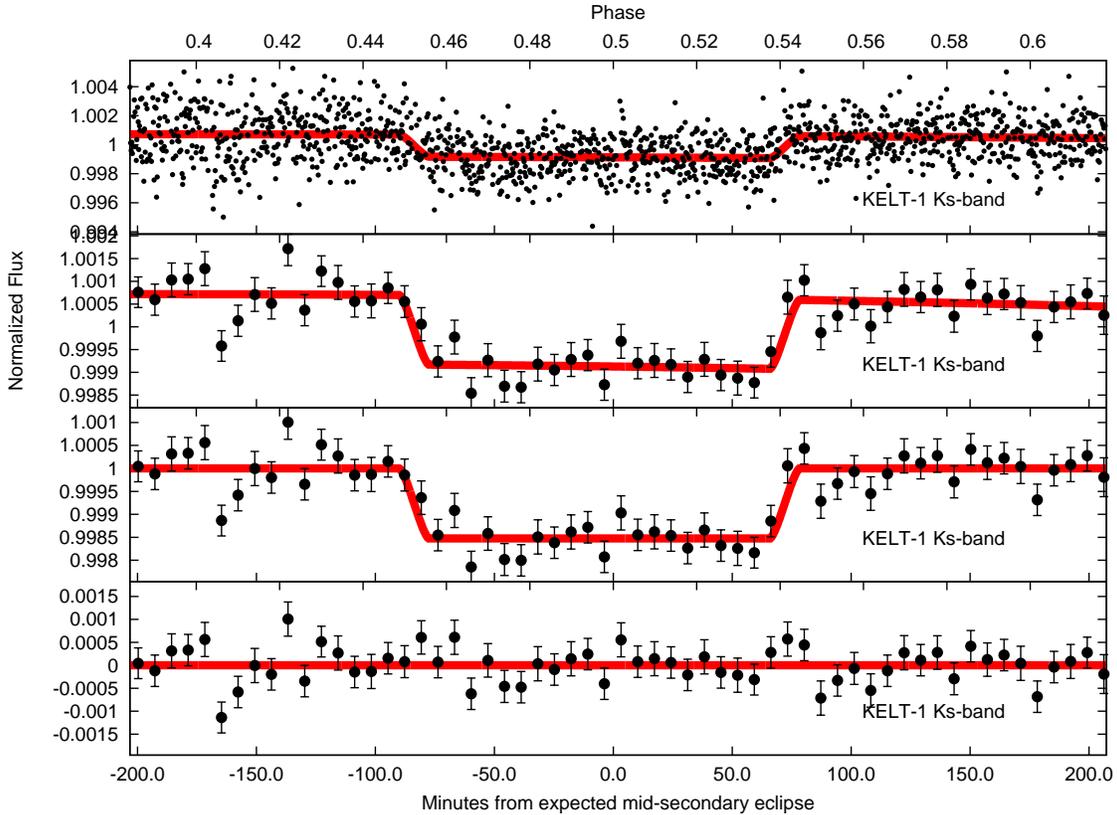}
\caption[KELT-1 Eclipse]
	{
		CFHT/WIRCam photometry of the secondary eclipse of KELT-1b observed in the Ks band on 
		2012 October 11. Otherwise, the figure layout is identical to Figure \ref{FigWASPTwelveEclipseKsOne}.
	}
\label{FigKELTOne}
\end{figure*}

\begin{figure*}
\centering
\includegraphics[scale=0.52,angle=270]{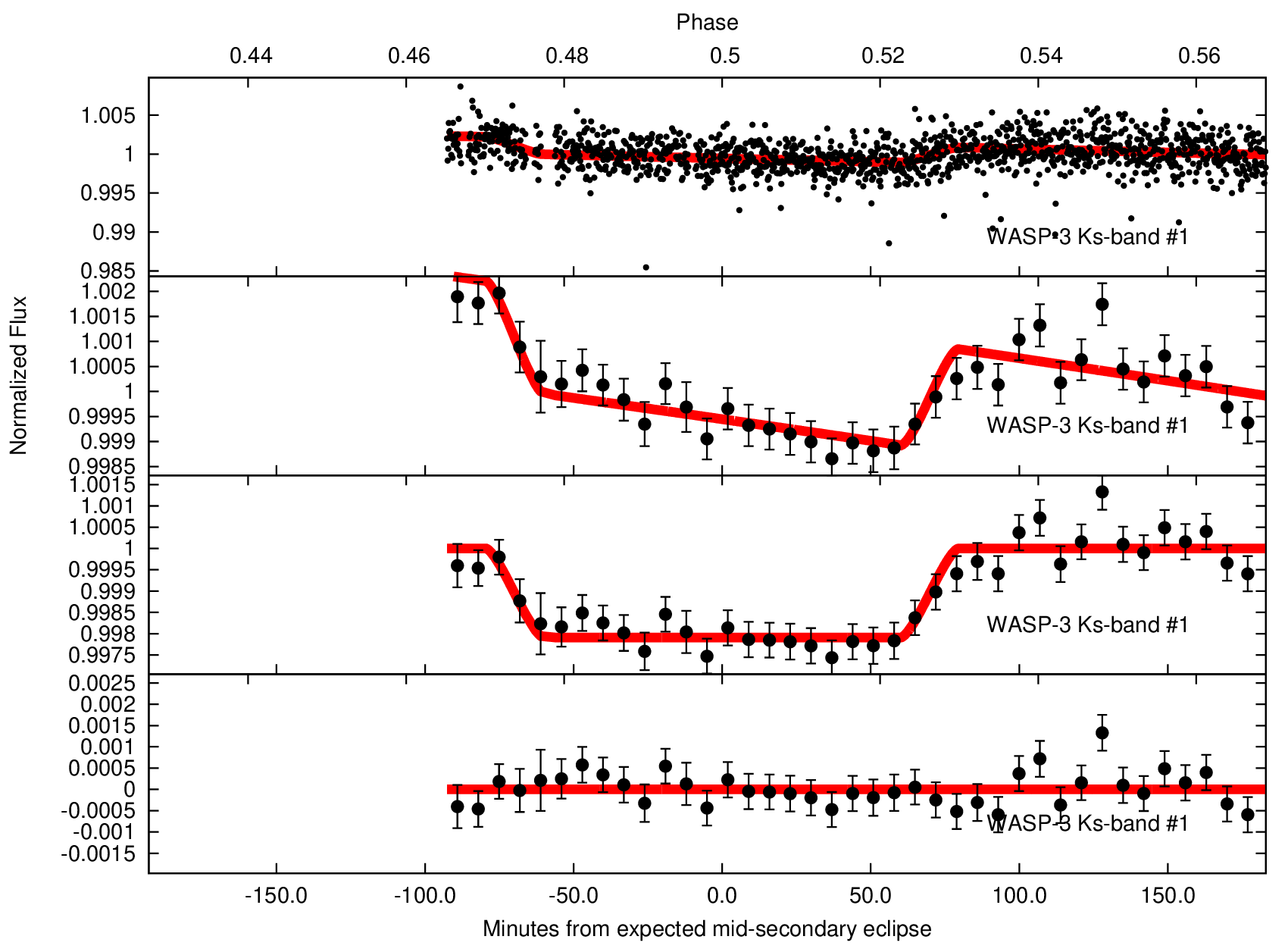}
\caption[Photometry of our first WASP-3b Ks-band secondary eclipse]
	{	CFHT/WIRCam photometry of our first Ks-band secondary eclipse of WASP-3b
		observed on 2009 June 3.
		Otherwise, the figure layout is identical to Figure \ref{FigWASPTwelveEclipseKsOne}.
	}
\label{FigWASPThree}
\end{figure*}

\begin{figure*}
\centering
\includegraphics[scale=0.52,angle=270]{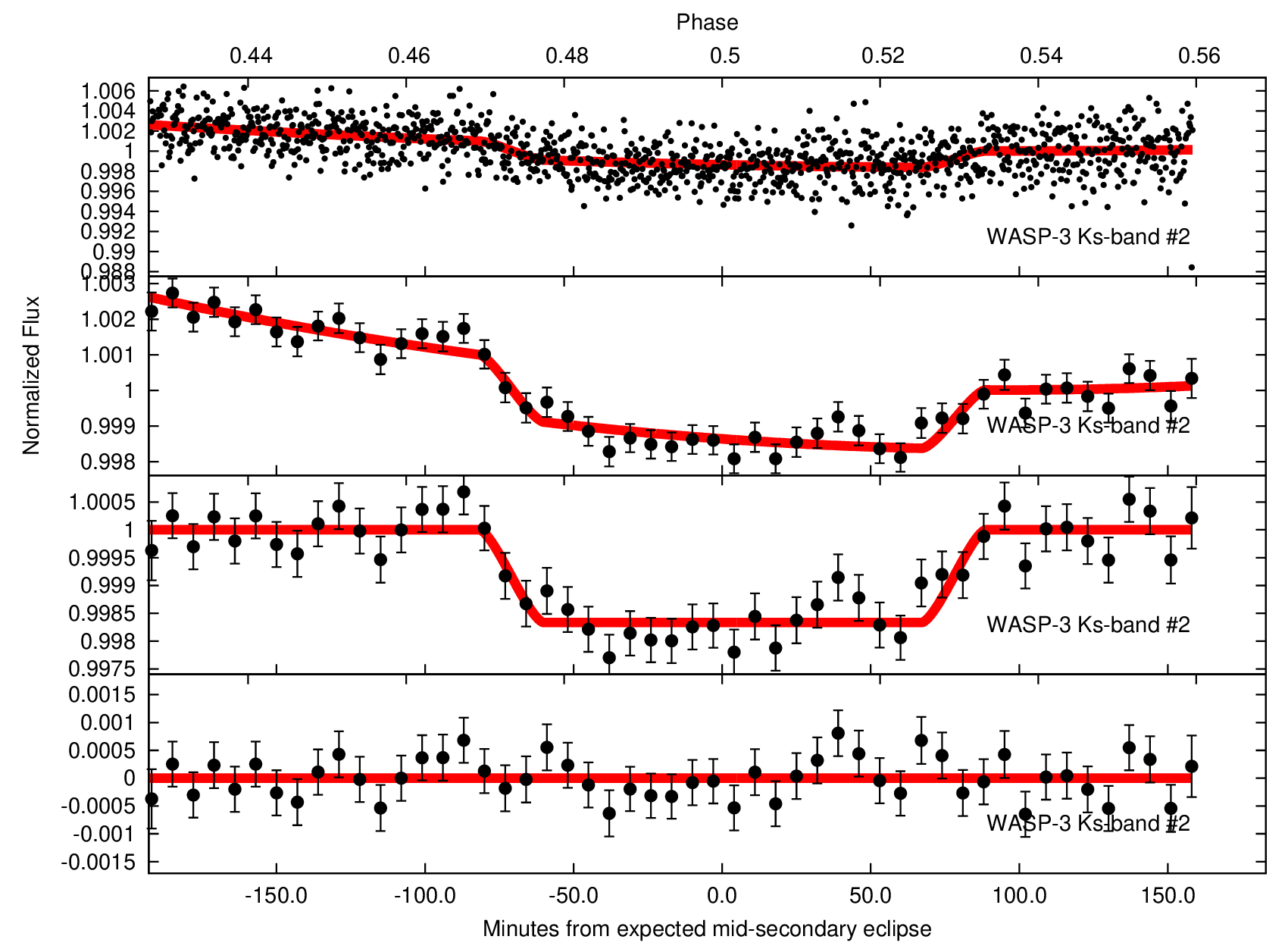}
\caption[Photometry of our second WASP-3b Ks-band secondary eclipses]
	{	CFHT/WIRCam photometry of our second Ks-band secondary eclipse of WASP-3b
		observed on 2009 June 15. 
		Otherwise, the figure layout is identical to Figure \ref{FigWASPTwelveEclipseKsOne}.
	}
\label{FigWASPThreeTwo}
\end{figure*}

\begin{deluxetable*}{cccccc}
\tablecaption{WASP-12 MCMC Eclipse Parameters}
\tablehead{
\colhead{Parameter} 	& \colhead{First Ks-band} 	& \colhead{Second Ks-band}	& \colhead{Third Ks-band}	& \colhead{Y-band}	& \colhead{$K_{CONT}$-band}	\\
\colhead{}		& \colhead{Eclipse}		& \colhead{Eclipse}		& \colhead{Eclipse}		& \colhead{Eclipse}	& \colhead{Eclipse} 		\\
}
\startdata
\multicolumn{6}{c}{Single Aperture Size and Reference Star Ensemble Fit}	\\
reduced $\chi^{2}$					& \ChiWASPTwelveKsBand$^{+\ChiPlusWASPTwelveKsBand}_{-\ChiMinusWASPTwelveKsBand}$										& \ChiWASPTwelveKsBandTwentyEleven$^{+\ChiPlusWASPTwelveKsBandTwentyEleven}_{-\ChiMinusWASPTwelveKsBandTwentyEleven}$										& \ChiWASPTwelveKsBandDecTwentyEleven$^{+\ChiPlusWASPTwelveKsBandDecTwentyEleven}_{-\ChiMinusWASPTwelveKsBandDecTwentyEleven}$										& \ChiWASPTwelveYBandTwentyEleven$^{+\ChiPlusWASPTwelveYBandTwentyEleven}_{-\ChiMinusWASPTwelveYBandTwentyEleven}$										& \ChiWASPTwelveKContBandJanTwentyTwelve$^{+\ChiPlusWASPTwelveKContBandJanTwentyTwelve}_{-\ChiMinusWASPTwelveKContBandJanTwentyTwelve}$										\\
$c_{1}$							& \cOneWASPTwelveKsBand$^{+\cOnePlusWASPTwelveKsBand}_{-\cOneMinusWASPTwelveKsBand}$										& \cOneWASPTwelveKsBandTwentyEleven$^{+\cOnePlusWASPTwelveKsBandTwentyEleven}_{-\cOneMinusWASPTwelveKsBandTwentyEleven}$ 									& \cOneWASPTwelveKsBandDecTwentyEleven$^{+\cOnePlusWASPTwelveKsBandDecTwentyEleven}_{-\cOneMinusWASPTwelveKsBandDecTwentyEleven}$ 									& \cOneWASPTwelveYBandTwentyEleven$^{+\cOnePlusWASPTwelveYBandTwentyEleven}_{-\cOneMinusWASPTwelveYBandTwentyEleven}$ 										& \cOneWASPTwelveKContBandJanTwentyTwelve$^{+\cOnePlusWASPTwelveKContBandJanTwentyTwelve}_{-\cOneMinusWASPTwelveKContBandJanTwentyTwelve}$ 									\\
$c_{2}$							& \cTwoWASPTwelveKsBand$^{+\cTwoPlusWASPTwelveKsBand}_{-\cTwoMinusWASPTwelveKsBand}$										& \cTwoWASPTwelveKsBandTwentyEleven$^{+\cTwoPlusWASPTwelveKsBandTwentyEleven}_{-\cTwoMinusWASPTwelveKsBandTwentyEleven}$ 									& \cTwoWASPTwelveKsBandDecTwentyEleven$^{+\cTwoPlusWASPTwelveKsBandDecTwentyEleven}_{-\cTwoMinusWASPTwelveKsBandDecTwentyEleven}$ 									& \cTwoWASPTwelveYBandTwentyEleven$^{+\cTwoPlusWASPTwelveYBandTwentyEleven}_{-\cTwoMinusWASPTwelveYBandTwentyEleven}$ 										&  \cTwoWASPTwelveKContBandJanTwentyTwelve$^{+\cTwoPlusWASPTwelveKContBandJanTwentyTwelve}_{-\cTwoMinusWASPTwelveKContBandJanTwentyTwelve}$ 									\\
$c_{3}$							& \cThreeWASPTwelveKsBand$^{+\cThreePlusWASPTwelveKsBand}_{-\cThreeMinusWASPTwelveKsBand}$									& \cThreeWASPTwelveKsBandTwentyEleven$^{+\cThreePlusWASPTwelveKsBandTwentyEleven}_{-\cThreeMinusWASPTwelveKsBandTwentyEleven}$									& \cThreeWASPTwelveKsBandDecTwentyEleven$^{+\cThreePlusWASPTwelveKsBandDecTwentyEleven}_{-\cThreeMinusWASPTwelveKsBandDecTwentyEleven}$ 								& \cThreeWASPTwelveYBandTwentyEleven$^{+\cThreePlusWASPTwelveYBandTwentyEleven}_{-\cThreeMinusWASPTwelveYBandTwentyEleven}$ 									&  \cThreeWASPTwelveKContBandJanTwentyTwelve$^{+\cThreePlusWASPTwelveKContBandJanTwentyTwelve}_{-\cThreeMinusWASPTwelveKContBandJanTwentyTwelve}$ 								\\
$F_{Ap}/F_{*}$	\tablenotemark{c}			& \FpOverFStarPercentAbstractWASPTwelveKsBand$^{+\FpOverFStarPercentAbstractPlusWASPTwelveKsBand}_{-\FpOverFStarPercentAbstractMinusWASPTwelveKsBand}$\%	& \FpOverFStarPercentAbstractWASPTwelveKsBandTwentyEleven$^{+\FpOverFStarPercentAbstractPlusWASPTwelveKsBandTwentyEleven}_{-\FpOverFStarPercentAbstractMinusWASPTwelveKsBandTwentyEleven}$\%	& \FpOverFStarPercentAbstractWASPTwelveKsBandDecTwentyEleven$^{+\FpOverFStarPercentAbstractPlusWASPTwelveKsBandDecTwentyEleven}_{-\FpOverFStarPercentAbstractMinusWASPTwelveKsBandDecTwentyEleven}$\% 	& \FpOverFStarPercentAbstractWASPTwelveYBandTwentyEleven$^{+\FpOverFStarPercentAbstractPlusWASPTwelveYBandTwentyEleven}_{-\FpOverFStarPercentAbstractMinusWASPTwelveYBandTwentyEleven}$\%	& \FpOverFStarPercentAbstractWASPTwelveKContBandJanTwentyTwelve$^{+\FpOverFStarPercentAbstractPlusWASPTwelveKContBandJanTwentyTwelve}_{-\FpOverFStarPercentAbstractMinusWASPTwelveKContBandJanTwentyTwelve}$\% 	\\
$t_{offset}$ ($min$)\tablenotemark{a}			& \TOffsetWASPTwelveKsBand$^{+\TOffsetPlusWASPTwelveKsBand}_{-\TOffsetMinusWASPTwelveKsBand}$									& \TOffsetWASPTwelveKsBandTwentyEleven$^{+\TOffsetPlusWASPTwelveKsBandTwentyEleven}_{-\TOffsetMinusWASPTwelveKsBandTwentyEleven}$ 								& \TOffsetWASPTwelveKsBandDecTwentyEleven$^{+\TOffsetPlusWASPTwelveKsBandDecTwentyEleven}_{-\TOffsetMinusWASPTwelveKsBandDecTwentyEleven}$								& \TOffsetWASPTwelveYBandTwentyEleven$^{+\TOffsetPlusWASPTwelveYBandTwentyEleven}_{-\TOffsetMinusWASPTwelveYBandTwentyEleven}$ 									& \TOffsetWASPTwelveKContBandJanTwentyTwelve$^{+\TOffsetPlusWASPTwelveKContBandJanTwentyTwelve}_{-\TOffsetMinusWASPTwelveKContBandJanTwentyTwelve}$ 								\\
\hline
\multicolumn{6}{c}{Combined Aperture Sizes and Reference Star Ensembles Fit} 	\\
$F_{Ap}/F_{*}$ \tablenotemark{c}			& \FpOverFStarPercentAbstractCorrWASPTwelveKsBand$^{+\FpOverFStarPercentAbstractCorrPlusWASPTwelveKsBand}_{-\FpOverFStarPercentAbstractCorrMinusWASPTwelveKsBand}$\%							& \FpOverFStarPercentAbstractCorrWASPTwelveKsBandTwentyEleven$^{+\FpOverFStarPercentAbstractCorrPlusWASPTwelveKsBandTwentyEleven}_{-\FpOverFStarPercentAbstractCorrMinusWASPTwelveKsBandTwentyEleven}$\%	& \FpOverFStarPercentAbstractCorrWASPTwelveKsBandDecTwentyEleven$^{+\FpOverFStarPercentAbstractCorrPlusWASPTwelveKsBandDecTwentyEleven}_{-\FpOverFStarPercentAbstractCorrMinusWASPTwelveKsBandDecTwentyEleven}$\% 	& \FpOverFStarPercentAbstractCorrWASPTwelveYBandTwentyEleven$^{+\FpOverFStarPercentAbstractCorrPlusWASPTwelveYBandTwentyEleven}_{-\FpOverFStarPercentAbstractCorrMinusWASPTwelveYBandTwentyEleven}$\%		& \FpOverFStarPercentAbstractCorrWASPTwelveKContBandJanTwentyTwelve$^{+\FpOverFStarPercentAbstractCorrPlusWASPTwelveKContBandJanTwentyTwelve}_{-\FpOverFStarPercentAbstractCorrMinusWASPTwelveKContBandJanTwentyTwelve}$\% 	\\
$t_{offset}$ ($min$)\tablenotemark{a}			& \TOffsetCorrWASPTwelveKsBand$^{+\TOffsetCorrPlusWASPTwelveKsBand}_{-\TOffsetCorrMinusWASPTwelveKsBand}$								& \TOffsetCorrWASPTwelveKsBandTwentyEleven$^{+\TOffsetCorrPlusWASPTwelveKsBandTwentyEleven}_{-\TOffsetCorrMinusWASPTwelveKsBandTwentyEleven}$ 									& \TOffsetCorrWASPTwelveKsBandDecTwentyEleven$^{+\TOffsetCorrPlusWASPTwelveKsBandDecTwentyEleven}_{-\TOffsetCorrMinusWASPTwelveKsBandDecTwentyEleven}$									& \TOffsetCorrWASPTwelveYBandTwentyEleven$^{+\TOffsetCorrPlusWASPTwelveYBandTwentyEleven}_{-\TOffsetCorrMinusWASPTwelveYBandTwentyEleven}$ 									& \TOffsetCorrWASPTwelveKContBandJanTwentyTwelve$^{+\TOffsetCorrPlusWASPTwelveKContBandJanTwentyTwelve}_{-\TOffsetCorrMinusWASPTwelveKContBandJanTwentyTwelve}$ 								\\
$t_{eclipse}$ \tablenotemark{b}				& \JDOffsetCorrWASPTwelveKsBand$^{+\JDOffsetCorrPlusWASPTwelveKsBand}_{-\JDOffsetCorrMinusWASPTwelveKsBand}$								& \JDOffsetCorrWASPTwelveKsBandTwentyEleven$^{+\JDOffsetCorrPlusWASPTwelveKsBandTwentyEleven}_{-\JDOffsetCorrMinusWASPTwelveKsBandTwentyEleven}$ 								& \JDOffsetCorrWASPTwelveKsBandDecTwentyEleven$^{+\JDOffsetCorrPlusWASPTwelveKsBandDecTwentyEleven}_{-\JDOffsetCorrMinusWASPTwelveKsBandDecTwentyEleven}$ 								& \JDOffsetCorrWASPTwelveYBandTwentyEleven$^{+\JDOffsetCorrPlusWASPTwelveYBandTwentyEleven}_{-\JDOffsetCorrMinusWASPTwelveYBandTwentyEleven}$ 									& \JDOffsetCorrWASPTwelveKContBandJanTwentyTwelve$^{+\JDOffsetCorrPlusWASPTwelveKContBandJanTwentyTwelve}_{-\JDOffsetCorrMinusWASPTwelveKContBandJanTwentyTwelve}$	 					\\
$\phi$							& \PhaseAbstractCorrWASPTwelveKsBand$^{+\PhaseAbstractCorrPlusWASPTwelveKsBand}_{-\PhaseAbstractCorrMinusWASPTwelveKsBand}$						& \PhaseAbstractCorrWASPTwelveKsBandTwentyEleven$^{+\PhaseAbstractCorrPlusWASPTwelveKsBandTwentyEleven}_{-\PhaseAbstractCorrMinusWASPTwelveKsBandTwentyEleven}$ 						& \PhaseAbstractCorrWASPTwelveKsBandDecTwentyEleven$^{+\PhaseAbstractCorrPlusWASPTwelveKsBandDecTwentyEleven}_{-\PhaseAbstractCorrMinusWASPTwelveKsBandDecTwentyEleven}$						& \PhaseAbstractCorrWASPTwelveYBandTwentyEleven$^{+\PhaseAbstractCorrPlusWASPTwelveYBandTwentyEleven}_{-\PhaseAbstractCorrMinusWASPTwelveYBandTwentyEleven}$							& \PhaseAbstractCorrWASPTwelveKContBandJanTwentyTwelve$^{+\PhaseAbstractCorrPlusWASPTwelveKContBandJanTwentyTwelve}_{-\PhaseAbstractCorrMinusWASPTwelveKContBandJanTwentyTwelve}$ 						\\
$e \cos(\omega)$ \tablenotemark{a} 			& \ECosOmegaCorrWASPTwelveKsBand$^{+\ECosOmegaCorrPlusWASPTwelveKsBand}_{-\ECosOmegaCorrMinusWASPTwelveKsBand}$								& \ECosOmegaCorrWASPTwelveKsBandTwentyEleven$^{+\ECosOmegaCorrPlusWASPTwelveKsBandTwentyEleven}_{-\ECosOmegaCorrMinusWASPTwelveKsBandTwentyEleven}$ 								& \ECosOmegaCorrWASPTwelveKsBandDecTwentyEleven$^{+\ECosOmegaCorrPlusWASPTwelveKsBandDecTwentyEleven}_{-\ECosOmegaCorrMinusWASPTwelveKsBandDecTwentyEleven}$								& \ECosOmegaCorrWASPTwelveYBandTwentyEleven$^{+\ECosOmegaCorrPlusWASPTwelveYBandTwentyEleven}_{-\ECosOmegaCorrMinusWASPTwelveYBandTwentyEleven}$								& \ECosOmegaCorrWASPTwelveKContBandJanTwentyTwelve$^{+\ECosOmegaCorrPlusWASPTwelveKContBandJanTwentyTwelve}_{-\ECosOmegaCorrMinusWASPTwelveKContBandJanTwentyTwelve}$ 							\\
\enddata
\tablenotetext{a}{We account for the increased light travel-time in the system \citep{Loeb05}.}
\tablenotetext{b}{$t_{eclipse}$ is the barycentric Julian Date of the mid-eclipse of the secondary eclipse
calculated using the terrestrial time standard (BJD-2440000; as calculated using the routines of \citealt{Eastman10}).}
\tablenotetext{c}{We reiterate that due to the presence of the nearby M-dwarf binary companion for WASP-12, the diluted
apparent eclipse depth, $F_{Ap}/F_{*}$, is not equivalent to the true eclipse depths, $F_{p}/F_{*}$, which are given in Table \ref{TableCorr}.}
\label{TableParams}
\end{deluxetable*}

\begin{deluxetable*}{ccccc}
\centering
\tablecaption{WASP-3, Qatar-1 \& KELT-1 best-fit secondary eclipse parameters} 
\tablehead{
\colhead{Parameter} 	& \colhead{Qatar-1 MCMC} 		& \colhead{WASP-3 Eclipse \#1}	& \colhead{WASP-3 Eclipse \#2}	& \colhead{KELT-1 MCMC}		\\
\colhead{}		& \colhead{eclipse solution}		& \colhead{MCMC solution}	& \colhead{MCMC solution}	& \colhead{eclipse solution}		\\
}
\startdata
\multicolumn{5}{c}{Single Aperture Size and Reference Star Ensemble Fit} 	\\
reduced $\chi^{2}$					& \ChiQatarOneKsBand$^{+\ChiPlusQatarOneKsBand}_{-\ChiMinusQatarOneKsBand}$										& \ChiWASPThreeKsBandEclipseOne$^{+\ChiPlusWASPThreeKsBandEclipseOne}_{-\ChiMinusWASPThreeKsBandEclipseOne}$										& \ChiWASPThreeKsBandEclipseTwo$^{+\ChiPlusWASPThreeKsBandEclipseTwo}_{-\ChiMinusWASPThreeKsBandEclipseTwo}$										& \ChiKELTOneKsBand$^{+\ChiPlusKELTOneKsBand}_{-\ChiMinusKELTOneKsBand}$									\\
$c_1$							& \cOneQatarOneKsBand$^{+\cOnePlusQatarOneKsBand}_{-\cOneMinusQatarOneKsBand}$										& \cOneWASPThreeKsBandEclipseOne$^{+\cOnePlusWASPThreeKsBandEclipseOne}_{-\cOneMinusWASPThreeKsBandEclipseOne}$										& \cOneWASPThreeKsBandEclipseTwo$^{+\cOnePlusWASPThreeKsBandEclipseTwo}_{-\cOneMinusWASPThreeKsBandEclipseTwo}$										& \cOneKELTOneKsBand$^{+\cOnePlusKELTOneKsBand}_{-\cOneMinusKELTOneKsBand}$									\\
$c_2$							& \cTwoQatarOneKsBand$^{+\cTwoPlusQatarOneKsBand}_{-\cTwoMinusQatarOneKsBand}$										& \cTwoWASPThreeKsBandEclipseOne$^{+\cTwoPlusWASPThreeKsBandEclipseOne}_{-\cTwoMinusWASPThreeKsBandEclipseOne}$										& \cTwoWASPThreeKsBandEclipseTwo$^{+\cTwoPlusWASPThreeKsBandEclipseTwo}_{-\cTwoMinusWASPThreeKsBandEclipseTwo}$										& \cTwoKELTOneKsBand$^{+\cTwoPlusKELTOneKsBand}_{-\cTwoMinusKELTOneKsBand}$									\\
$c_3$							& \cThreeQatarOneKsBand$^{+\cThreePlusQatarOneKsBand}_{-\cThreeMinusQatarOneKsBand}$									& \cThreeWASPThreeKsBandEclipseOne$^{+\cThreePlusWASPThreeKsBandEclipseOne}_{-\cThreeMinusWASPThreeKsBandEclipseOne}$									& \cThreeWASPThreeKsBandEclipseTwo$^{+\cThreePlusWASPThreeKsBandEclipseTwo}_{-\cThreeMinusWASPThreeKsBandEclipseTwo}$									& \cThreeKELTOneKsBand$^{+\cThreePlusKELTOneKsBand}_{-\cThreeMinusKELTOneKsBand}$								\\
$F_{Ap}/F_{*}$						& \FpOverFStarPercentAbstractQatarOneKsBand$^{+\FpOverFStarPercentAbstractPlusQatarOneKsBand}_{-\FpOverFStarPercentAbstractMinusQatarOneKsBand} $	& \FpOverFStarPercentAbstractWASPThreeKsBandEclipseOne$^{+\FpOverFStarPercentAbstractPlusWASPThreeKsBandEclipseOne}_{-\FpOverFStarPercentAbstractMinusWASPThreeKsBandEclipseOne} $	& \FpOverFStarPercentAbstractWASPThreeKsBandEclipseTwo$^{+\FpOverFStarPercentAbstractPlusWASPThreeKsBandEclipseTwo}_{-\FpOverFStarPercentAbstractMinusWASPThreeKsBandEclipseTwo} $	& \FpOverFStarPercentAbstractKELTOneKsBand$^{+\FpOverFStarPercentAbstractPlusKELTOneKsBand}_{-\FpOverFStarPercentAbstractMinusKELTOneKsBand} $	\\
$t_{offset}$ ($min$)\tablenotemark{a}			& \TOffsetQatarOneKsBand$^{+\TOffsetPlusQatarOneKsBand}_{-\TOffsetMinusQatarOneKsBand}$									& \TOffsetWASPThreeKsBandEclipseOne \tablenotemark{c}																			& \TOffsetWASPThreeKsBandEclipseTwo$^{+\TOffsetPlusWASPThreeKsBandEclipseTwo}_{-\TOffsetMinusWASPThreeKsBandEclipseTwo}$								& \TOffsetKELTOneKsBand$^{+\TOffsetPlusKELTOneKsBand}_{-\TOffsetMinusKELTOneKsBand}$								\\ 
\hline
\multicolumn{5}{c}{Combined Aperture Sizes and Reference Star Ensembles Fit}	\\
$F_{Ap}/F_{*}$						& \FpOverFStarPercentAbstractCorrQatarOneKsBandAll$^{+\FpOverFStarPercentAbstractCorrPlusQatarOneKsBandAll}_{-\FpOverFStarPercentAbstractCorrMinusQatarOneKsBandAll}$				& \FpOverFStarPercentAbstractCorrWASPThreeKsBandEclipseOneAll$^{+\FpOverFStarPercentAbstractCorrPlusWASPThreeKsBandEclipseOneAll}_{-\FpOverFStarPercentAbstractCorrMinusWASPThreeKsBandEclipseOneAll}$ 				& \FpOverFStarPercentAbstractCorrWASPThreeKsBandEclipseTwoAll$^{+\FpOverFStarPercentAbstractCorrPlusWASPThreeKsBandEclipseTwoAll}_{-\FpOverFStarPercentAbstractCorrMinusWASPThreeKsBandEclipseTwoAll}$				& \FpOverFStarPercentAbstractCorrKELTOneKsBandAll$^{+\FpOverFStarPercentAbstractCorrPlusKELTOneKsBandAll}_{-\FpOverFStarPercentAbstractCorrMinusKELTOneKsBandAll}$ 				\\
$t_{offset}$ ($min$)\tablenotemark{a}			& \TOffsetCorrQatarOneKsBandAll$^{+\TOffsetCorrPlusQatarOneKsBandAll}_{-\TOffsetCorrMinusQatarOneKsBandAll}$						& \TOffsetCorrWASPThreeKsBandEclipseOneAll \tablenotemark{c}																		& \TOffsetCorrWASPThreeKsBandEclipseTwoAll$^{+\TOffsetCorrPlusWASPThreeKsBandEclipseTwoAll}_{-\TOffsetCorrMinusWASPThreeKsBandEclipseTwoAll}$						& \TOffsetCorrKELTOneKsBandAll$^{+\TOffsetCorrPlusKELTOneKsBandAll}_{-\TOffsetCorrMinusKELTOneKsBandAll}$					\\	
$t_{eclipse}$ \tablenotemark{b}	 	 		& \JDOffsetCorrQatarOneKsBandAll$^{+\JDOffsetCorrPlusQatarOneKsBandAll}_{-\JDOffsetCorrMinusQatarOneKsBandAll}$						& \JDOffsetCorrWASPThreeKsBandEclipseOneAll \tablenotemark{c}																		& \JDOffsetCorrWASPThreeKsBandEclipseTwoAll$^{+\JDOffsetCorrPlusWASPThreeKsBandEclipseTwoAll}_{-\JDOffsetCorrMinusWASPThreeKsBandEclipseTwoAll}$ 					& \JDOffsetCorrKELTOneKsBandAll$^{+\JDOffsetCorrPlusKELTOneKsBandAll}_{-\JDOffsetCorrMinusKELTOneKsBandAll}$ 					\\	
$\phi$							& \PhaseAbstractCorrQatarOneKsBandAll$^{+\PhaseAbstractCorrPlusQatarOneKsBandAll}_{-\PhaseAbstractCorrMinusQatarOneKsBandAll}$				& \PhaseAbstractCorrWASPThreeKsBandEclipseOneAll \tablenotemark{c}																	& \PhaseAbstractCorrWASPThreeKsBandEclipseTwoAll$^{+\PhaseAbstractCorrPlusWASPThreeKsBandEclipseTwoAll}_{-\PhaseAbstractCorrMinusWASPThreeKsBandEclipseTwoAll}$				& \PhaseAbstractCorrKELTOneKsBandAll$^{+\PhaseAbstractCorrPlusKELTOneKsBandAll}_{-\PhaseAbstractCorrMinusKELTOneKsBandAll}$			\\	
$e \cos(\omega)$ \tablenotemark{a}			& \ECosOmegaCorrQatarOneKsBandAll$^{+\ECosOmegaCorrPlusQatarOneKsBandAll}_{-\ECosOmegaCorrMinusQatarOneKsBandAll}$					& \ECosOmegaCorrWASPThreeKsBandEclipseOneAll \tablenotemark{c}																		& \ECosOmegaCorrWASPThreeKsBandEclipseTwoAll$^{+\ECosOmegaCorrPlusWASPThreeKsBandEclipseTwoAll}_{-\ECosOmegaCorrMinusWASPThreeKsBandEclipseTwoAll}$					& \ECosOmegaCorrKELTOneKsBandAll$^{+\ECosOmegaCorrPlusKELTOneKsBandAll}_{-\ECosOmegaCorrMinusKELTOneKsBandAll}$					\\	
\enddata
\tablenotetext{a}{We account for the increased light travel-time in the system \citep{Loeb05}.}
\tablenotetext{b}{$t_{eclipse}$ is the barycentric Julian Date of the mid-eclipse of the secondary eclipse
calculated using the terrestrial time standard (BJD-2440000; as calculated using the routines of \citealt{Eastman10}).}
\tablenotetext{c}{We do not fit for the mid-point of the eclipse, $t_{offset}$, for our first WASP-3 Ks-band eclipse.}
\label{TableParamsOther}
\end{deluxetable*}

As discussed in Section \ref{SecTechniques} we extensively consider the variations 
in the observed eclipse depth with the size of aperture used in our aperture photometry,
and the choice of reference star ensembles to correct the light curve.
For reasons detailed in Section \ref{SecAperture}, to quantify the best-fit light curve we often use the RMS
of the data following the subtraction of the best-fit light-curve multiplied by $\beta^2$.
The minimum RMS$\times$$\beta^2$ for these data-sets are found with an
aperture size and number of reference stars in the reference star ensemble,
as given in Table \ref{TableApertureNStar}.
It is with these aperture size and reference star ensemble choices that
we present our
best-fit MCMC eclipse fits for 
our various light curves in Figures \ref{FigWASPTwelveEclipseKsOne} -  \ref{FigWASPThreeTwo}.	
The associated best-fit parameters from our MCMC fits
for a single aperture size and reference star ensemble combination are listed in the top-half of Table \ref{TableParams}
for our WASP-12 eclipses, and at the top-half of Table \ref{TableParamsOther} for our other 
data-sets\footnote{In Tables \ref{TableParams} and \ref{TableParamsOther}, $\phi$
represents the orbital phase of the mid-point of the secondary eclipses (with $\phi$=0.0 denoting the transit). 
The associated inferred eccentricity and cosine of the argument of periastron, $e \cos(\omega)$, of the orbit are also presented for each eclipse.}.
The eclipse depths and errors once we've taken into account the effects of different
aperture sizes and reference stars, as detailed in Section \ref{SecHonestEclipses}, are given in
the bottom of Tables \ref{TableParams} and \ref{TableParamsOther}. 

\subsection{Correcting the diluted eclipse depths}
\label{SecTrueDepths}

\begin{deluxetable*}{ccc}
\tablecaption{Corrected Eclipse Depths}
\tablehead{
\colhead{Eclipse} 		& \colhead{Apparent Eclipse Depth} 	& \colhead{Actual Eclipse Depth}			\\
\colhead{Light curve}		& \colhead{$F_{Ap}/F_{*}$ (\%)}		& \colhead{$F_{p}/F_{*}$ (\%)}		\\
}
\startdata
First WASP-12 Ks-band		& \FpOverFStarPercentAbstractCorrWASPTwelveKsBand$^{+\FpOverFStarPercentAbstractCorrPlusWASPTwelveKsBand}_{-\FpOverFStarPercentAbstractCorrMinusWASPTwelveKsBand}$\%								& \FpOverFStarPercentAbstractUndilutedRealWASPTwelveKsBand$^{+\FpOverFStarPercentAbstractUndilutedRealPlusWASPTwelveKsBand}_{-\FpOverFStarPercentAbstractUndilutedRealMinusWASPTwelveKsBand}$\% 							\\
Second WASP-12 Ks-band		& \FpOverFStarPercentAbstractCorrWASPTwelveKsBandTwentyEleven$^{+\FpOverFStarPercentAbstractCorrPlusWASPTwelveKsBandTwentyEleven}_{-\FpOverFStarPercentAbstractCorrMinusWASPTwelveKsBandTwentyEleven}$\%			& \FpOverFStarPercentAbstractUndilutedRealWASPTwelveKsBandTwentyEleven$^{+\FpOverFStarPercentAbstractUndilutedRealPlusWASPTwelveKsBandTwentyEleven}_{-\FpOverFStarPercentAbstractUndilutedRealMinusWASPTwelveKsBandTwentyEleven}$\%			\\
Third WASP-12 Ks-band		& \FpOverFStarPercentAbstractCorrWASPTwelveKsBandDecTwentyEleven$^{+\FpOverFStarPercentAbstractCorrPlusWASPTwelveKsBandDecTwentyEleven}_{-\FpOverFStarPercentAbstractCorrMinusWASPTwelveKsBandDecTwentyEleven}$\%		& \FpOverFStarPercentAbstractUndilutedRealWASPTwelveKsBandDecTwentyEleven$^{+\FpOverFStarPercentAbstractUndilutedRealPlusWASPTwelveKsBandDecTwentyEleven}_{-\FpOverFStarPercentAbstractUndilutedRealMinusWASPTwelveKsBandDecTwentyEleven}$\%		\\
WASP-12 Y-band			& \FpOverFStarPercentAbstractCorrWASPTwelveYBandTwentyEleven$^{+\FpOverFStarPercentAbstractCorrPlusWASPTwelveYBandTwentyEleven}_{-\FpOverFStarPercentAbstractCorrMinusWASPTwelveYBandTwentyEleven}$\%				& \FpOverFStarPercentAbstractUndilutedRealWASPTwelveYBandTwentyEleven$^{+\FpOverFStarPercentAbstractUndilutedRealPlusWASPTwelveYBandTwentyEleven}_{-\FpOverFStarPercentAbstractUndilutedRealMinusWASPTwelveYBandTwentyEleven}$\%  			\\
WASP-12 $K_{CONT}$-band		& \FpOverFStarPercentAbstractCorrWASPTwelveKContBandJanTwentyTwelve$^{+\FpOverFStarPercentAbstractCorrPlusWASPTwelveKContBandJanTwentyTwelve}_{-\FpOverFStarPercentAbstractCorrMinusWASPTwelveKContBandJanTwentyTwelve}$\%	& \FpOverFStarPercentAbstractUndilutedRealWASPTwelveKContBandJanTwentyTwelve$^{+\FpOverFStarPercentAbstractUndilutedRealPlusWASPTwelveKContBandJanTwentyTwelve}_{-\FpOverFStarPercentAbstractUndilutedRealMinusWASPTwelveKContBandJanTwentyTwelve}$\%	\\
Qatar-1 Ks-band			& n/a \tablenotemark{a}																										& \FpOverFStarPercentAbstractCorrQatarOneKsBandAll$^{+\FpOverFStarPercentAbstractCorrPlusQatarOneKsBandAll}_{-\FpOverFStarPercentAbstractCorrMinusQatarOneKsBandAll}$\%							\\
KELT-1 Ks-band			& n/a \tablenotemark{a}																										& \FpOverFStarPercentAbstractCorrKELTOneKsBandAll$^{+\FpOverFStarPercentAbstractCorrPlusKELTOneKsBandAll}_{-\FpOverFStarPercentAbstractCorrMinusKELTOneKsBandAll}$\%							\\
First WASP-3 Ks-band Eclipse	& n/a \tablenotemark{a}																										& \FpOverFStarPercentAbstractCorrWASPThreeKsBandEclipseOneAll$^{+\FpOverFStarPercentAbstractCorrPlusWASPThreeKsBandEclipseOneAll}_{-\FpOverFStarPercentAbstractCorrMinusWASPThreeKsBandEclipseOneAll}$\%		\\
Second WASP-3 Ks-band Eclipse	& n/a \tablenotemark{a}																										& \FpOverFStarPercentAbstractCorrWASPThreeKsBandEclipseTwoAll$^{+\FpOverFStarPercentAbstractCorrPlusWASPThreeKsBandEclipseTwoAll}_{-\FpOverFStarPercentAbstractCorrMinusWASPThreeKsBandEclipseTwoAll}$\% 		\\
\hline																																										
WASP-3 Ks-band combined		& n/a \tablenotemark{a}																										& \FpOverFStarPercentAbstractWeatherPlotWASPThreeKsBandCombined \ $\pm$ \FpOverFStarPercentAbstractPlusMinusWeatherPlotWASPThreeKsBandCombined\% 	\\
WASP-12 Ks-band combined	& n/a 																												& \FpOverFStarPercentAbstractWeatherPlotWASPTwelveKsBandCombined \ $\pm$ \FpOverFStarPercentAbstractPlusMinusWeatherPlotWASPTwelveKsBandCombined\% 												 																	\\
\enddata
\label{TableCorr}
\tablenotetext{a}{As there are no nearby companions to dilute the secondary eclipses of these targets (or the companion is too faint as for KELT-1b; see Section \ref{SecTrueDepths}),
		the apparent eclipse depths, $F_{Ap}/F_{*}$, are equal to the actual eclipse depths, $F_{p}/F_{*}$.}
\end{deluxetable*}

Two of our target stars, KELT-1 and WASP-12, have stellar companions that dilute their
apparent eclipse depths, $F_{Ap}/F_{*}$.
Here we present our method to correct for this dilution and 
determine the true eclipse depths, $F_{p}/F_{*}$, for these stars.

WASP-12 is in fact a triple system with an M-dwarf binary, WASP-12 BC, separated from
WASP-12 by 1\arcsec, and is approximately 4 magnitudes 
fainter in the i-band \citep{Bergfors13,Crossfield12,Bechter14,Sing13}.
This M-dwarf binary is completely enclosed within our defocused apertures.
To determine the actual eclipse depths, $F_{p}/F_{*}$, 
we correct the diluted depths, $F_{Ap}/F_{*}$, by using the calculated factors given in \citet{Stevenson14Second}
that these nearby stellar companions dilute our previous near-infrared WIRCam eclipse depths of WASP-12b \citep{CrollWASPTwelve}.
These dilution factors were calculated 
using Kurucz stellar atmospheric models \citep{CastelliKurucz04} and determining the flux ratios of the models
of WASP-12 to its nearby binary companion WASP-12 BC \citep{Stevenson14} in the JHK and z'-bands; 
for our Y-band, and $K_{cont}$-band secondary eclipses
we simply use the \citet{Stevenson14Second} 
dilution factor given for the z'-band, and K-band, respectively.
The actual eclipse depths, once corrected for the diluting effects of the nearby M-dwarf
binary for WASP-12 using this method, are given in 
Table \ref{TableCorr}.



In addition to KELT-1b, KELT-1 has a nearby companion\footnote{Proper common proper motion has not been confirmed for this object,
but \citet{Siverd12} report that the likelihood of a chance alignment is minute.}
that is likely an M-dwarf;
it is separated from KELT-1 by $\sim$0.6\arcsec \ and is fainter than KELT-1 in the K-band by 
$\Delta K$=5.59 magnitudes \citep{Siverd12}.
Due to the faintness of the companion compared to the target,
we do not correct the secondary eclipses we report for the flux of this nearby faint companion,
as the difference in the resulting eclipse depth, $F_{p}/F_{*}$, is negligible.

\section{Optimal techniques for ground-based, near-infrared, differential photometry}
\label{SecTechniques}

  In our CFHT/WIRCam photometry of the eclipses and transits of exoplanets we have occasionally noticed
that the transit/eclipse depths we measure vary by a significant amount if we choose different aperture sizes
or with different reference star combinations; perhaps more troublingly, occasionally these different aperture size and 
reference star combinations result in very similar goodness of 
fits\footnote{To determine the goodness of fits we employ the RMS of the residuals
multiplied by a factor to account for the correlated noise, as discussed in Section \ref{SecAperture}.}.
As the precision we are able to reach from the ground in
the near-infrared relies solely on the precision of our differential photometry,
it is essential that our best-fit and error-bars account for the variations due to 
different reference star ensembles and aperture sizes.
In Section \ref{SecAperture} we discuss how to choose the optimal aperture size 
for our ground-based, aperture photometry.
In Section \ref{SecCircularApertures} we discuss the importance of taking
into account the fractional contribution of pixels at the edge of the aperture in different photometry, even for large aperture
sizes. In Section \ref{SecOptimalReferenceStars} we discuss how to choose the optimal reference star combination, and discuss the properties of
these reference stars. Finally, in Section \ref{SecHonestEclipses} we present how to return
the eclipse depth and the associated error,
even if there are correlations of the eclipse depth with the choice of aperture size 
or the choice of reference star ensemble.

\subsection{Optimal Aperture Radii Choices}
\label{SecAperture}

\begin{figure*}
\centering
\includegraphics[scale=0.44, angle = 270]{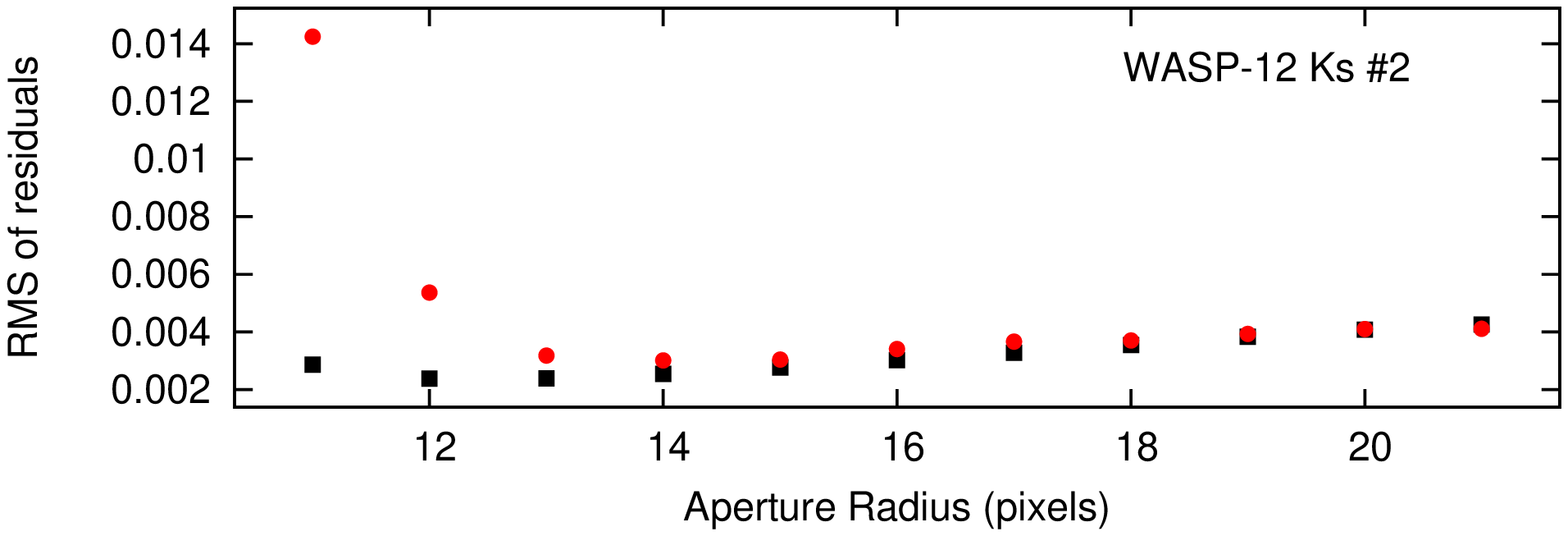}
\includegraphics[scale=0.44, angle = 270]{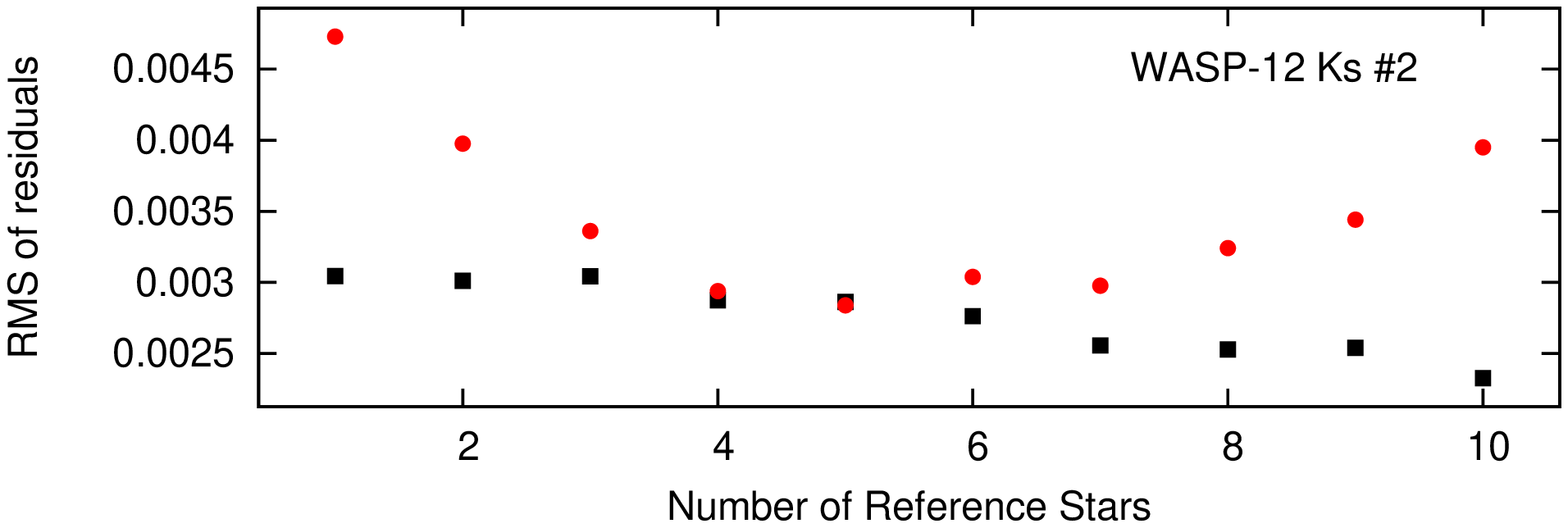}
\includegraphics[scale=0.44, angle = 270]{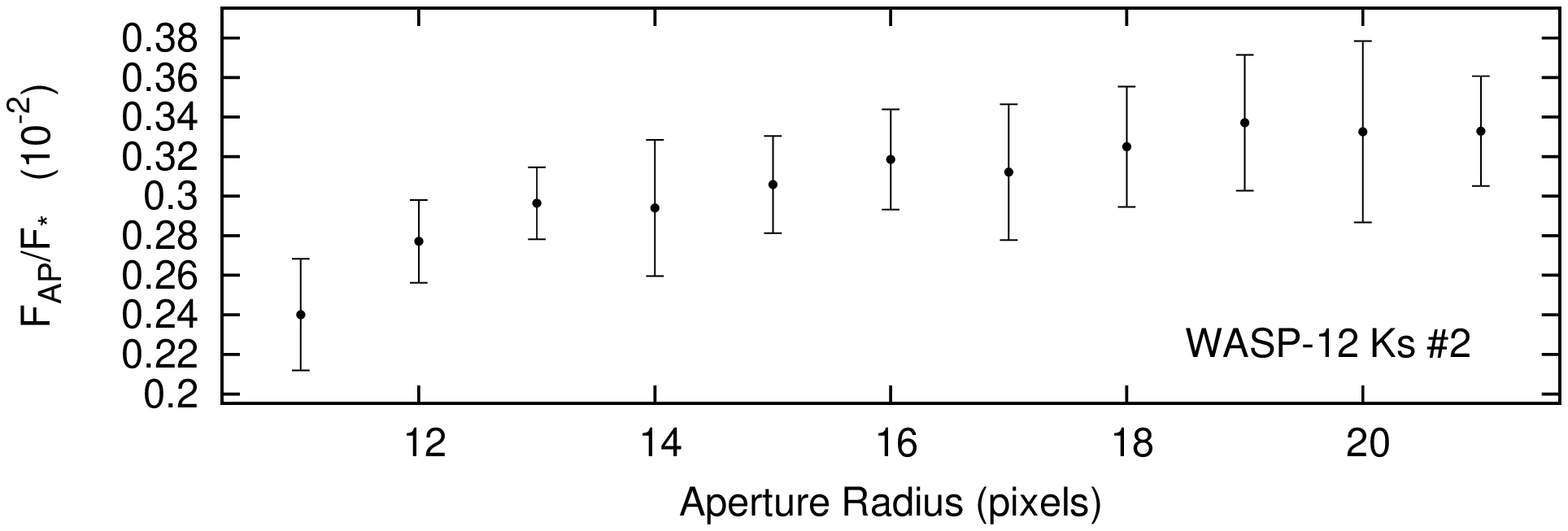}
\includegraphics[scale=0.44, angle = 270]{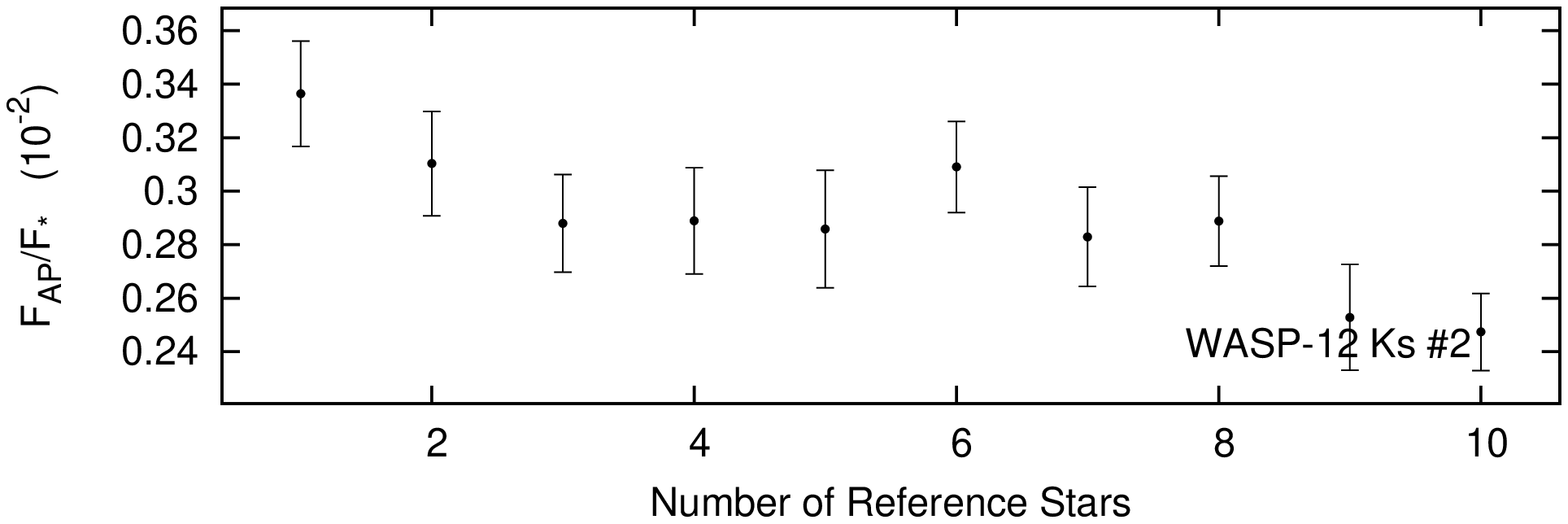}
\caption[ApertureSizeNumStar]
	{
	Top panels: The precision of the data for various aperture sizes (left; for a six star reference star ensemble),	
	and for the different number of reference stars for our second WASP-12 Ks-band eclipse (right; for a 15-pixel aperture);
	the precision of the data is indicated by the RMS of the residuals from the best-fit eclipse model (black squares),
	and the RMS of the residuals multiplied by the relevant $\beta^2$ factor (red circles).		
	Bottom panels: The associated measured MCMC eclipse depths for our
	our second WASP-12 Ks-band eclipse.
	}
\label{FigApertureNumStar}
\end{figure*}

Selecting the optimal aperture radius for aperture photometry 
is a topic that has been receiving increasing attention by those performing 
high precision photometry (e.g. \citealt{Gillon07,Blecic13, Beatty14}).
In the near-infrared, where the sky background is notoriously high and contributes
a large fraction of the noise budget (see Section \ref{SecNoise}),
aperture size choices present a delicate balancing act between 
favouring small aperture sizes to minimize the impact of the high sky background, and
large aperture sizes to ensure that the aperture catches all, or the overwhelming majority,
of the light from the star.

In our previous analyses of the thermal emission
of hot Jupiters \citep{CrollTrESTwo,CrollTrESThree,CrollWASPTwelve},
the optimal aperture choice was determined by selecting the aperture 
that minimized the RMS of the out-of-eclipse photometry, while clearly capturing the vast majority of the light from the star.
Further investigation revealed that, due to the high sky background in the near-infrared,
that if we attempted to simply
minimize the RMS of the out-of-eclipse photometry, this technique, on occasion, favoured too small of aperture choices.
During occasions of poor seeing or guiding, these small apertures resulted in a small amount of light 
near the edge of the aperture to be lost; this is apparent 
as time-correlated noise in many of our light-curves analyzed with small aperture (e.g. 
our WASP-12 Y-band photometry, or our KELT-1 Ks-band photometry; the left set of panels of Figure	
\ref{FigWASPTwelveYbandFidelityMany} or Figure \ref{FigKeltOneKsbandFidelityMany}). 

Our preferred metric for determining the optimal aperture size, and the optimal reference star combination 
(see Section \ref{SecOptimalReferenceStars}), is to minimize
the RMS$\times$$\beta^2$ of the residuals of the photometry once
the best-fit model is subtracted. 
We frequently observe that the minimum RMS 
is reached for relatively small aperture radii (for small apertures 
one is able to reduce the impact of the high sky background),
while a lack of time-correlated noise ($\beta$$\sim$1) is achieved
only for sufficiently large aperture values (where one is able to ensure that even during moments of poor
seeing and guiding the aperture captures the vast majority of the light
from the target and the reference stars);
an example of the impact of aperture sizes has on
eclipse depths and the precision of the light curve is displayed for our second 
WASP-12 Ks-band eclipse in the left panel of Figure \ref{FigApertureNumStar}.
In most cases, the minimum of the RMS$\times$$\beta^2$ does a reasonable job of balancing these two competing pressures,
of avoiding the high sky background that come along with large apertures, and mitigating the presence
of time-correlated noise that comes along with small apertures;
therefore, we 
select our optimal aperture choice
by identifying the aperture with the minimum RMS$\times$$\beta^2$.
We note that in a previous application of this technique to near-infrared photometry \citep{CrollKIC}, we used 
a metric of RMS$\times$$\beta$; further investigation revealed that 
this metric did not provide a sufficiently high penalty against time-correlated noise, and 
occasionally resulted
in light-curves with obvious time-correlated noise being favoured as the best-fit light curves.


\subsection{The importance of accounting for the fractional flux at the edge of circular apertures}
\label{SecCircularApertures}

In this subsection, we highlight the importance for our precise, differential photometry of taking into account
the fractional contribution of pixels at the edge of the circular aperture.
Taking this fractional contribution of pixels into account, has become common-place in Spitzer/IRAC 
analyses (e.g. \citealt{Agol10}),
due to the fact the aperture sizes are frequently small (3-5 pixels in radius); for such small apertures, a
large percentage of the pixels are near the edge of the aperture, and therefore it is imperative to take
into account these edge effects.

 However, the much larger apertures used in many ground-based applications, and generally in our 
photometry\footnote{In our previous analyses of the 
thermal emission and transmission spectroscopy of hot Jupiters and super-Earths 
\citep{CrollTrESTwo,CrollTrESThree,CrollWASPTwelve,CrollGJ}, aperture sizes were typically 
$\sim$14 - 20 pixels in radius. Our recent analysis of KIC 12557458b \citep{CrollKIC} is the exception,
where we consider aperture sizes as small as 5-10 pixels; as a result, we did take into account
the impact of fractional pixels at the edge of the aperture.},
would appear to mitigate this issue.
Nonetheless, the accuracy of precise, ground-based photometry typically relies 
heavily on the use of differential photometry, and the assumption
that the flux in a single exposure from one star is intimately associated with the flux of a reference star.
Unfortunately, imperfect guiding commonly leads to small shifts
in the centroid of the target and reference star PSF;
this is a problem, especially for differential photometry, as the centroid of the 
target star may lie on one edge of a pixel, while the centroid of the reference stars
may fall on different edges of their pixels. Therefore,
as the target and reference star apertures shift around, the decomposition of the circular aperture
into square pixels often leads to a slightly different number of pixels, and different pixels,
in each aperture
from one exposure to the next.
In the optical, which typically features a very low sky background,
this is generally not a significant problem, as apertures many times the full-width-half-maximum (FWHM)
of the PSF are used and therefore at the edge 
of apertures there is no contribution other than the sky; however, in the near-infrared, which typically
features high sky background, the lowest RMS is often achieved for apertures
only fractionally larger than a few times the FWHM of the PSF (as discussed in Section \ref{SecAperture}),
or just larger than the flux annulus for highly defocused
PSFs, such as what we use in our photometry here.
To mitigate this issue, we take
into account the fractional contribution of the square pixels at the edge of our circular aperture,
by multiplying the flux of these pixels by the fraction of the pixel that falls within our circular aperture;
to do this, we utilize the GSFC Astronomy LIbrary IDL procedure {\it pixwt.pro}.
We have noticed that after applying this technique our eclipse/transit depths, and our photometry,
are much more robust and consistent even when varying the aperture sizes, as discussed in \ref{SecAperture}.

\subsection{Optimal Reference Star Choices}
\label{SecOptimalReferenceStars}

\begin{figure*}
\centering
\includegraphics[scale=0.6, angle = 270]{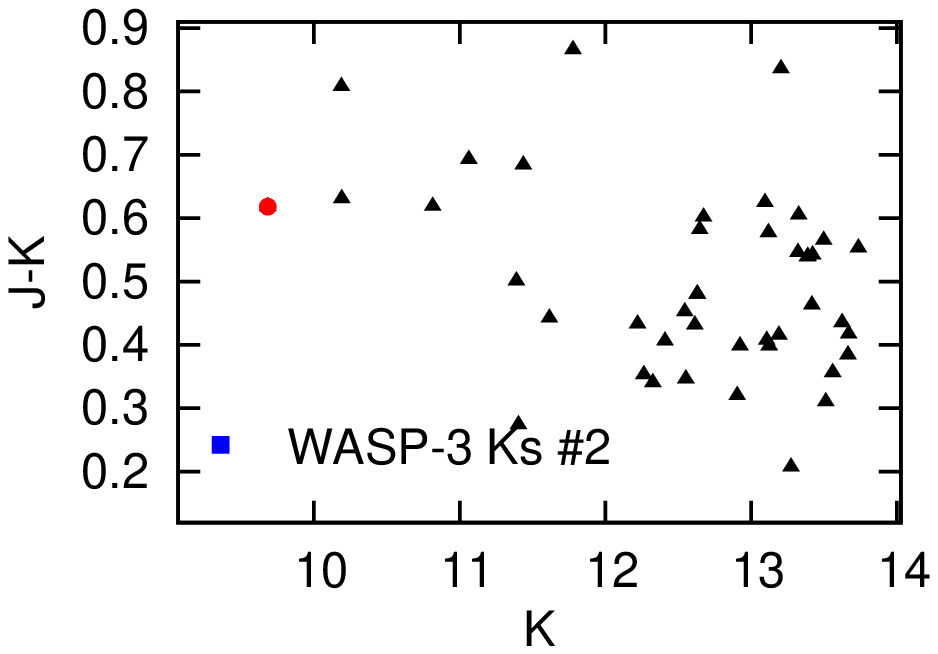}
\includegraphics[scale=0.6, angle = 270]{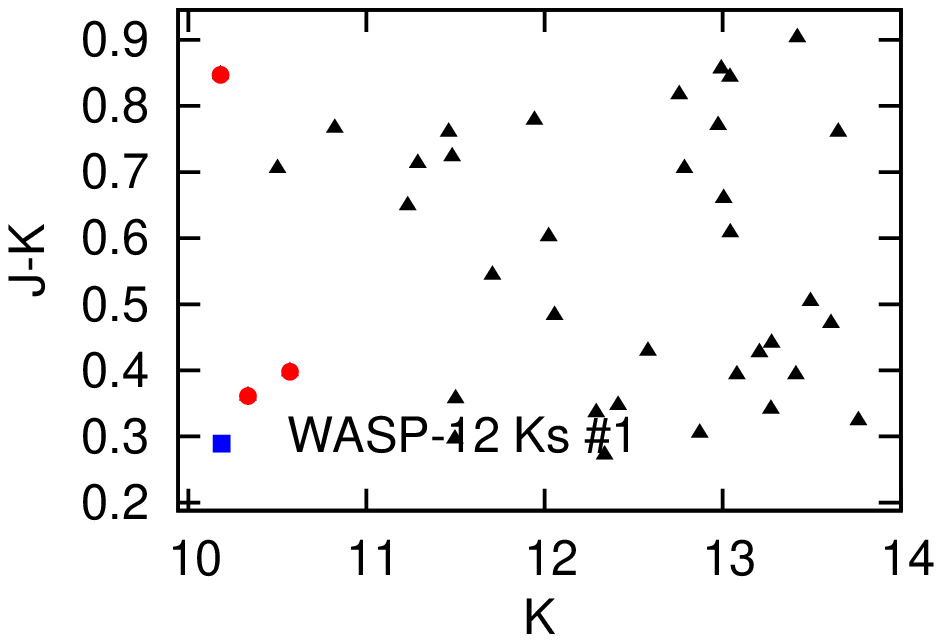}
\includegraphics[scale=0.6, angle = 270]{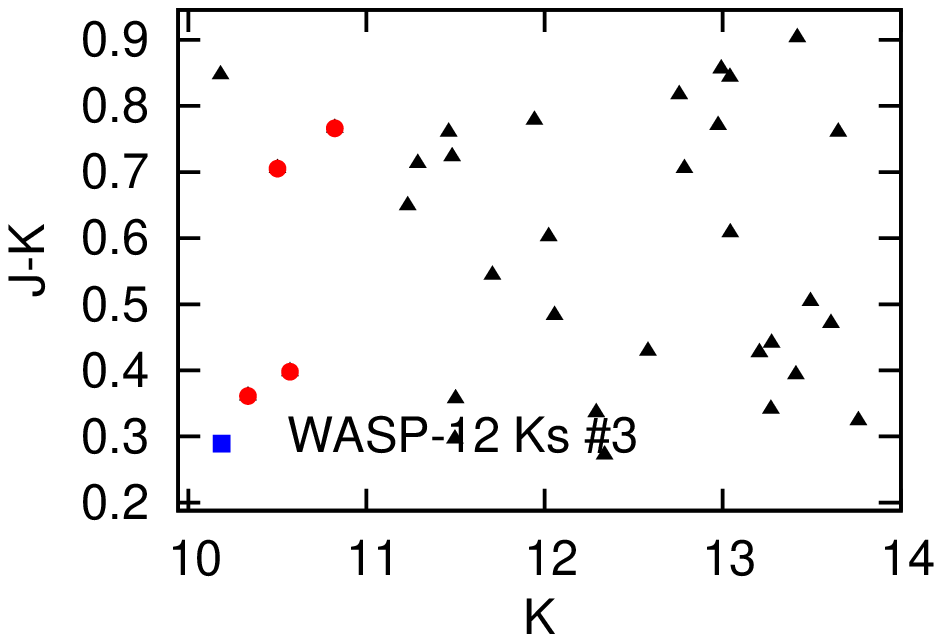}
\includegraphics[scale=0.6, angle = 270]{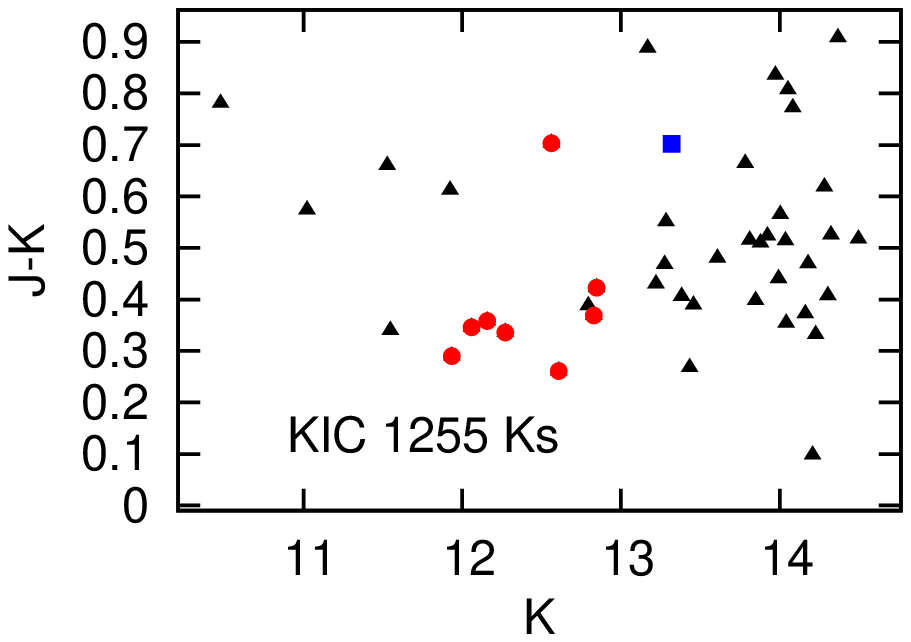}
\includegraphics[scale=0.6, angle = 270]{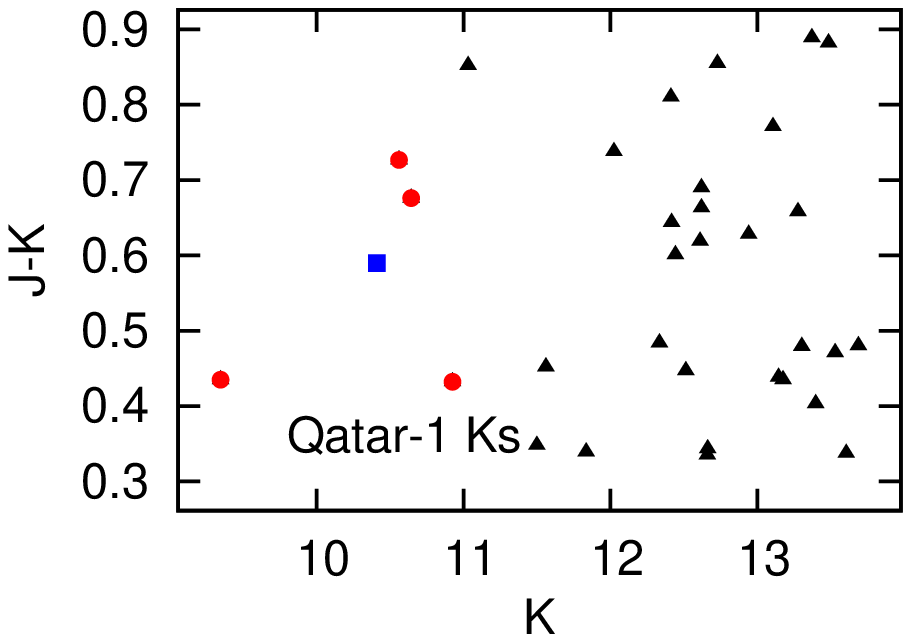}
\includegraphics[scale=0.6, angle = 270]{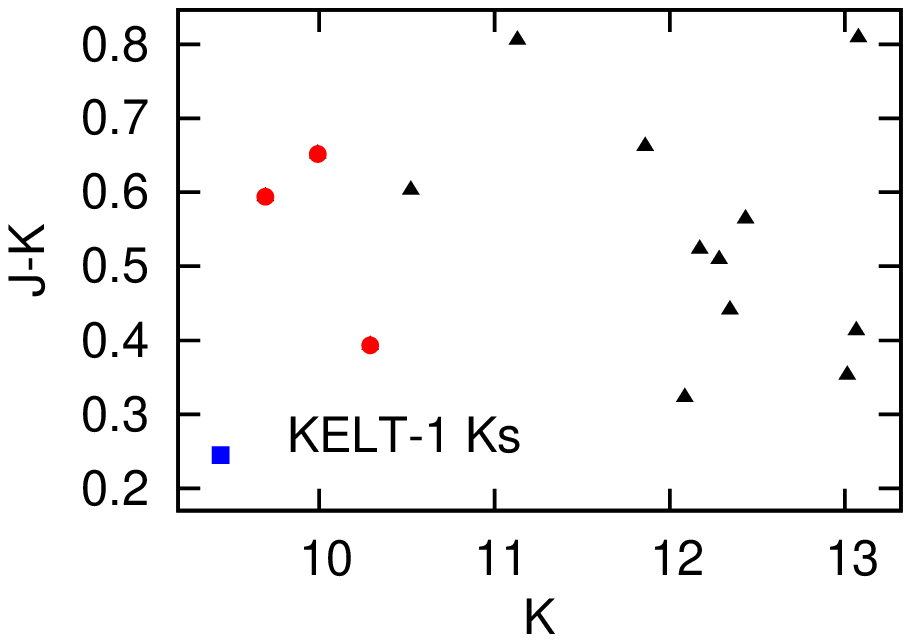}
\caption[Colour Magnitude Plots]
	{
		The 2Mass K-band magnitude and colour (2Mass J-band minus K-band magnitude)
		of the reference stars compared to the target stars
		for -- clockwise from top left -- our second WASP-3 Ks-band eclipse, WASP-12 Ks-band first eclipse,
		WASP-12 Ks-band third eclipse,
		KELT-1 Ks-band eclipse, Qatar-1 Ks-band eclipse, 
		and our KIC 12557548 Ks-band transit photometry.
		Our target stars, for each eclipse/transit, are displayed with a blue square,
		while the optimal reference star ensemble
		are displayed with red circles, and our reference stars that are unused
		are displayed with the black triangles.
		Optimal reference stars appear to be similar in magnitude or slightly brighter than the target star,
		but not necessarily similar in colour.
	}
\label{FigColourMagnitude}
\end{figure*}

The precision of our ground-based near-infrared photometry is based solely on the accuracy
imparted by our differential photometry; our 10\arcmin$\times$10\arcmin \ field-of-view
from a single WIRCam chip often offers $\sim$10-50 suitable reference stars. 
Although this is usually an asset, it can cause complications to arise 
if different reference star combinations,
or different number of reference stars, lead to different eclipse depths,
as has been observed for several of our eclipse/transit light curves (e.g. the right panels of Figure \ref{FigApertureNumStar}).
As with our aperture size choices, we choose the optimal reference star ensemble 
by minimizing the RMS$\times$$\beta^2$ of our residuals to our best-fit light curve.
In general, our experience suggests that adding additional reference
stars (up to $\sim$4-10 reference stars) usually reduces the RMS$\times$$\beta^2$, if the reference stars
do not exhibit correlated noise. We therefore discuss our optimal method for choosing a
reference star ensemble here.

 Our most fool-proof method for choosing a reference star ensemble has been to
perform and repeat our analysis -- fitting for the best-fit eclipse, and background trend, as discussed in Section \ref{SecAnalysis} --
for reference star light curves consisting of each individual reference star.
The reference star light curves with the lowest RMS$\times$$\beta^2$ are then ranked,
and a reference star ensemble is composed by adding in one-by-one the best ranked (RMS$\times$$\beta^2$)
reference stars. Our full analysis is repeated until the 
RMS$\times$$\beta^2$ no longer improves.
This process can be viewed in the right panels of Figure \ref{FigApertureNumStar} for our second WASP-12 Ks-band eclipse.


 In some cases, the use of additional reference stars does not improve the photometry.  
Our photometry of the second eclipse of the hot Jupiter WASP-3b 
is best corrected by a single reference star only, and additional reference stars only contribute correlated noise
(see Figure \ref{FigWASPThreeKsbandFidelityManyTwo}). 
The other reference
stars for our WASP-3 observation are not significantly dissimilar in colour, but are slightly fainter than the target star
and the one suitable reference star (see the top-left panel of Figure \ref{FigColourMagnitude}).
The reason that these other reference stars contribute significant correlated noise,
may be due to an imperfect non-linearity correction. 



\subsubsection{Lessons for reference star selection for other programs}

WIRCam \citep{Puget04} has an enviable 21\arcmin$\times$21\arcmin \ field-of-view
that many near-infrared imagers are unable to match. 
Even though we restrict our field-of-view in the analysis presented here 
to a single WIRCam chip (10\arcmin$\times$10\arcmin), we are still able to perform differential photometry
on a great many more reference stars than most other infrared imagers. Our program may, therefore,
be able to provide lessons 
for the selection of reference stars that may be applicable to others attempting to perform
precise, photometry from the ground in the near-infrared. 

We display the magnitude and colour of our target star and the reference stars that we use 
to correct the flux of our target, and the reference stars
that we reject for this task, for a number of our hot Jupiter light curves in Figure \ref{FigColourMagnitude}.
We note that the optimal reference star ensembles appear to feature stars similar in brightness to the target star.
Although these stars are often fainter than the target,
this is largely due to the relative dearth of stars brighter than our typical K$\sim$9-10 exoplanet host 
star\footnote{There are occasionally stars brighter than our typical exoplanet host star (K$\sim$9-10) on our WIRCam field-of-view, but,
as we often optimize 
our exposure times and defocus amounts for our target, stars significantly brighter than our target often saturate,
or enter into the highly non-linear regime.}.
KIC 12557548
is an obvious counter-example; the faintness of the target star ($K$$\sim$13.3) results
in there being a wealth of reference
stars brighter than the target that are useful for correcting the target's flux.
Stars significantly brighter than KIC 12557548 end up not being used in our analysis, likely
due to the fact that these stars suffer from significant non-linearity, or saturate for some exposures. 
For our goal of high precision photometry, we
note that colour of the reference stars compared to the target seems to be a secondary consideration;
colour differences between the target and the reference stars do not appear to be as important
as the reference stars being similar in brightness to the target star for our ground-based Ks-band photometry. 

\subsection{Honest Near-infrared Eclipse Depths}
\label{SecHonestEclipses}

\begin{figure*}
\centering
\includegraphics[scale=0.44, angle = 270]{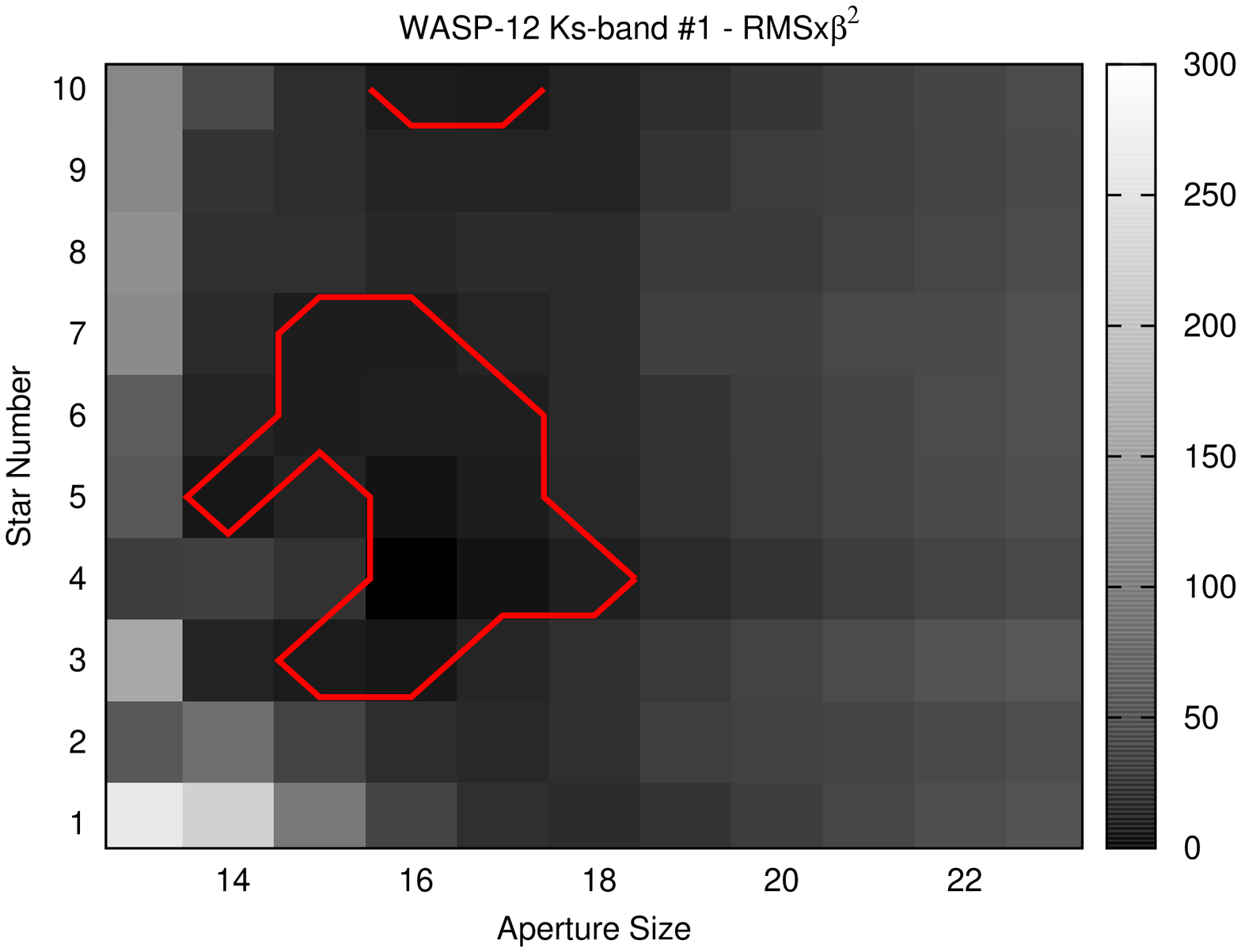}
\includegraphics[scale=0.44, angle = 270]{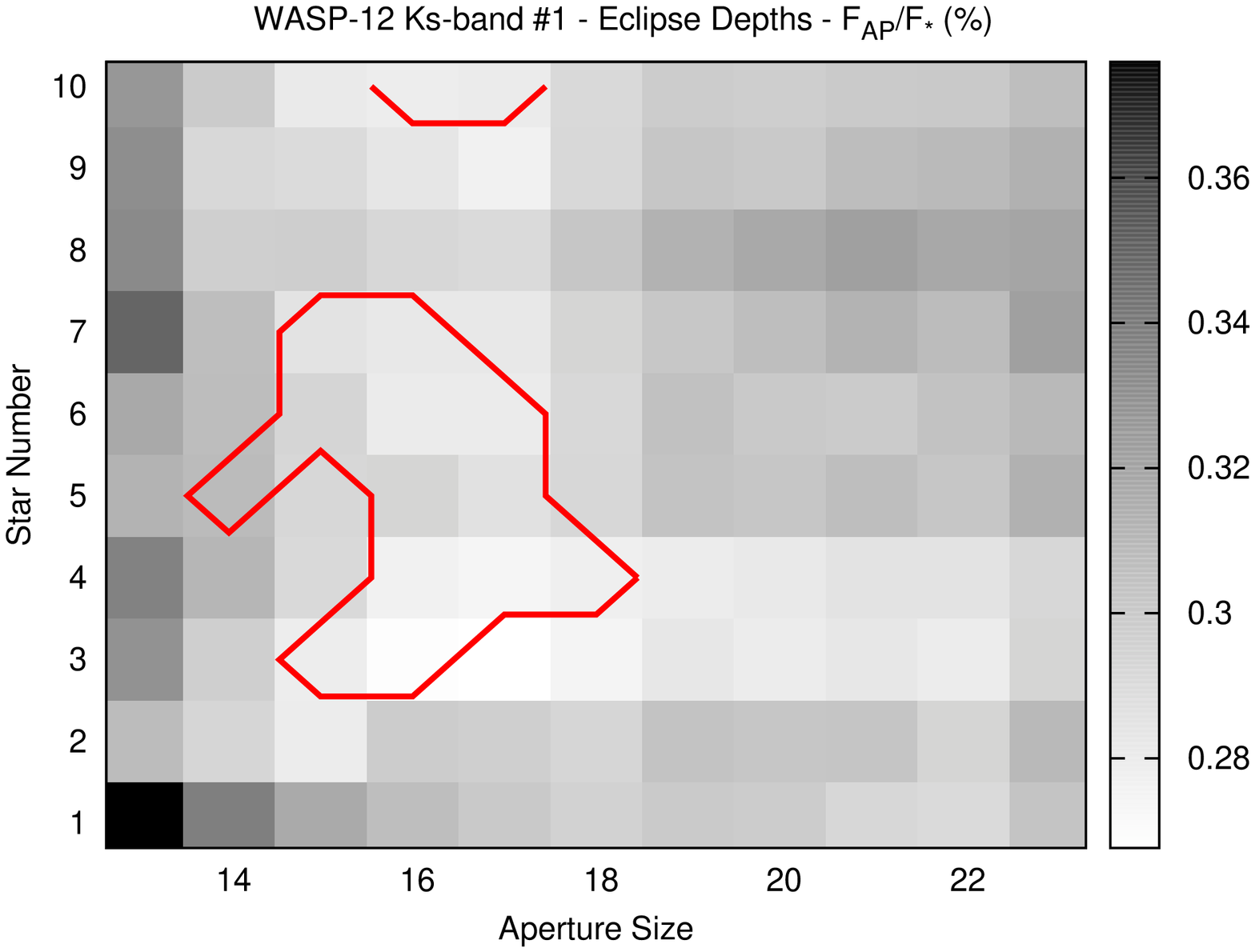}

\includegraphics[scale=0.28, angle = 270]{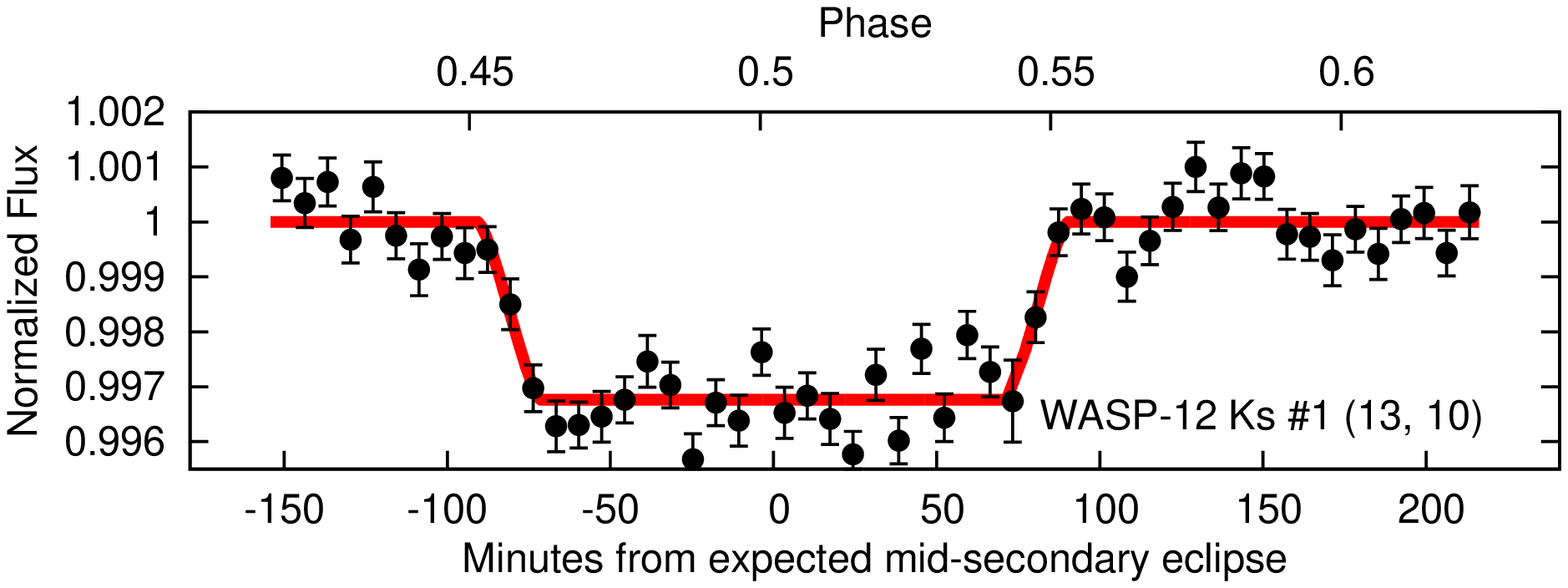}
\includegraphics[scale=0.28, angle = 270]{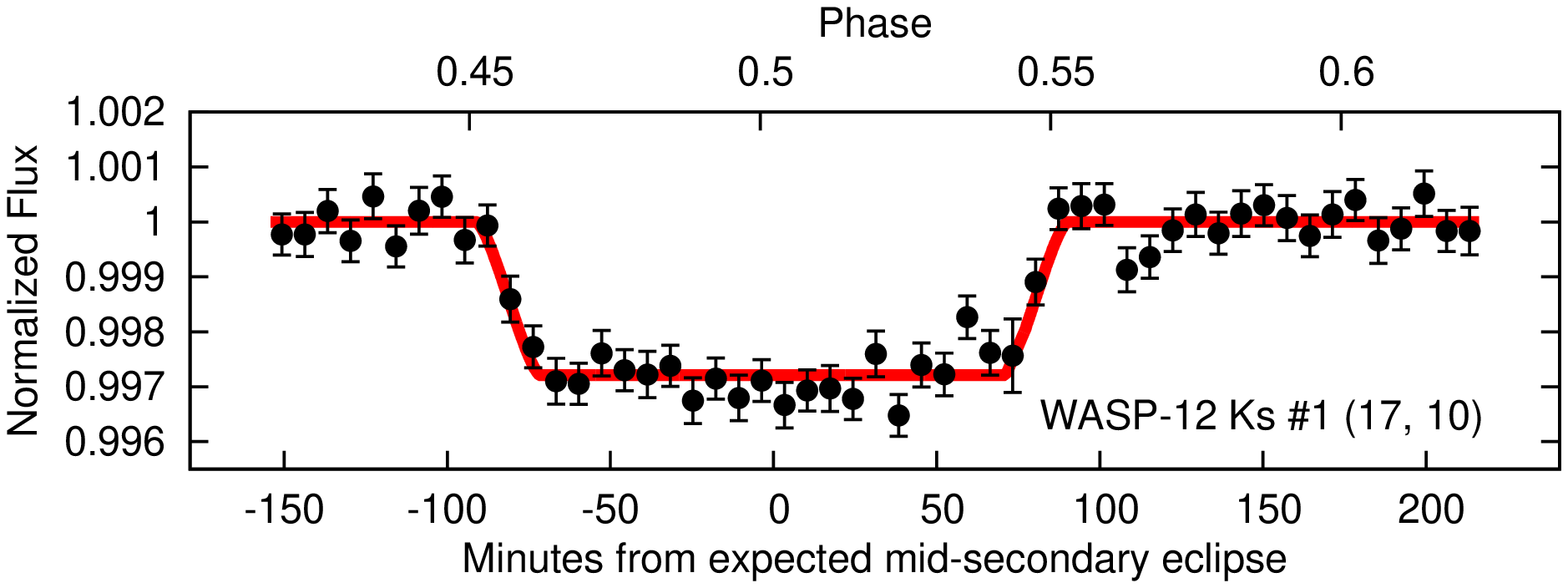}
\includegraphics[scale=0.28, angle = 270]{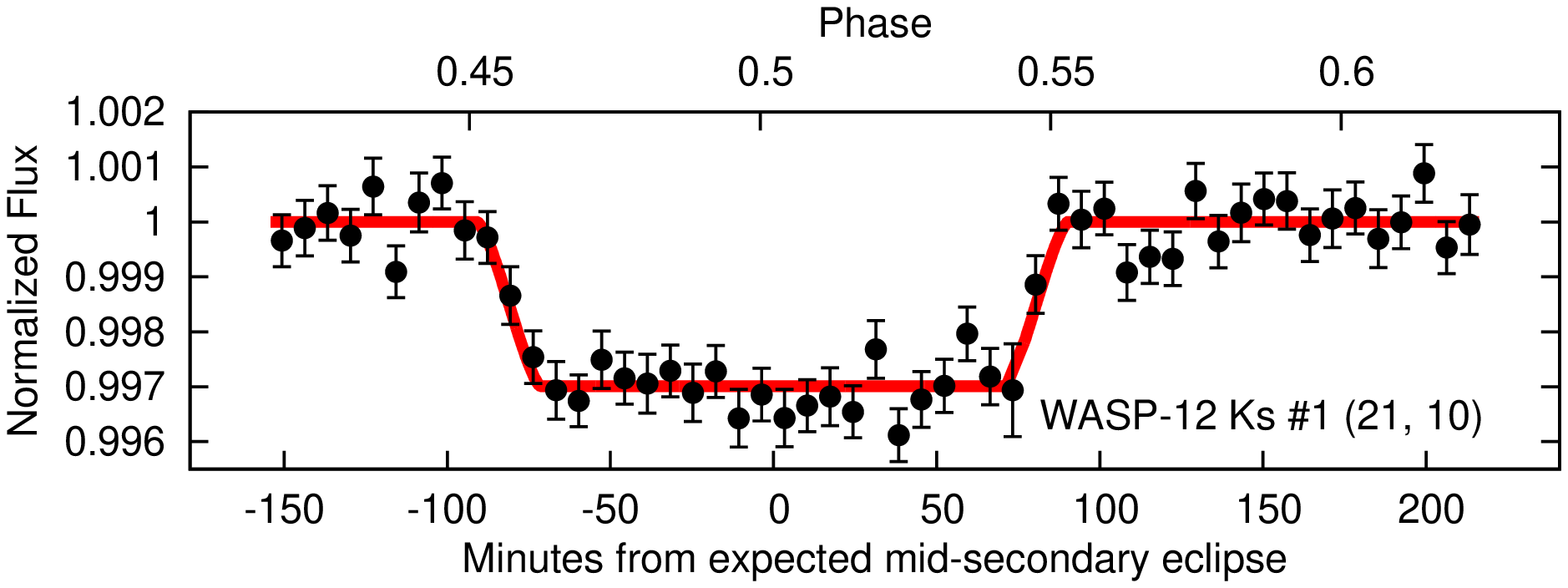}
 
\includegraphics[scale=0.28, angle = 270]{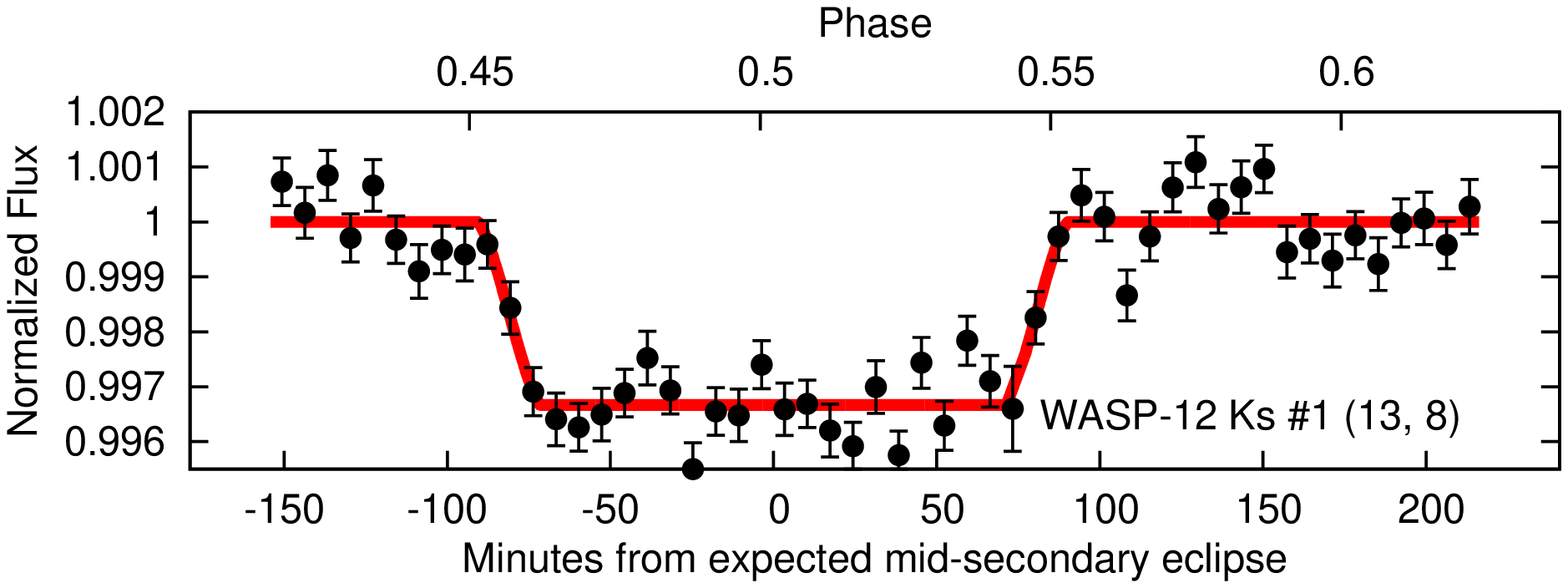}
\includegraphics[scale=0.28, angle = 270]{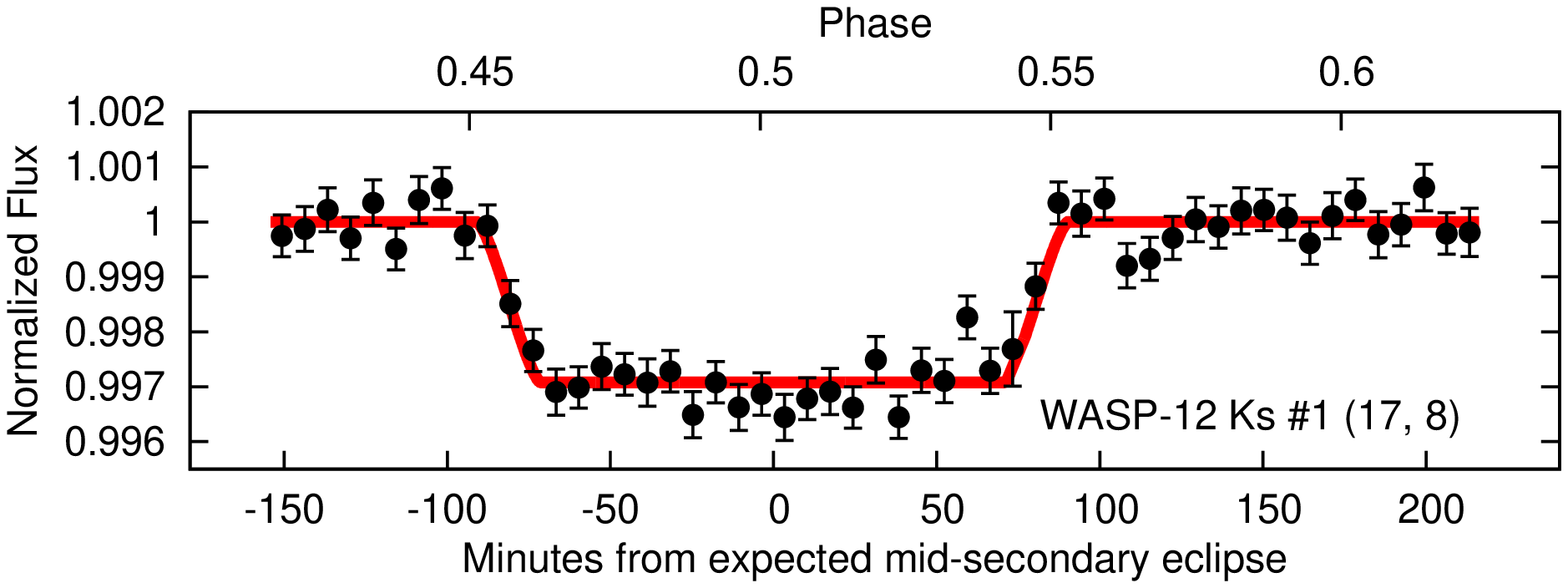}
\includegraphics[scale=0.28, angle = 270]{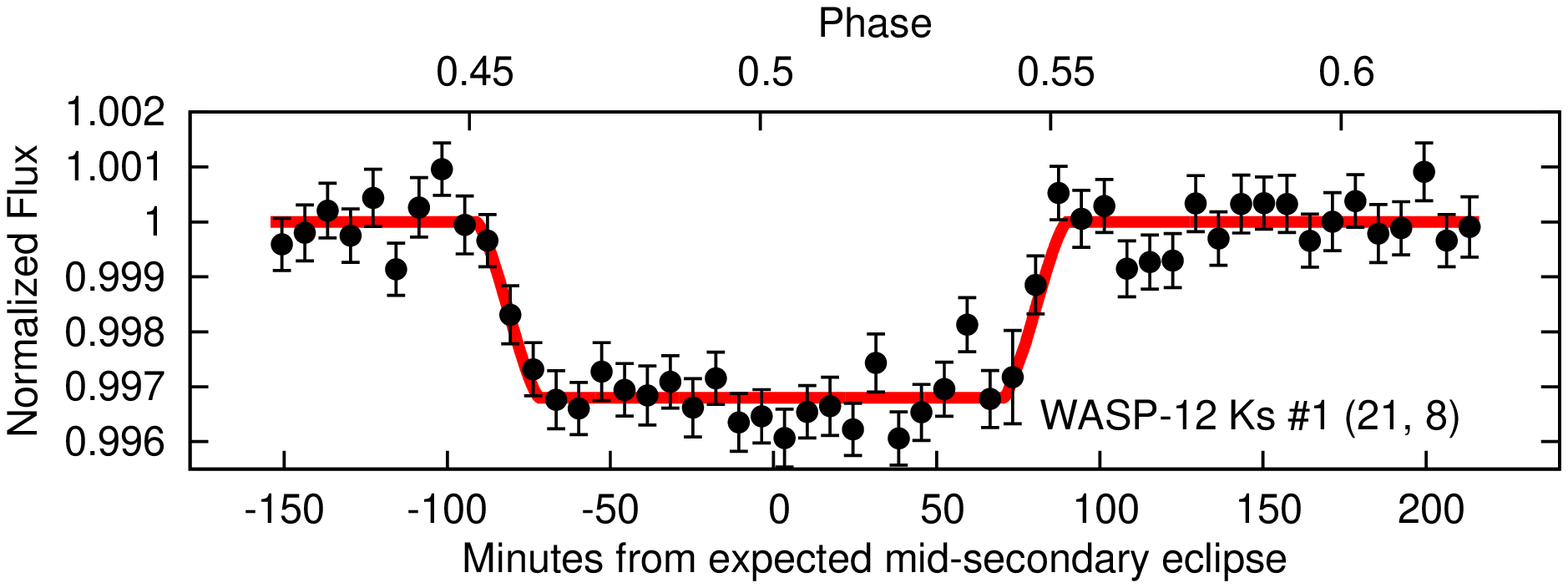}

\includegraphics[scale=0.28, angle = 270]{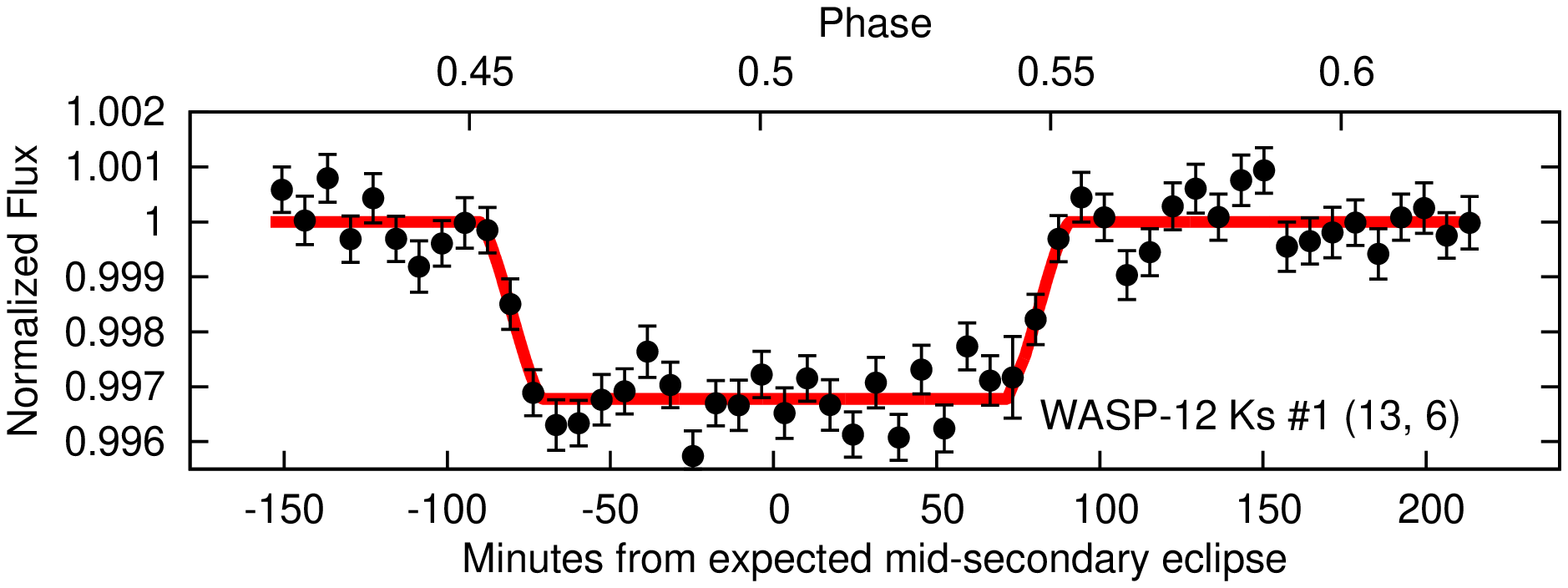}
\includegraphics[scale=0.28, angle = 270]{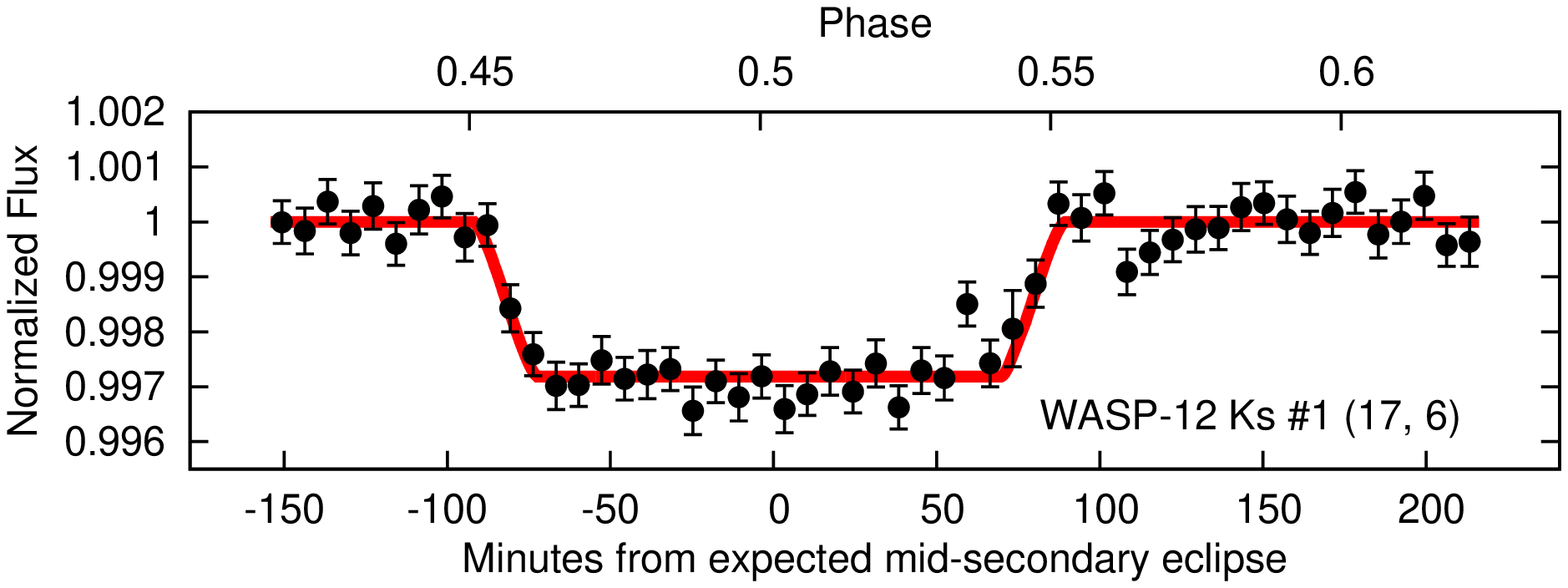}
\includegraphics[scale=0.28, angle = 270]{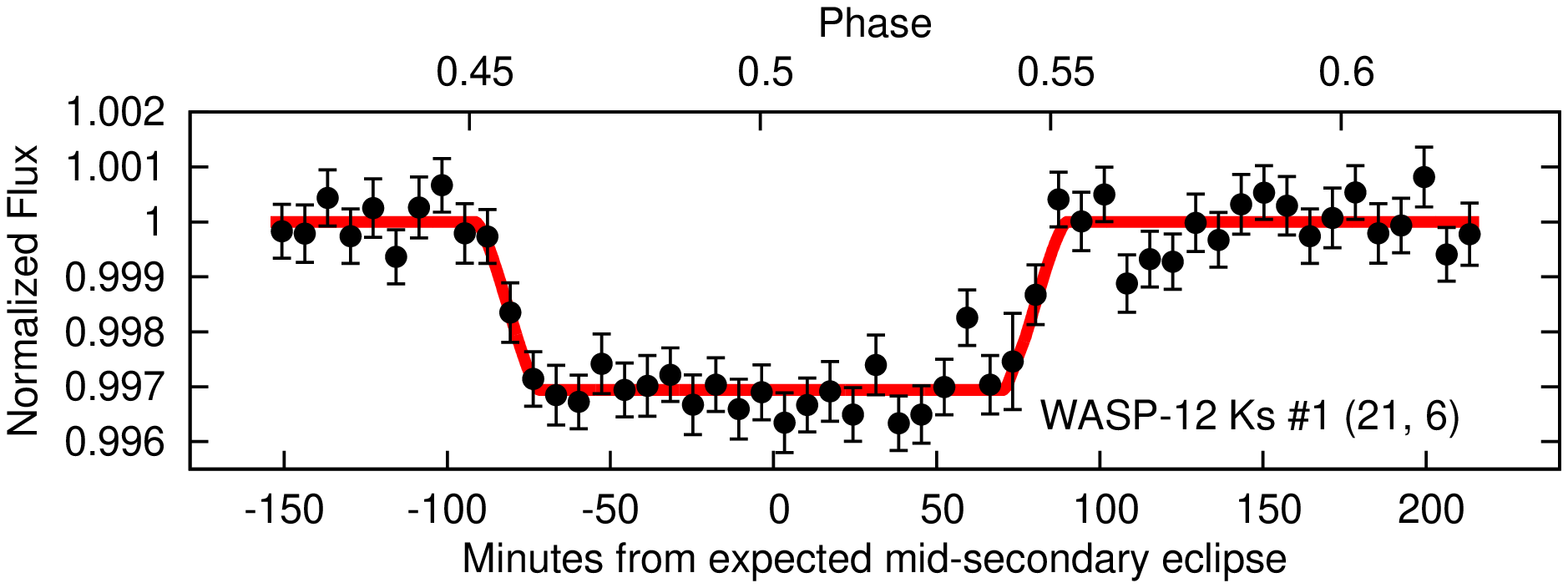}

\includegraphics[scale=0.28, angle = 270]{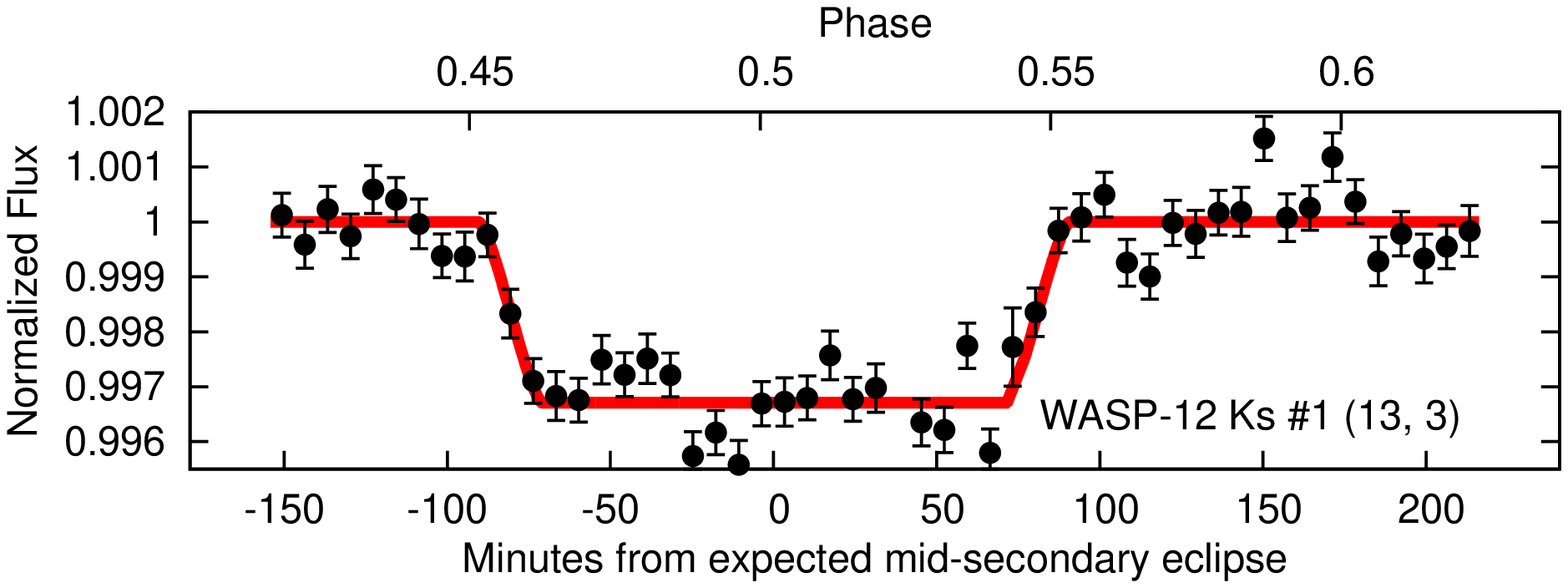}
\includegraphics[scale=0.28, angle = 270]{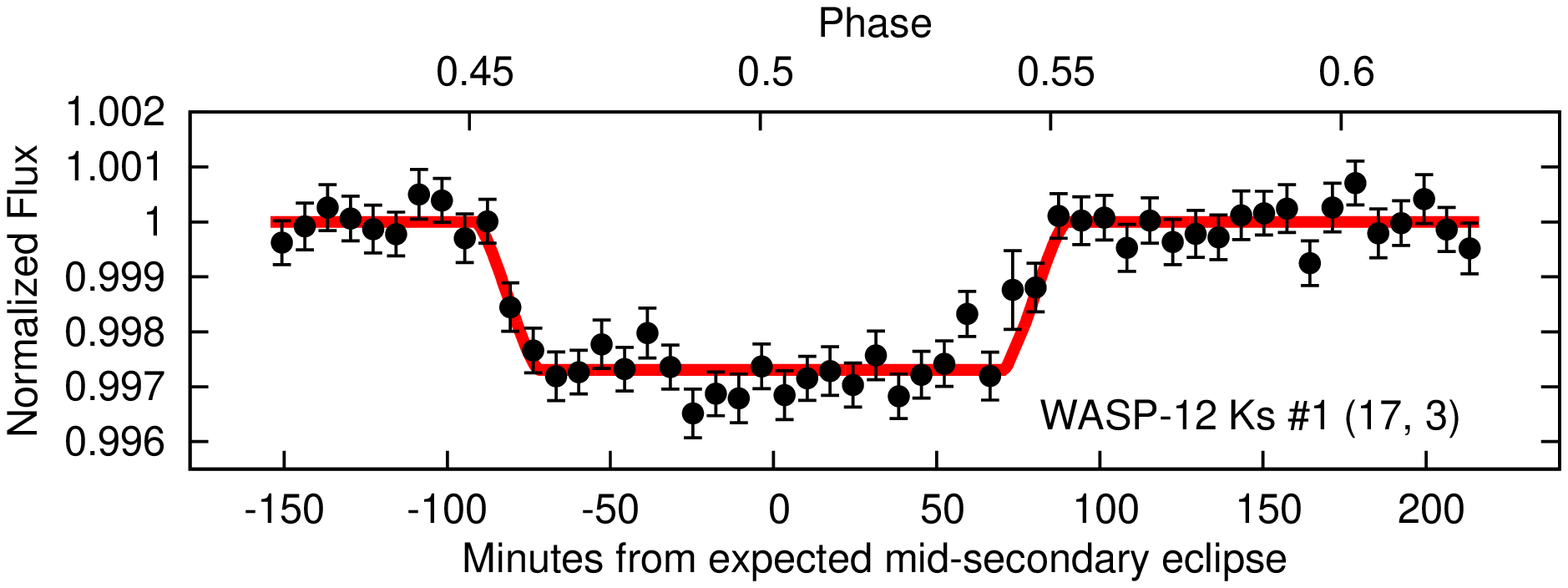}
\includegraphics[scale=0.28, angle = 270]{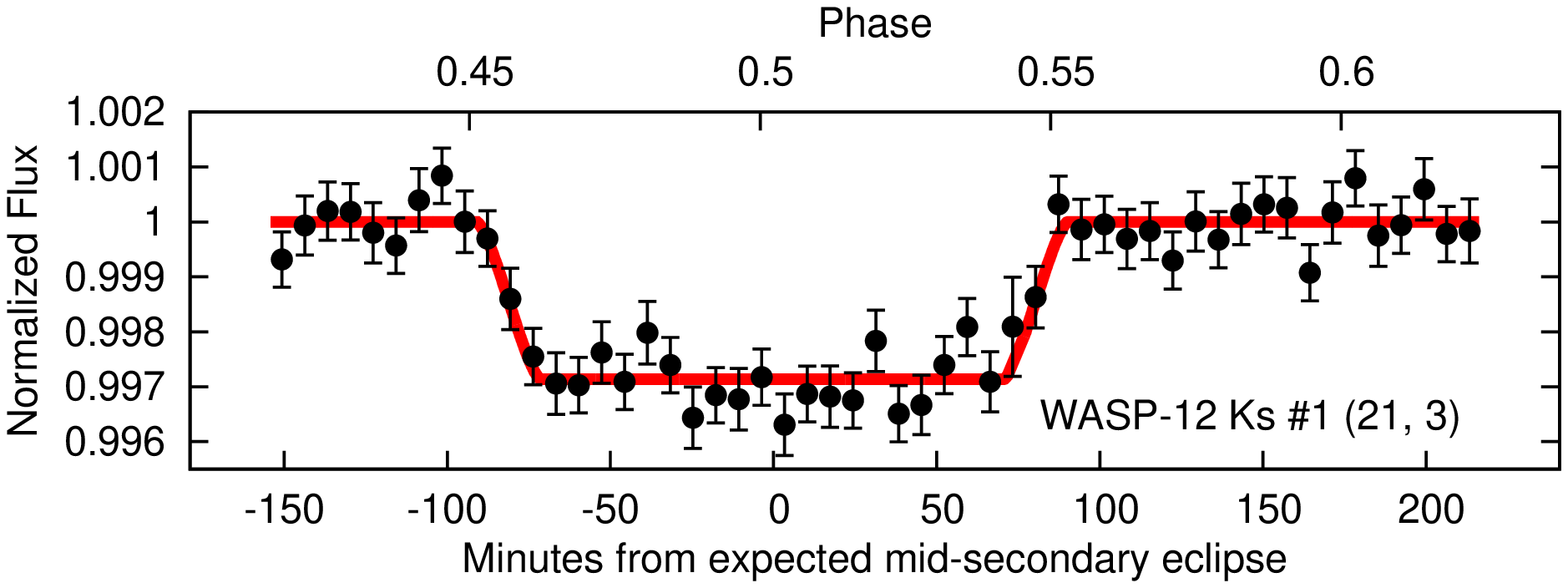}

\includegraphics[scale=0.28, angle = 270]{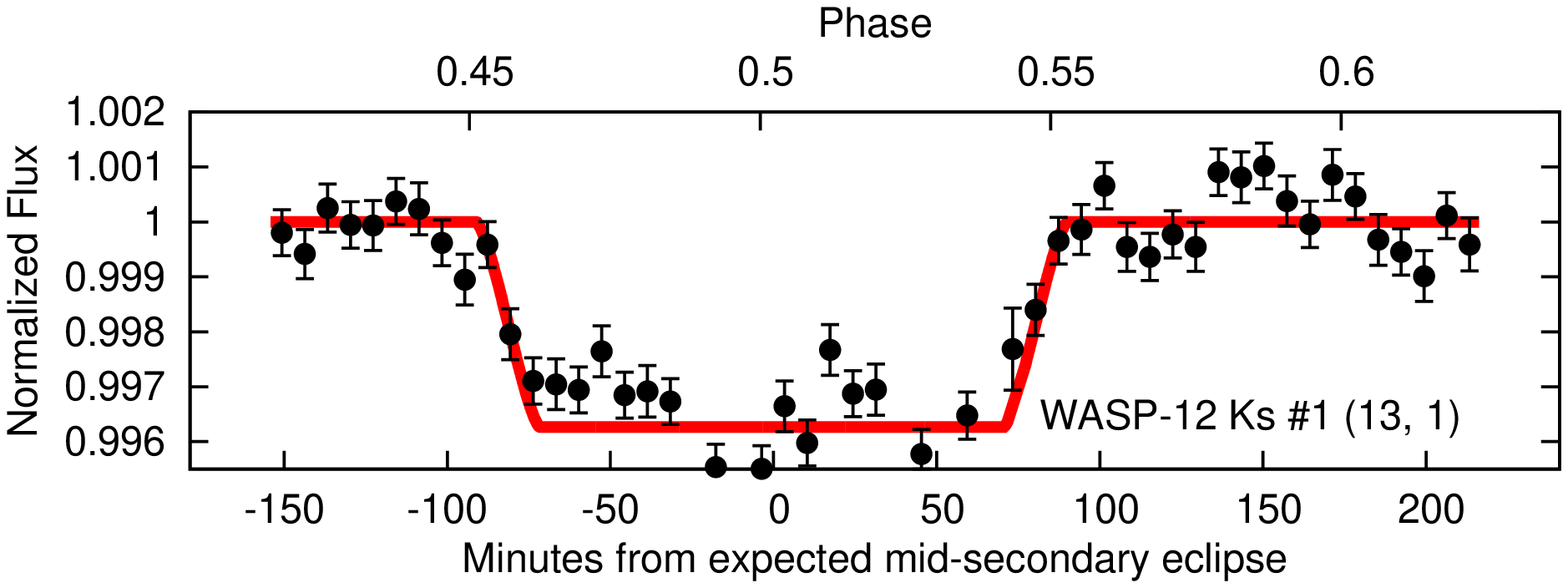}
\includegraphics[scale=0.28, angle = 270]{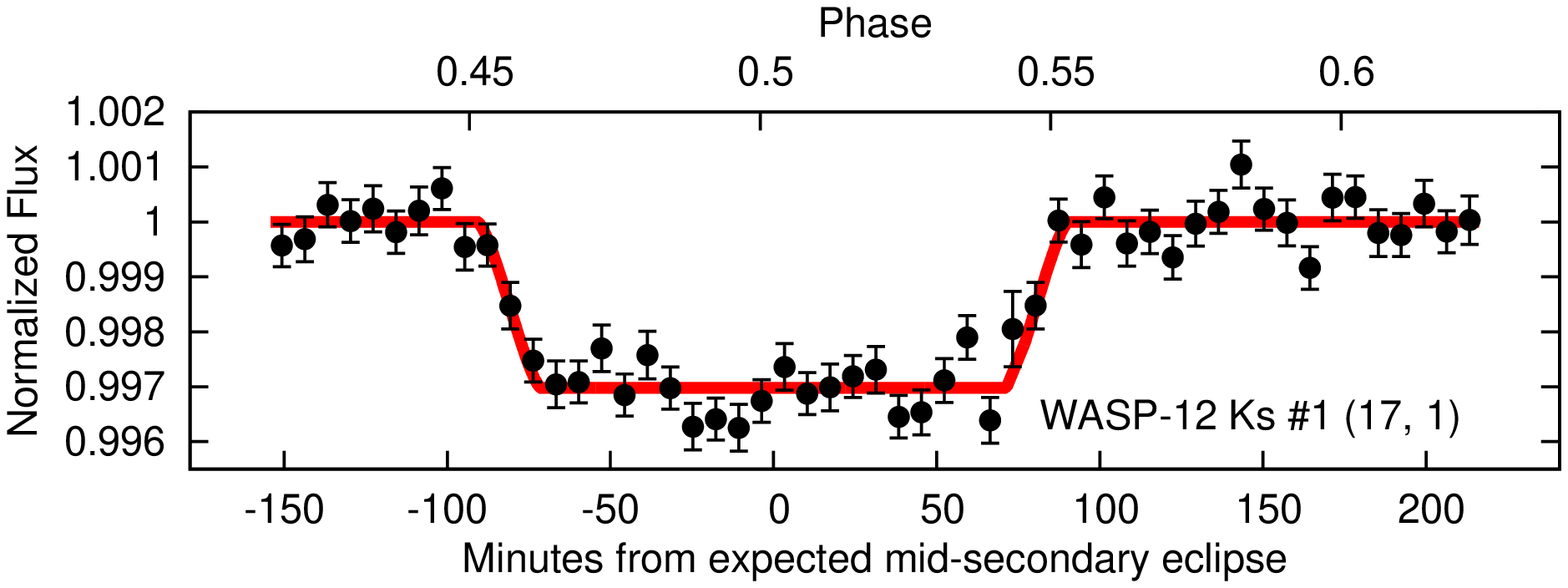}
\includegraphics[scale=0.28, angle = 270]{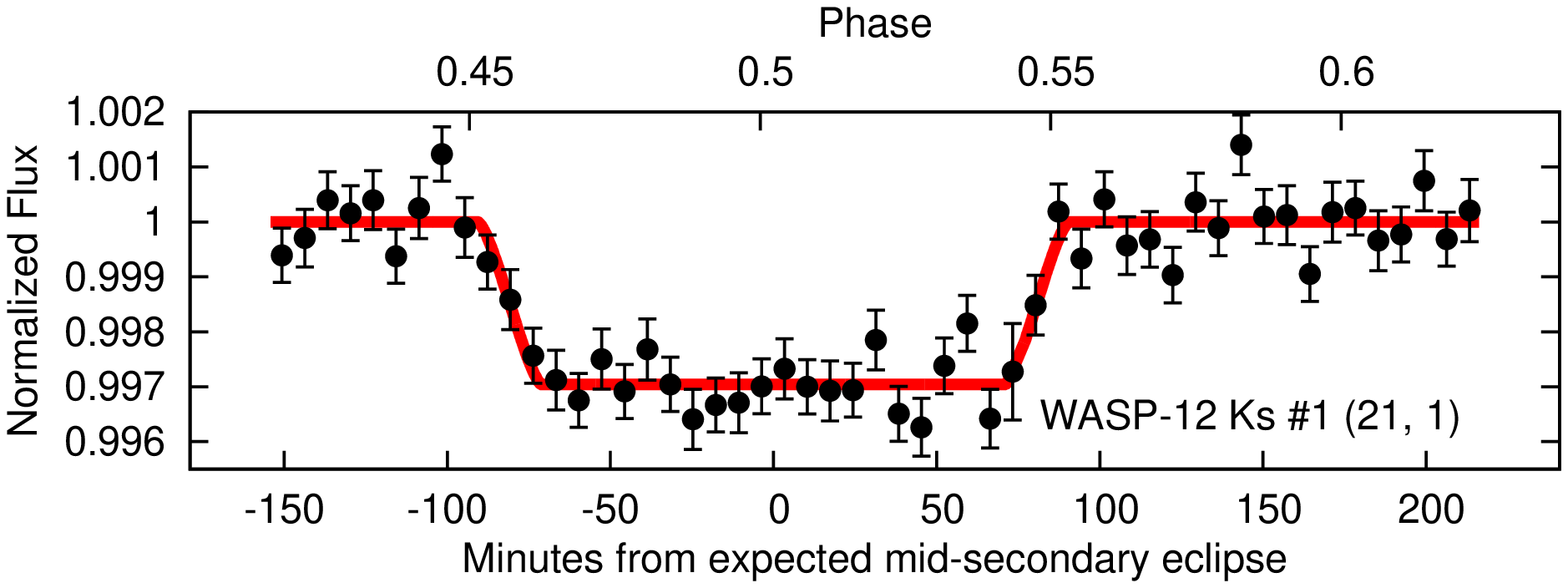}

\caption[WASP-12 One Fidelity Many]
	{	
 		Top panels: For various aperture sizes
		and number of reference stars in our reference star ensemble, in 
		the top left panel we display the precision of the data (the RMS$\times$$\beta^2$ of the residuals from our
		best-fit model; the intensity-bar on the right of the panel represents
		the percentage above the minimum RMS$\times$$\beta^2$ that we observe
		in our reference star ensemble and aperture size grid),
		while the top right panel displays 
		the depth of the secondary eclipse for our first WASP-12 Ks-band eclipse; in this case
		the intensity bar represents the percentage depth of the apparent secondary eclipse ($F_{Ap}/F_{*}$)
		in terms of the stellar flux. 
		Values within \NUMBERHERERMSLIMIT\% of the minimum recorded value of the RMS$\times$$\beta^2$ for various aperture sizes
		and number of reference stars are encircled by the red lines, and are used to determine
		the value of our eclipse depth and uncertainty.
		Other (bottom) panels: 	the binned light curves (every $\sim$7 minutes) after the subtraction
		of the best-fit background trend, $B_f$, for our first WASP-12 Ks-band eclipse
		for various aperture sizes and reference star ensembles,
		with the best-fit MCMC eclipse fit given with the solid red line.
		The aperture sizes increase from left to right, and the number of reference stars in the ensemble increase
		from bottom to top; the aperture size is denoted before the comma,
		and the number of reference stars is denoted after the comma,
		in the parenthetical comment in the bottom right of each of the bottom panels.
		For our first WASP-12 Ks-band eclipse the eclipse depths are relatively constant for various aperture
		sizes and reference star ensembles.
	}
\label{FigWASPTwelveKsbandFidelityManyOne}
\end{figure*}

\begin{figure*}
\centering

\includegraphics[scale=0.44, angle = 270]{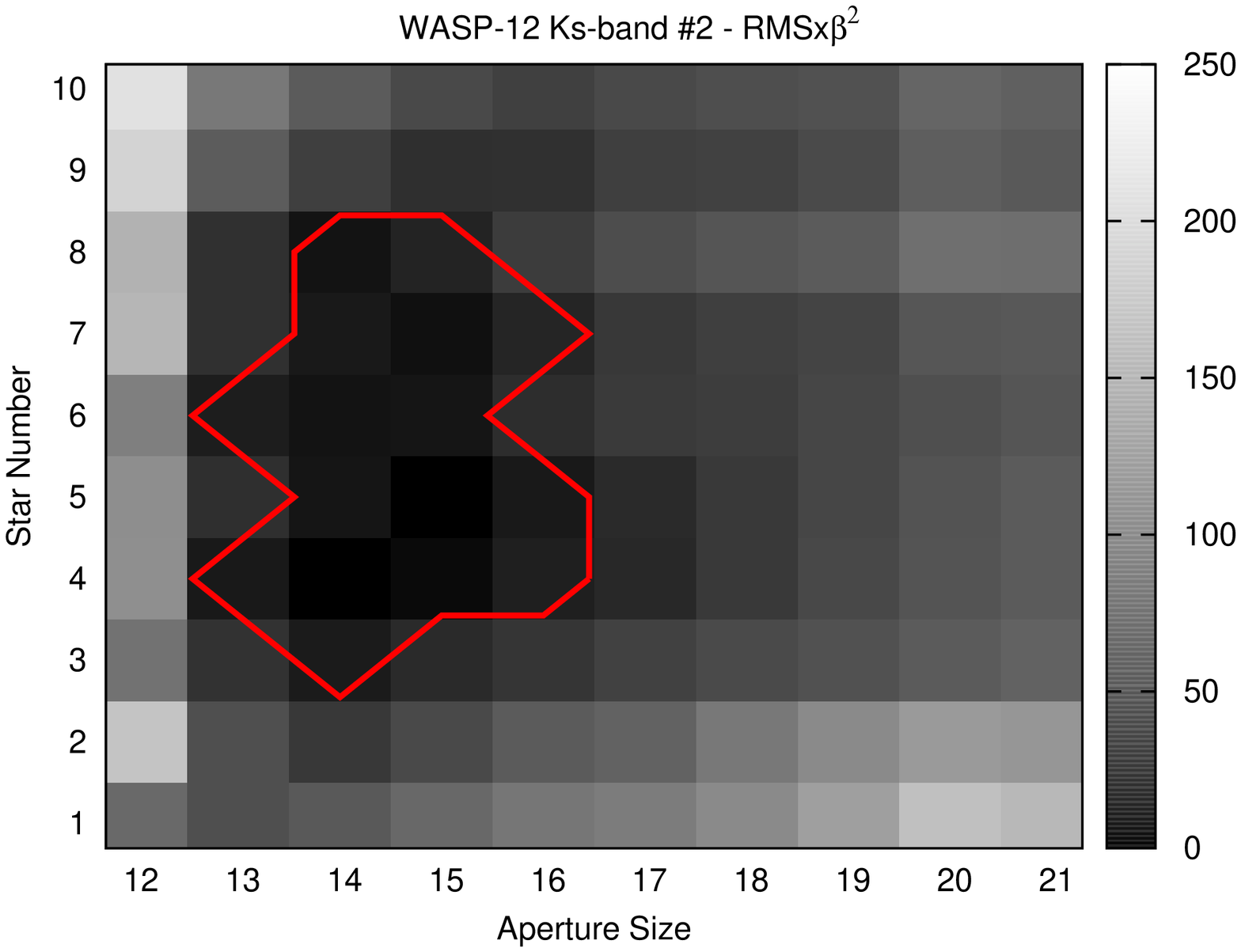}
\includegraphics[scale=0.44, angle = 270]{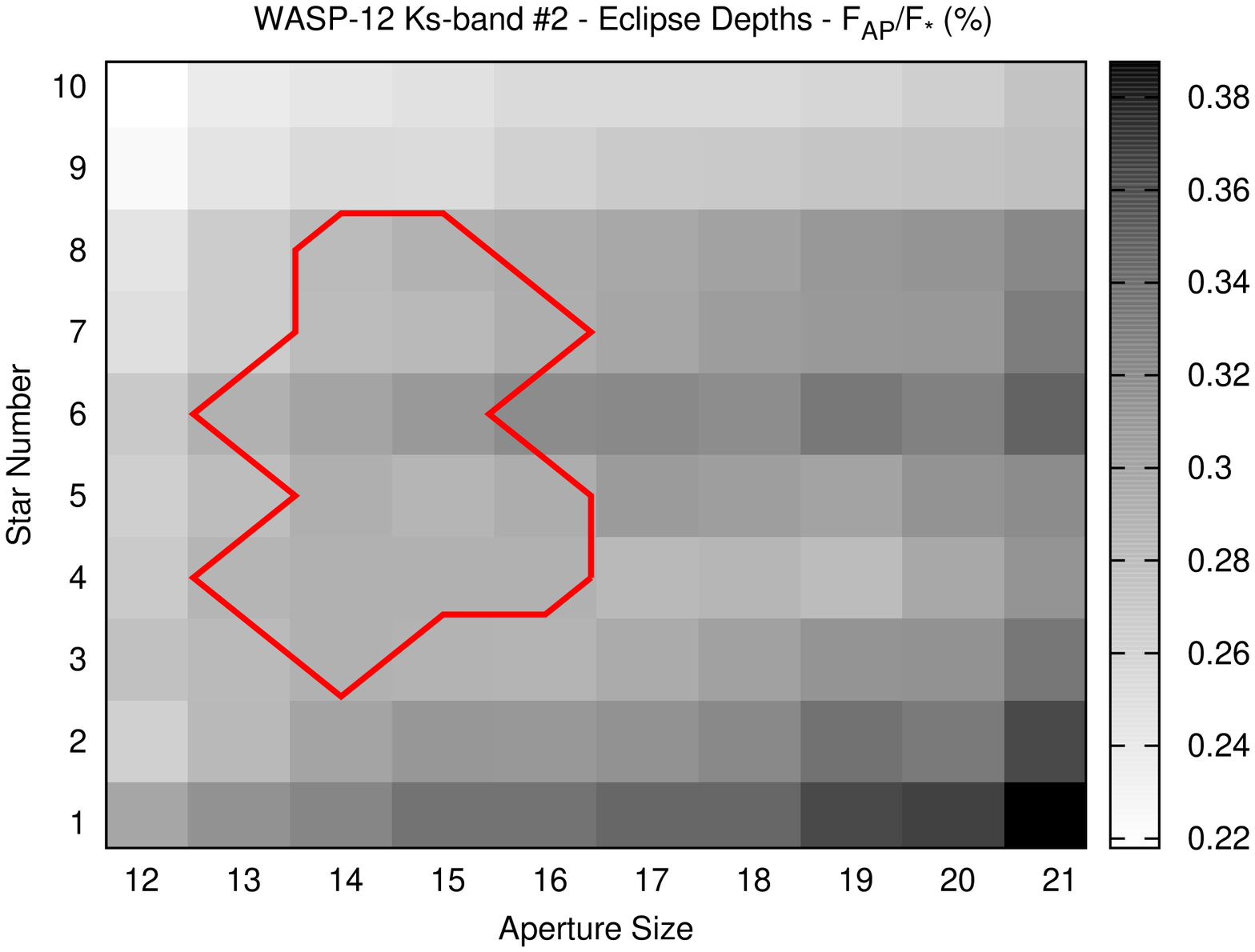}

\includegraphics[scale=0.27, angle = 270]{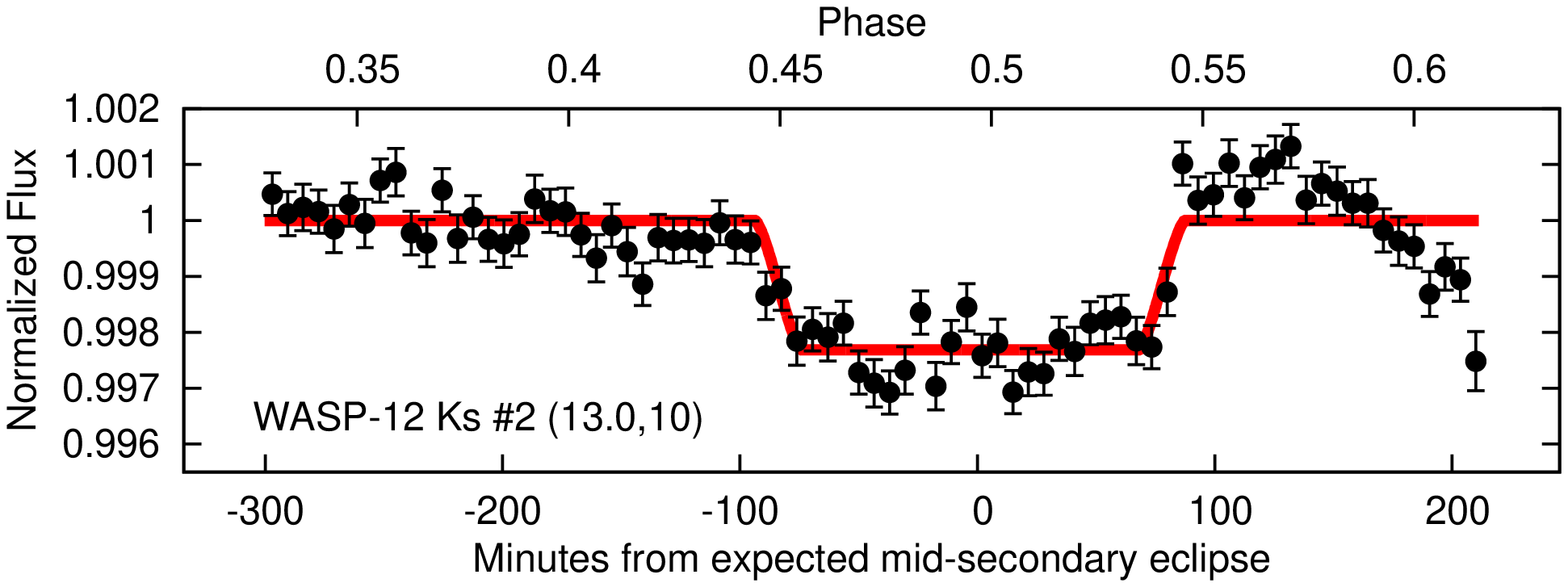}
\includegraphics[scale=0.27, angle = 270]{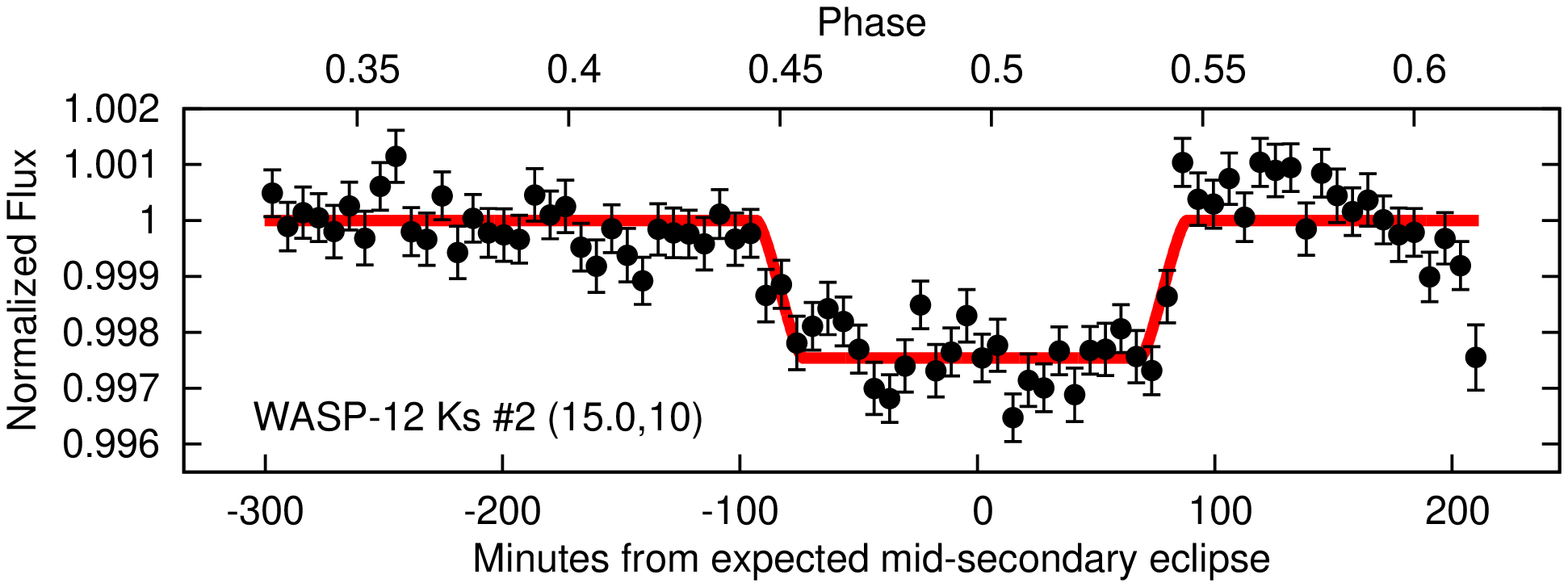}
\includegraphics[scale=0.27, angle = 270]{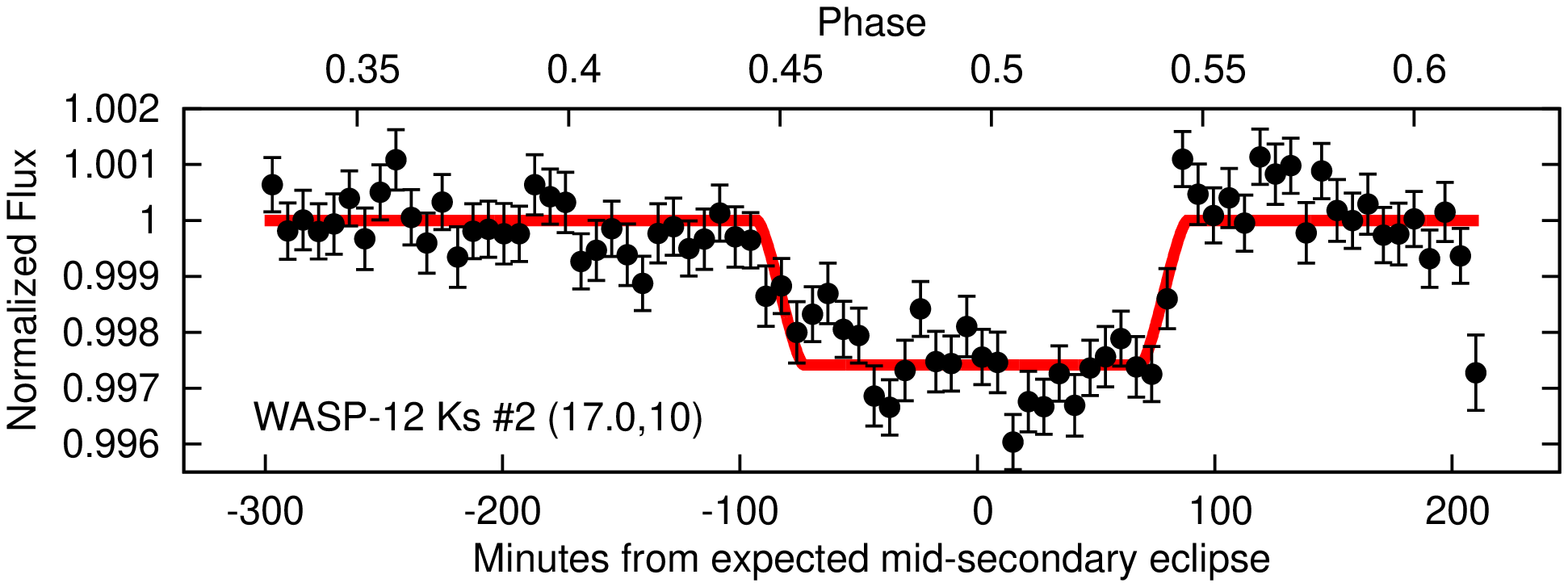}

\includegraphics[scale=0.27, angle = 270]{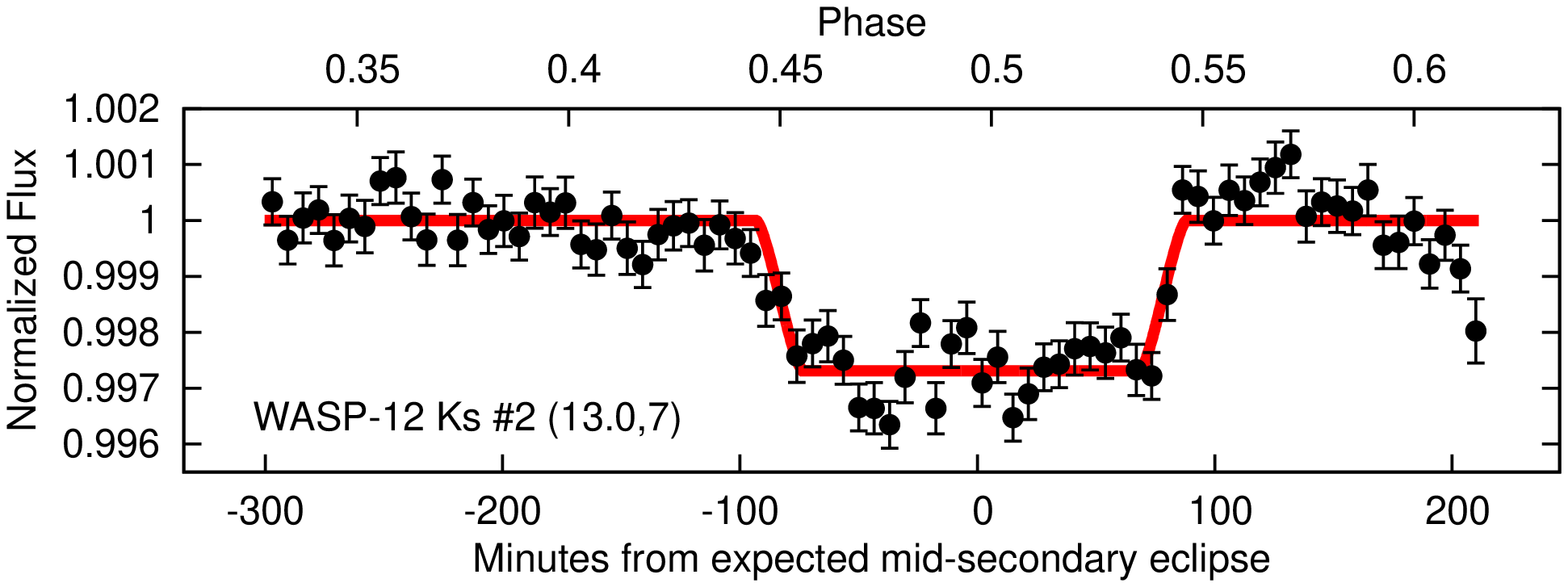}
\includegraphics[scale=0.27, angle = 270]{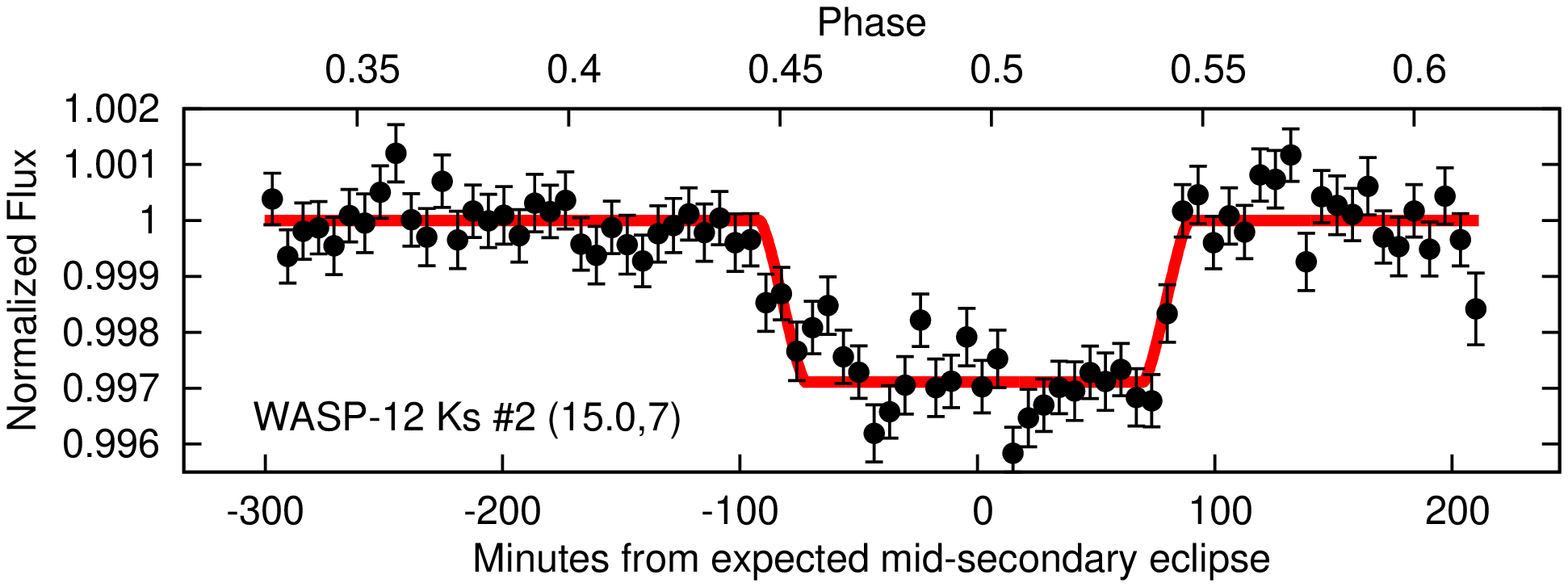}
\includegraphics[scale=0.27, angle = 270]{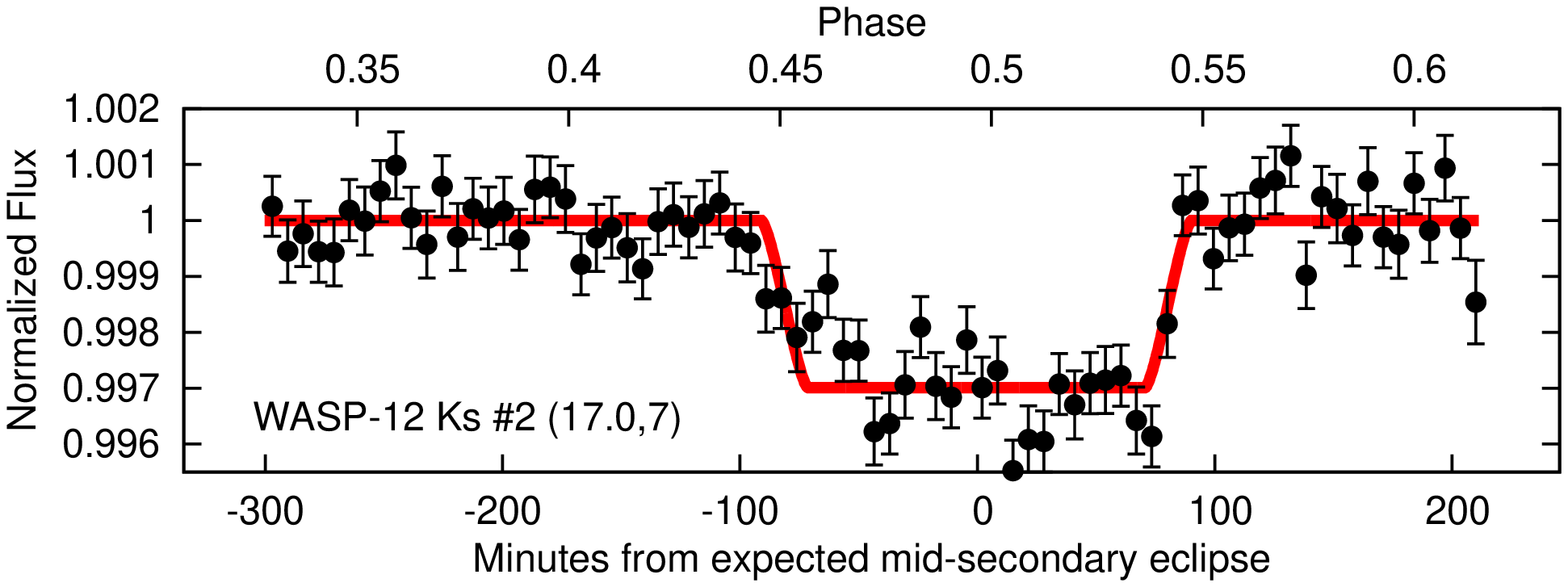}

\includegraphics[scale=0.27, angle = 270]{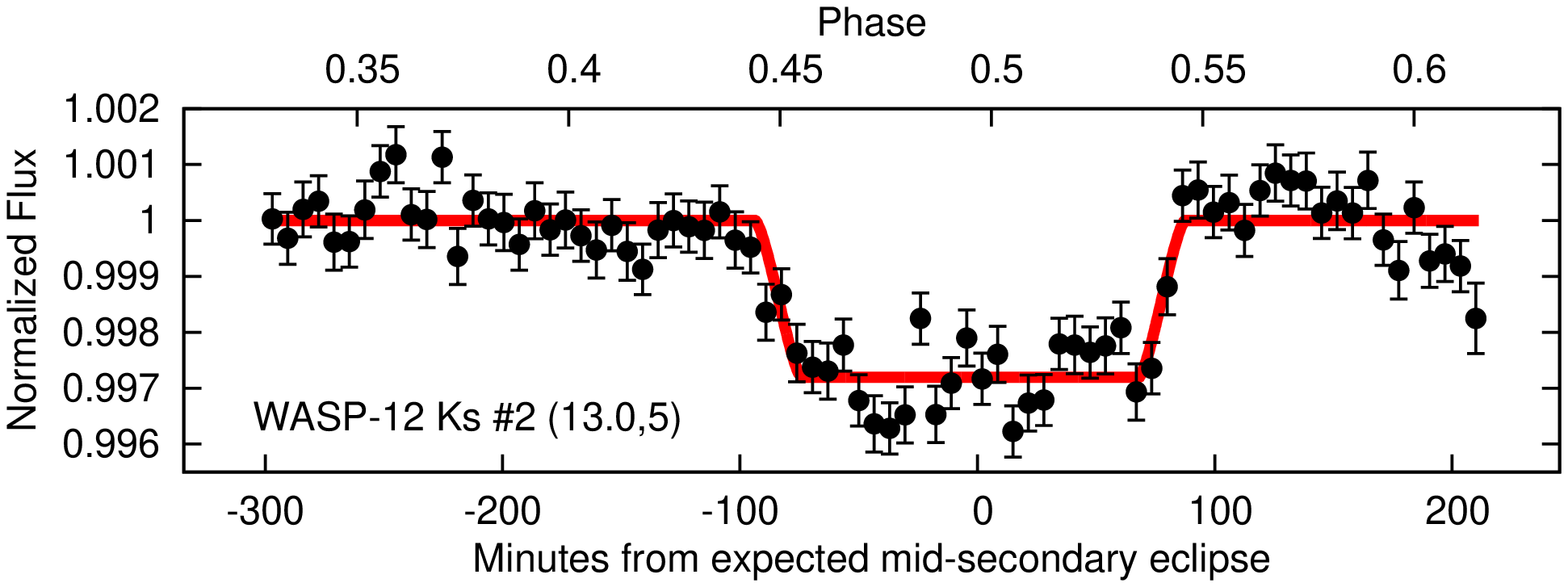}
\includegraphics[scale=0.27, angle = 270]{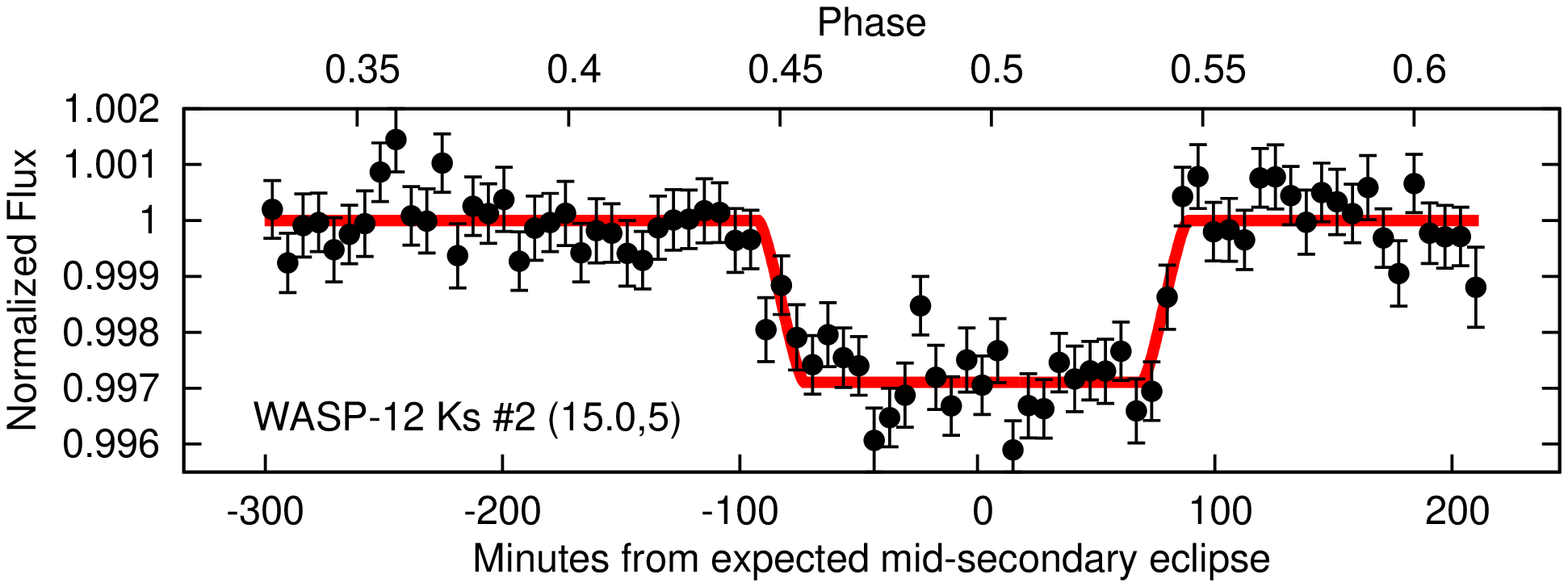}
\includegraphics[scale=0.27, angle = 270]{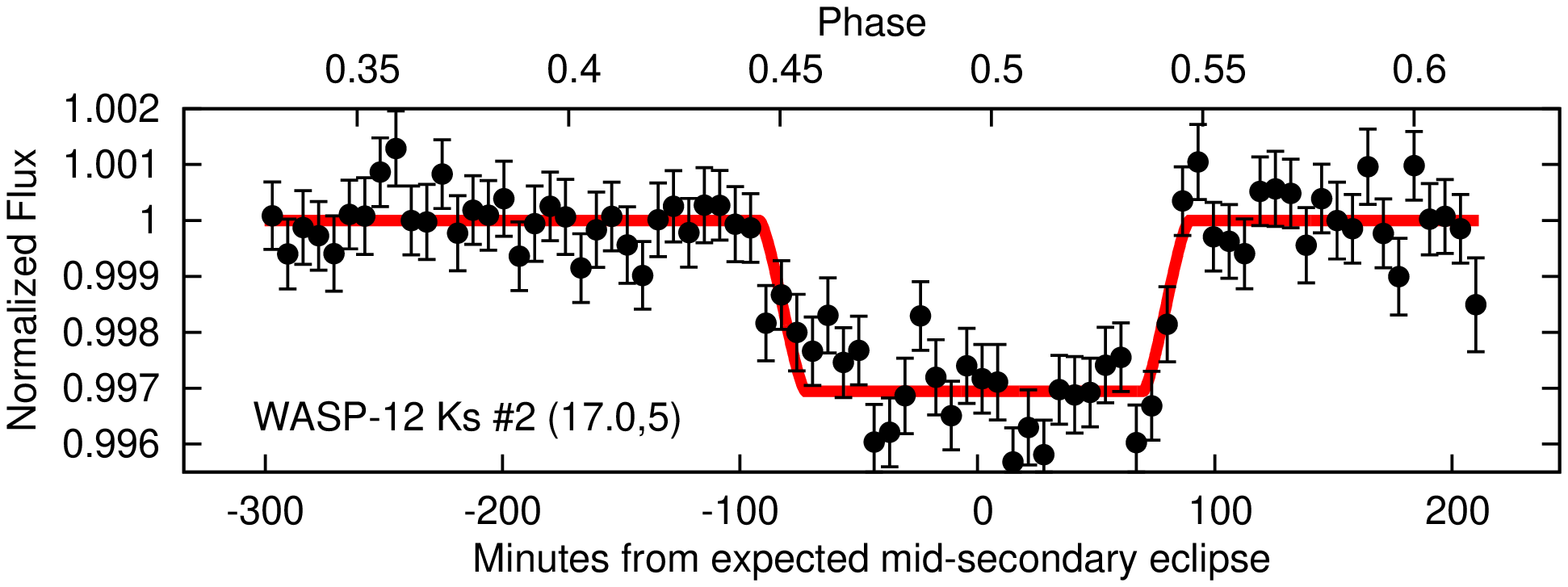}

\includegraphics[scale=0.27, angle = 270]{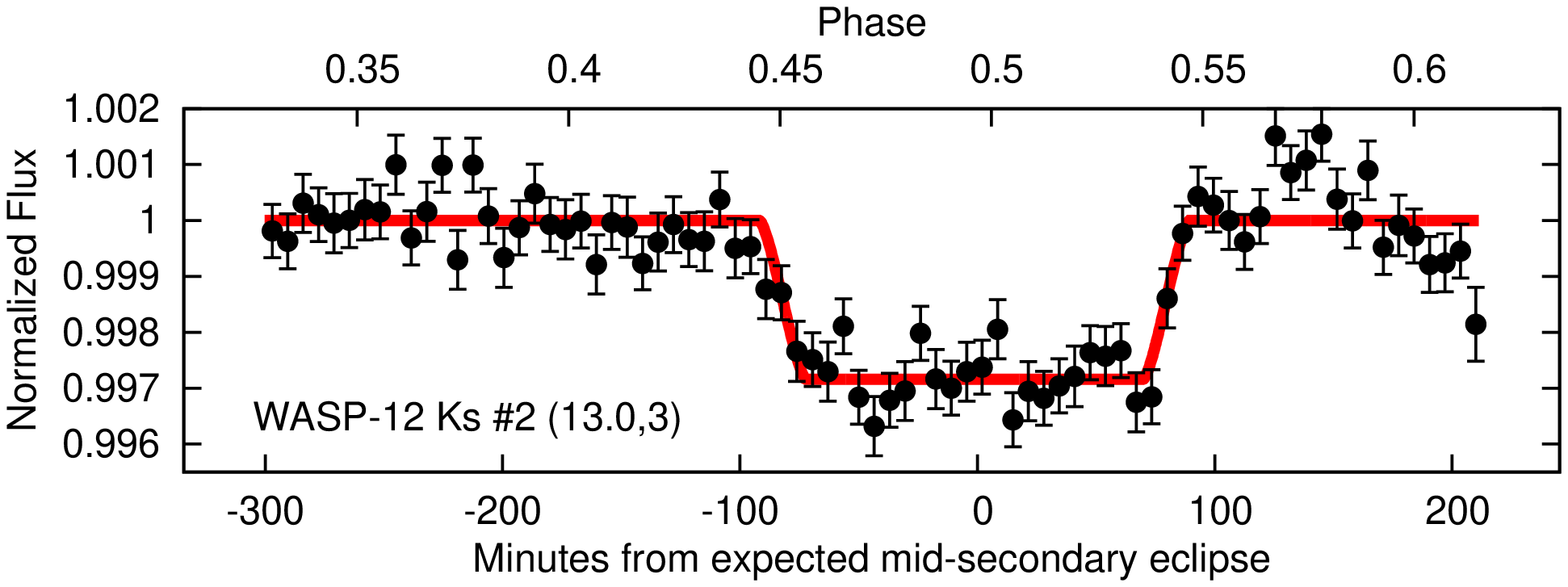}
\includegraphics[scale=0.27, angle = 270]{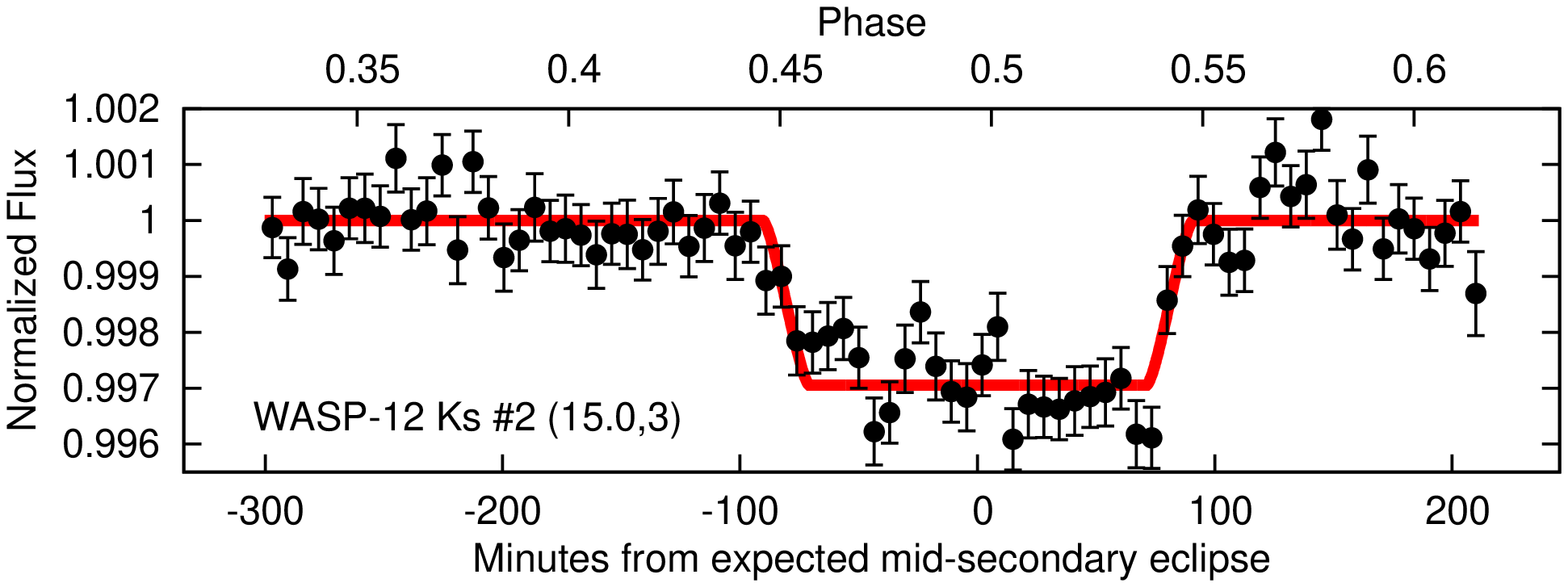}
\includegraphics[scale=0.27, angle = 270]{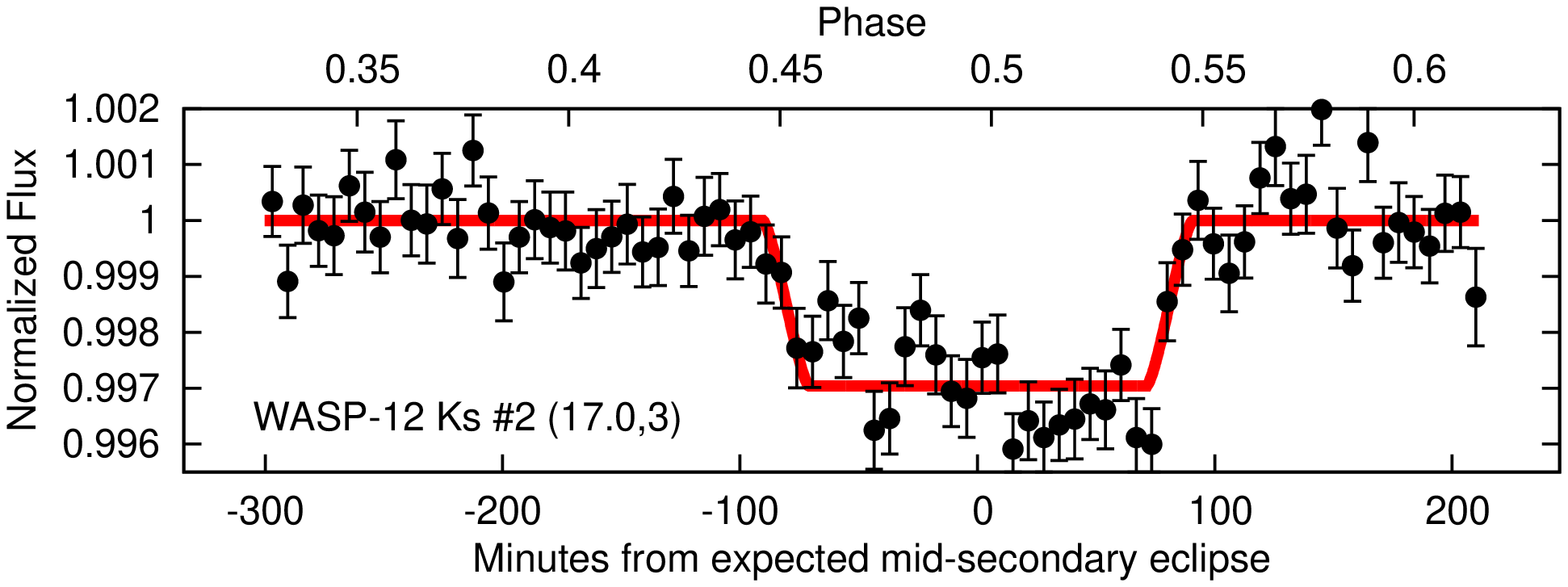}

\includegraphics[scale=0.27, angle = 270]{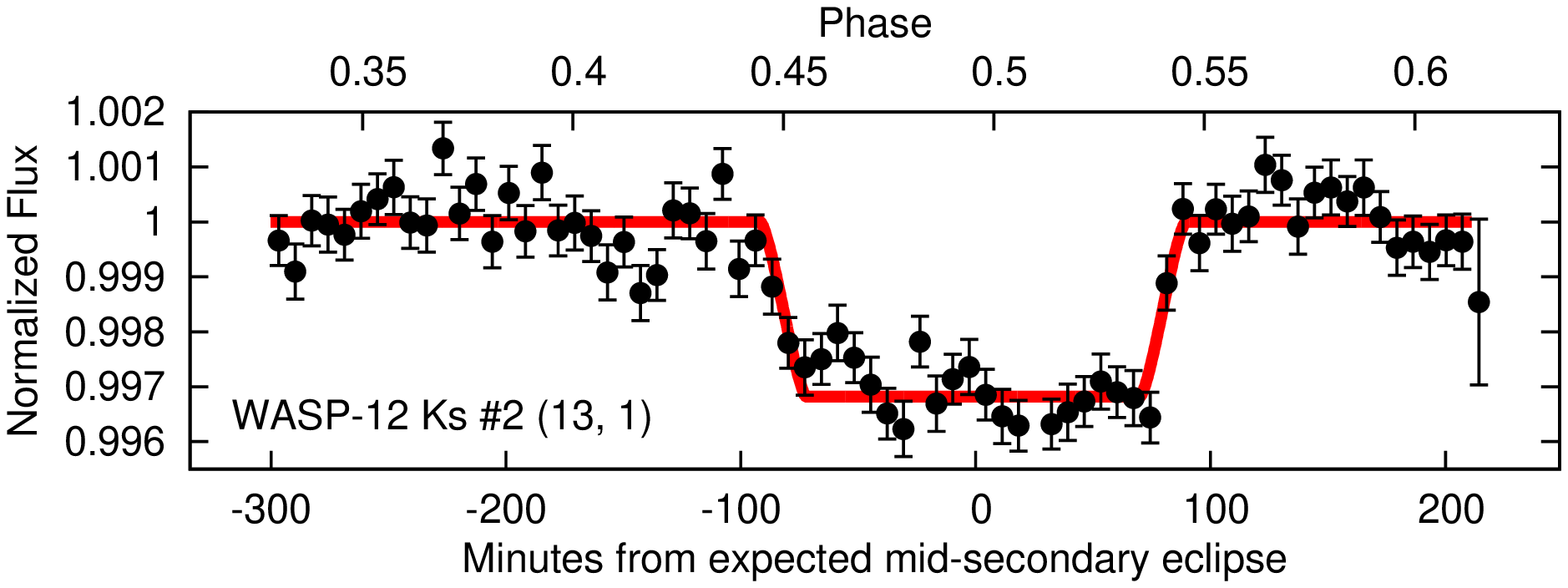}
\includegraphics[scale=0.27, angle = 270]{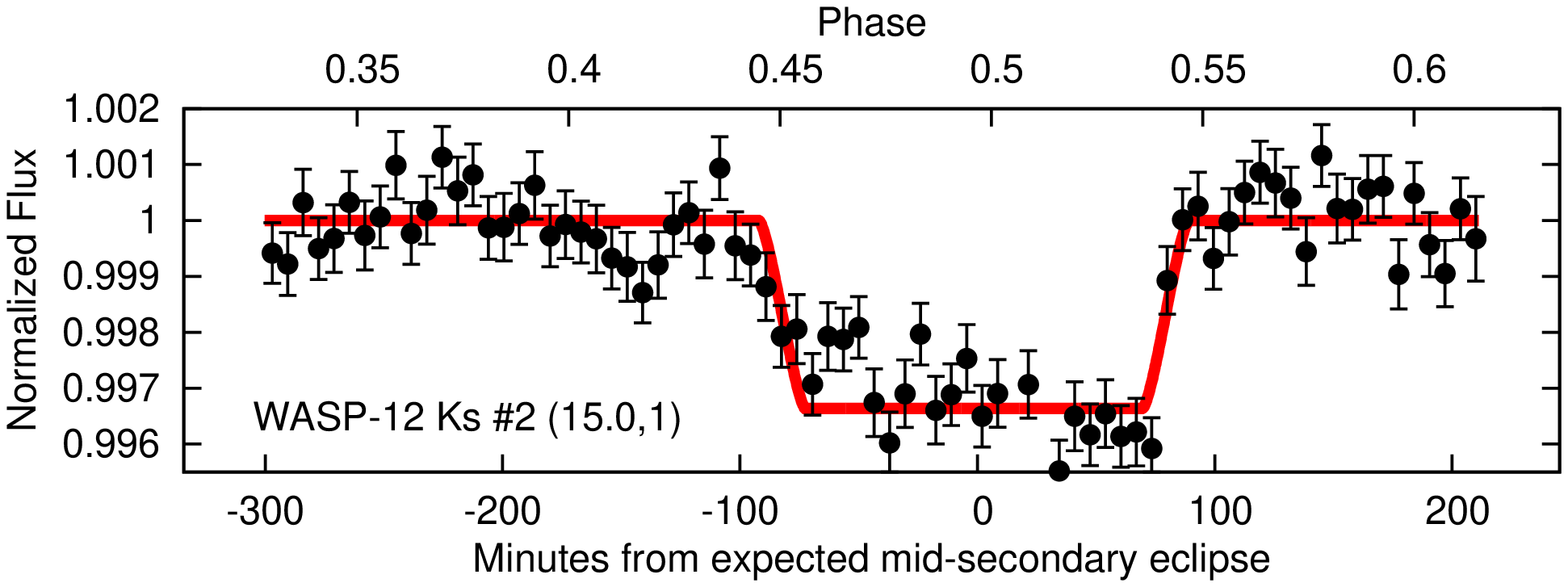}
\includegraphics[scale=0.27, angle = 270]{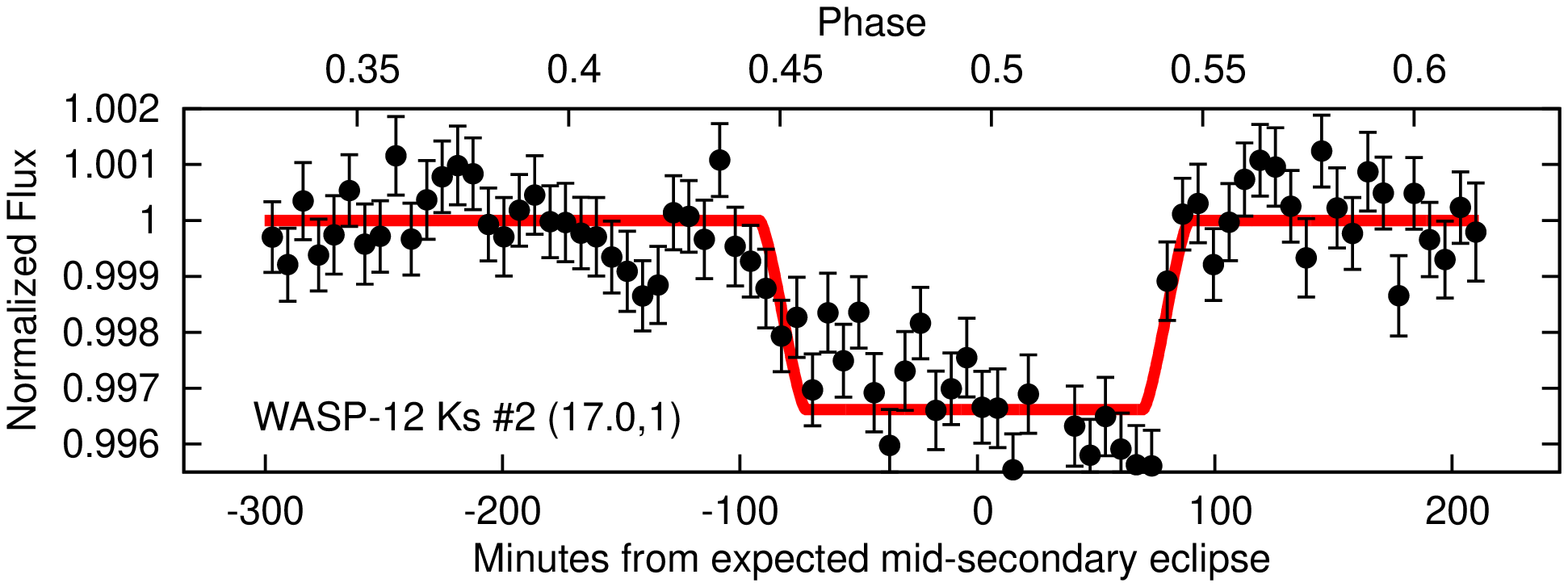}

\caption[WASP-12 Two Fidelity Many]
	{	
		Same as Figure \ref{FigWASPTwelveKsbandFidelityManyOne} except for our second WASP-12 Ks-band secondary eclipse,
		and the fact that we use a binning time of $\sim$6.5 minutes for our bottom panels.
		The scale of the bottom panels is identical to that of the other WASP-12 Ks-band
		eclipses (Figure \ref{FigWASPTwelveKsbandFidelityManyOne} \& \ref{FigWASPTwelveKsbandFidelityManyThree}).
	}
\label{FigWASPTwelveKsbandFidelityManyTwo}
\end{figure*}

\begin{figure*}
\centering
\includegraphics[scale=0.44, angle = 270]{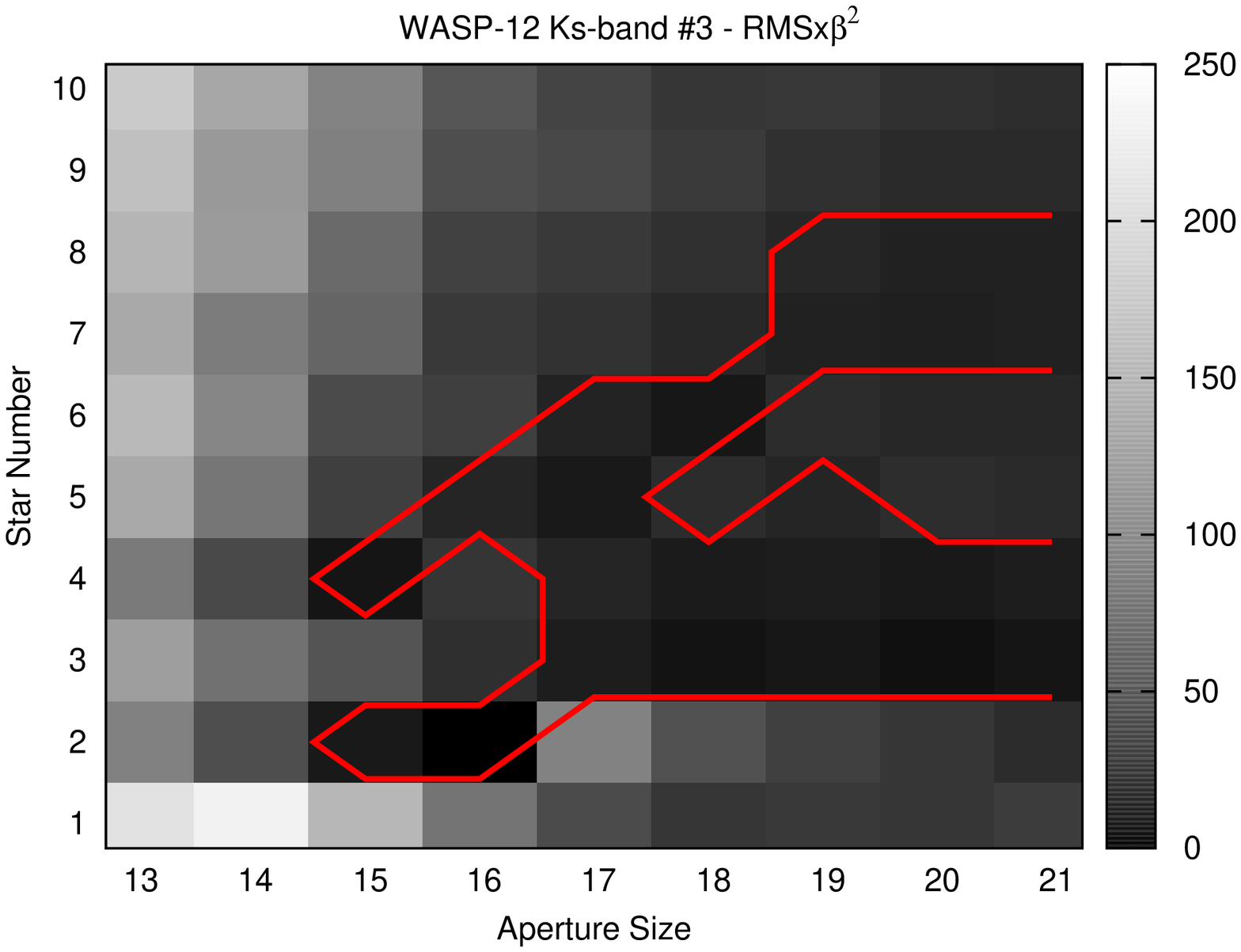}
\includegraphics[scale=0.44, angle = 270]{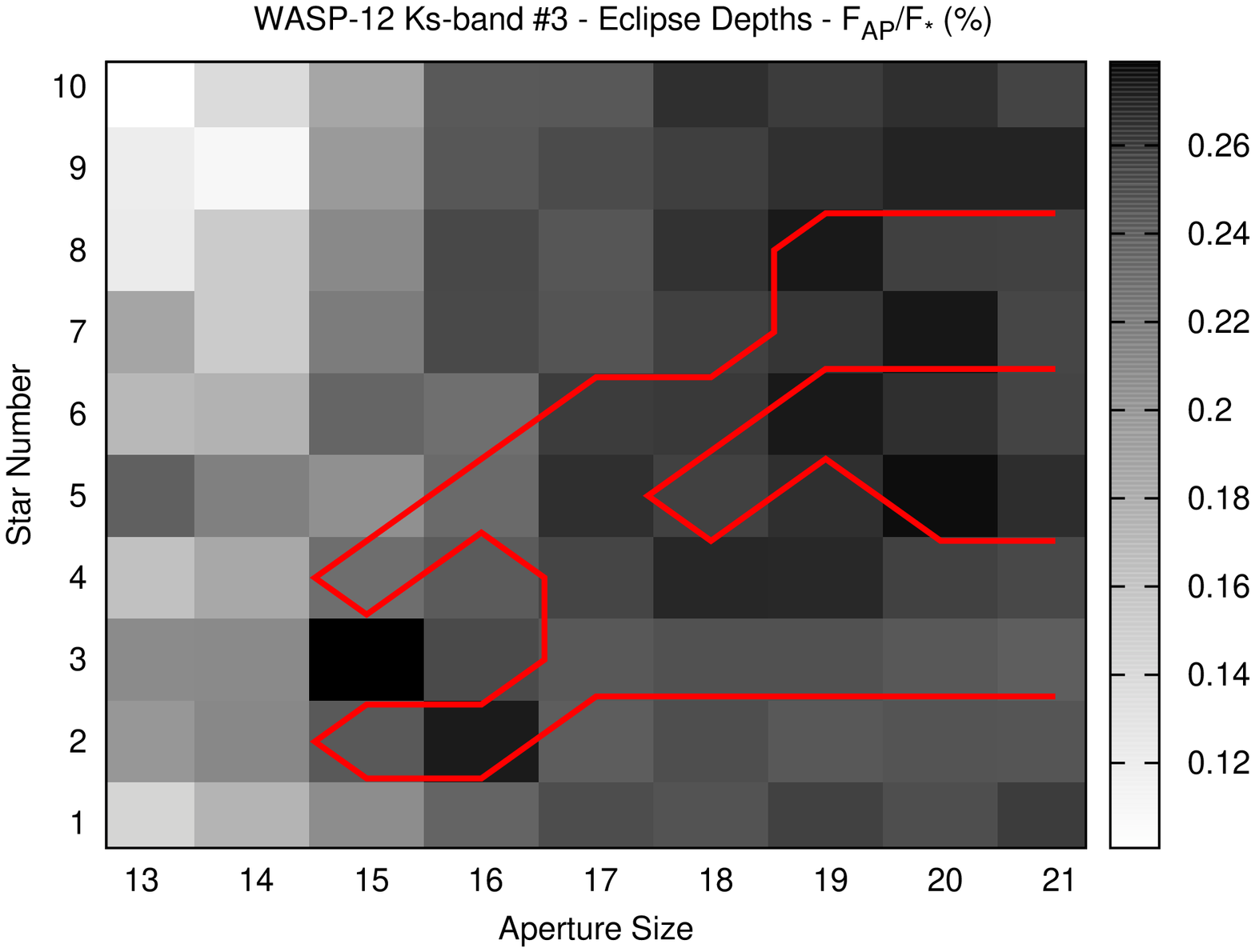}

\includegraphics[scale=0.27, angle = 270]{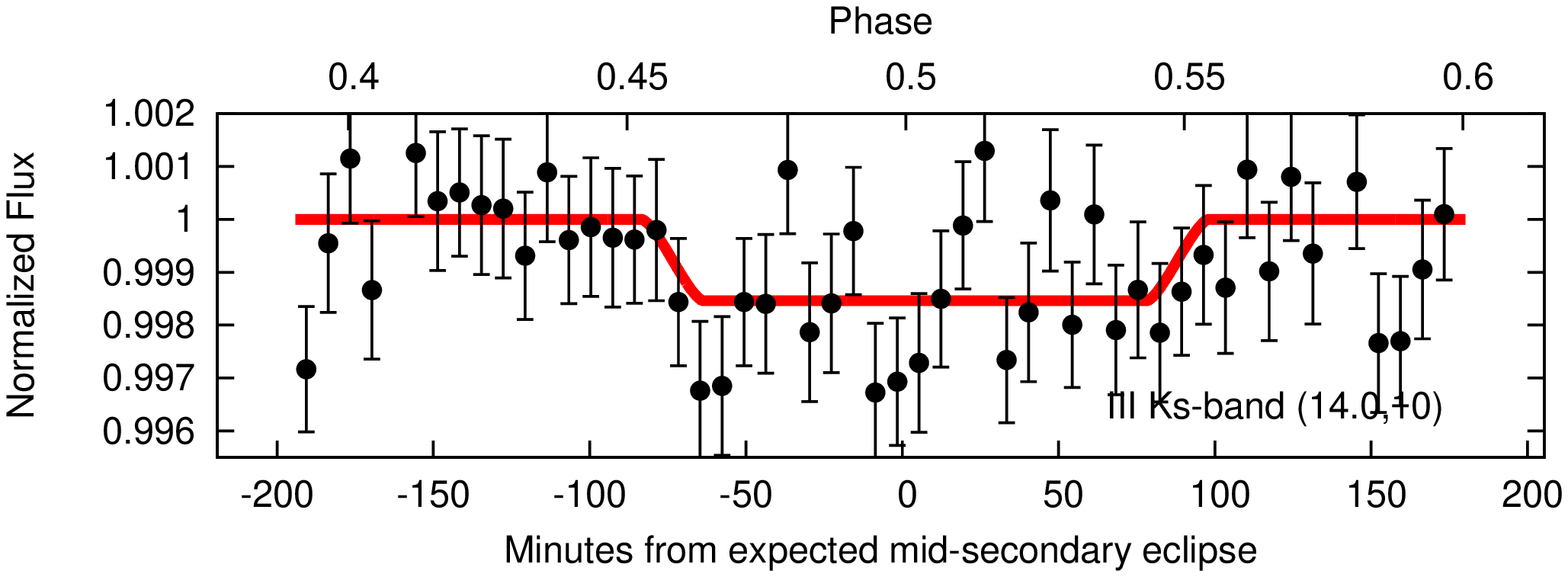}
\includegraphics[scale=0.27, angle = 270]{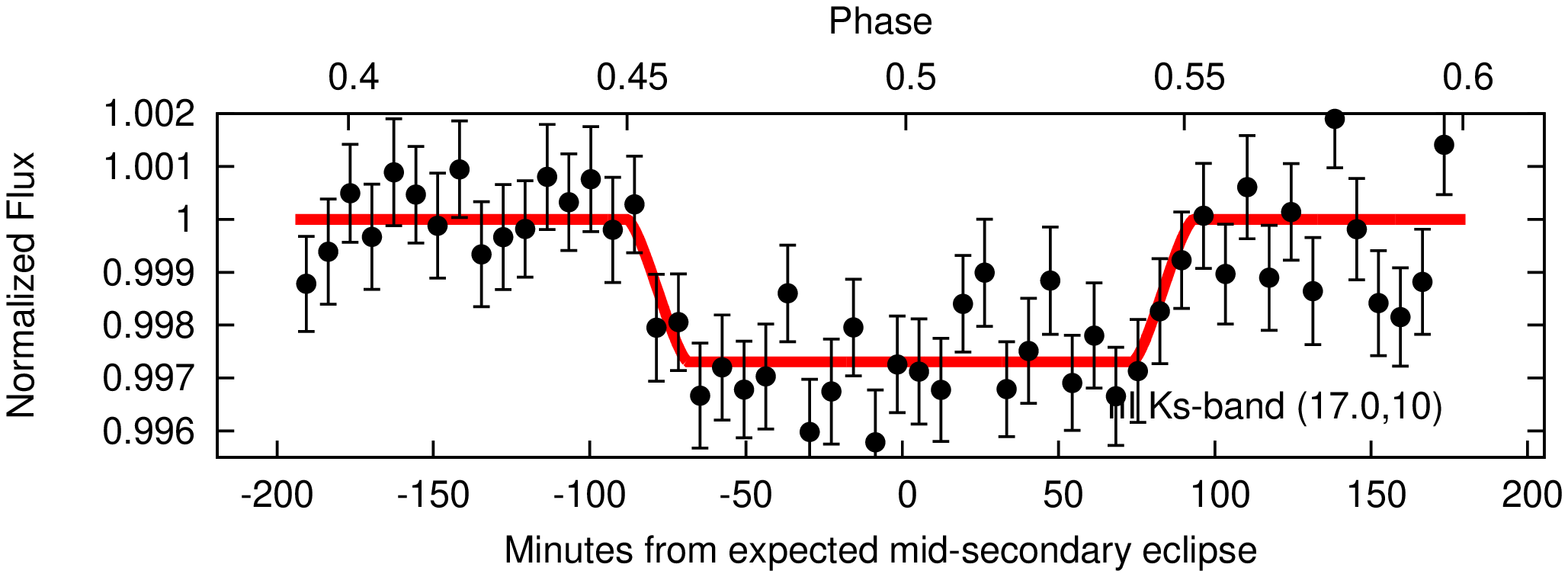}
\includegraphics[scale=0.27, angle = 270]{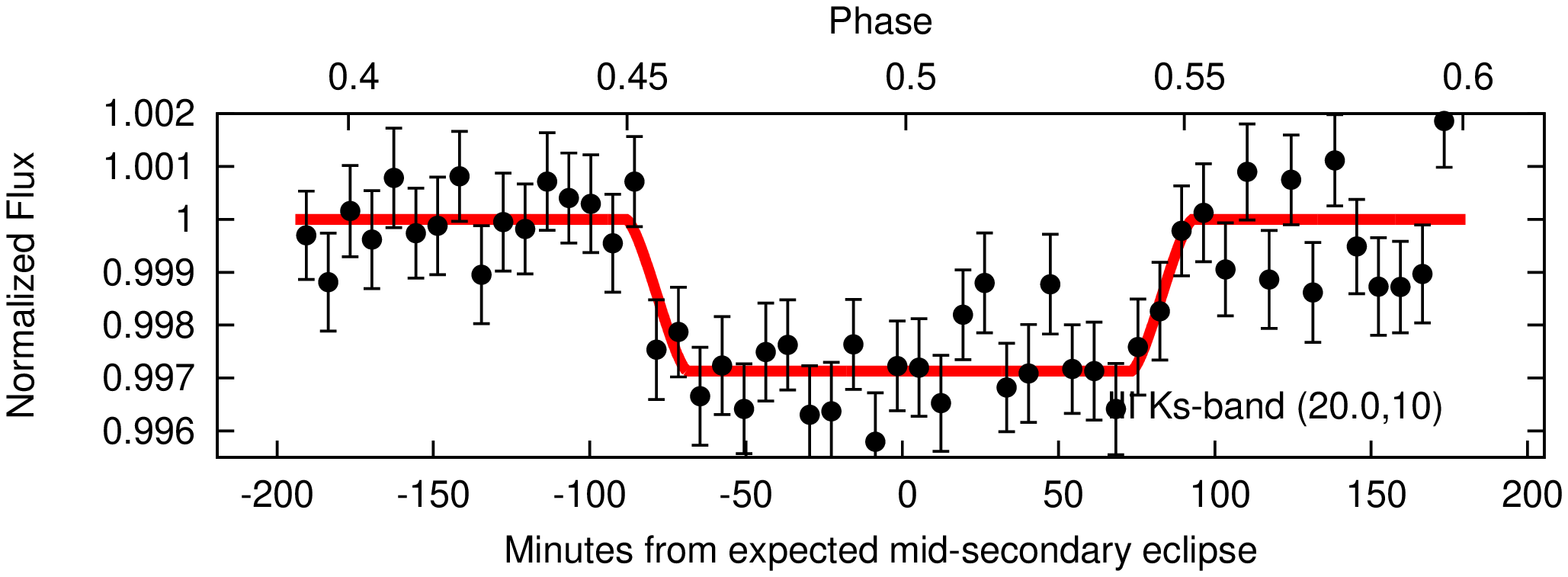}

\includegraphics[scale=0.27, angle = 270]{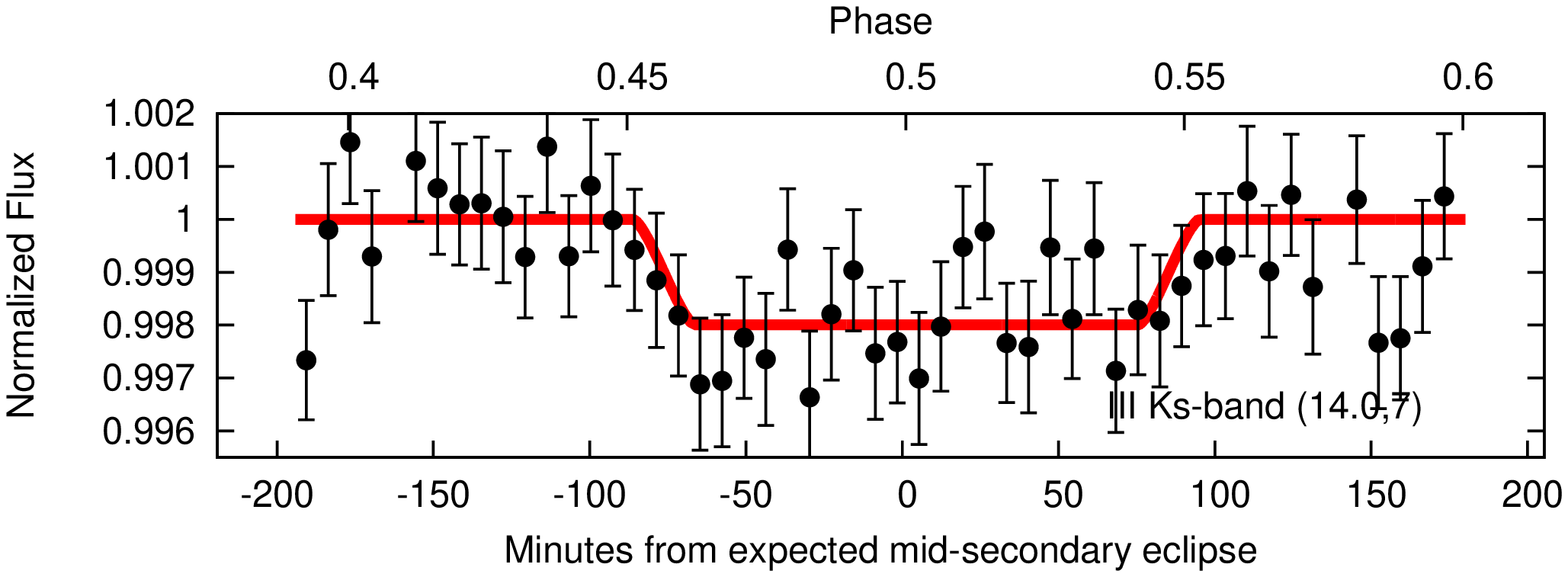}
\includegraphics[scale=0.27, angle = 270]{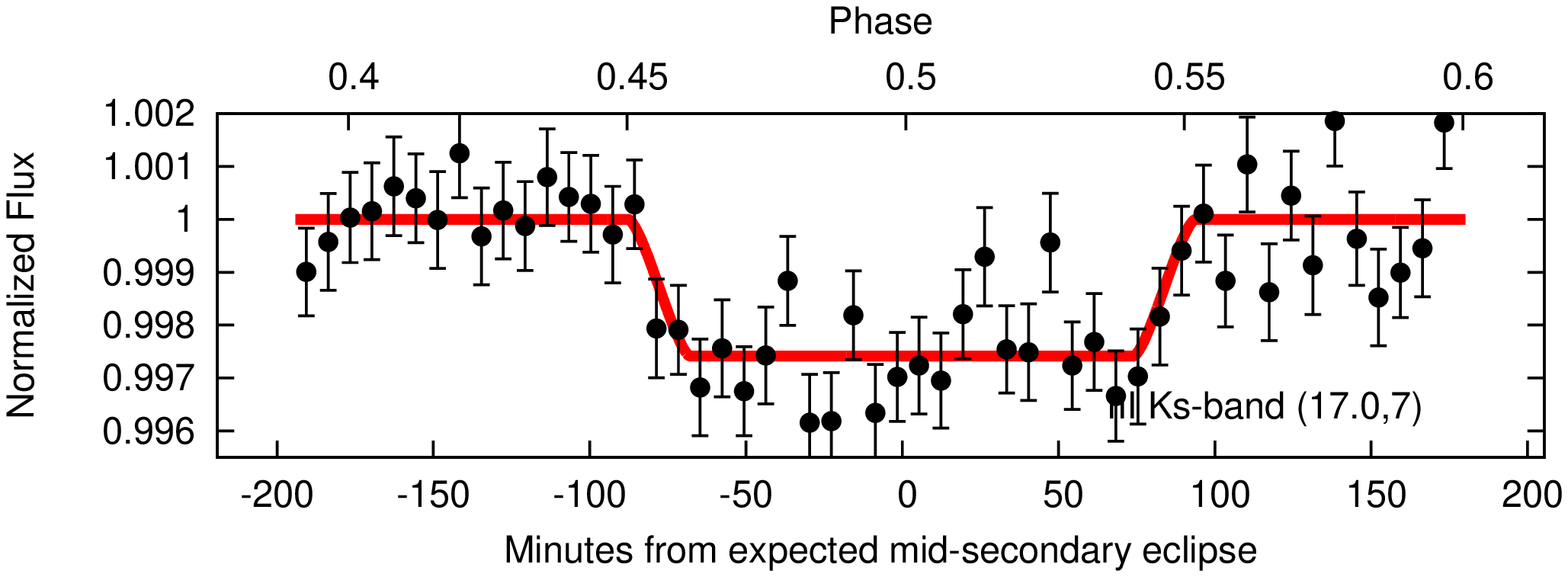}
\includegraphics[scale=0.27, angle = 270]{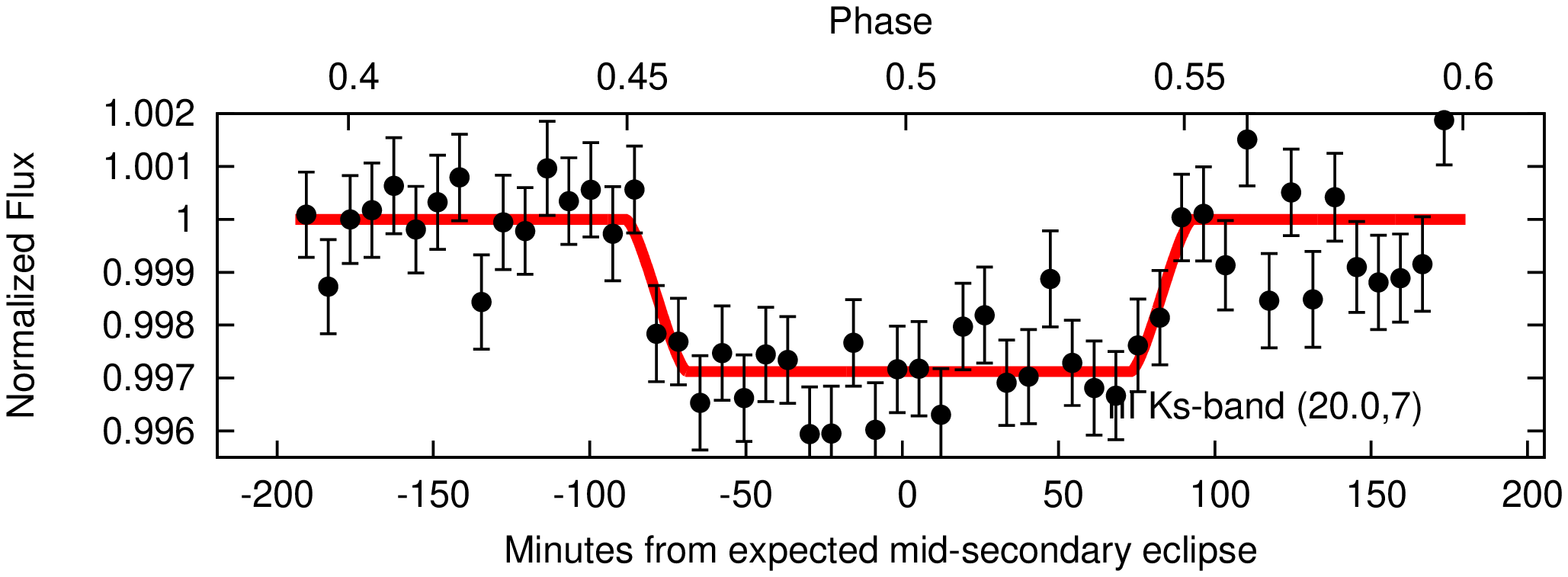}

\includegraphics[scale=0.27, angle = 270]{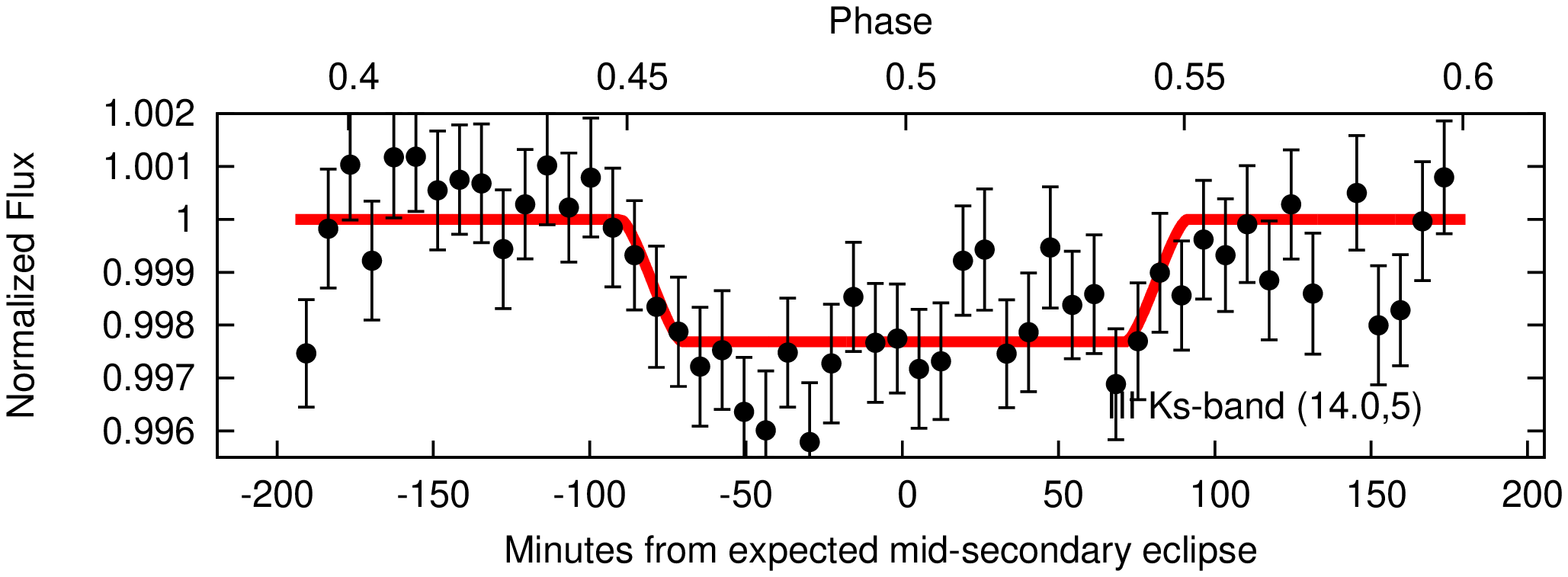}
\includegraphics[scale=0.27, angle = 270]{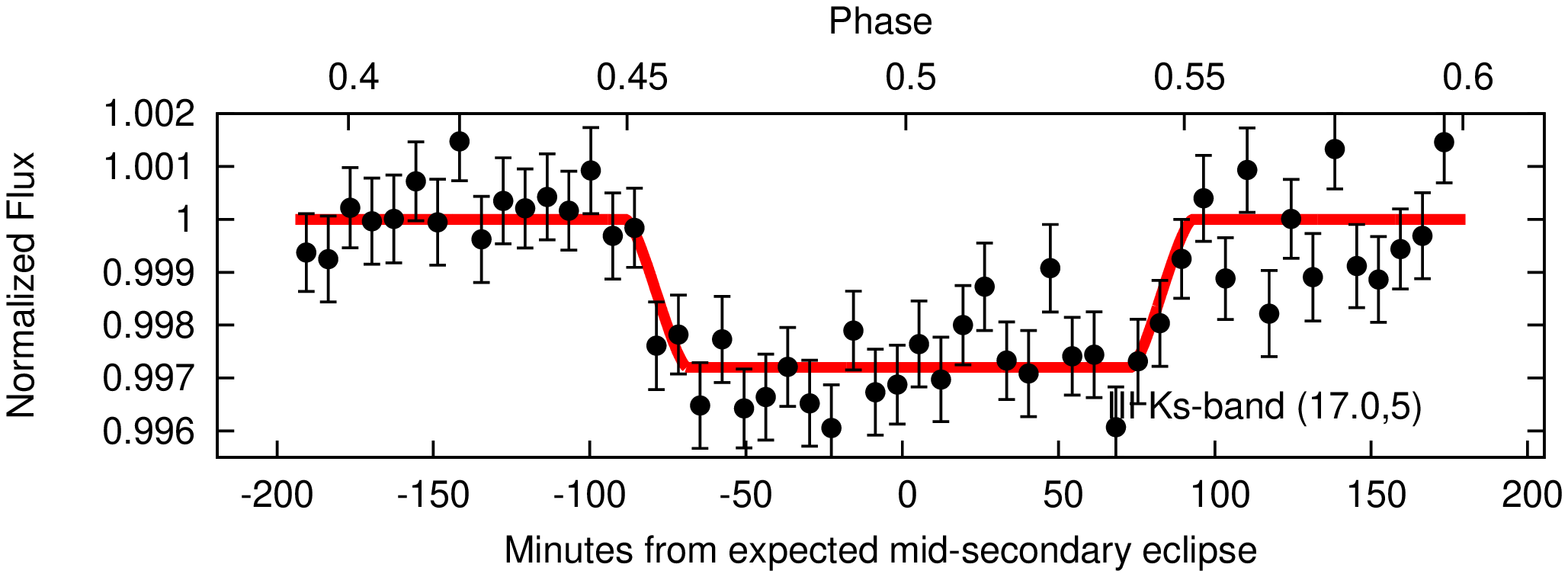}
\includegraphics[scale=0.27, angle = 270]{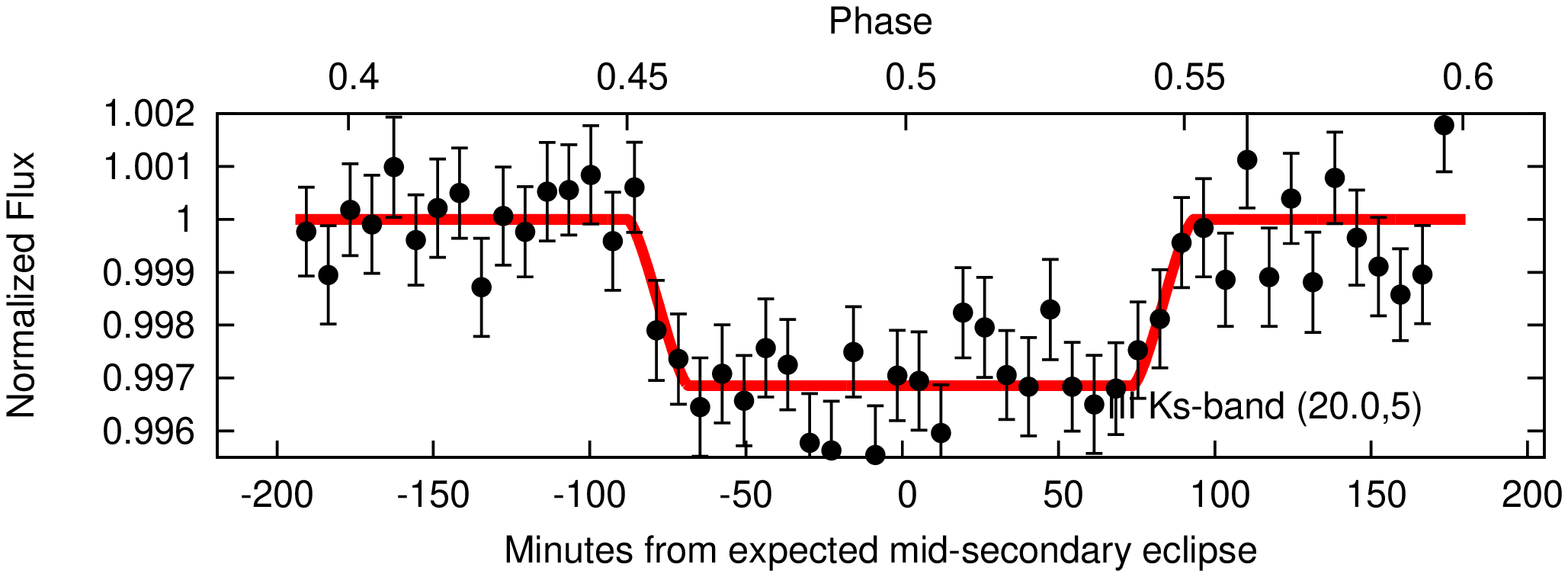}

\includegraphics[scale=0.27, angle = 270]{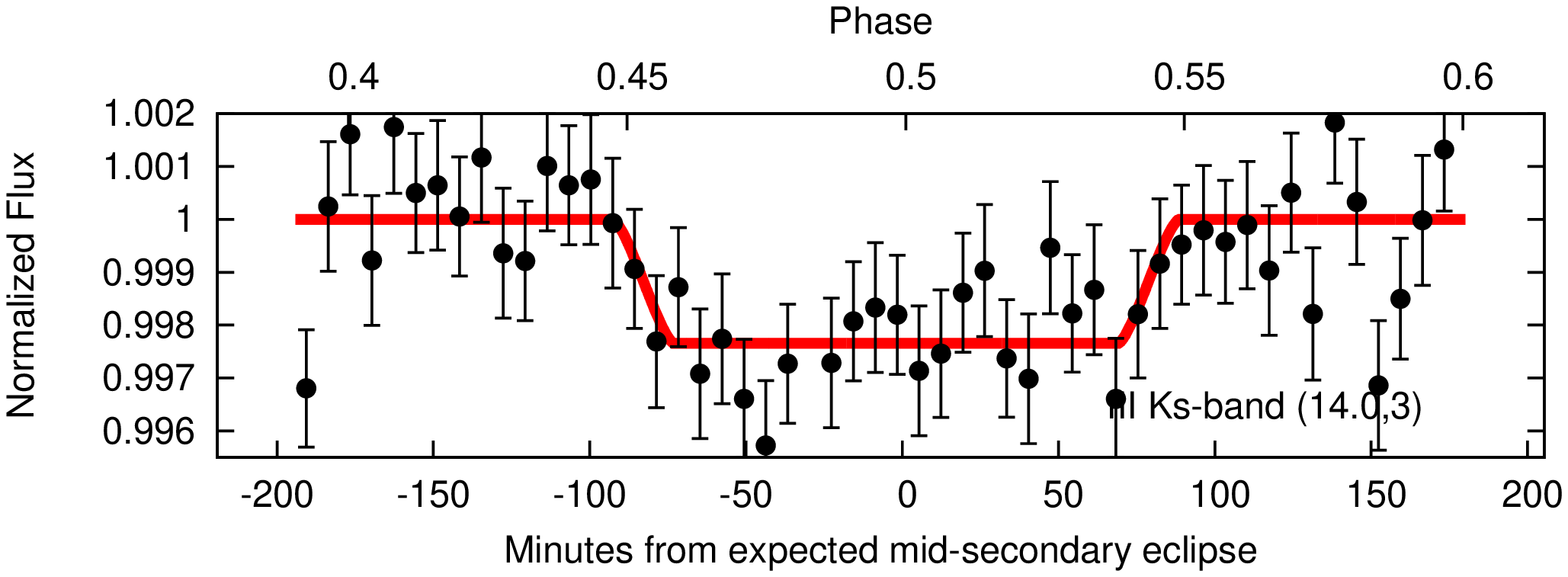}
\includegraphics[scale=0.27, angle = 270]{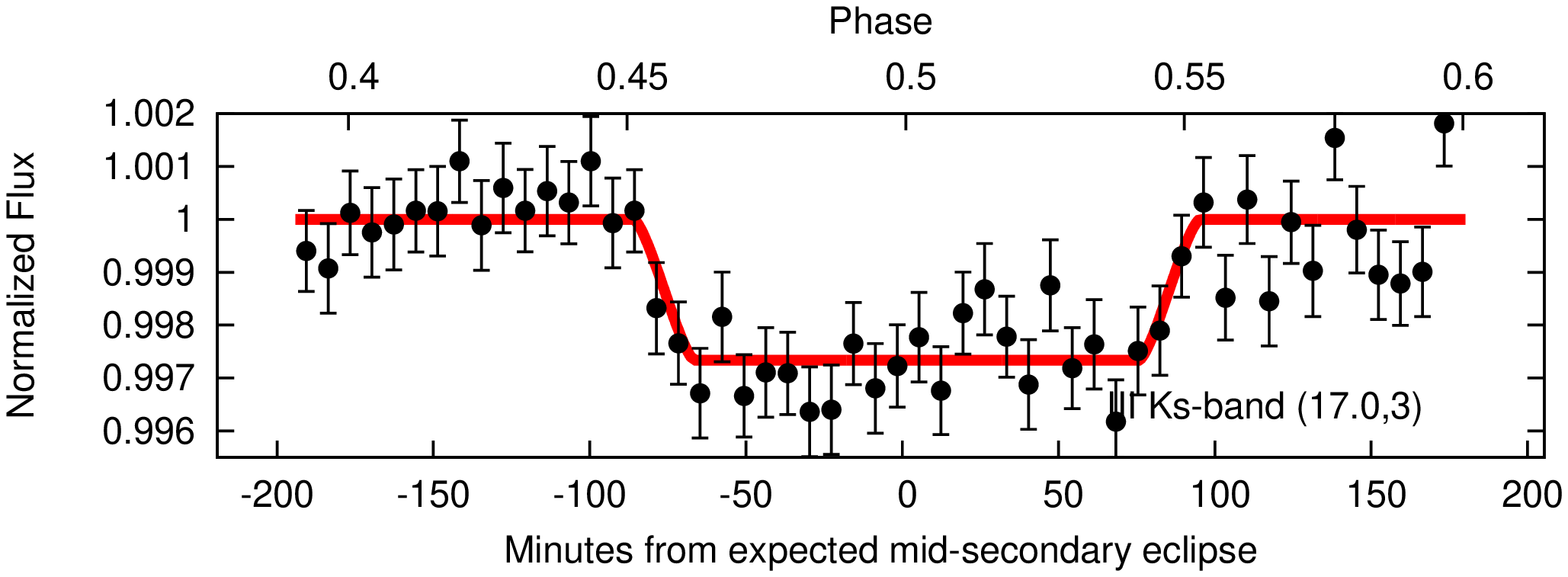}
\includegraphics[scale=0.27, angle = 270]{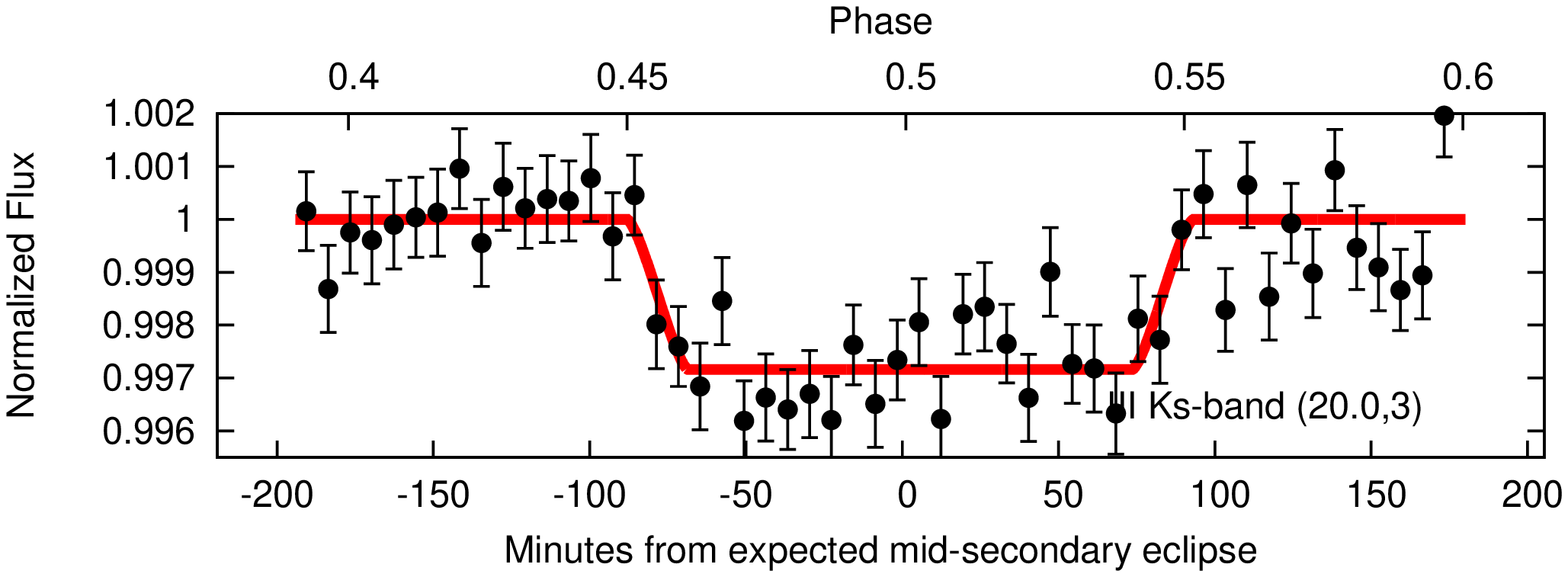}

\includegraphics[scale=0.27, angle = 270]{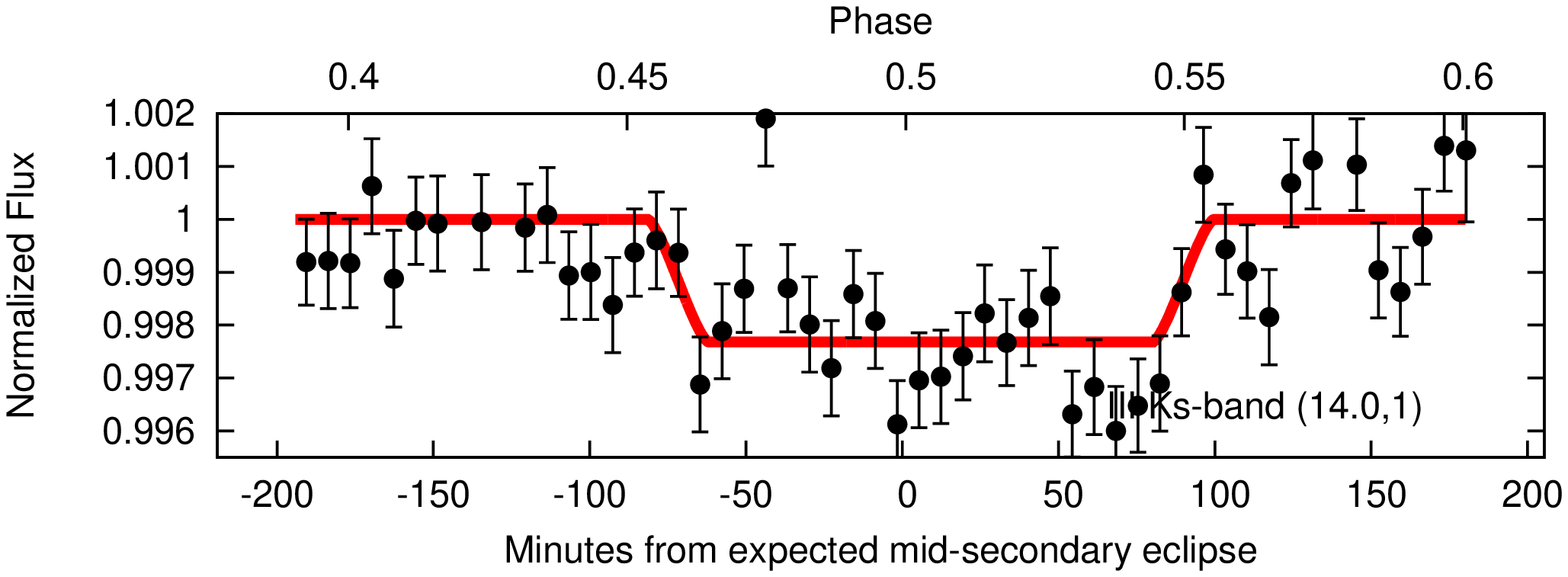}
\includegraphics[scale=0.27, angle = 270]{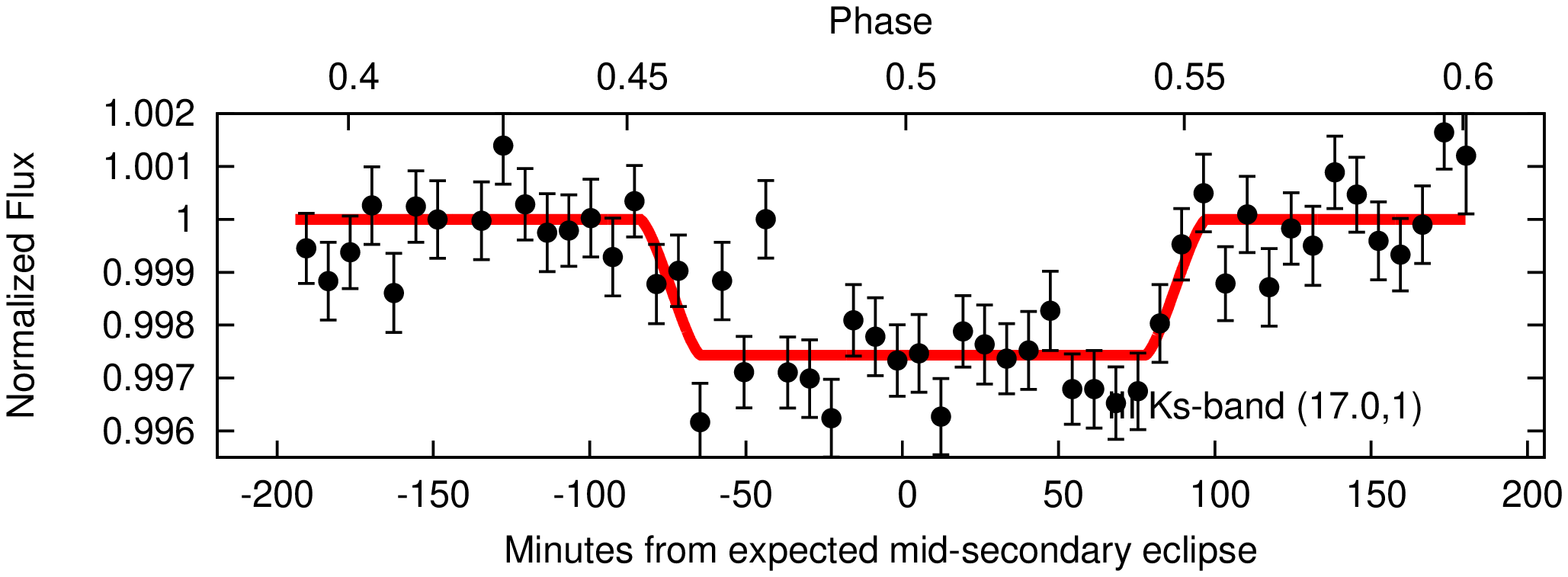}
\includegraphics[scale=0.27, angle = 270]{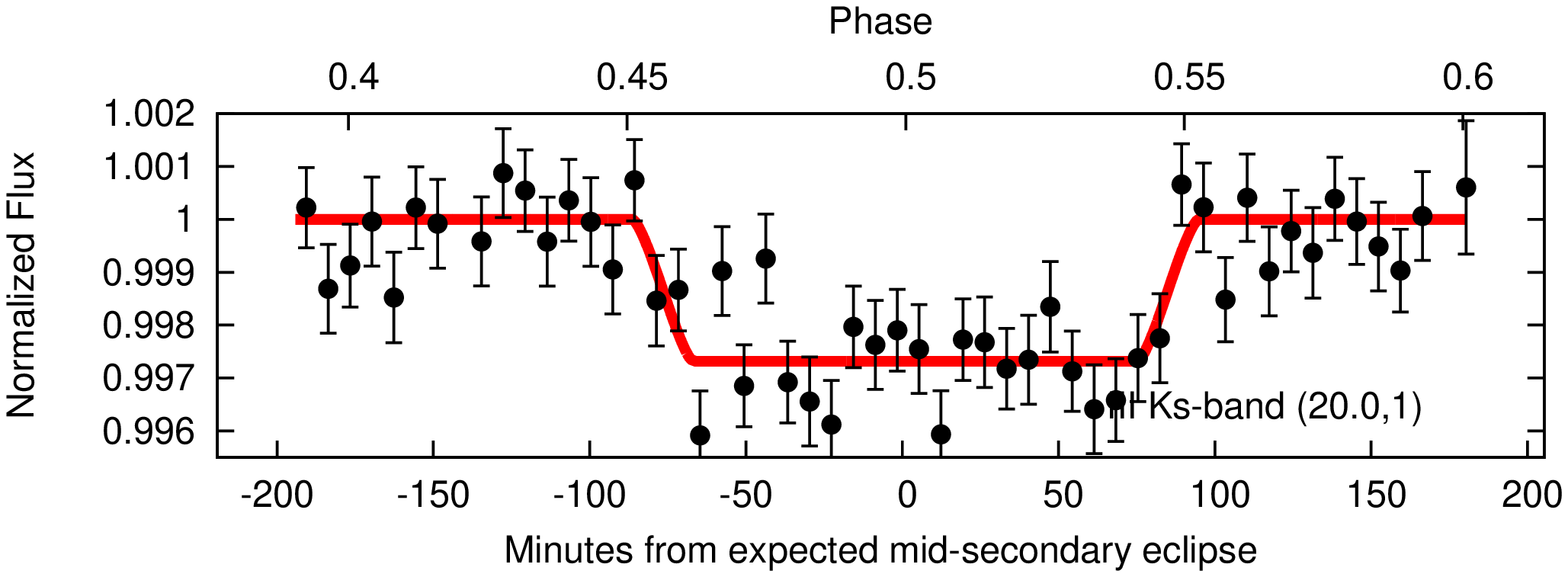}

\caption[WASP-12 Three Fidelity Many]
	{	
		Same as Figure \ref{FigWASPTwelveKsbandFidelityManyOne} except for our third WASP-12 Ks-band secondary eclipse.
		The scale of the bottom panels is identical to that of the other WASP-12 Ks-band
		eclipses (Figure \ref{FigWASPTwelveKsbandFidelityManyOne} \& \ref{FigWASPTwelveKsbandFidelityManyTwo}).
	}
\label{FigWASPTwelveKsbandFidelityManyThree}
\end{figure*}

\begin{figure*}
\centering

\includegraphics[scale=0.44, angle = 270]{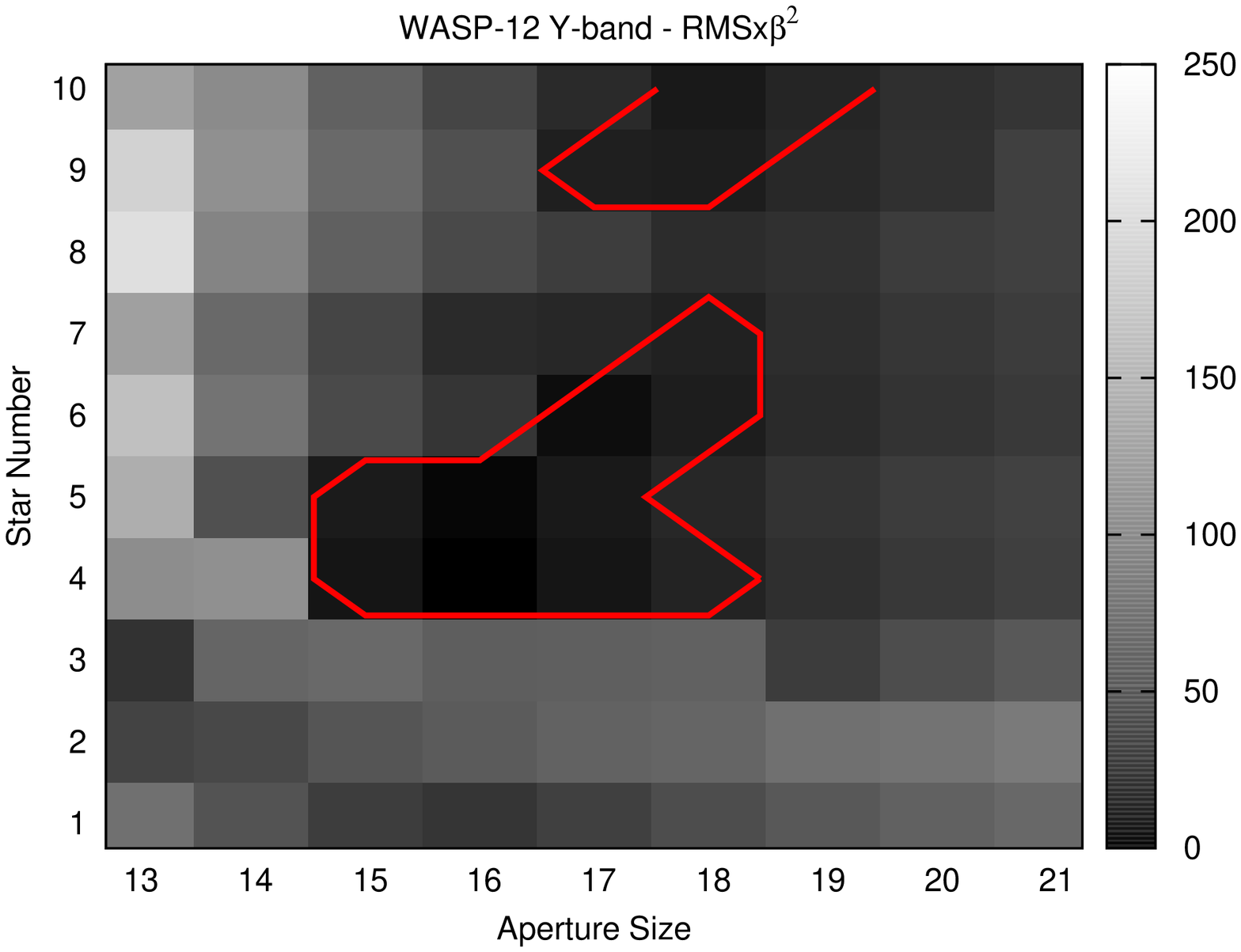}
\includegraphics[scale=0.44, angle = 270]{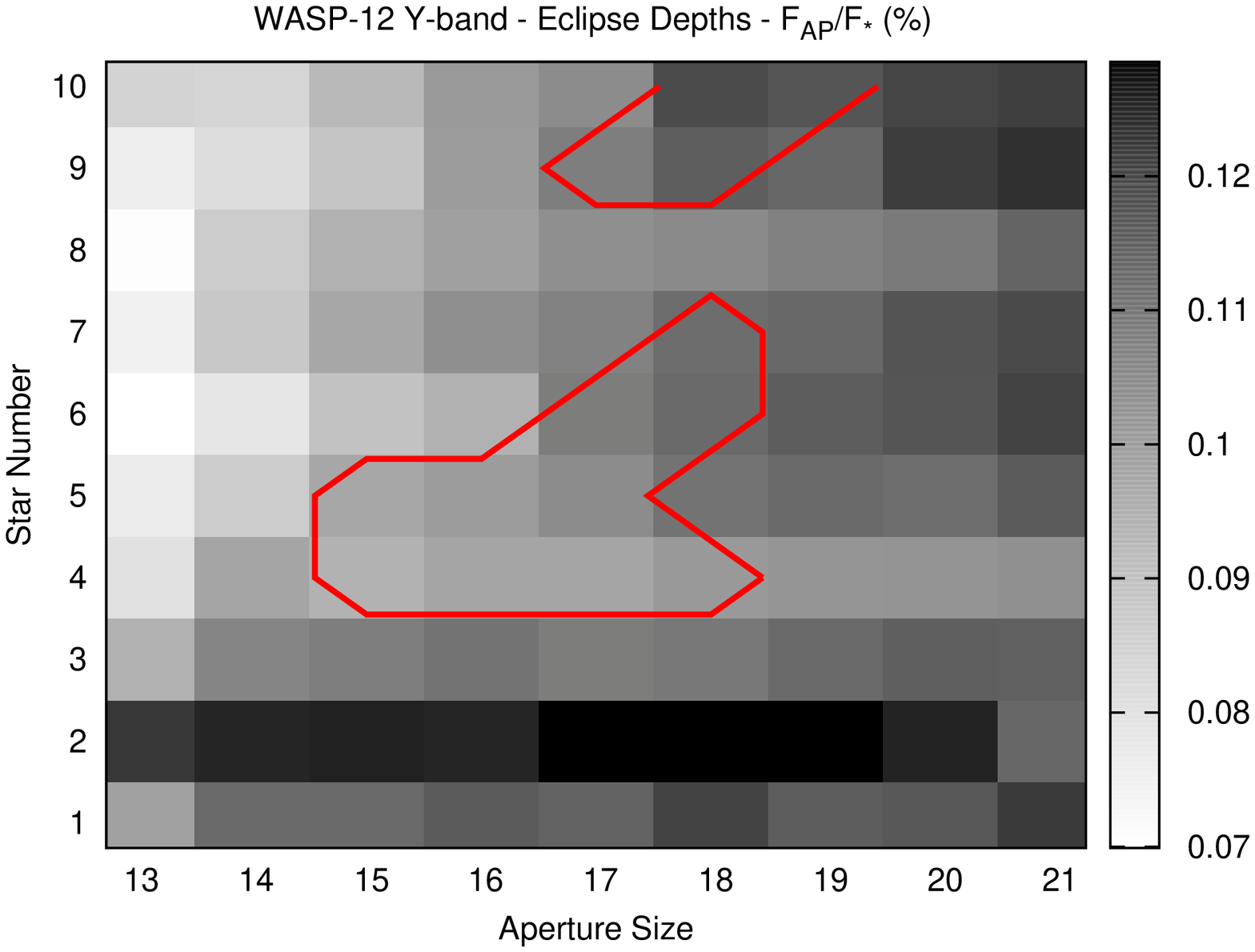}

\includegraphics[scale=0.27, angle = 270]{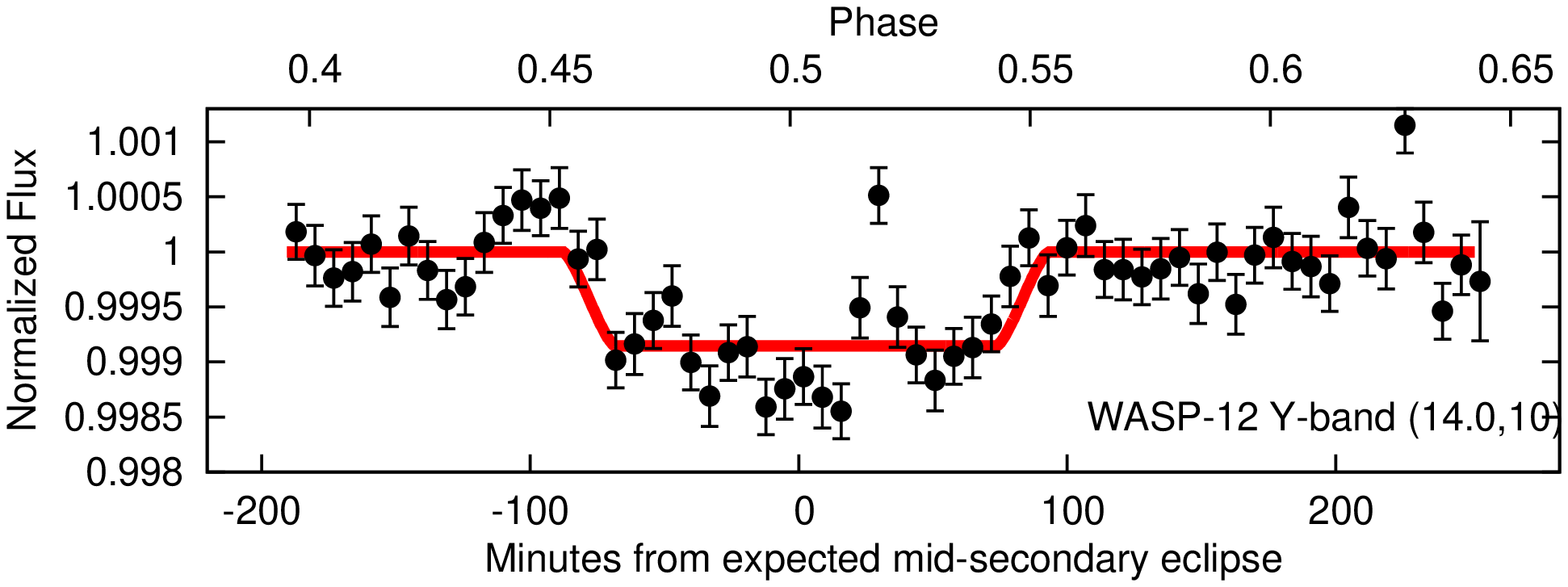}
\includegraphics[scale=0.27, angle = 270]{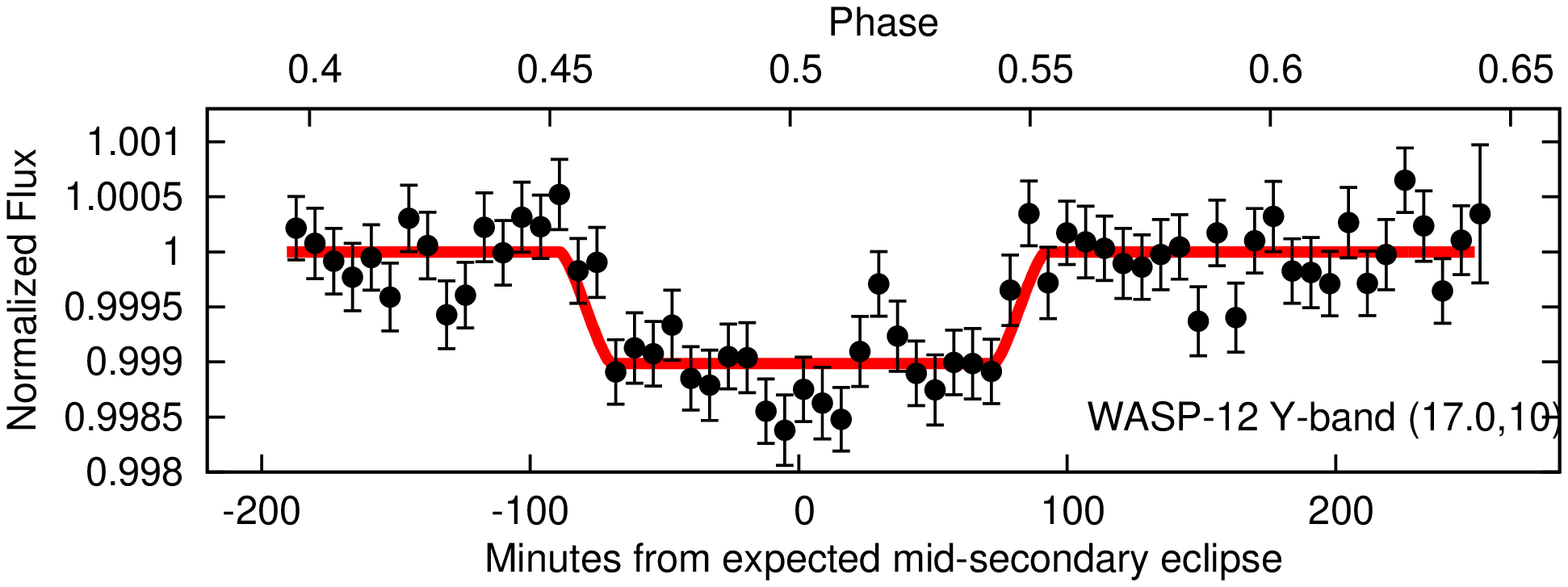}
\includegraphics[scale=0.27, angle = 270]{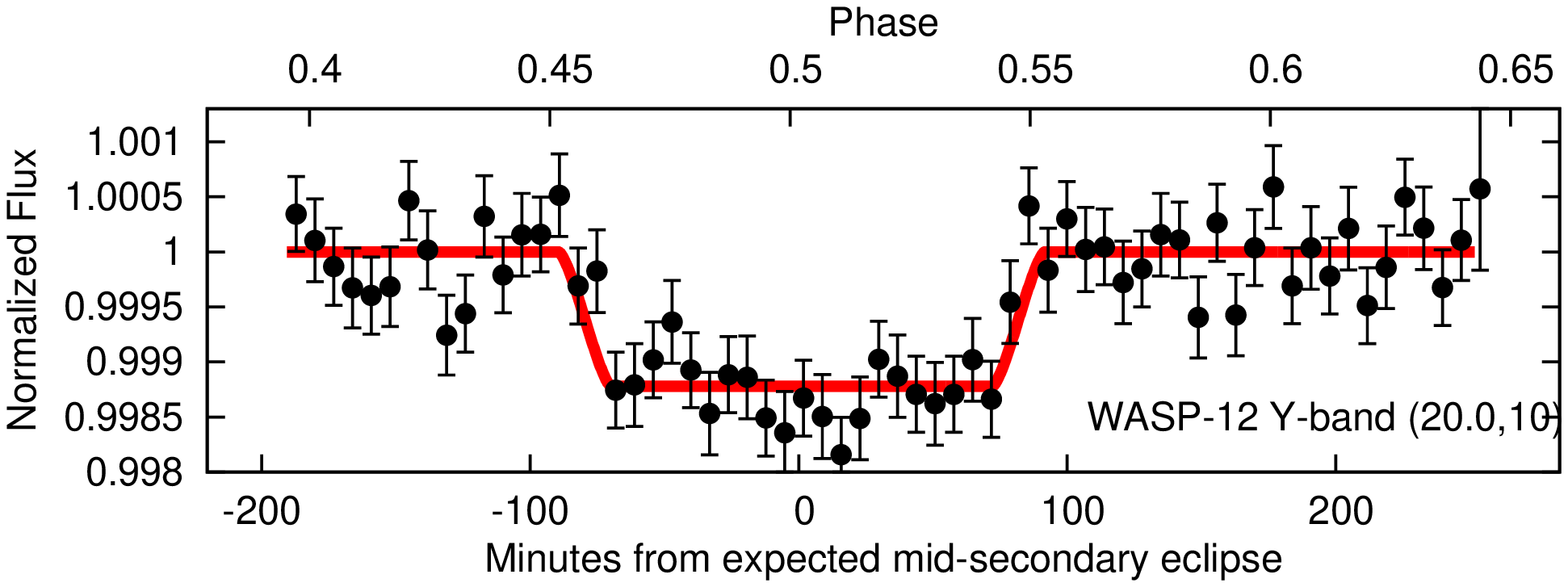}

\includegraphics[scale=0.27, angle = 270]{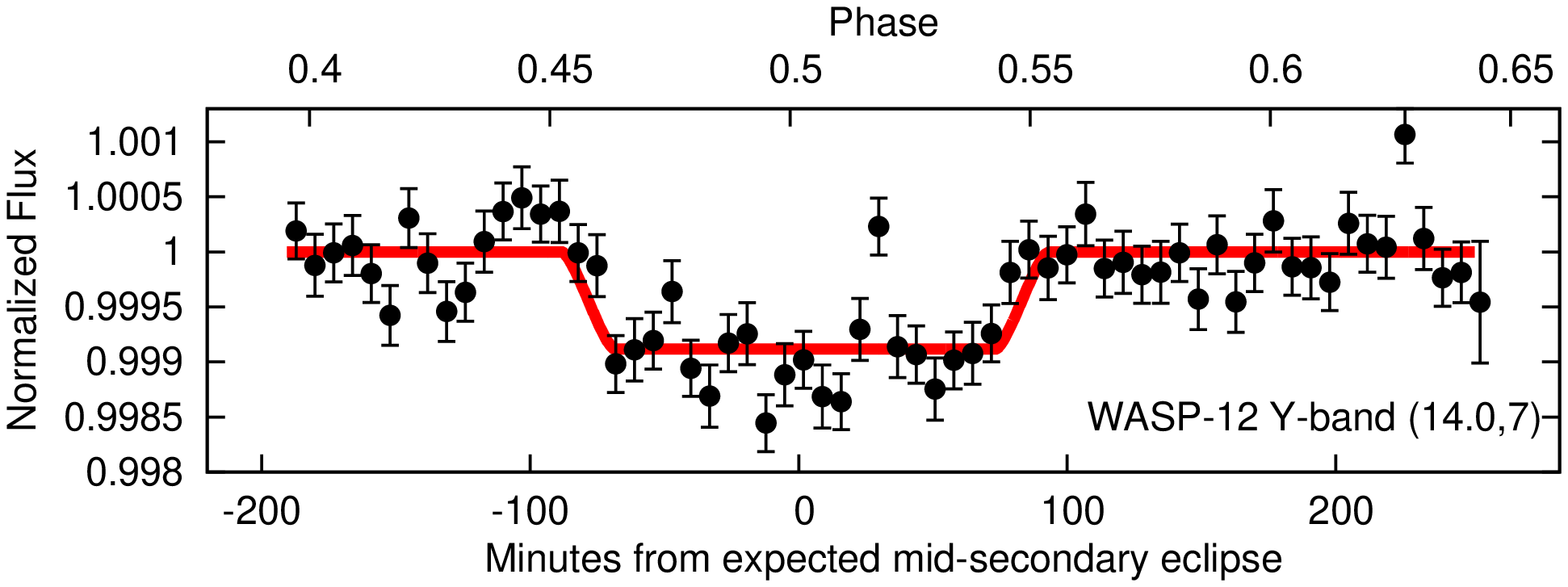}
\includegraphics[scale=0.27, angle = 270]{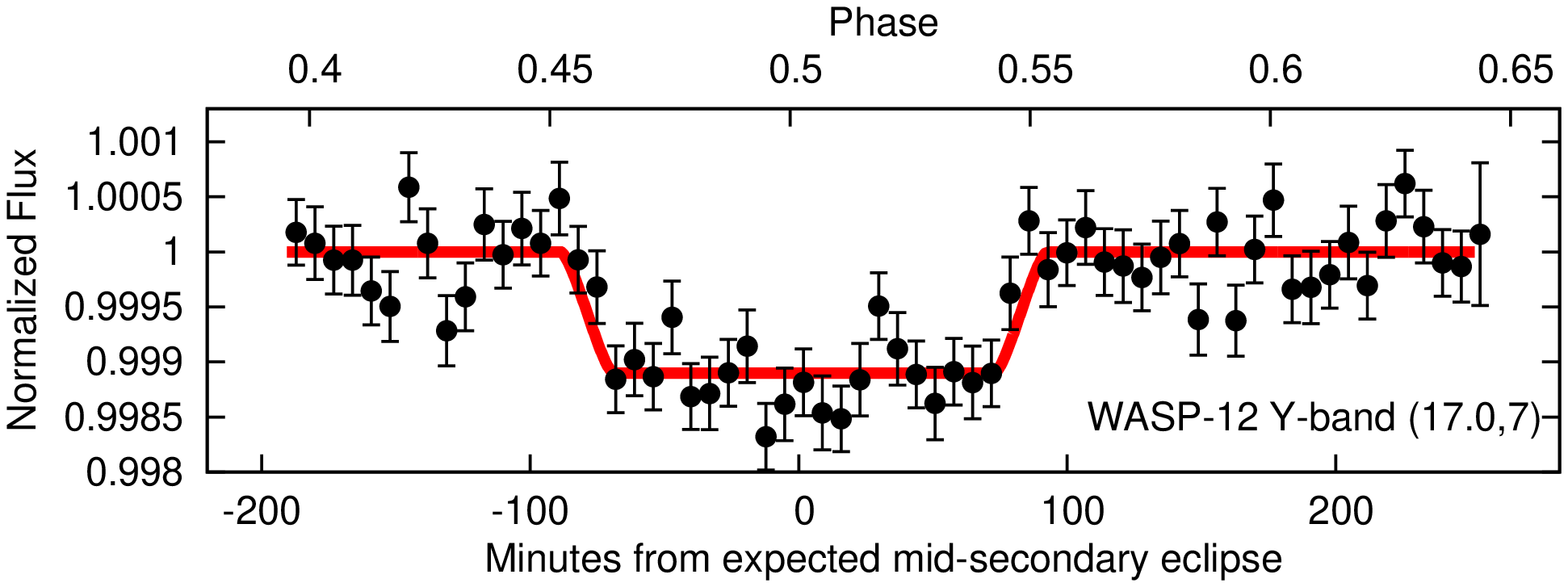}
\includegraphics[scale=0.27, angle = 270]{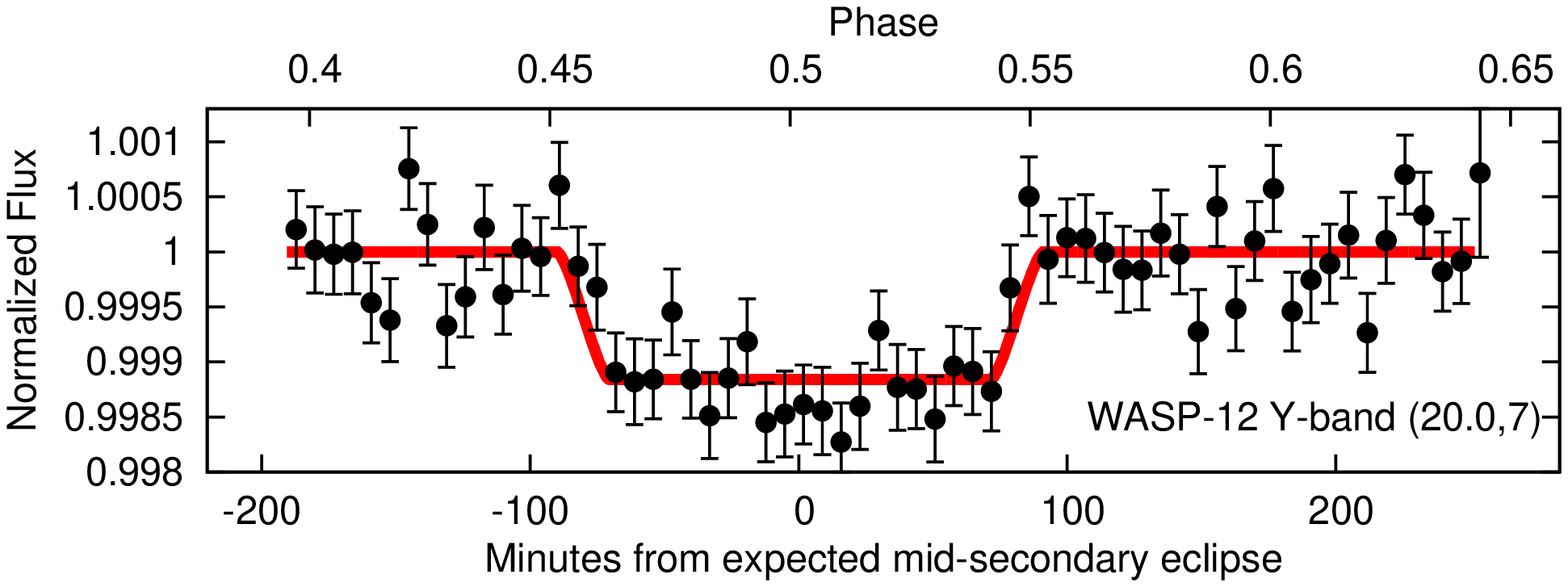}

\includegraphics[scale=0.27, angle = 270]{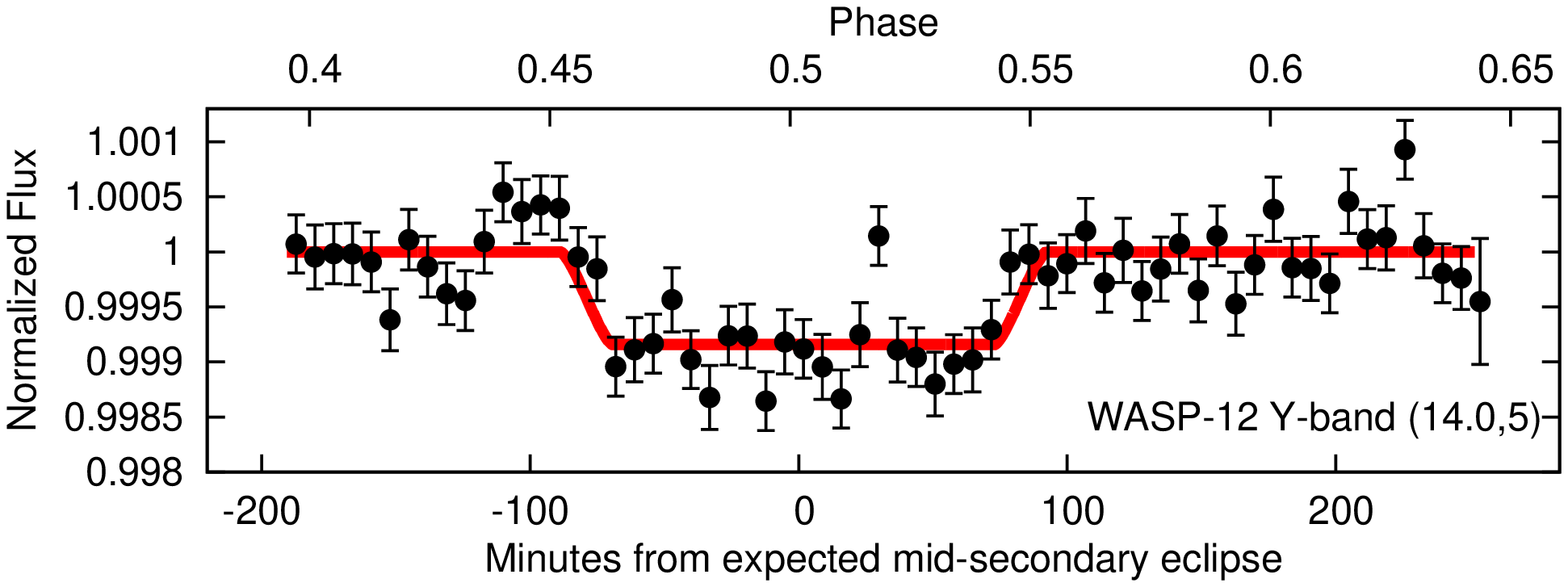}
\includegraphics[scale=0.27, angle = 270]{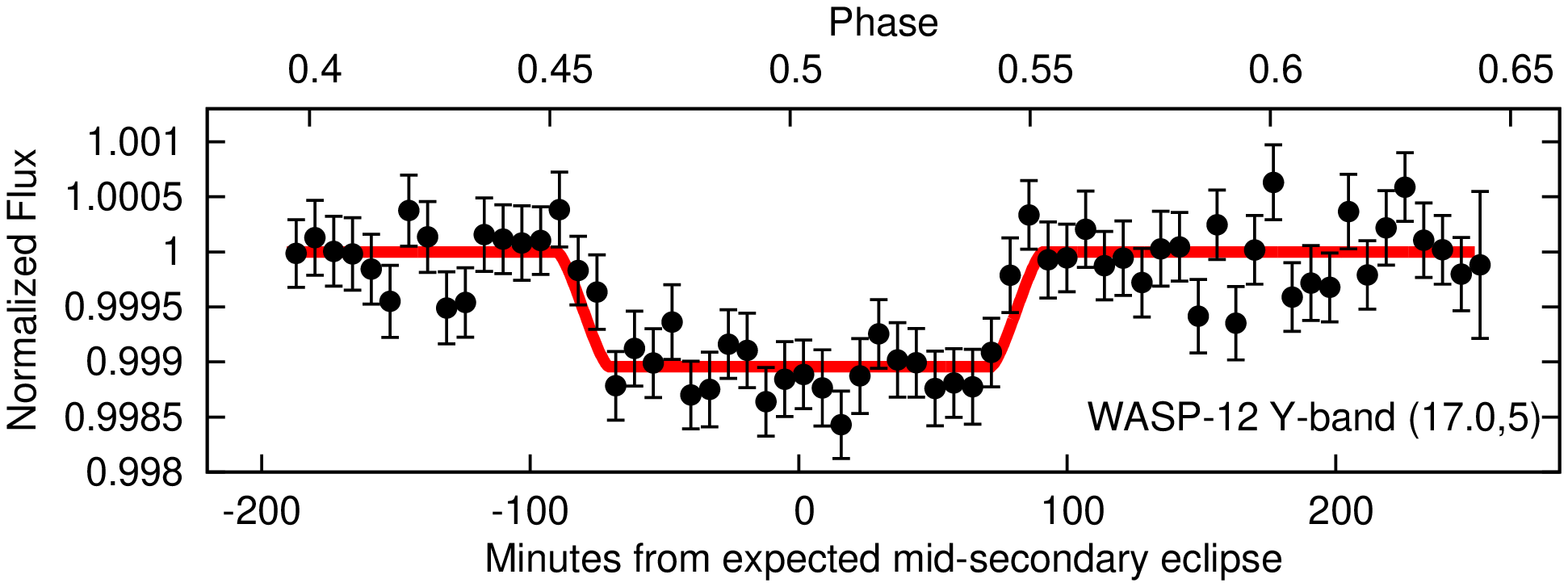}
\includegraphics[scale=0.27, angle = 270]{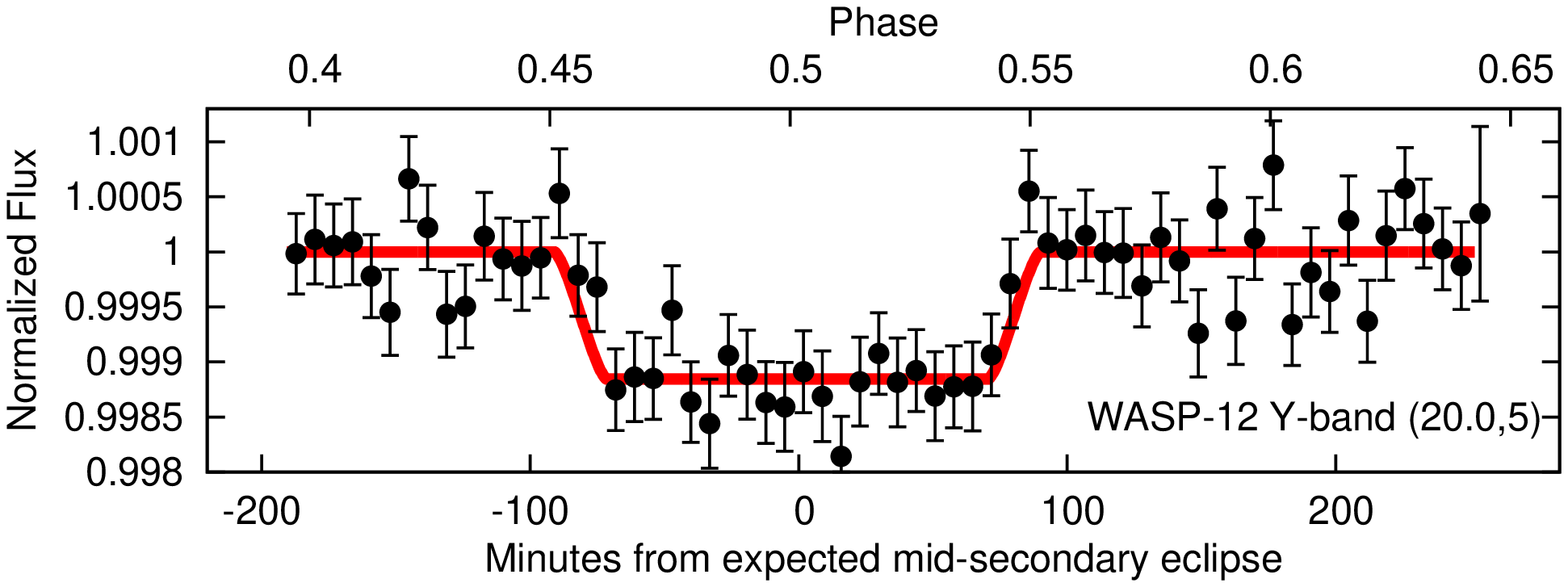}

\includegraphics[scale=0.27, angle = 270]{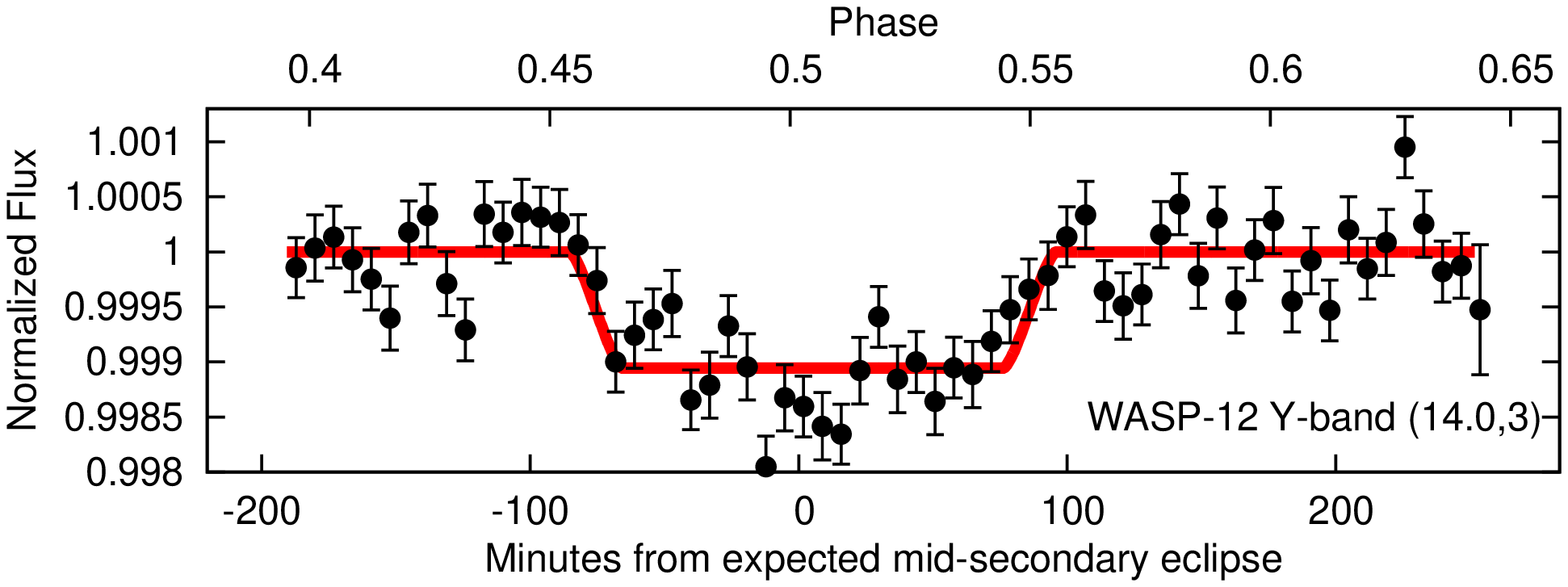}
\includegraphics[scale=0.27, angle = 270]{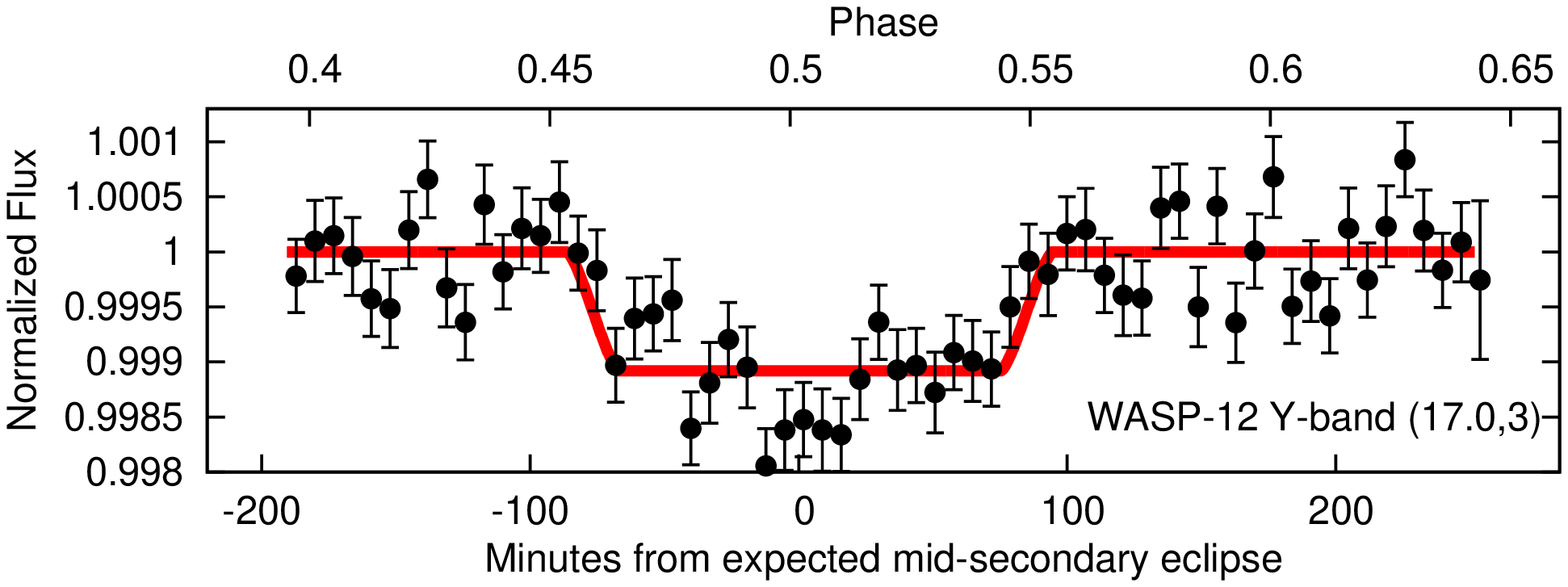}
\includegraphics[scale=0.27, angle = 270]{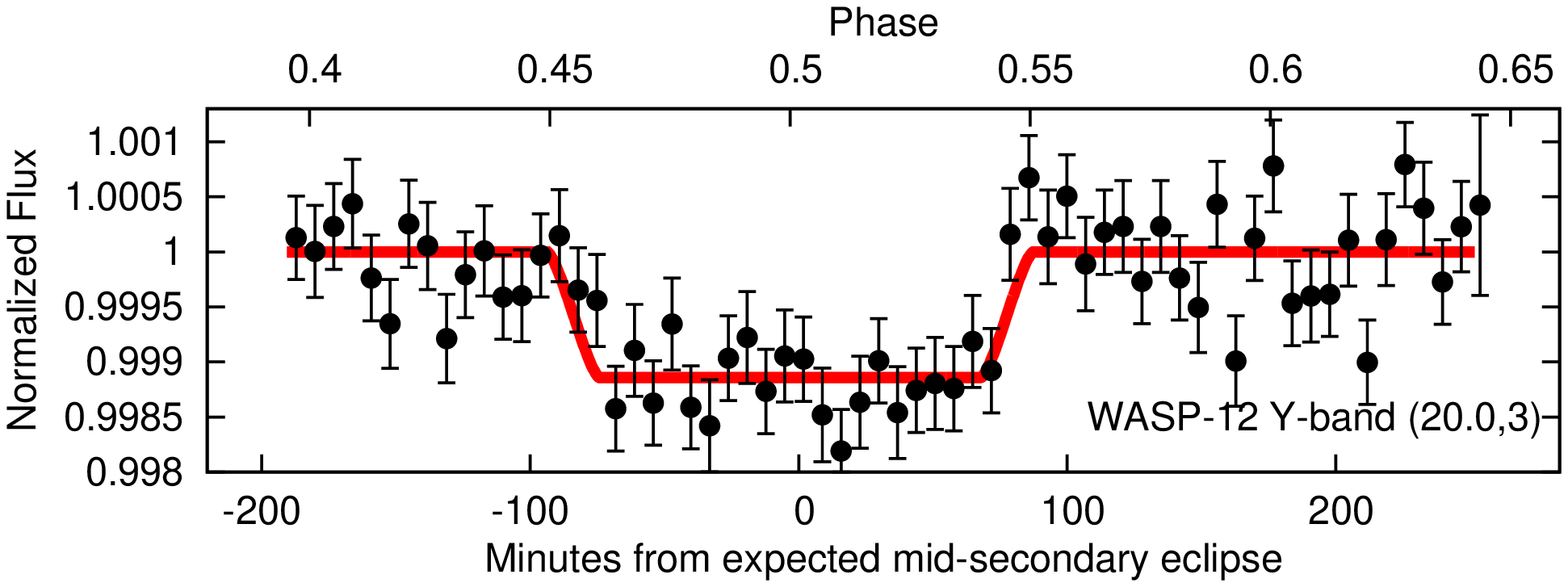}

\includegraphics[scale=0.27, angle = 270]{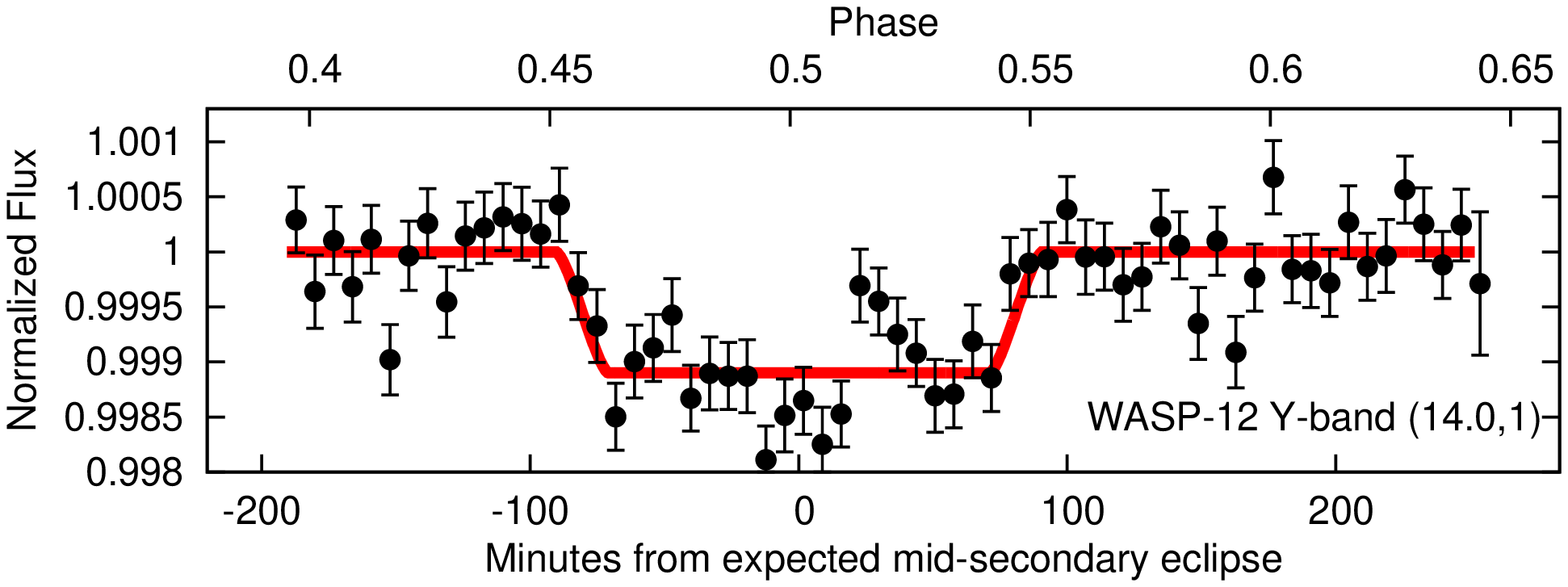}
\includegraphics[scale=0.27, angle = 270]{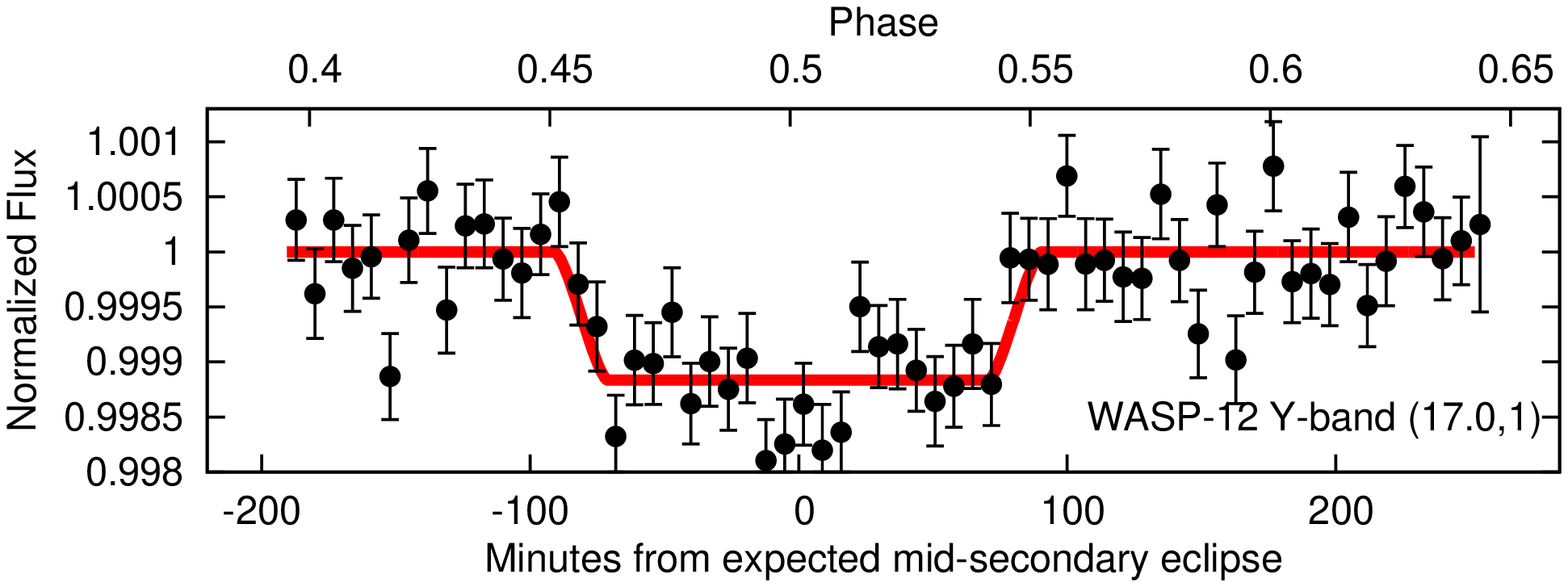}
\includegraphics[scale=0.27, angle = 270]{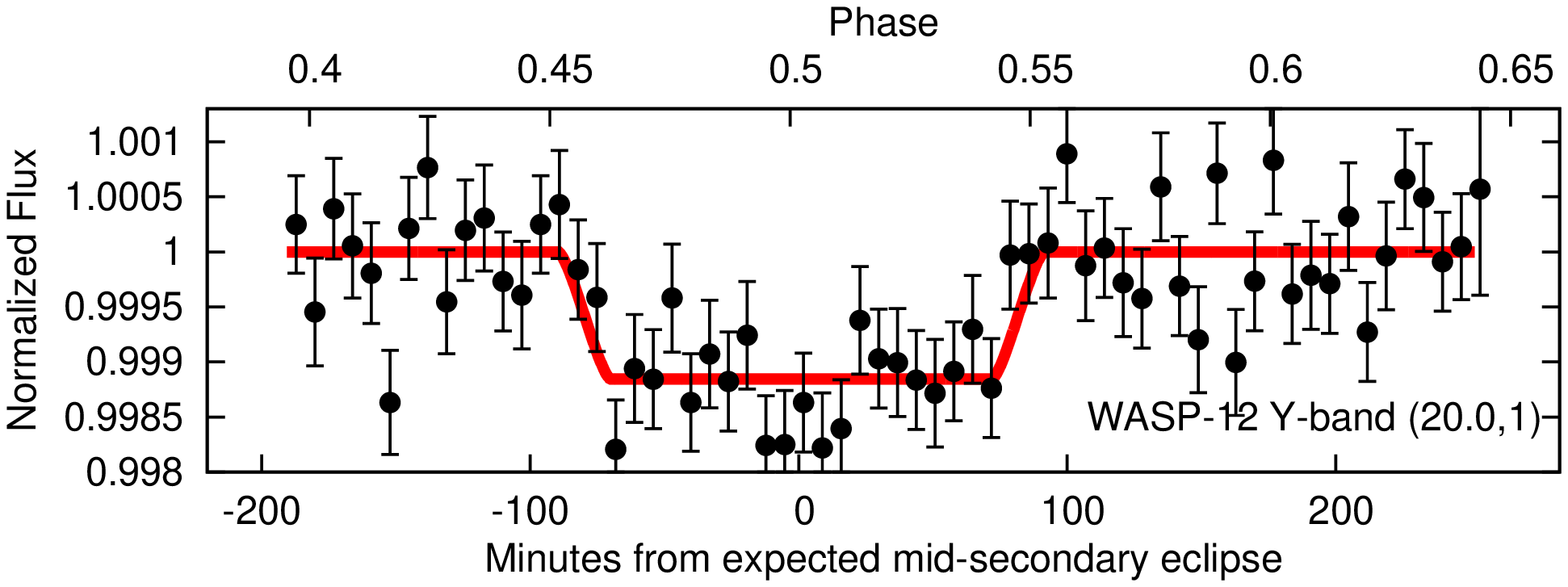}

\caption[WASP-12 Y Fidelity Many]
	{	
		Same as Figure \ref{FigWASPTwelveKsbandFidelityManyOne} except for our WASP-12 Y-band secondary eclipse.
		For our WASP-12 Y-band eclipse, the smallest apertures display correlated noise (the left set of panels for the bottom plots).	
	}
\label{FigWASPTwelveYbandFidelityMany}
\end{figure*}

\begin{figure*}
\centering

\includegraphics[scale=0.44, angle = 270]{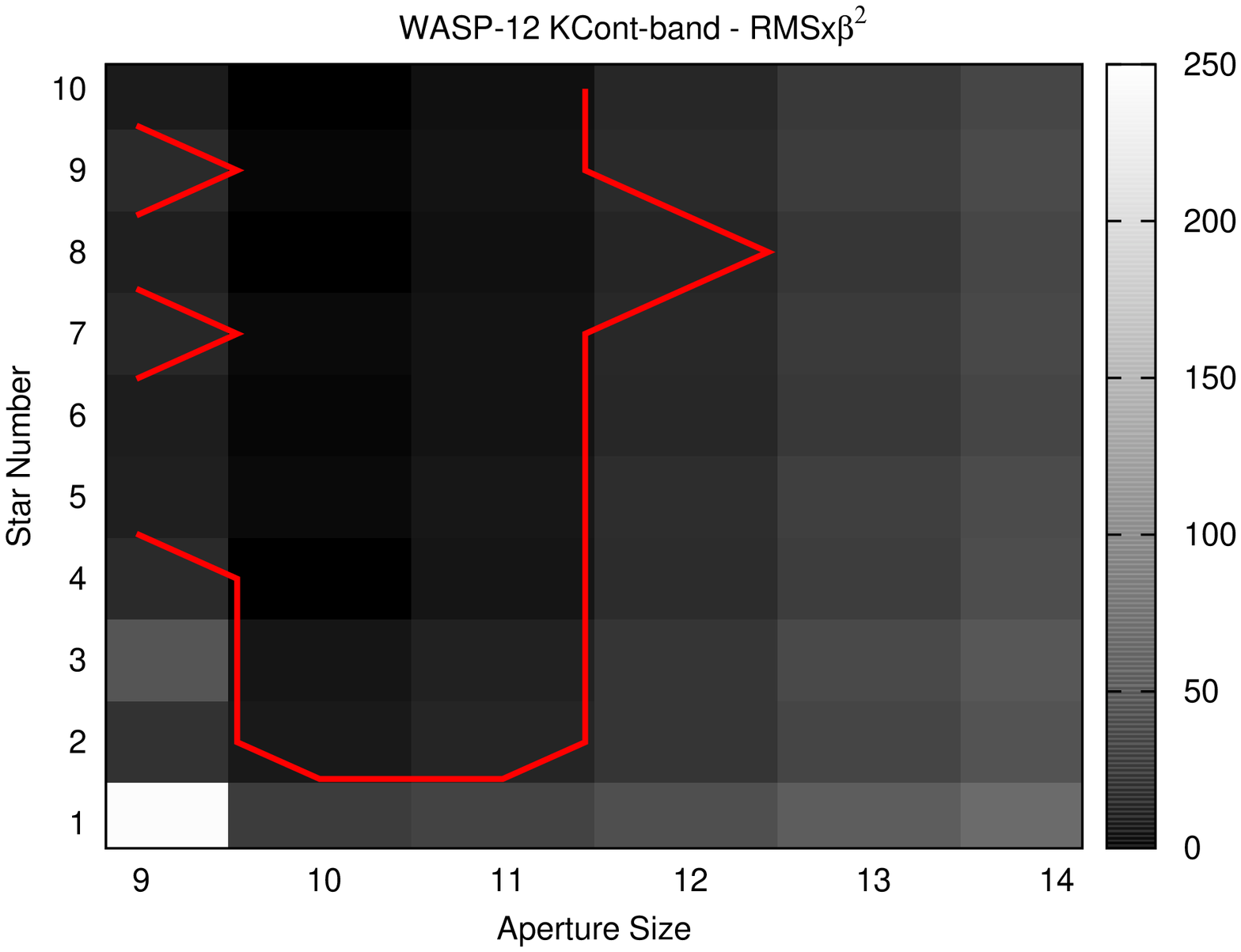}
\includegraphics[scale=0.44, angle = 270]{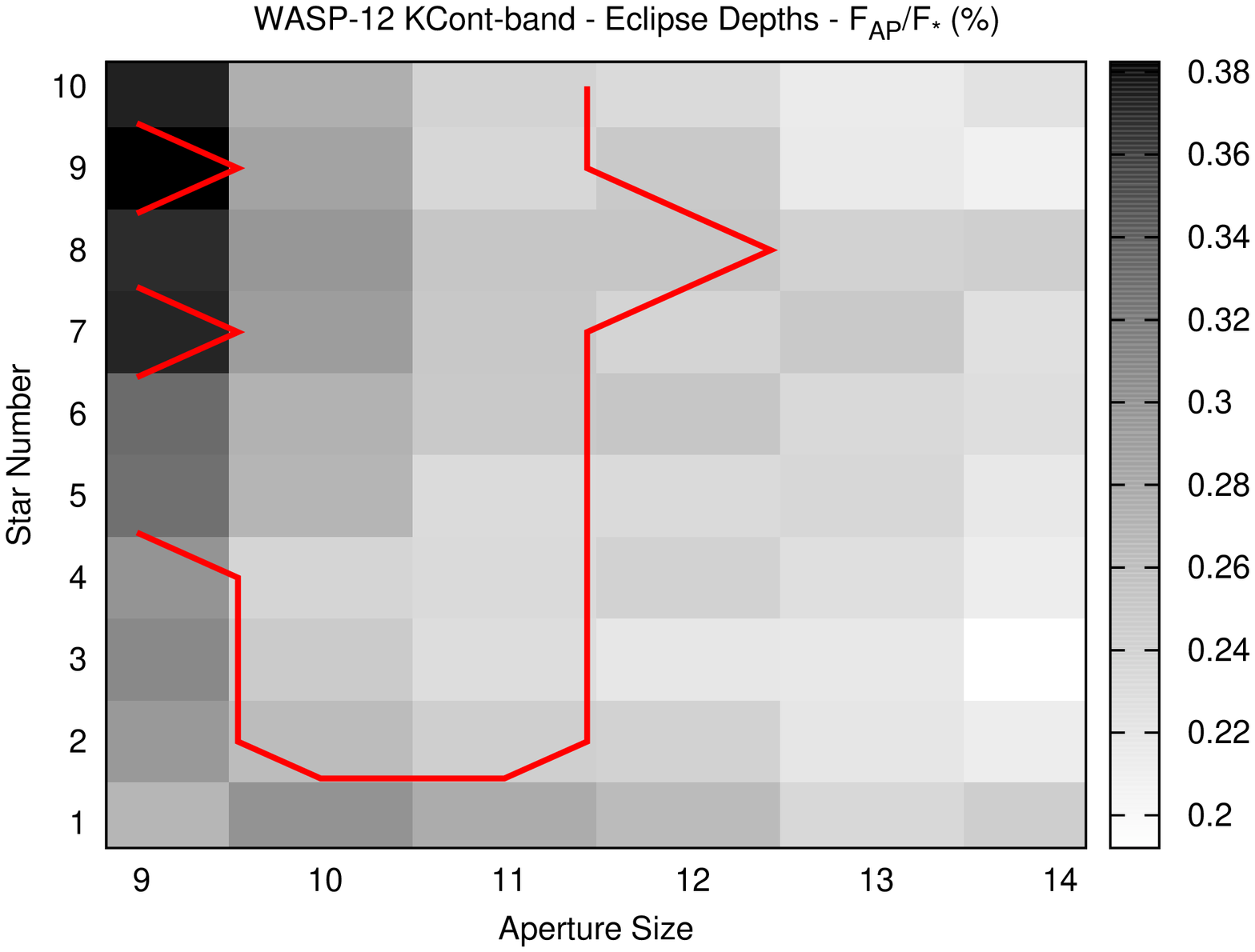}

\includegraphics[scale=0.27, angle = 270]{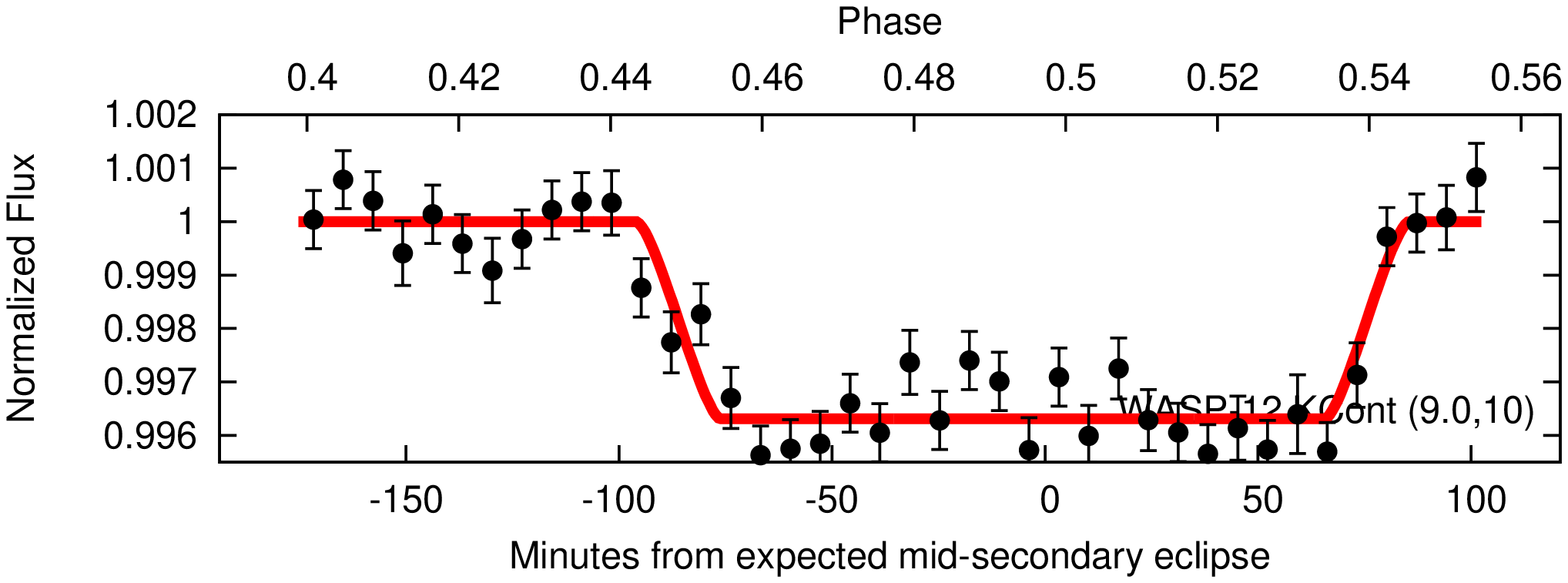}
\includegraphics[scale=0.27, angle = 270]{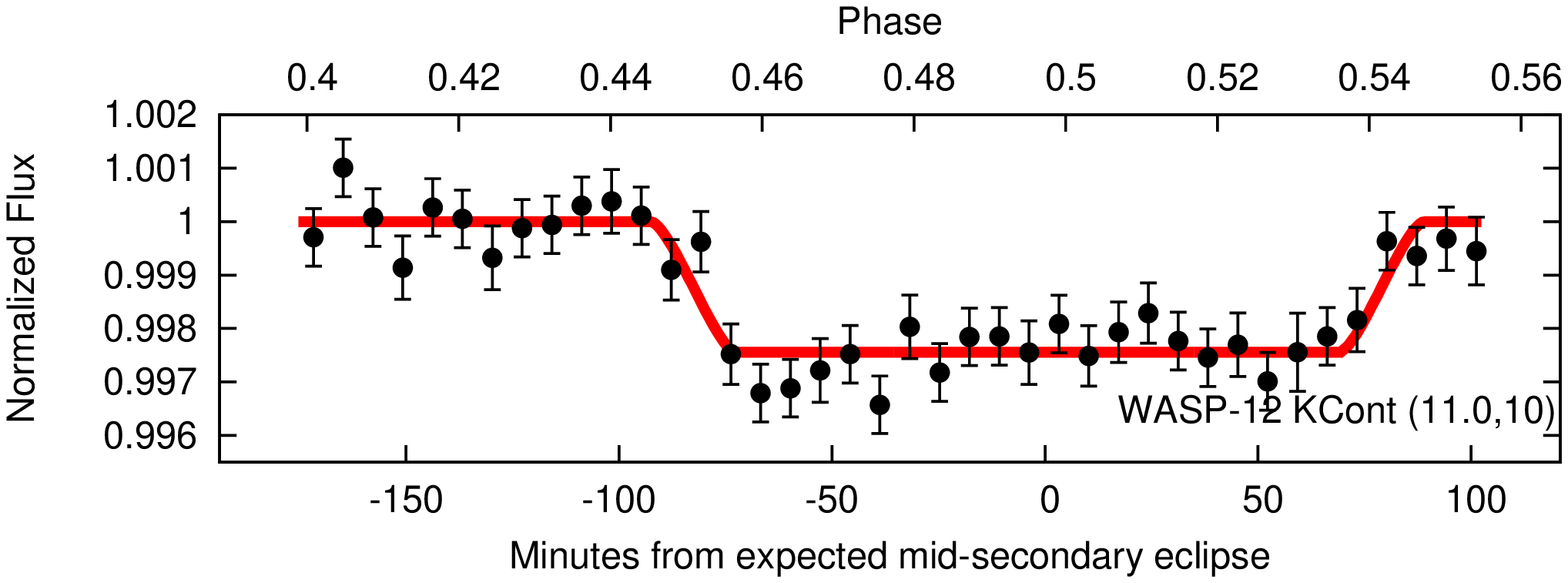}
\includegraphics[scale=0.27, angle = 270]{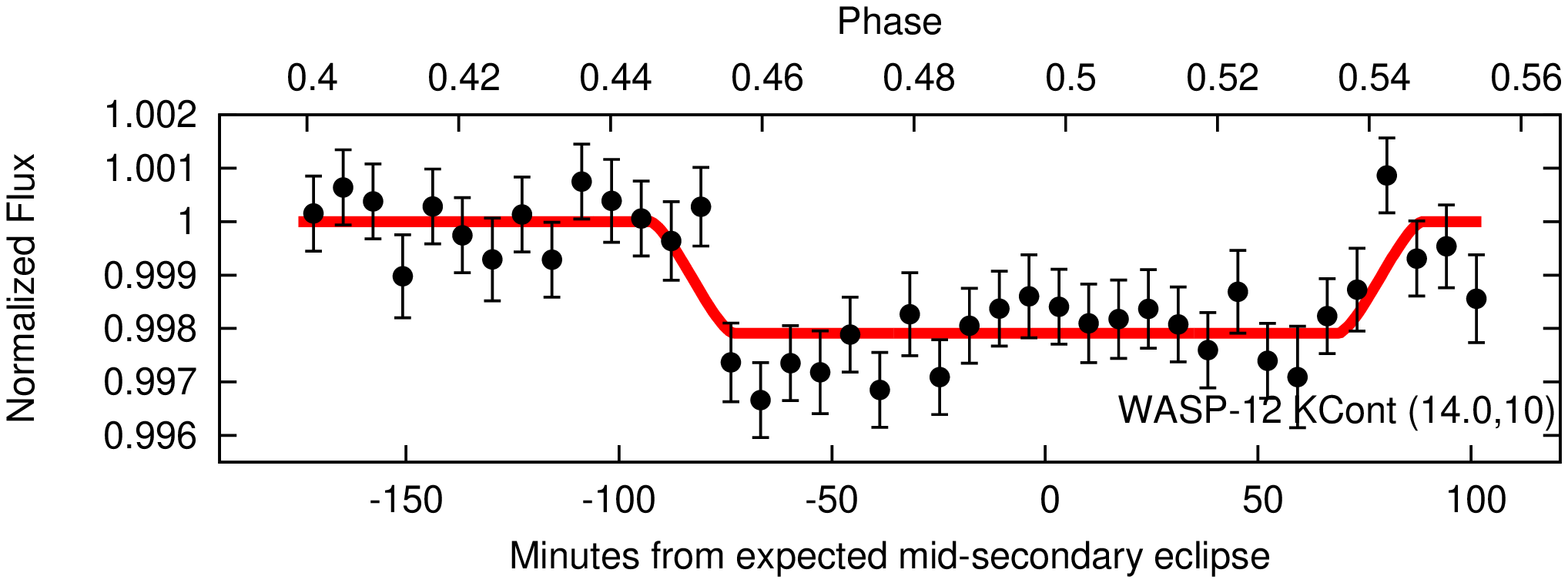}

\includegraphics[scale=0.27, angle = 270]{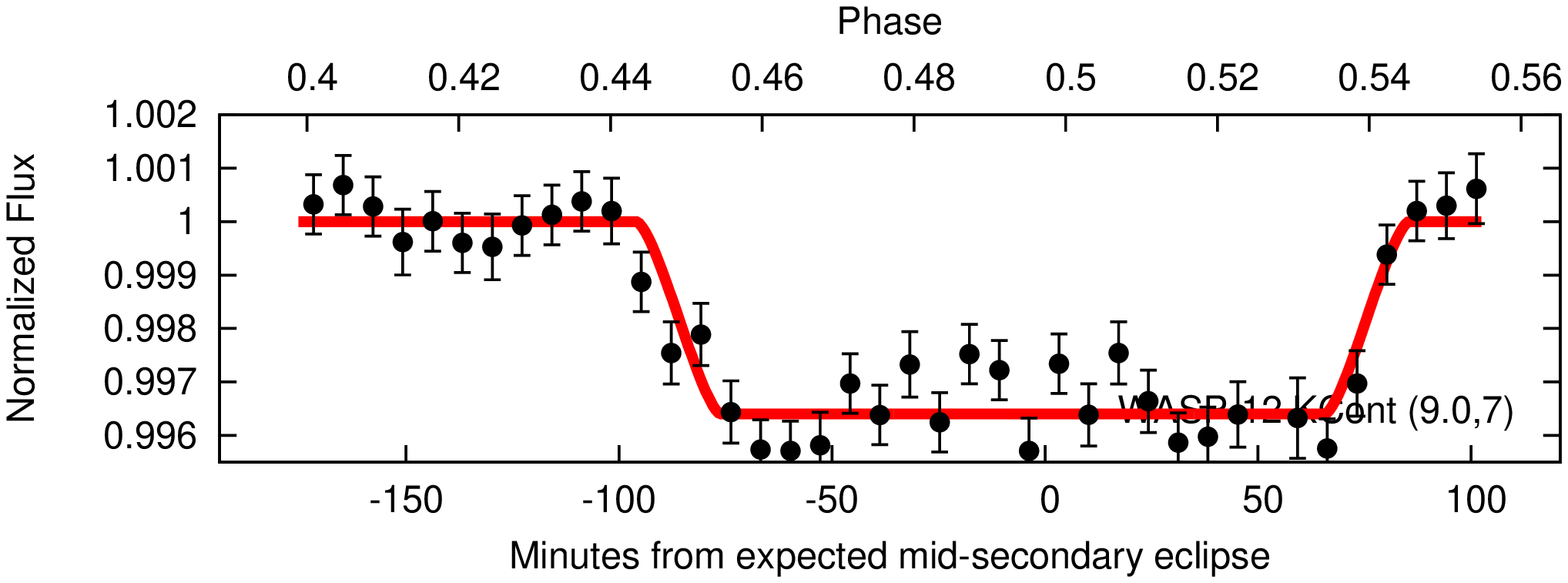}
\includegraphics[scale=0.27, angle = 270]{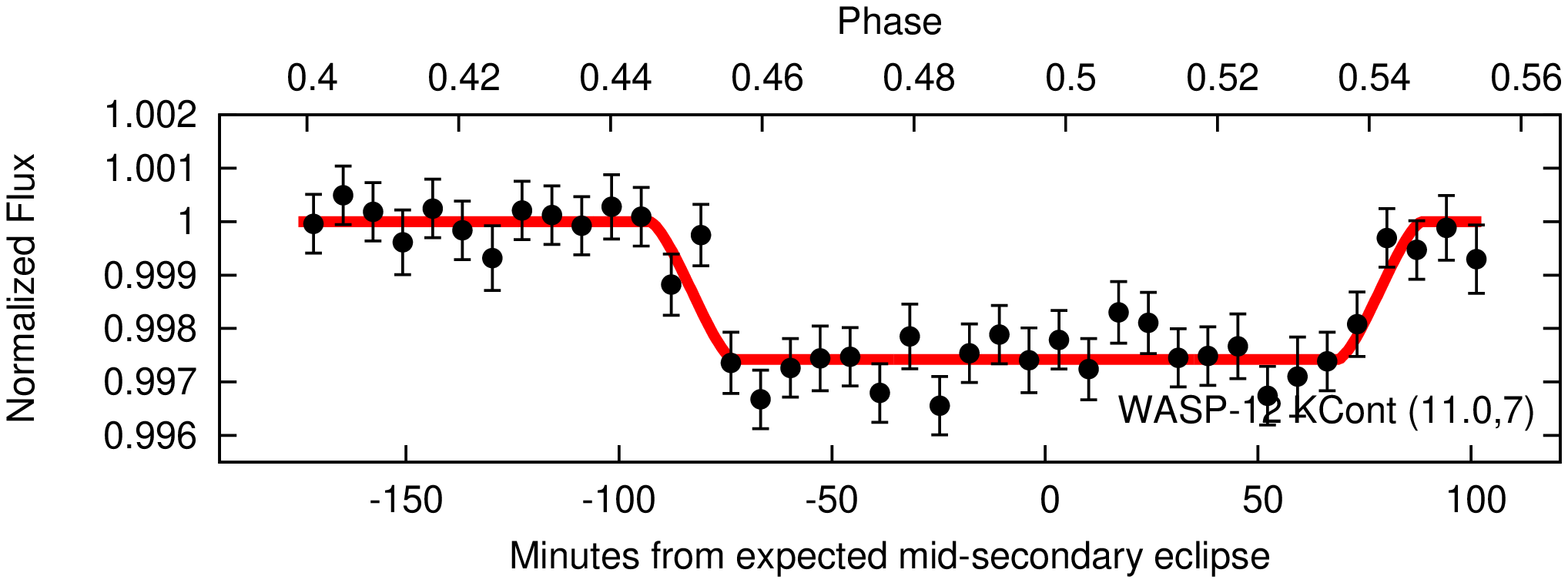}
\includegraphics[scale=0.27, angle = 270]{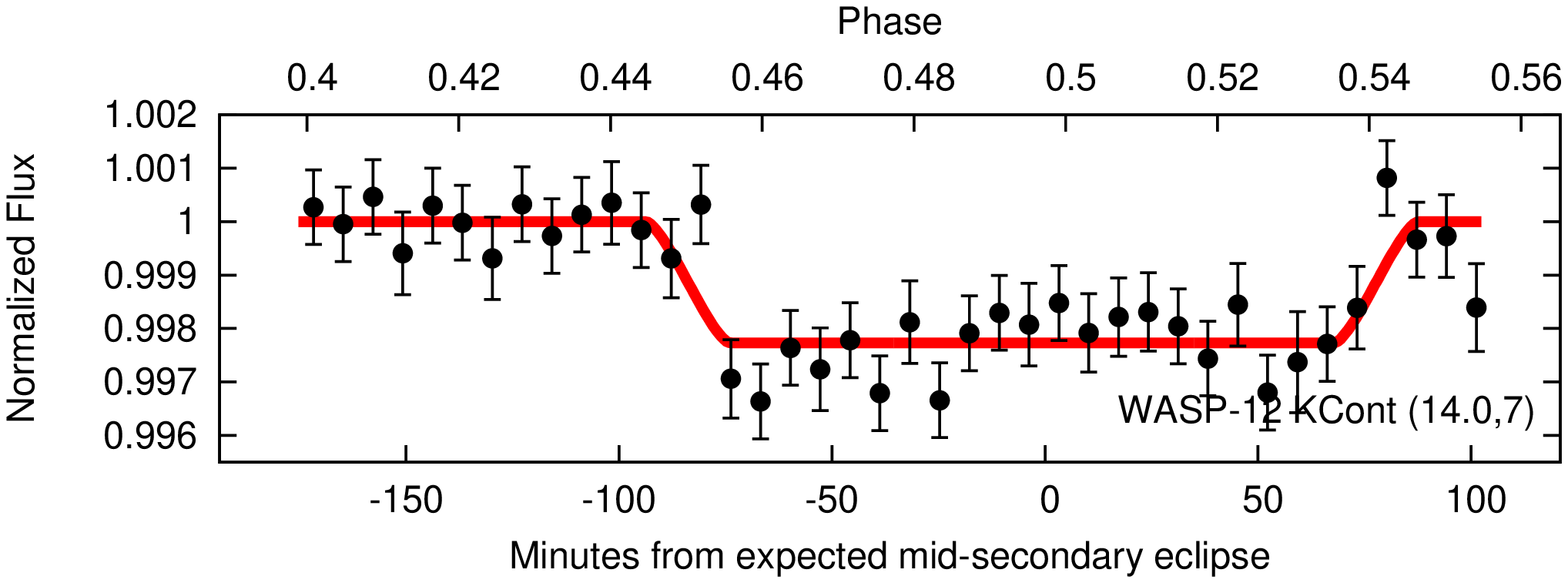}

\includegraphics[scale=0.27, angle = 270]{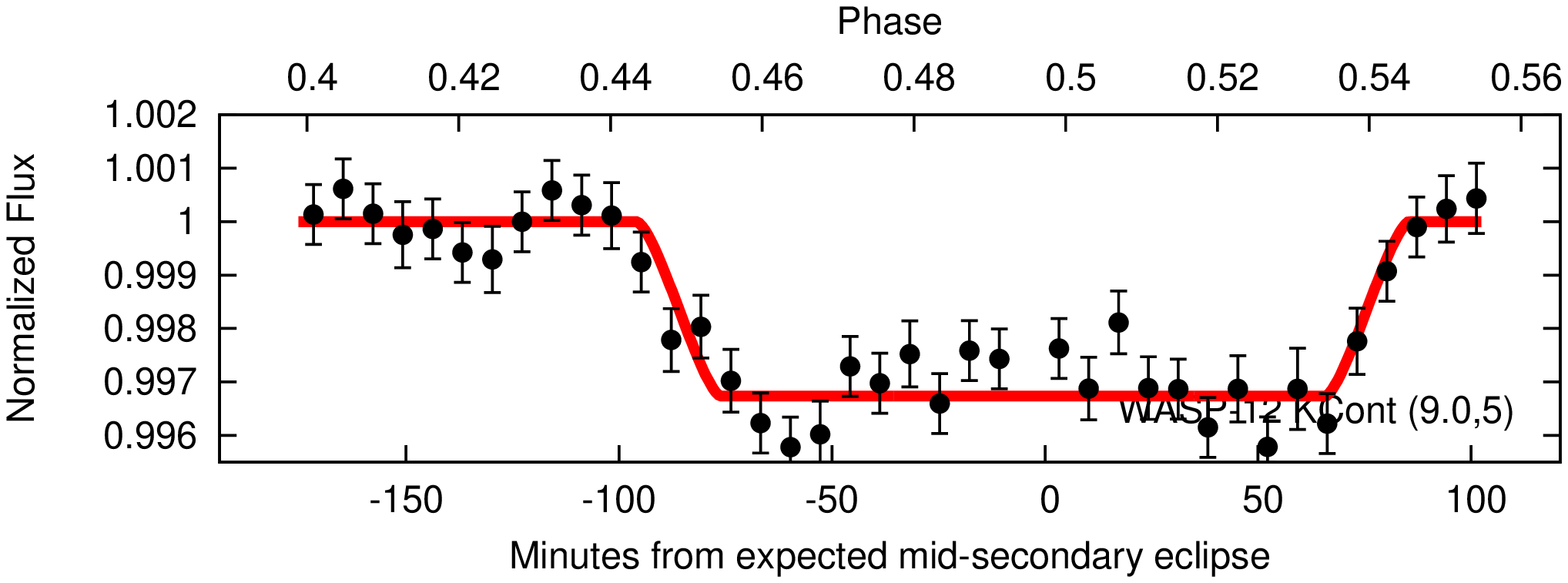}
\includegraphics[scale=0.27, angle = 270]{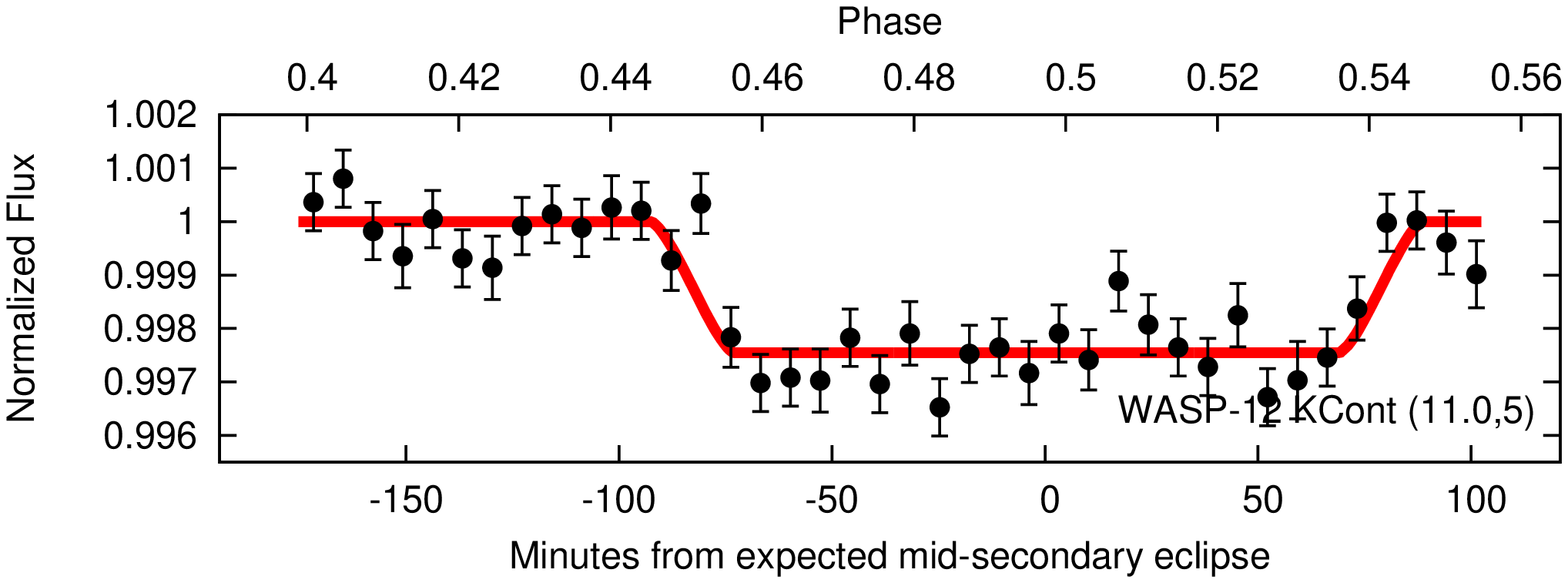}
\includegraphics[scale=0.27, angle = 270]{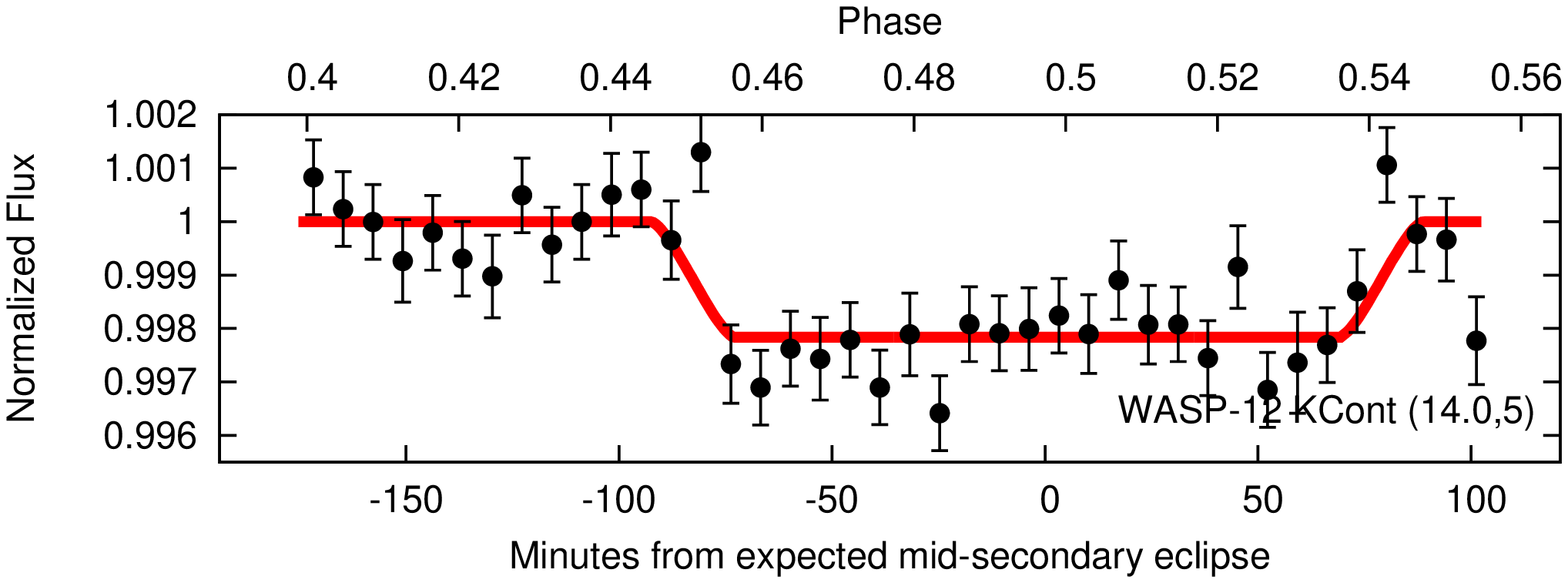}

\includegraphics[scale=0.27, angle = 270]{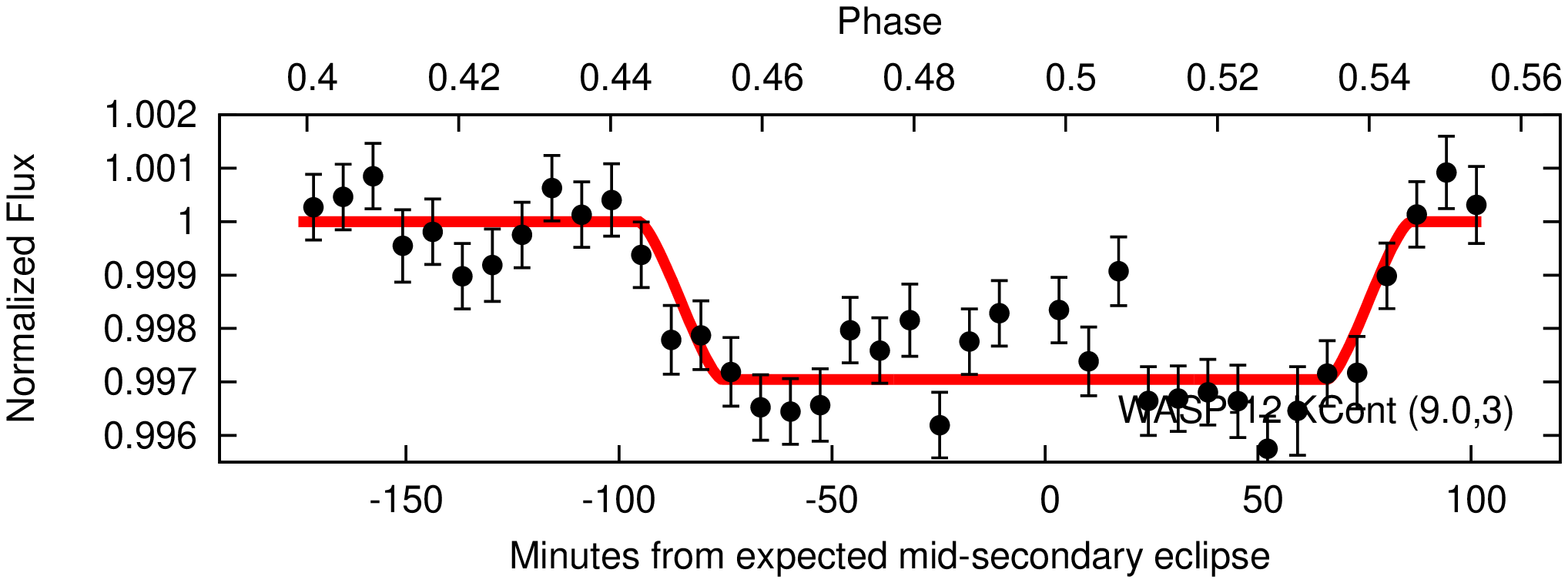}
\includegraphics[scale=0.27, angle = 270]{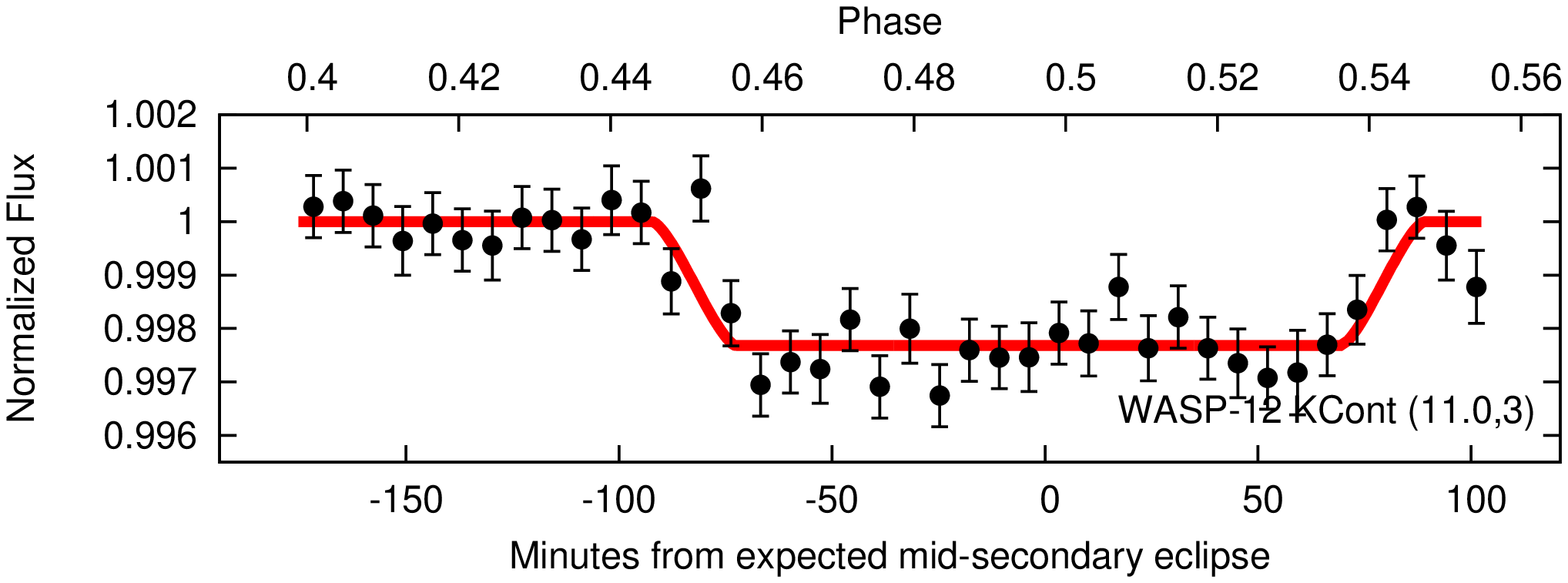}
\includegraphics[scale=0.27, angle = 270]{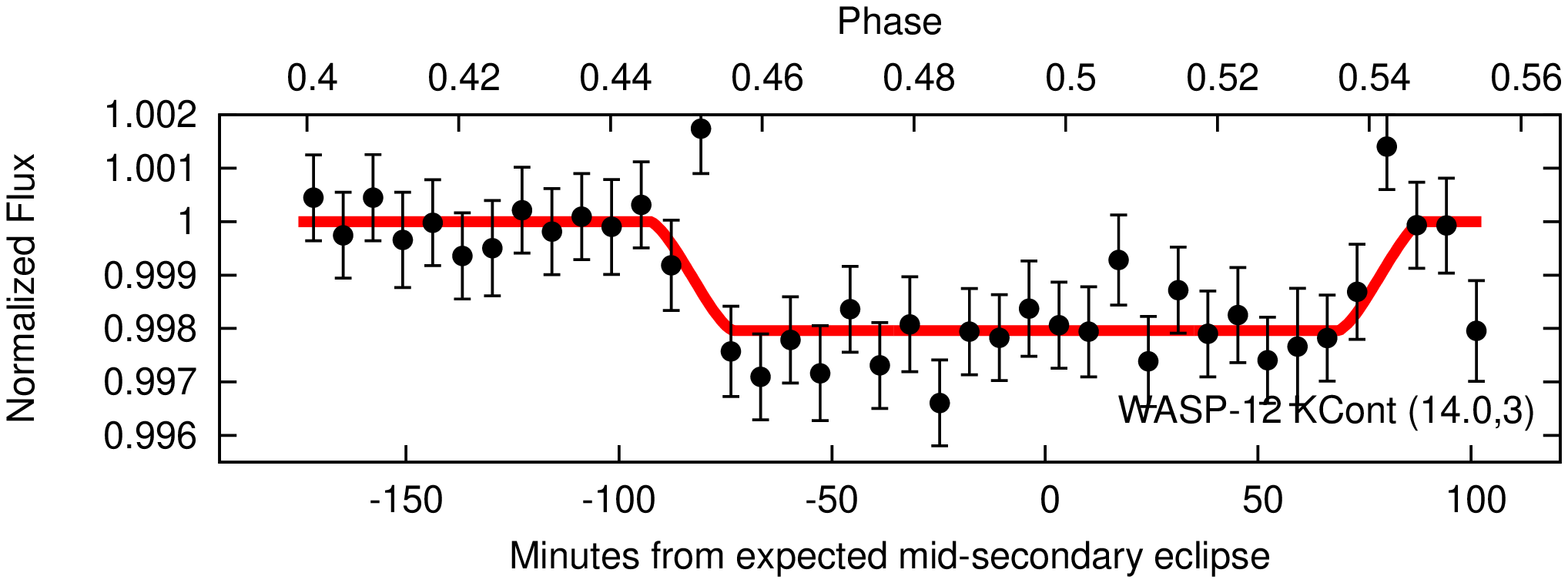}

\includegraphics[scale=0.27, angle = 270]{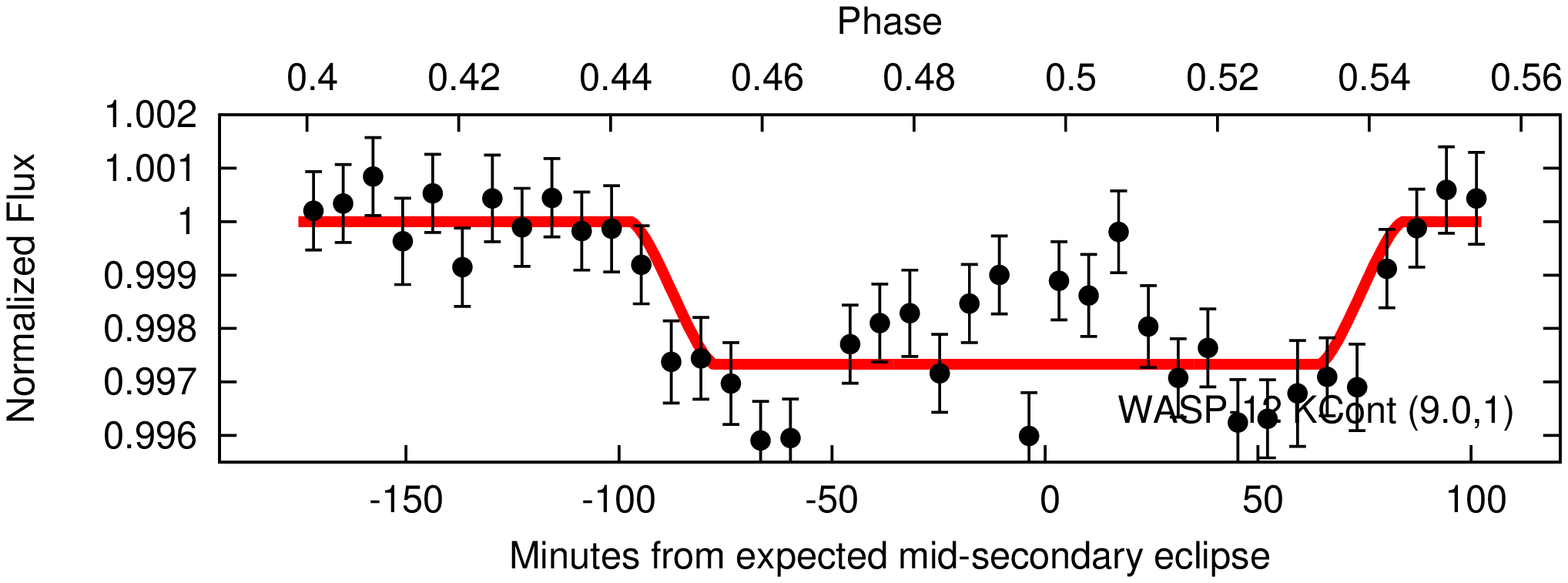}
\includegraphics[scale=0.27, angle = 270]{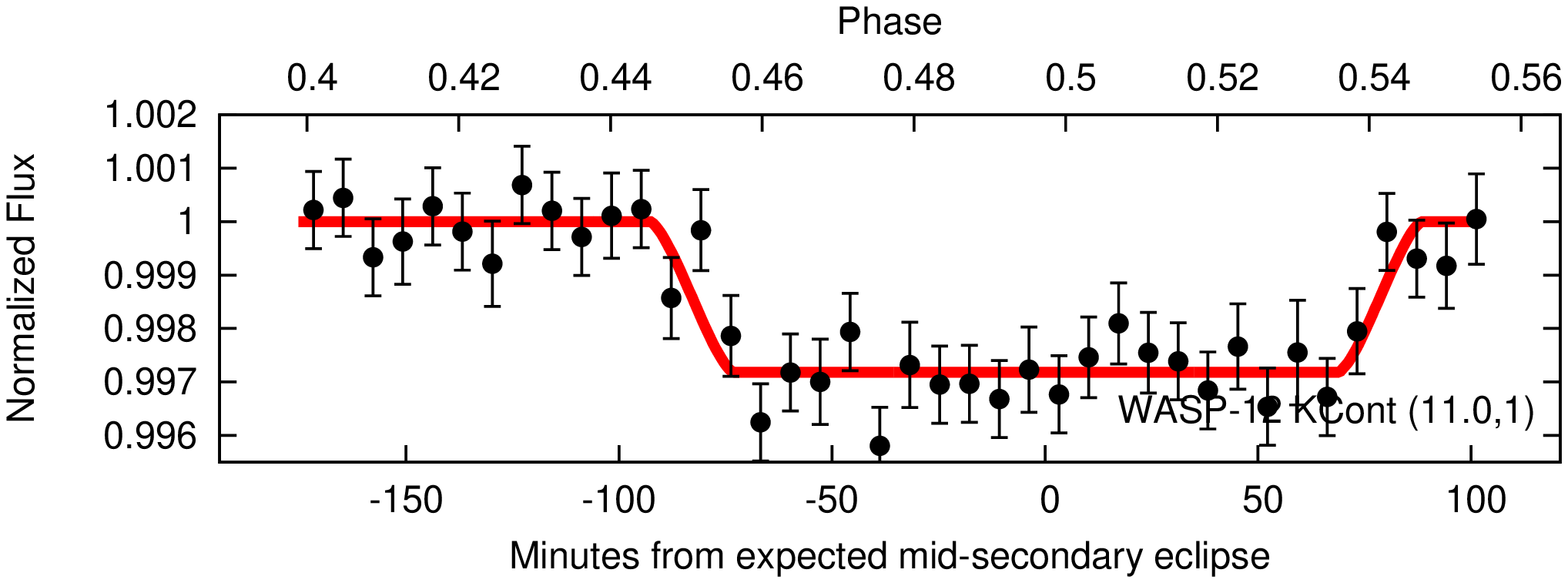}
\includegraphics[scale=0.27, angle = 270]{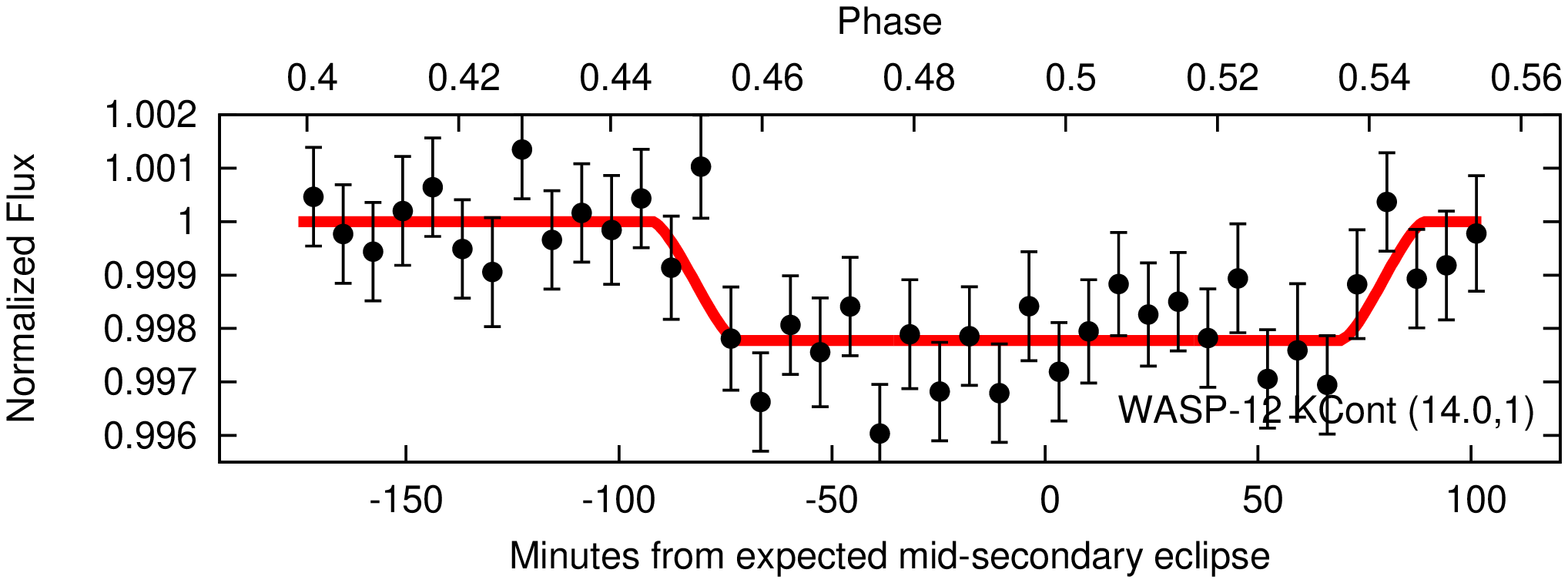}

\caption[WASP-12 $K_{CONT}$ Fidelity Many]
	{	
		Same as Figure \ref{FigWASPTwelveKsbandFidelityManyOne} except for our WASP-12 $K_{CONT}$-band secondary eclipse.
	}
\label{FigWASPTwelveKContbandFidelityMany}
\end{figure*}

\begin{figure*}
\centering

\includegraphics[scale=0.44, angle = 270]{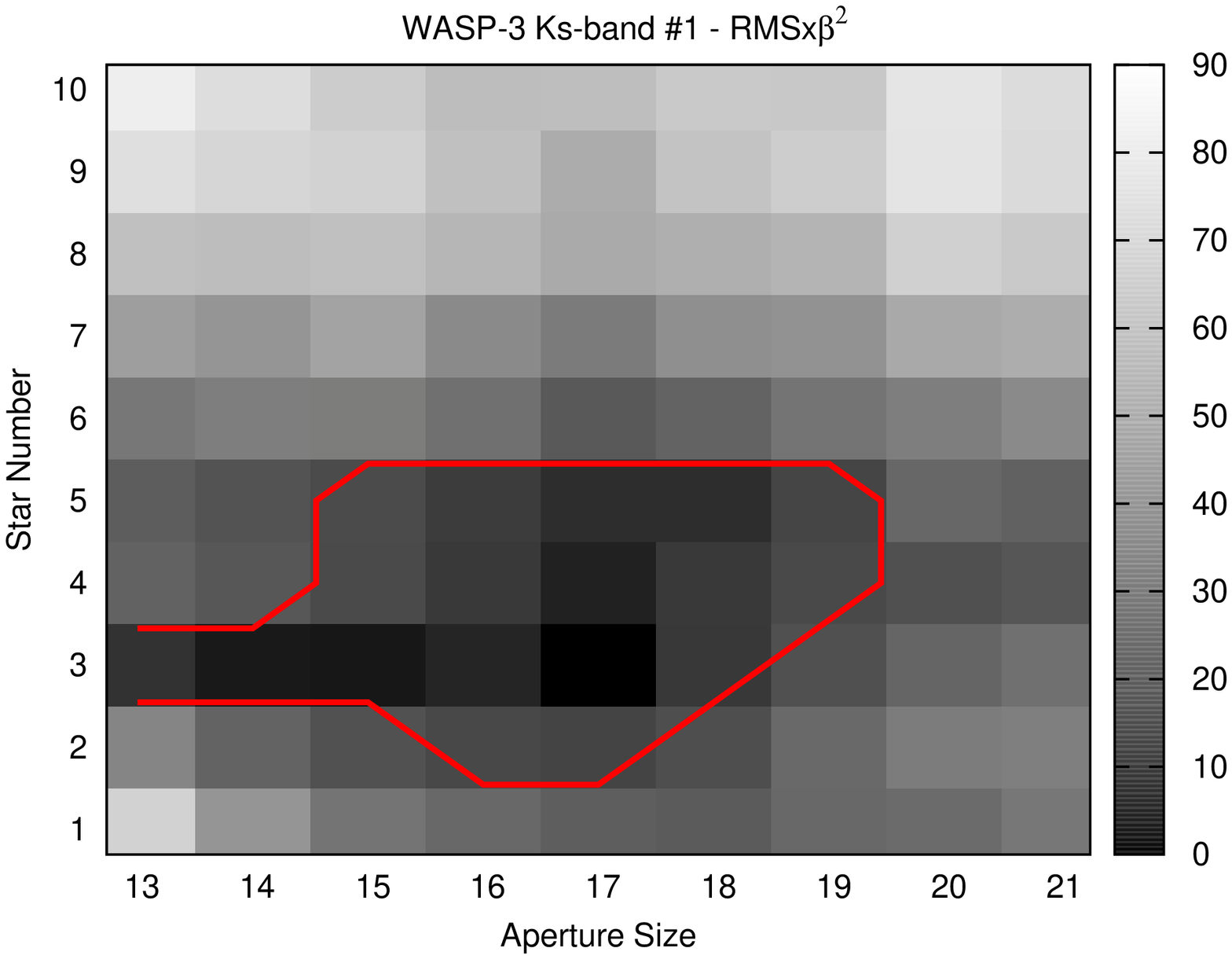}
\includegraphics[scale=0.44, angle = 270]{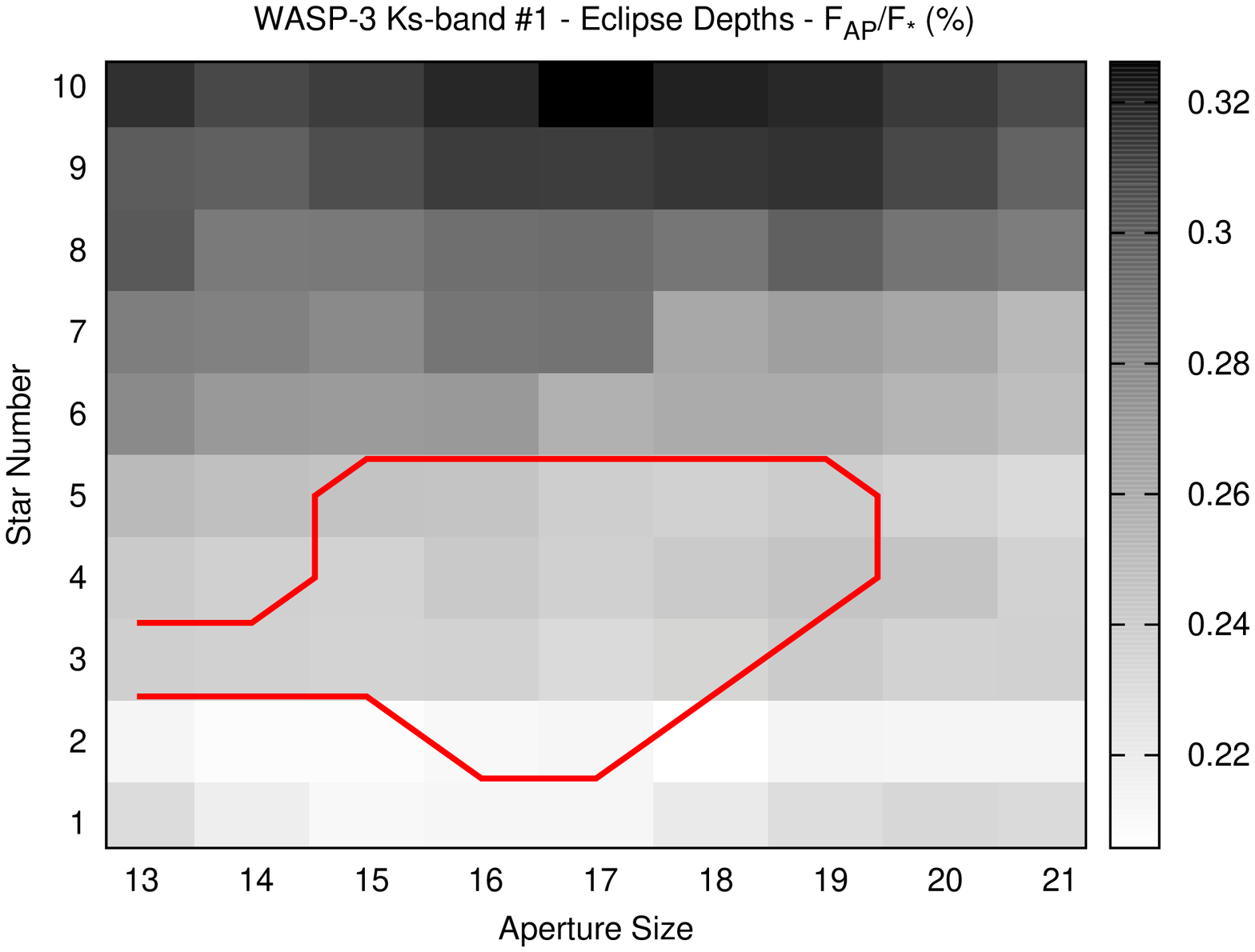}

\includegraphics[scale=0.27, angle = 270]{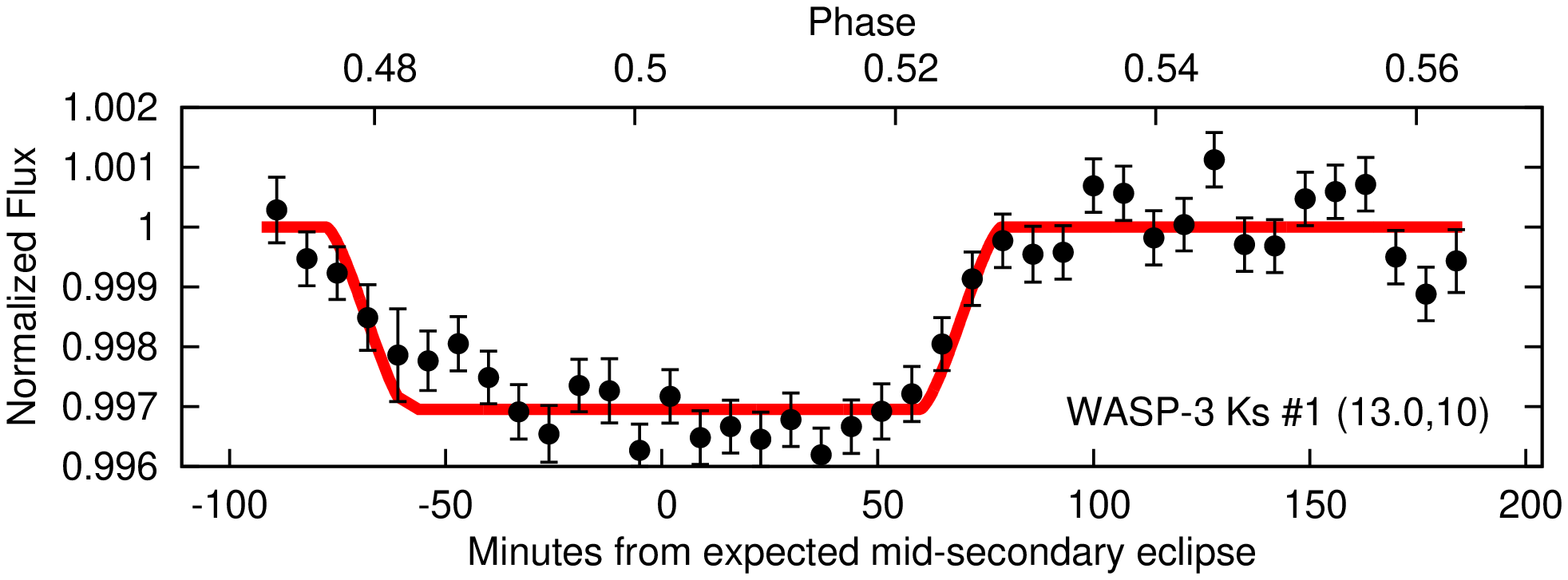}
\includegraphics[scale=0.27, angle = 270]{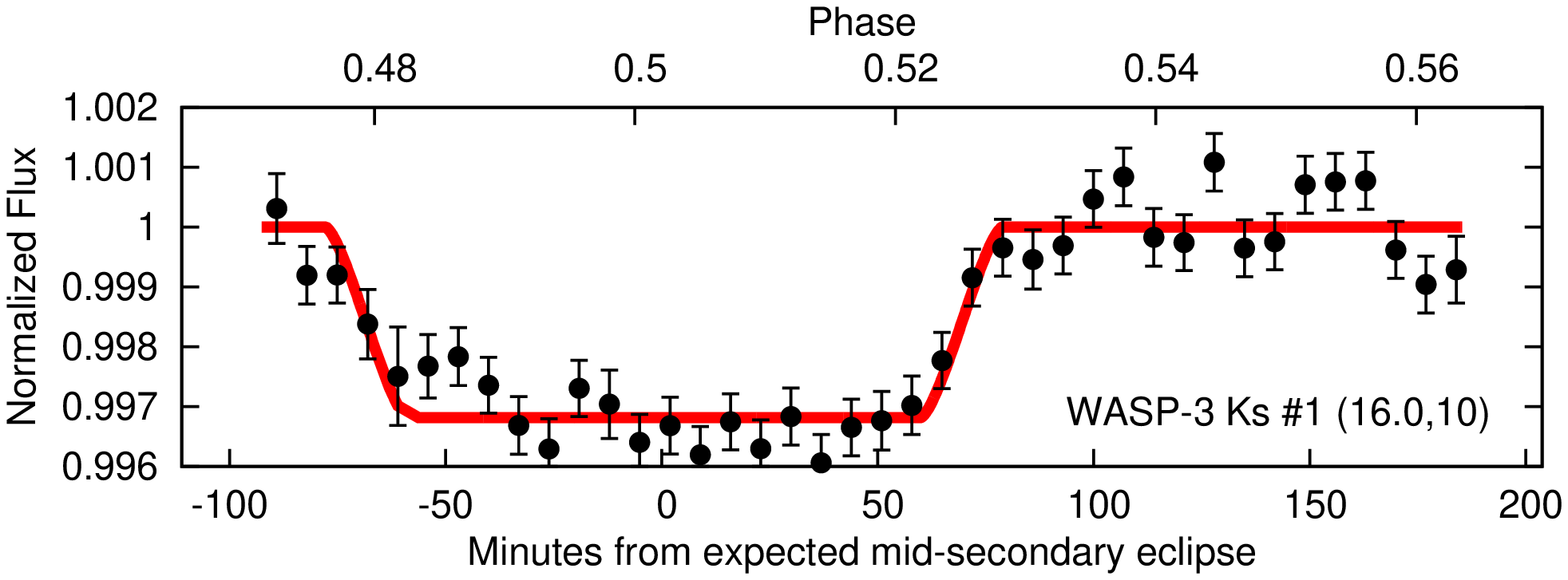}
\includegraphics[scale=0.27, angle = 270]{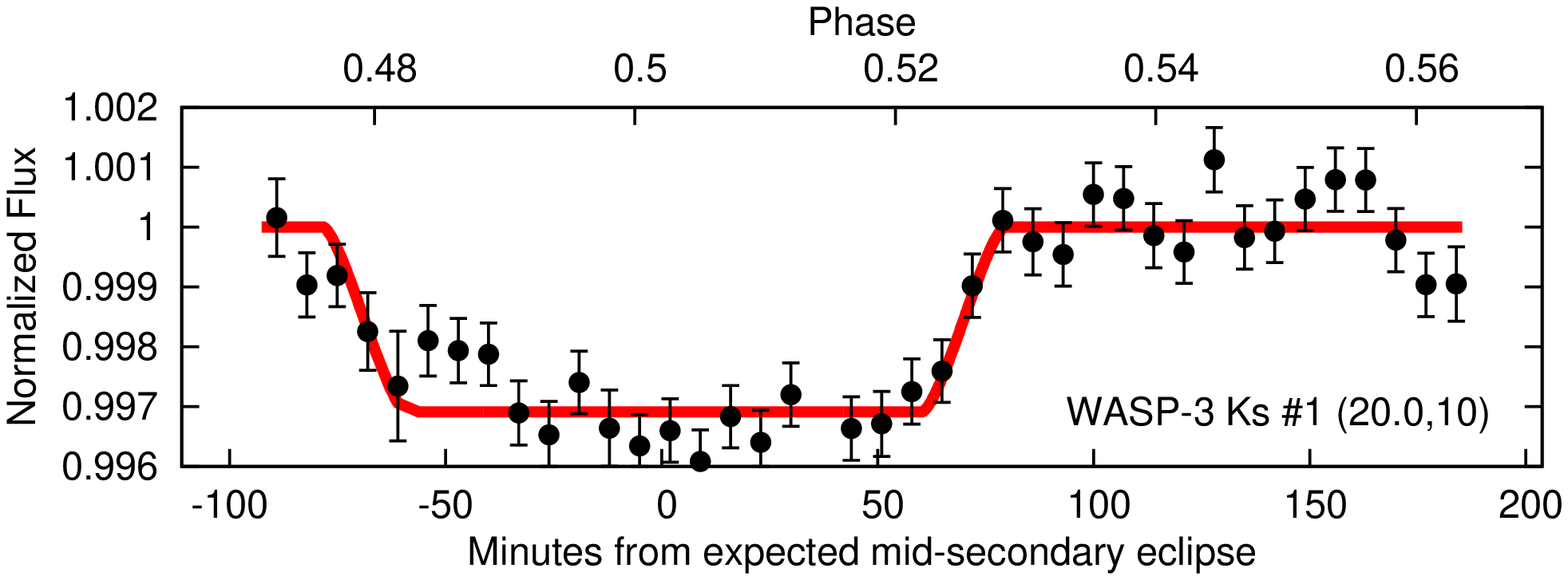}

\includegraphics[scale=0.27, angle = 270]{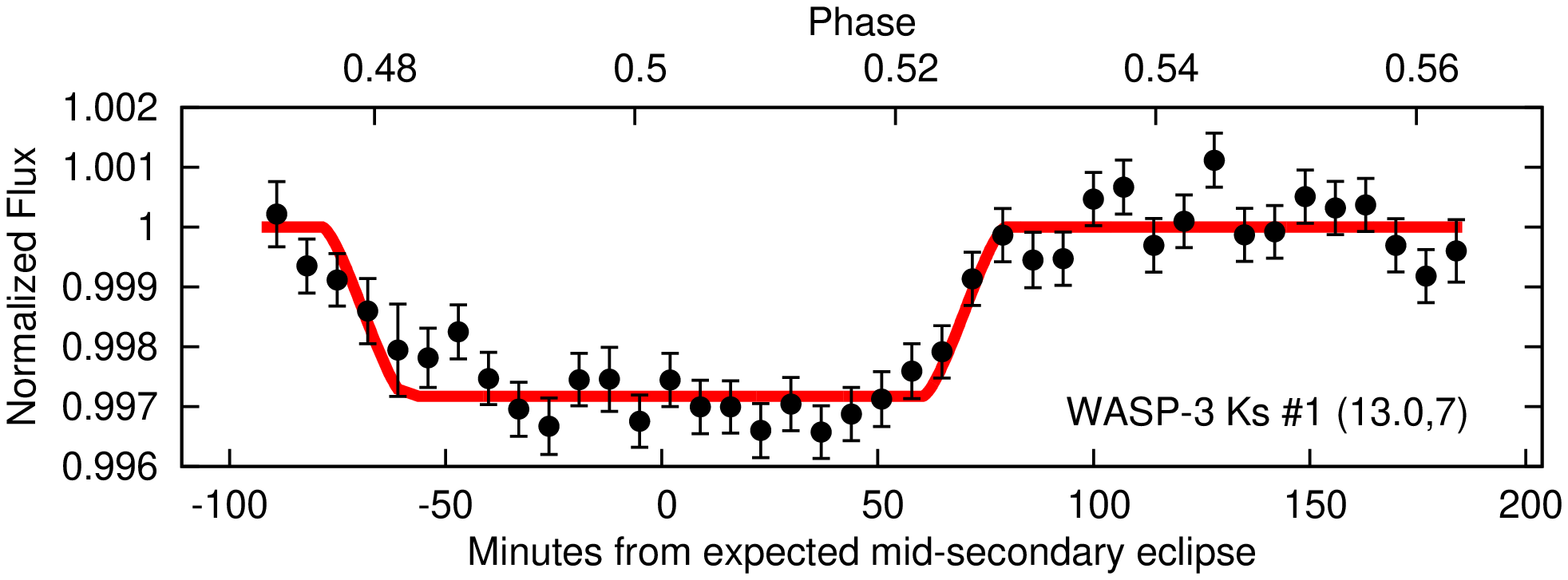}
\includegraphics[scale=0.27, angle = 270]{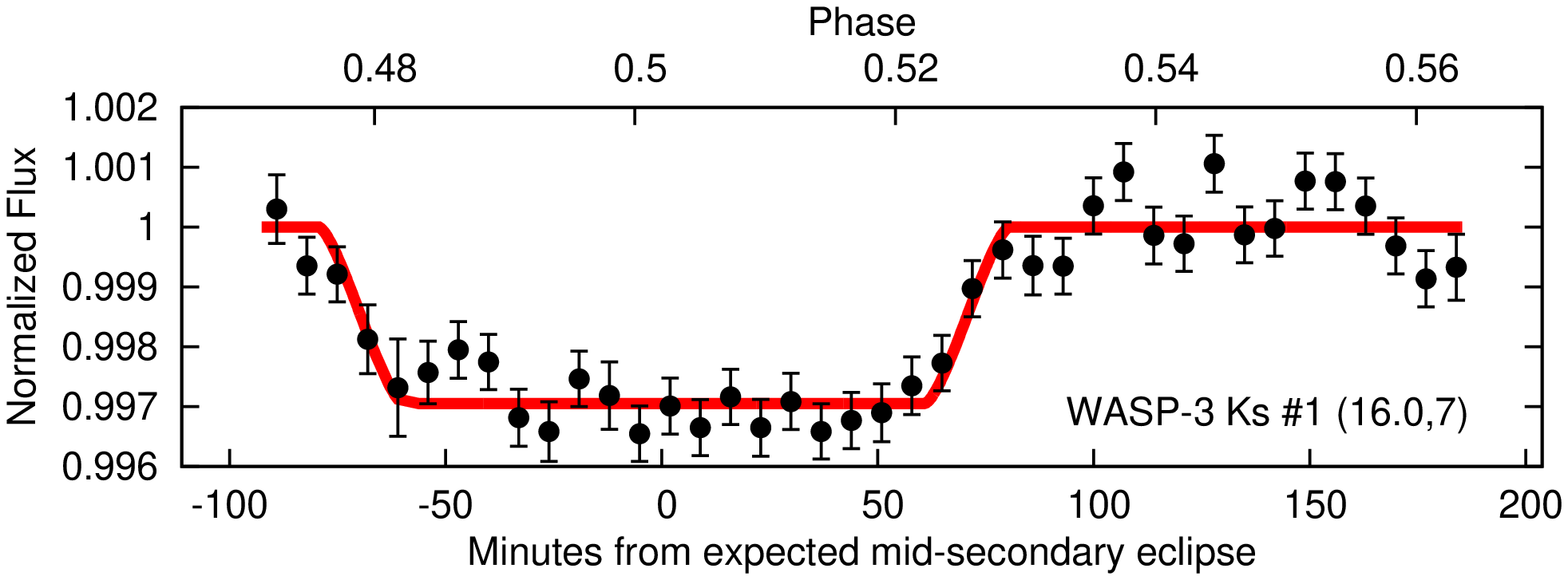}
\includegraphics[scale=0.27, angle = 270]{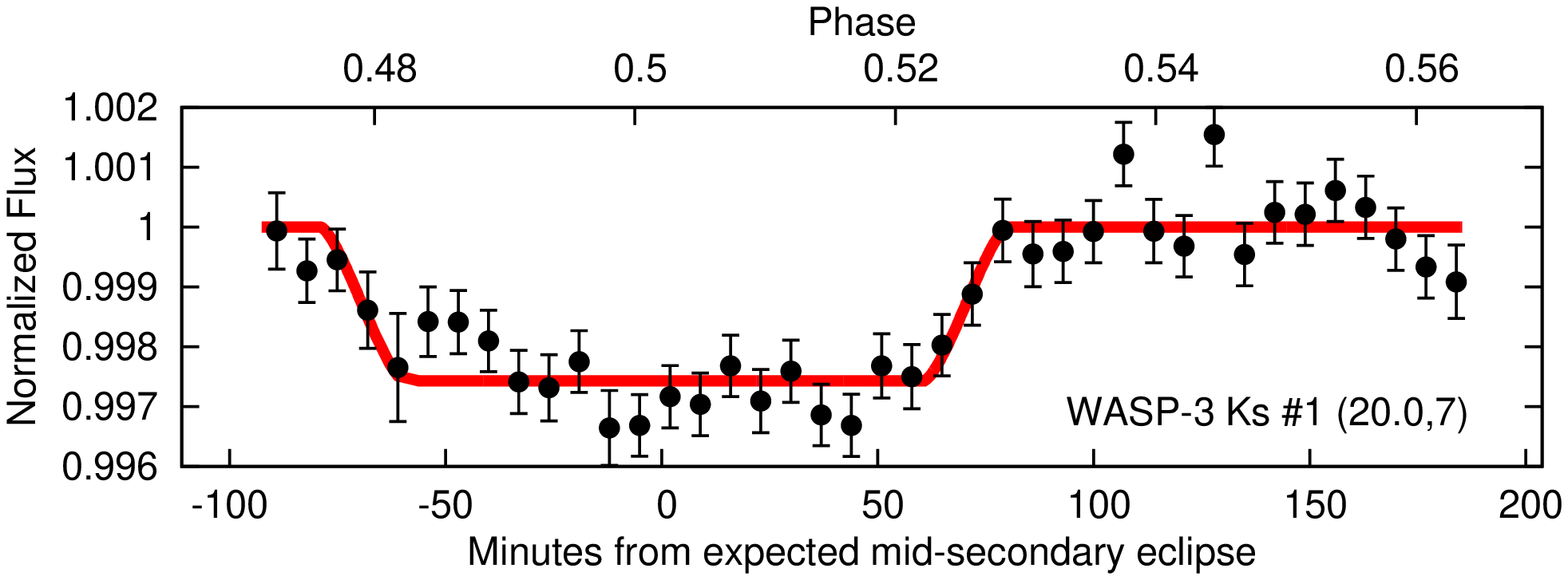}

\includegraphics[scale=0.27, angle = 270]{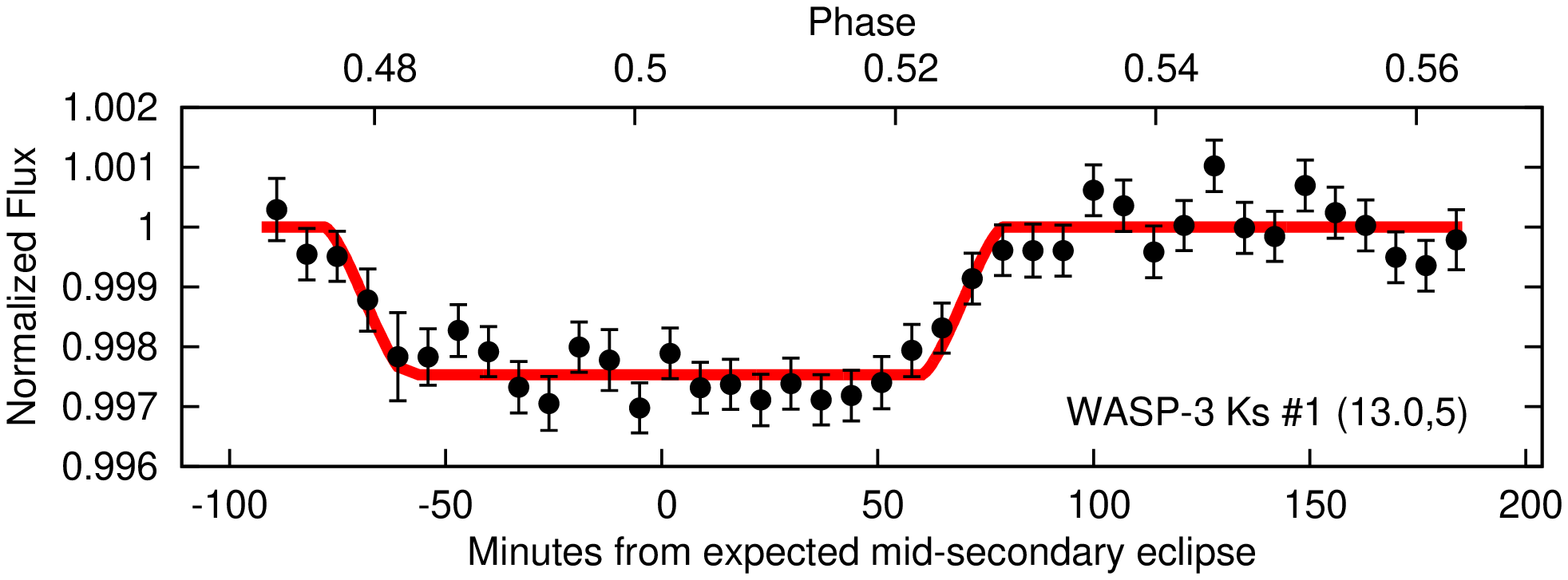}
\includegraphics[scale=0.27, angle = 270]{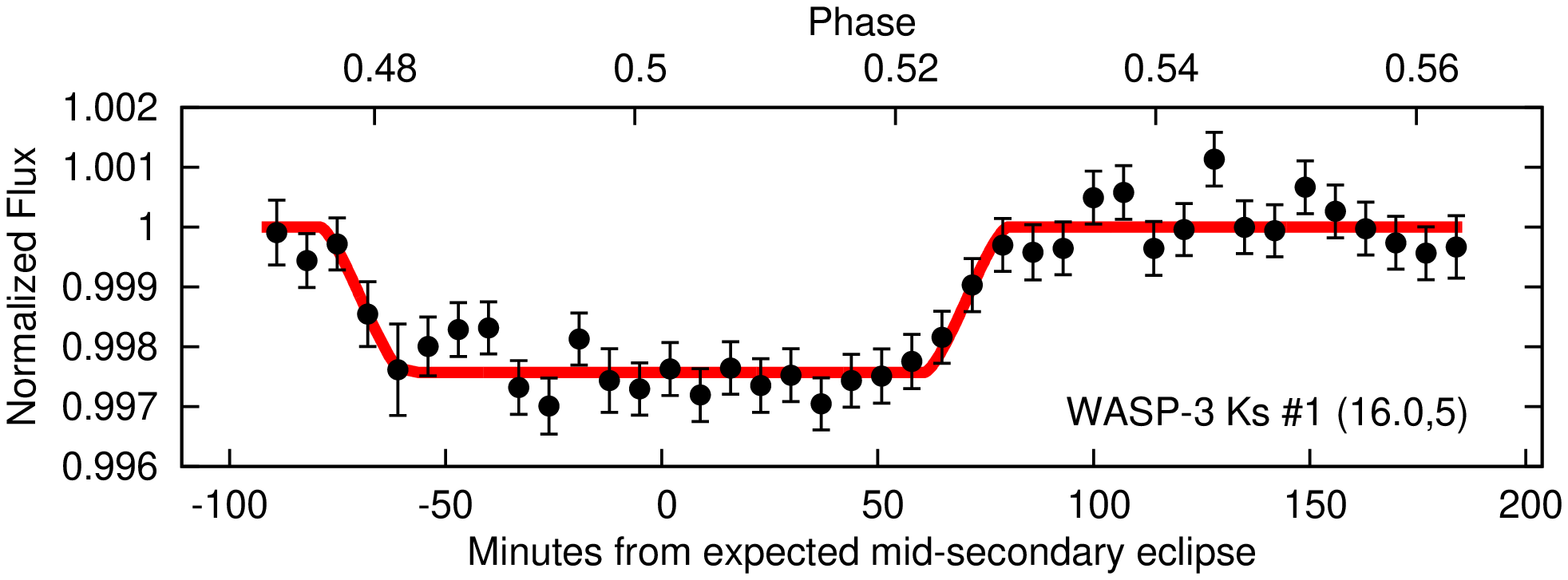}
\includegraphics[scale=0.27, angle = 270]{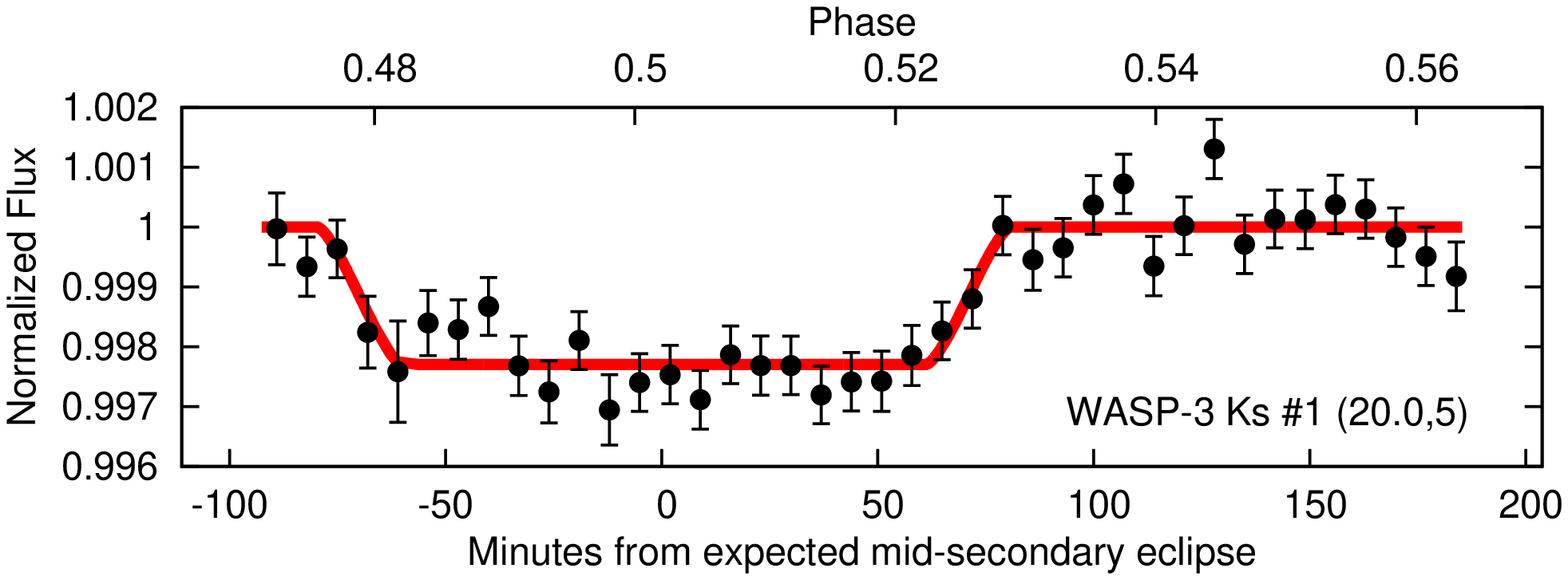}

\includegraphics[scale=0.27, angle = 270]{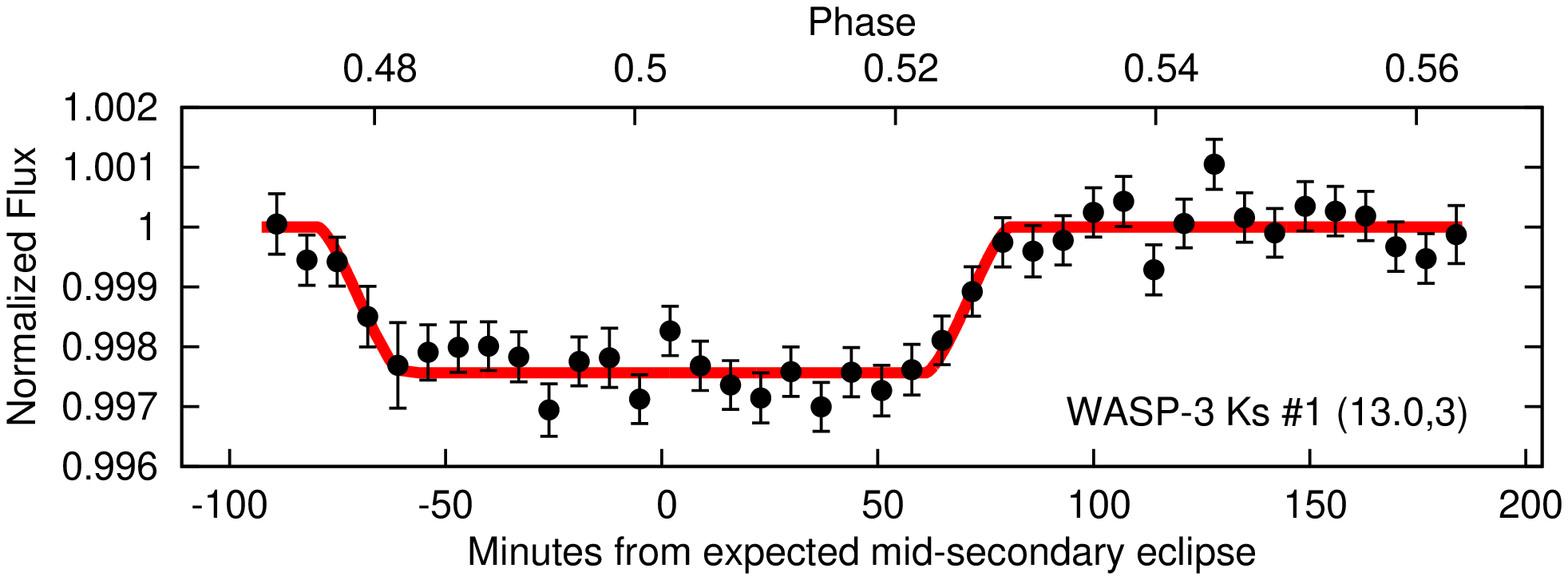}
\includegraphics[scale=0.27, angle = 270]{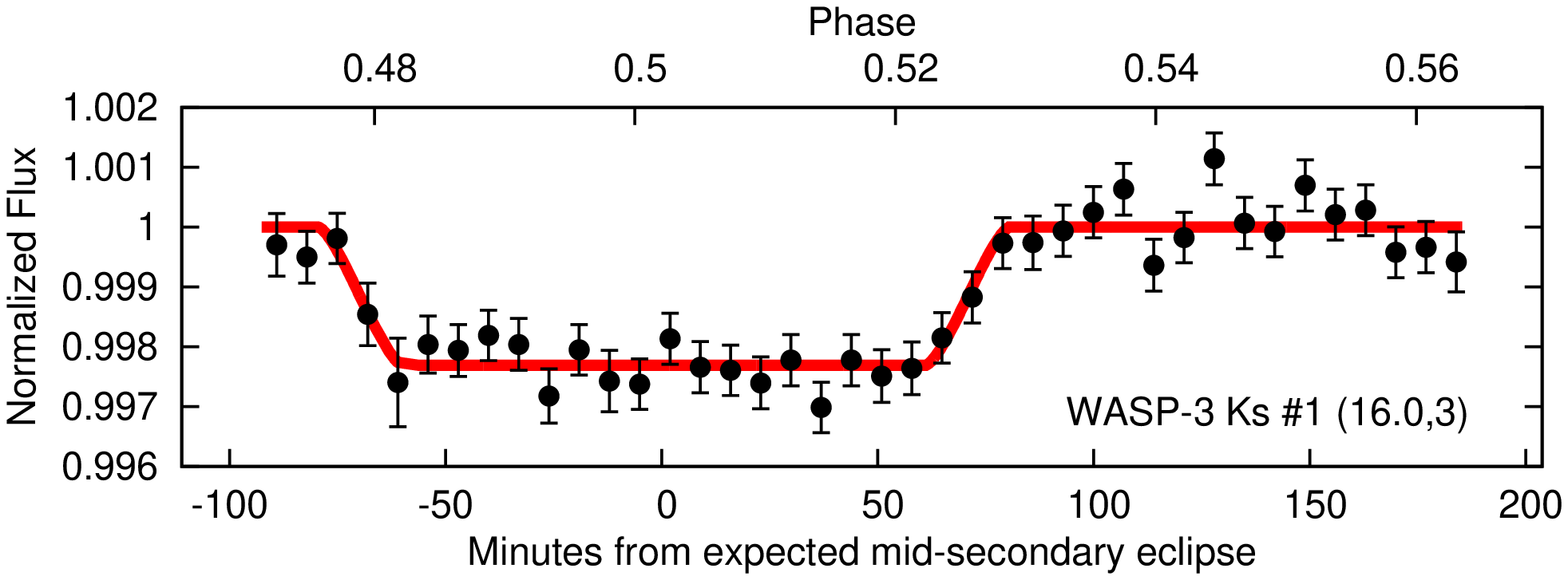}
\includegraphics[scale=0.27, angle = 270]{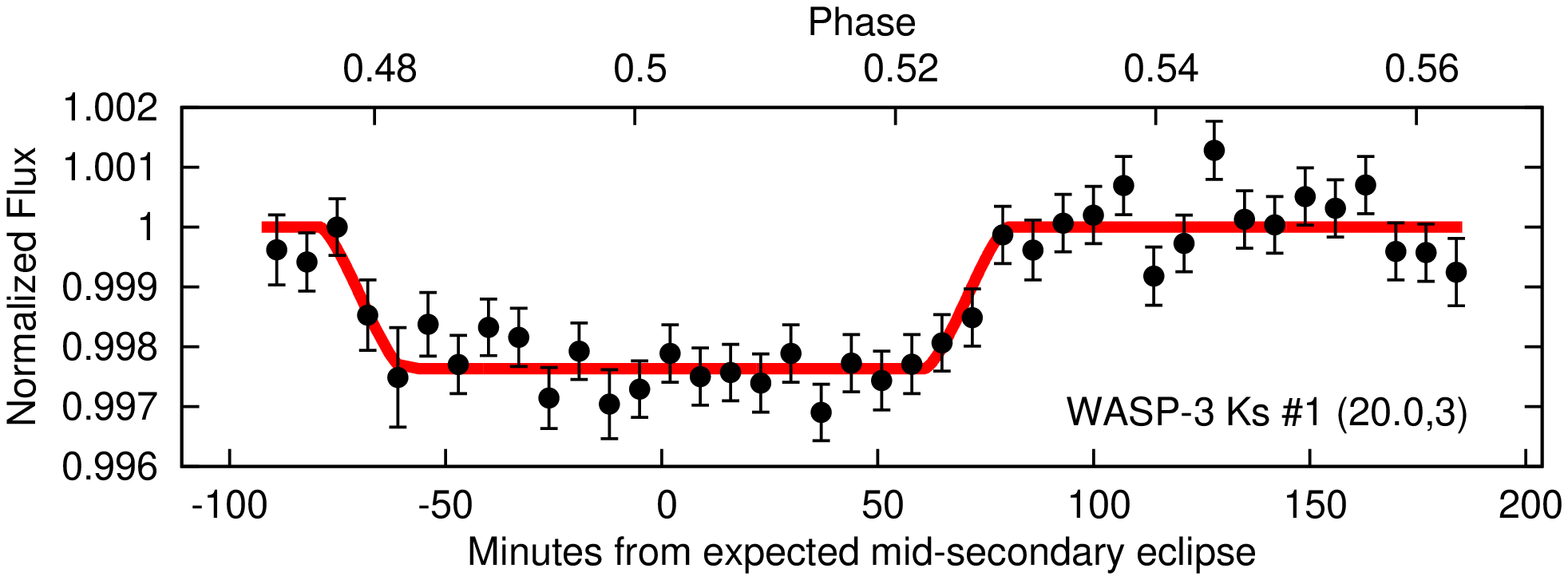}

\includegraphics[scale=0.27, angle = 270]{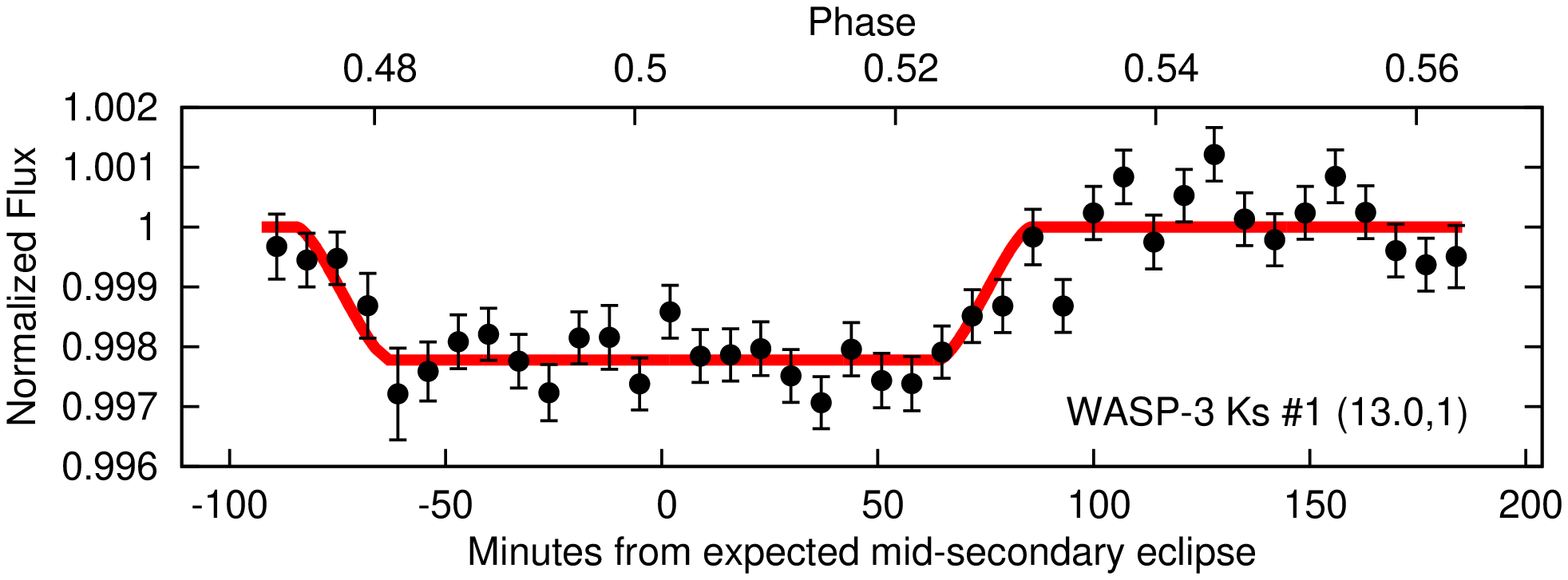}
\includegraphics[scale=0.27, angle = 270]{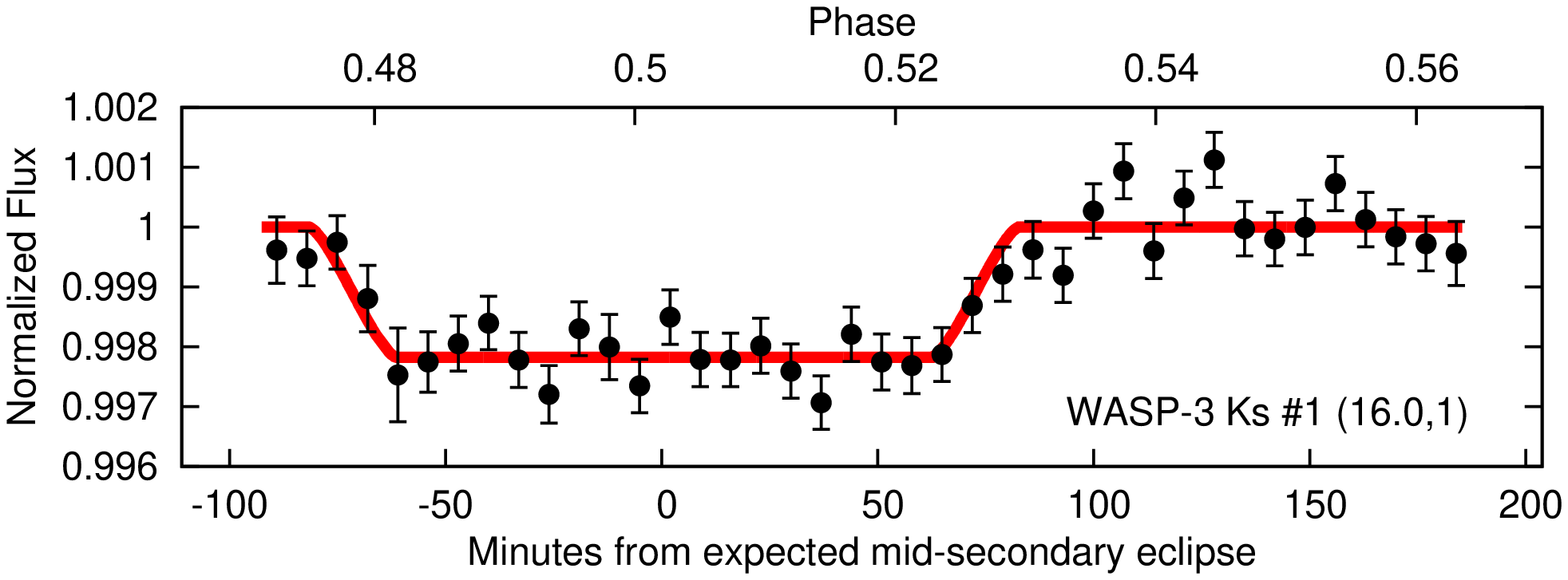}
\includegraphics[scale=0.27, angle = 270]{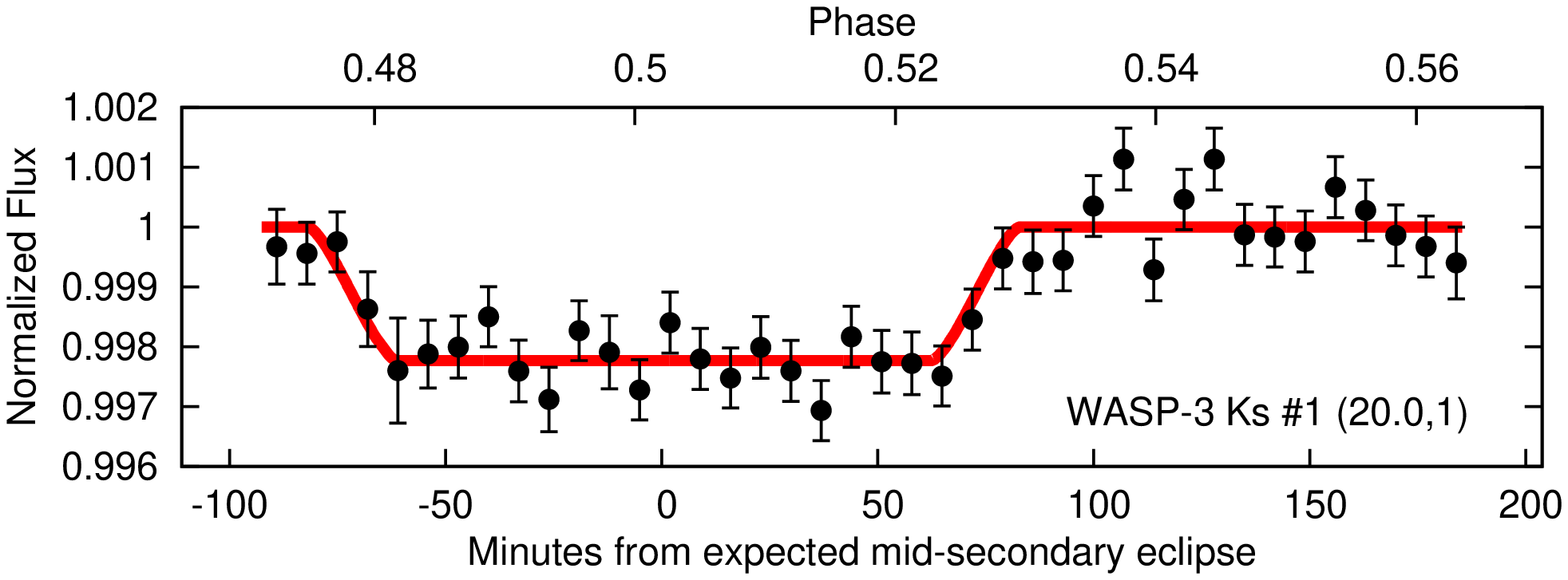}
\caption[BLAH]
	{	
		Same as Figure \ref{FigWASPTwelveKsbandFidelityManyOne} except for our second WASP-3 Ks-band secondary eclipse.
		The scale of the bottom panels is identical to that of the other WASP-3 Ks-band
		eclipse (Figure \ref{FigWASPThreeKsbandFidelityManyTwo}).
		For our first WASP-3 Ks-band eclipse, utilizing more than a few reference stars introduces correlated noise.
	}
\label{FigWASPThreeKsbandFidelityManyOne}
\end{figure*}

\begin{figure*}
\centering
\includegraphics[scale=0.44, angle = 270]{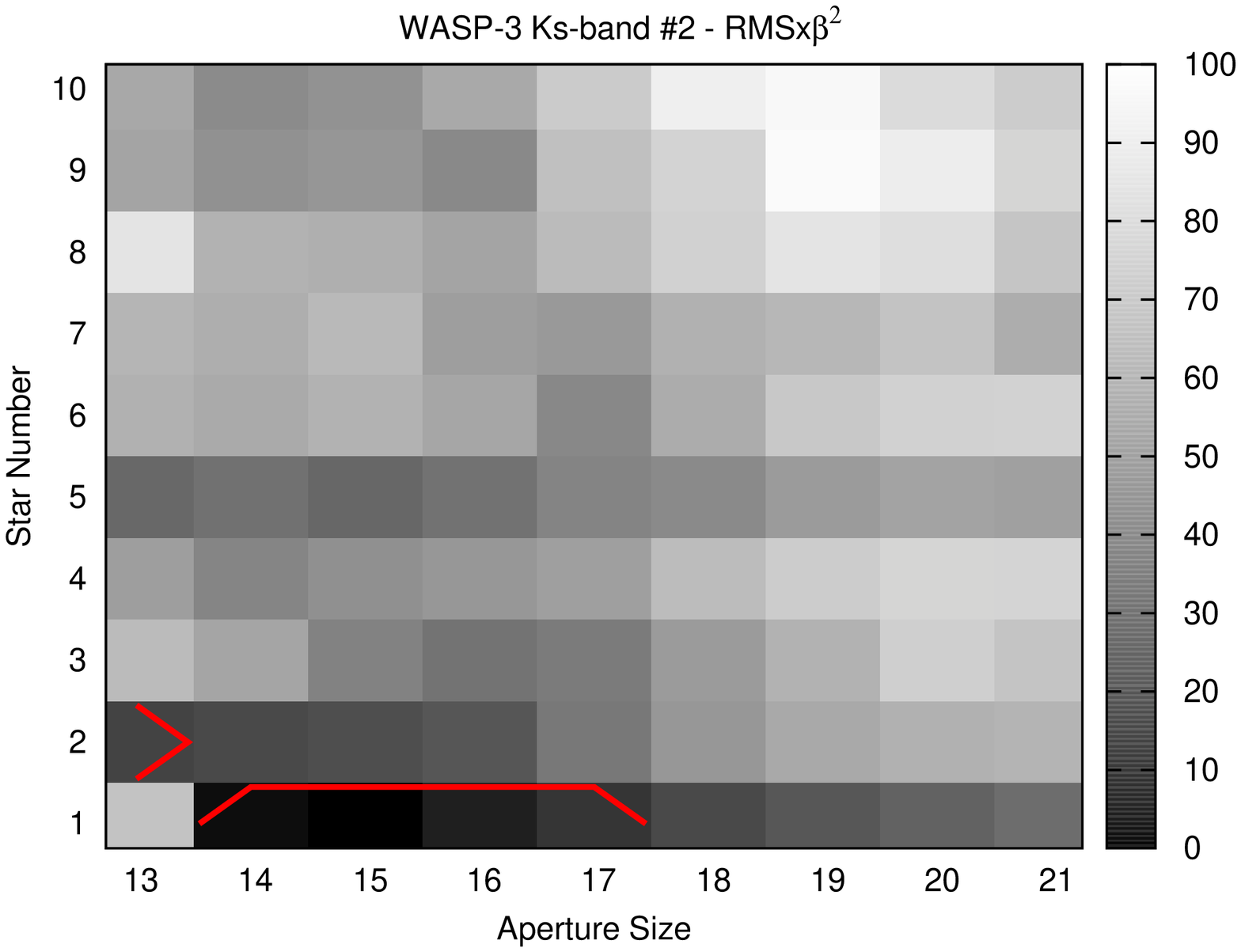}
\includegraphics[scale=0.44, angle = 270]{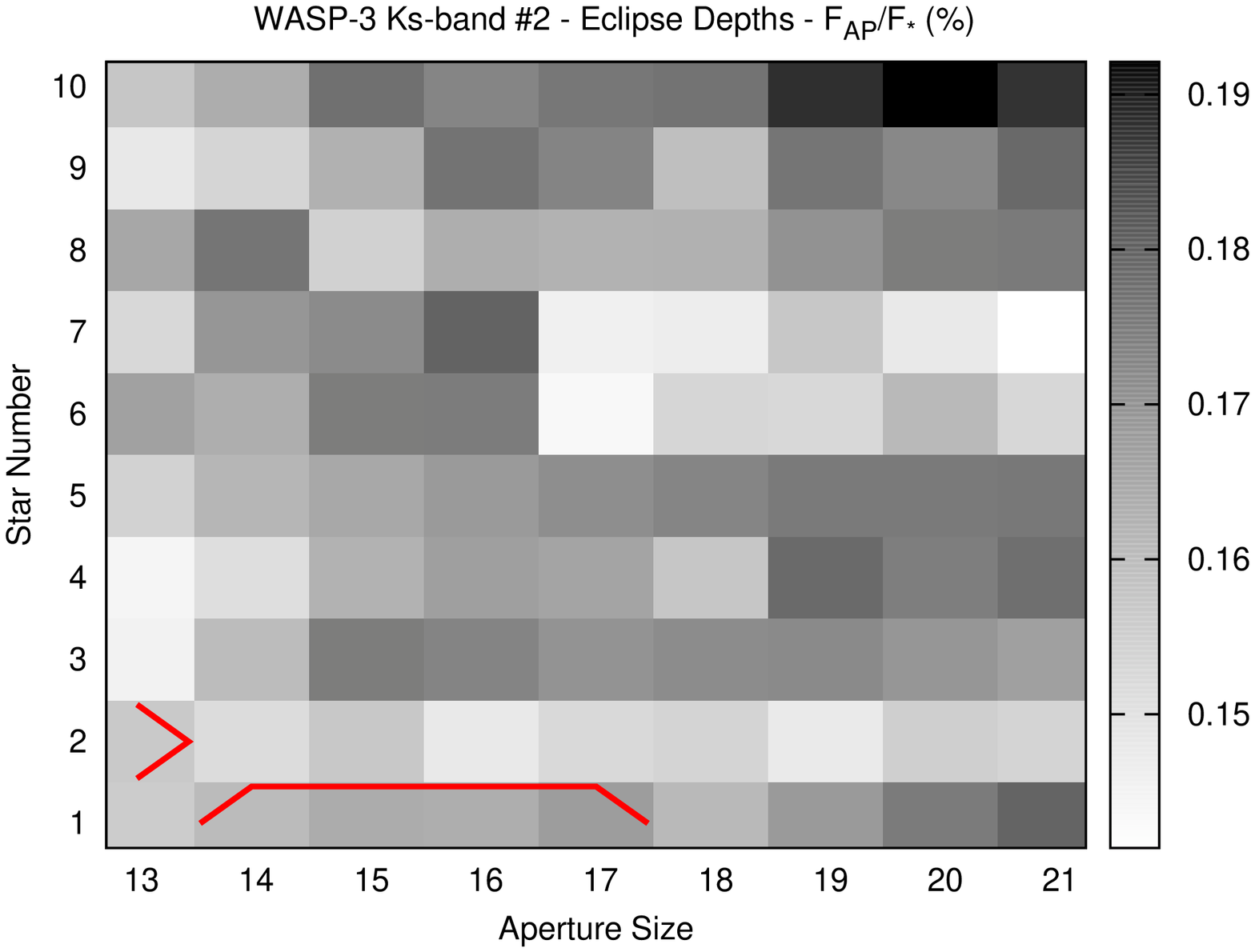}

\includegraphics[scale=0.27, angle = 270]{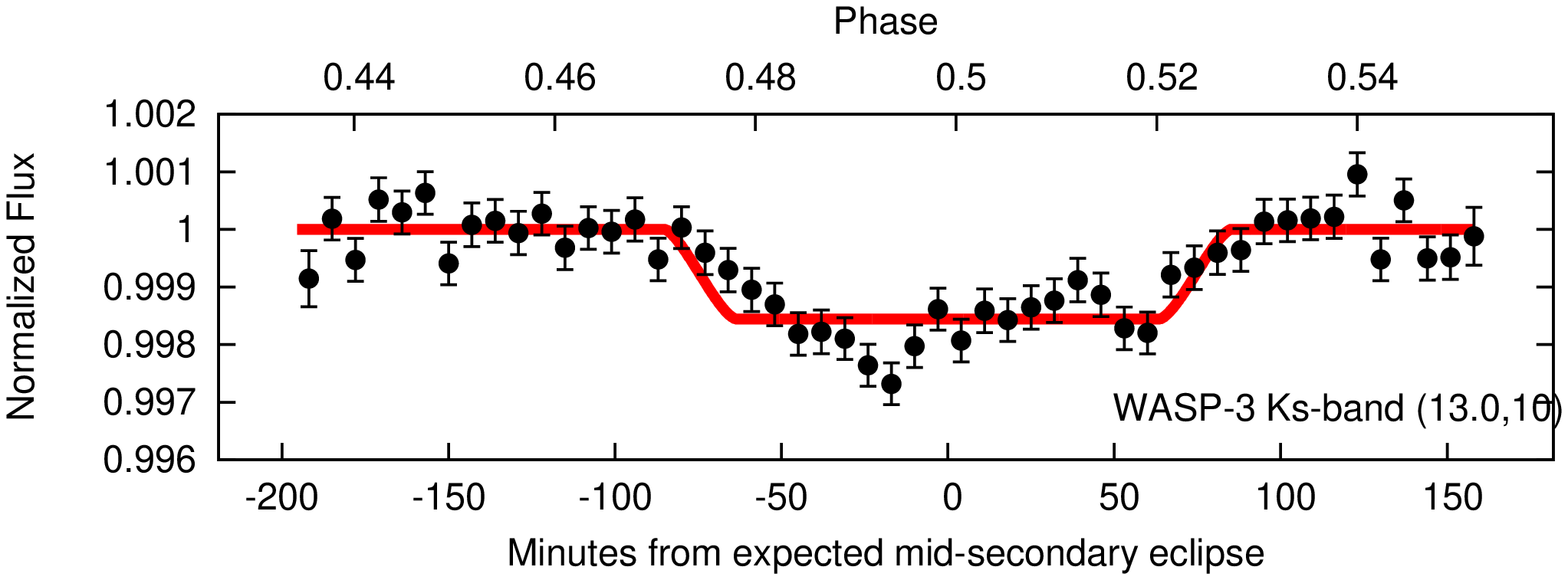}
\includegraphics[scale=0.27, angle = 270]{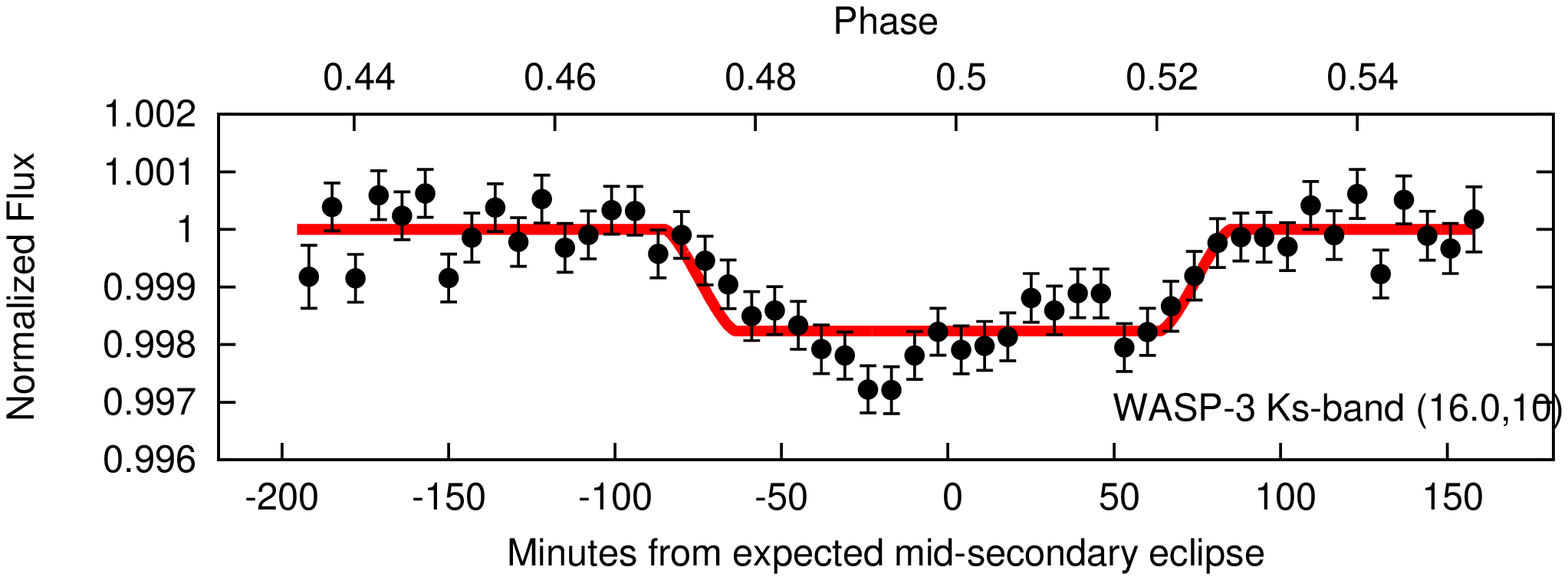}
\includegraphics[scale=0.27, angle = 270]{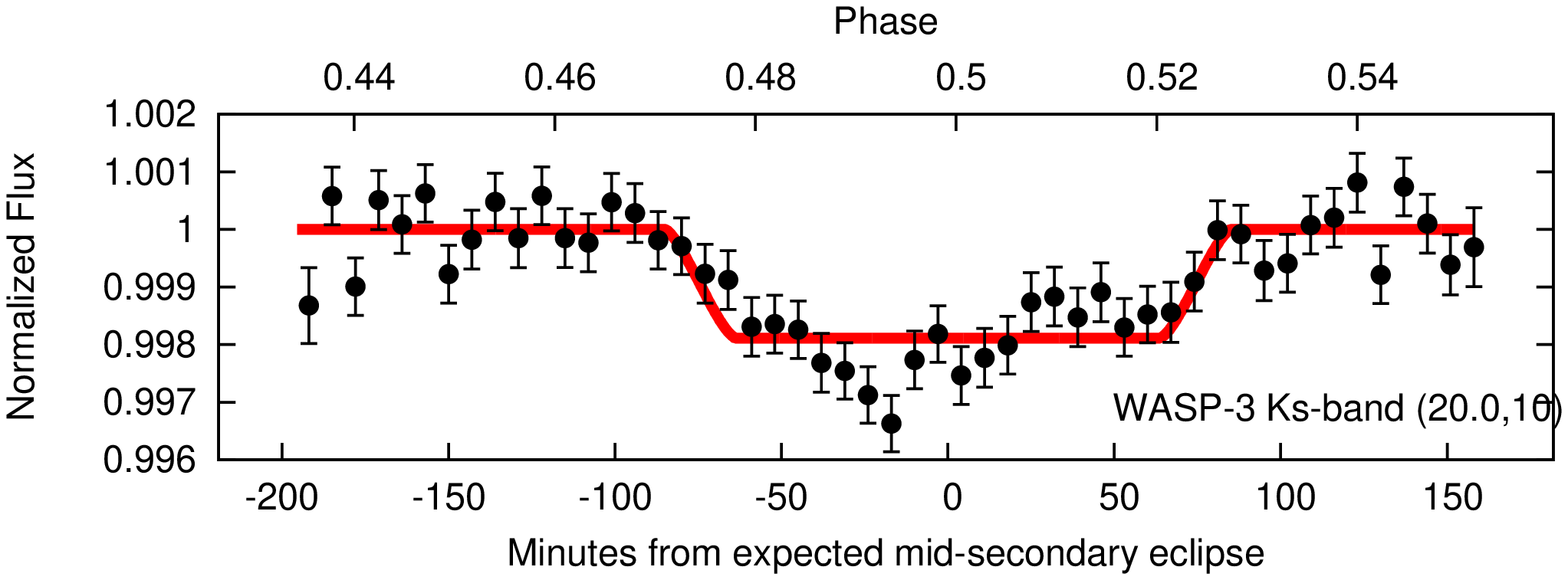}

\includegraphics[scale=0.27, angle = 270]{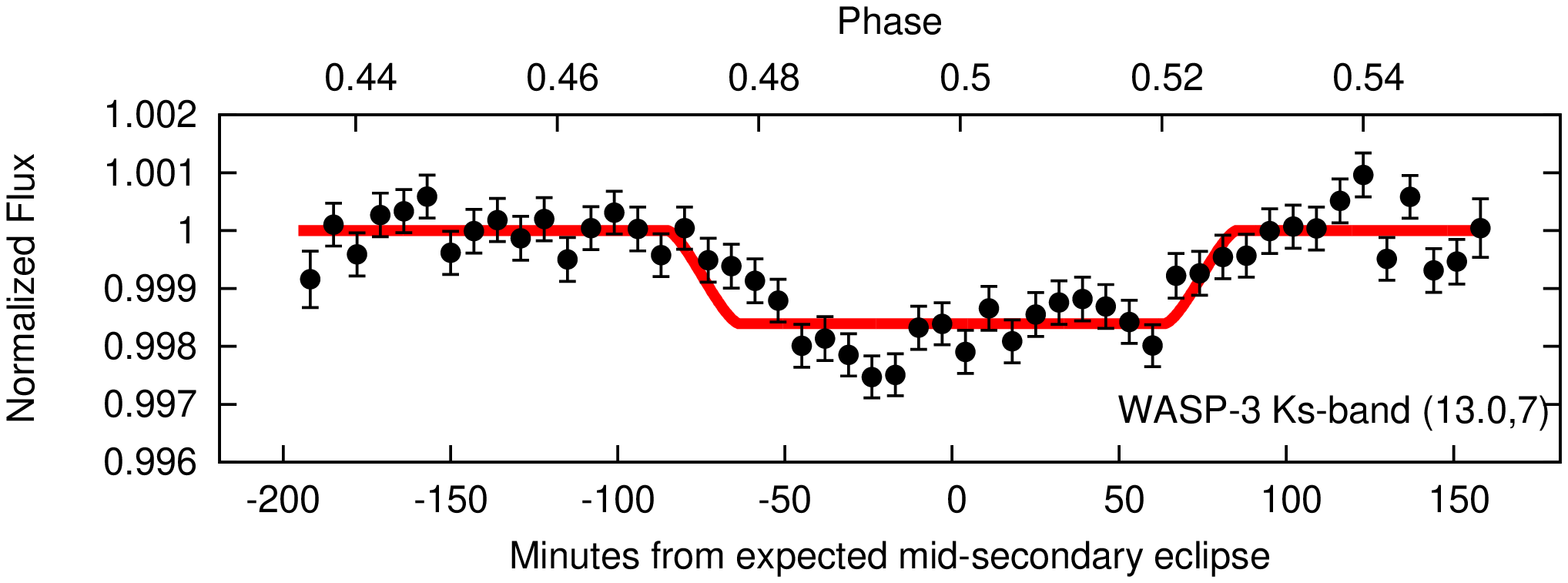}
\includegraphics[scale=0.27, angle = 270]{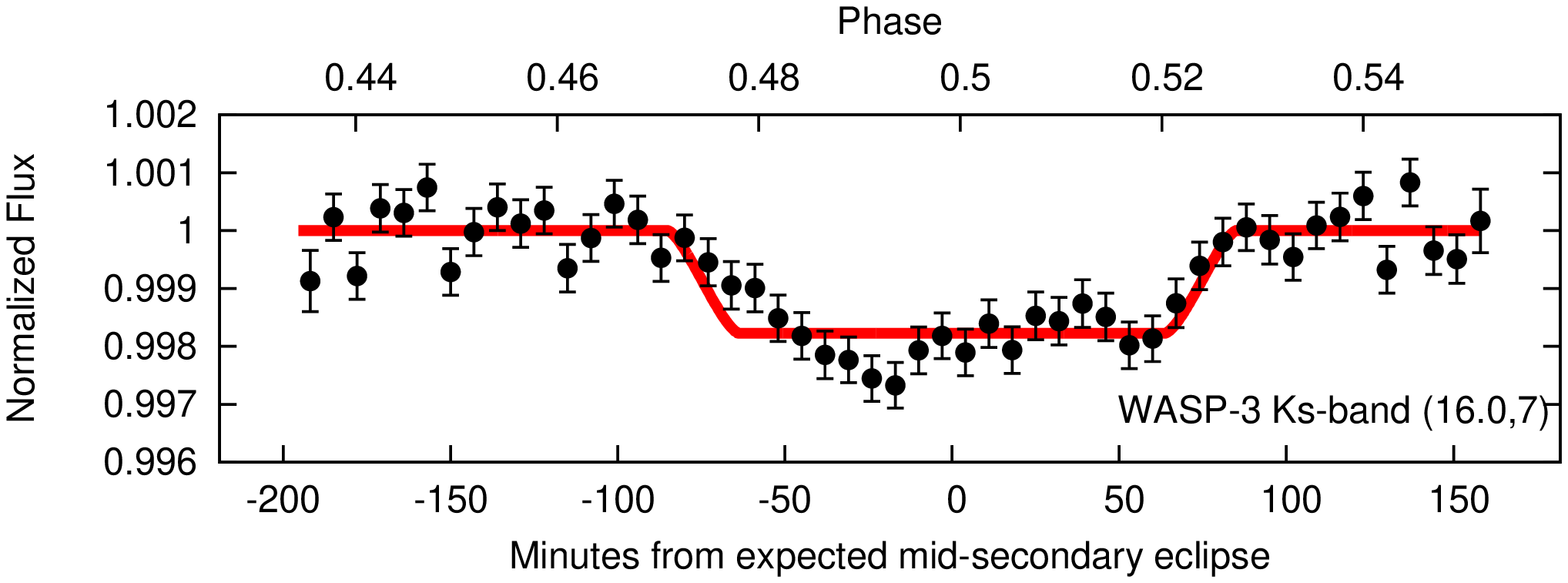}
\includegraphics[scale=0.27, angle = 270]{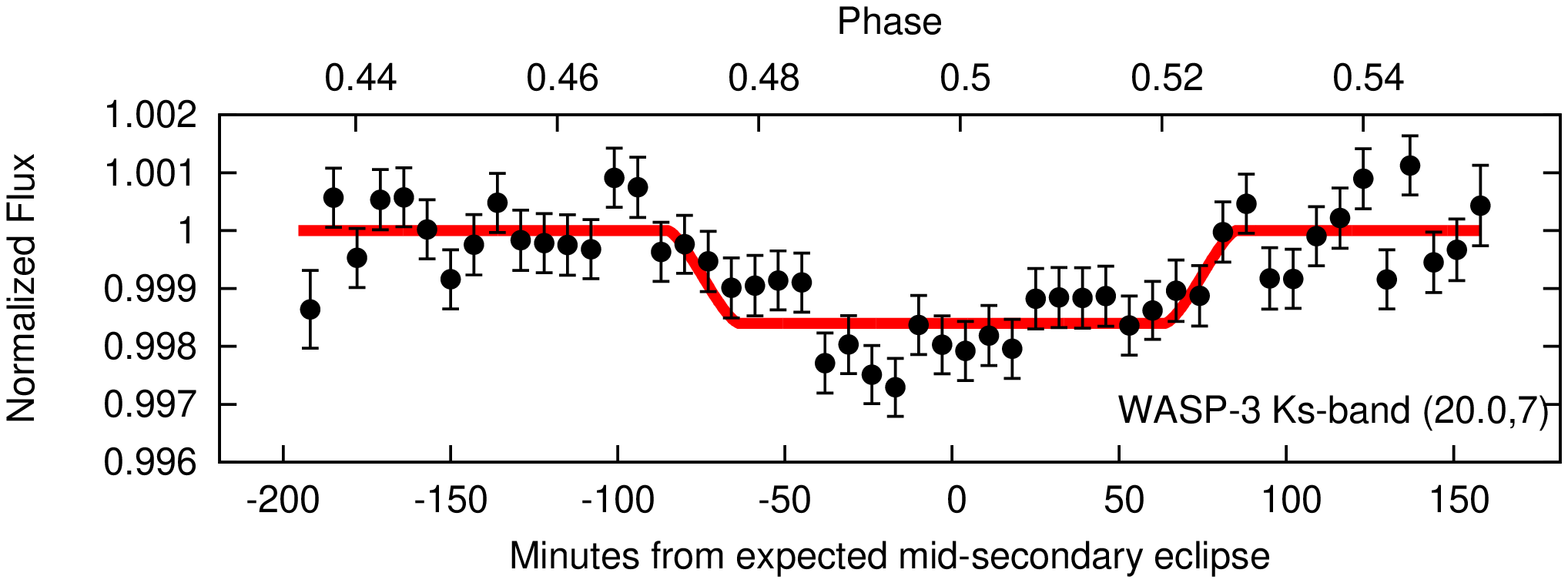}

\includegraphics[scale=0.27, angle = 270]{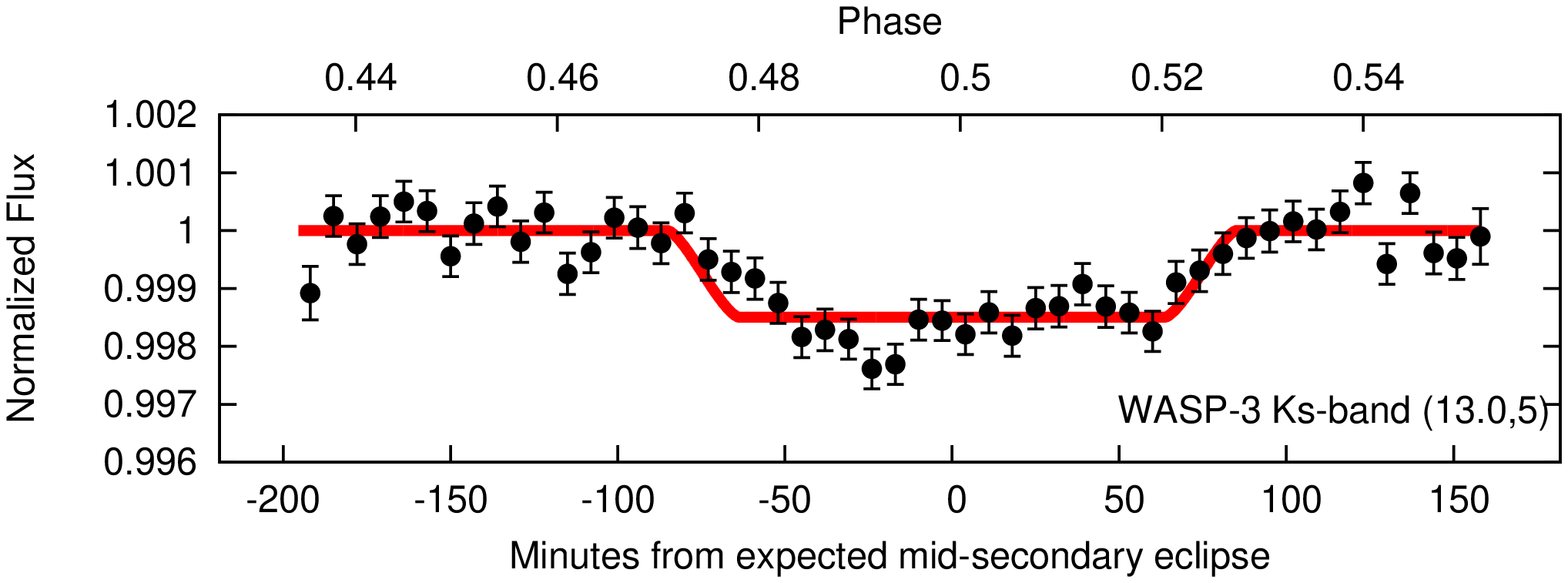}
\includegraphics[scale=0.27, angle = 270]{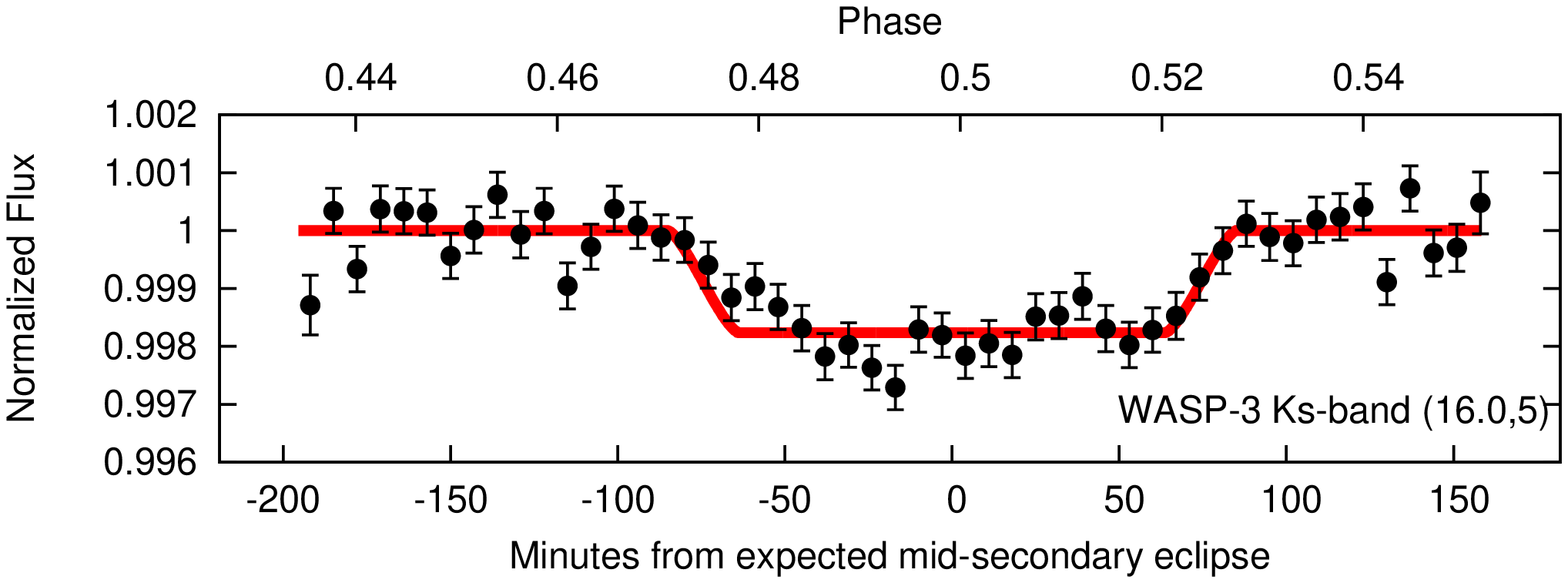}
\includegraphics[scale=0.27, angle = 270]{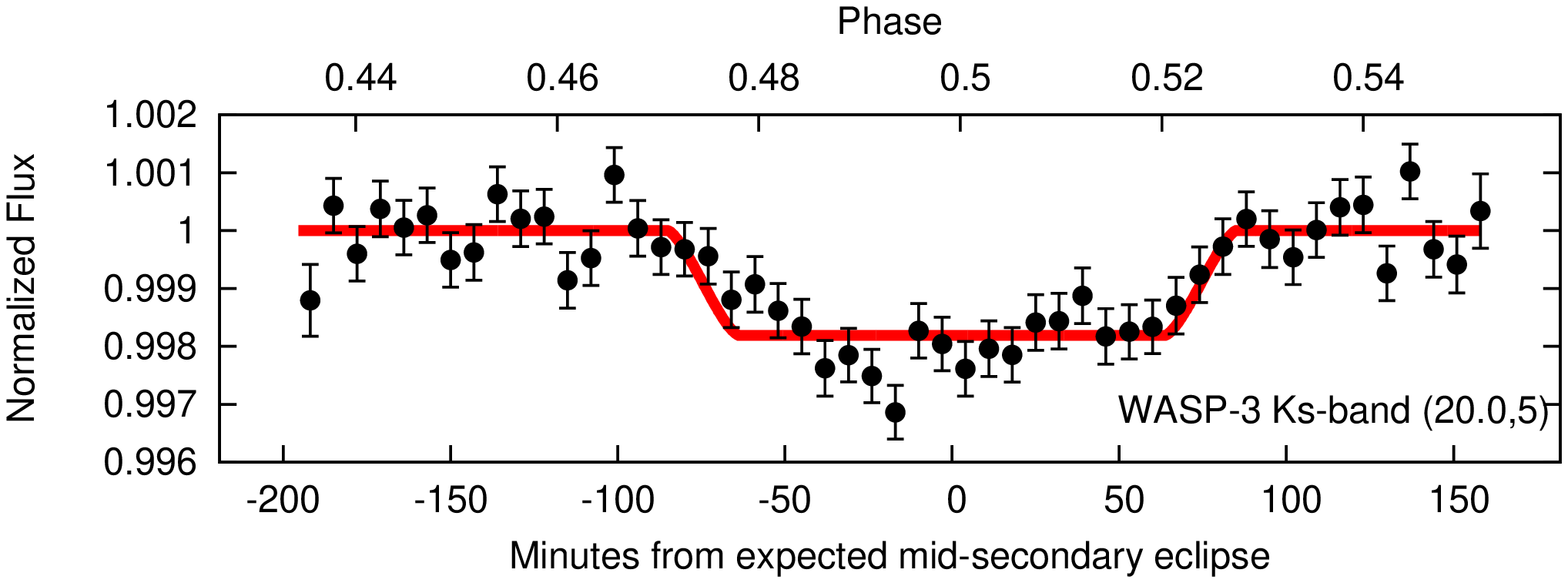}

\includegraphics[scale=0.27, angle = 270]{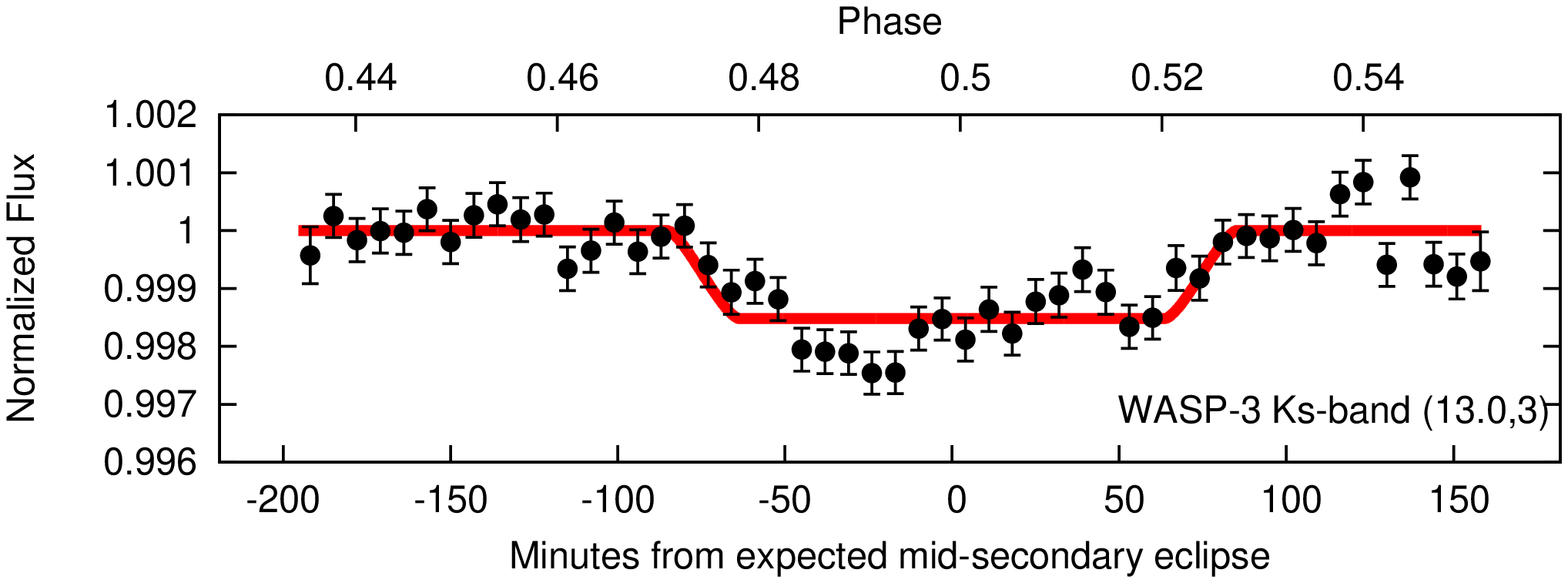}
\includegraphics[scale=0.27, angle = 270]{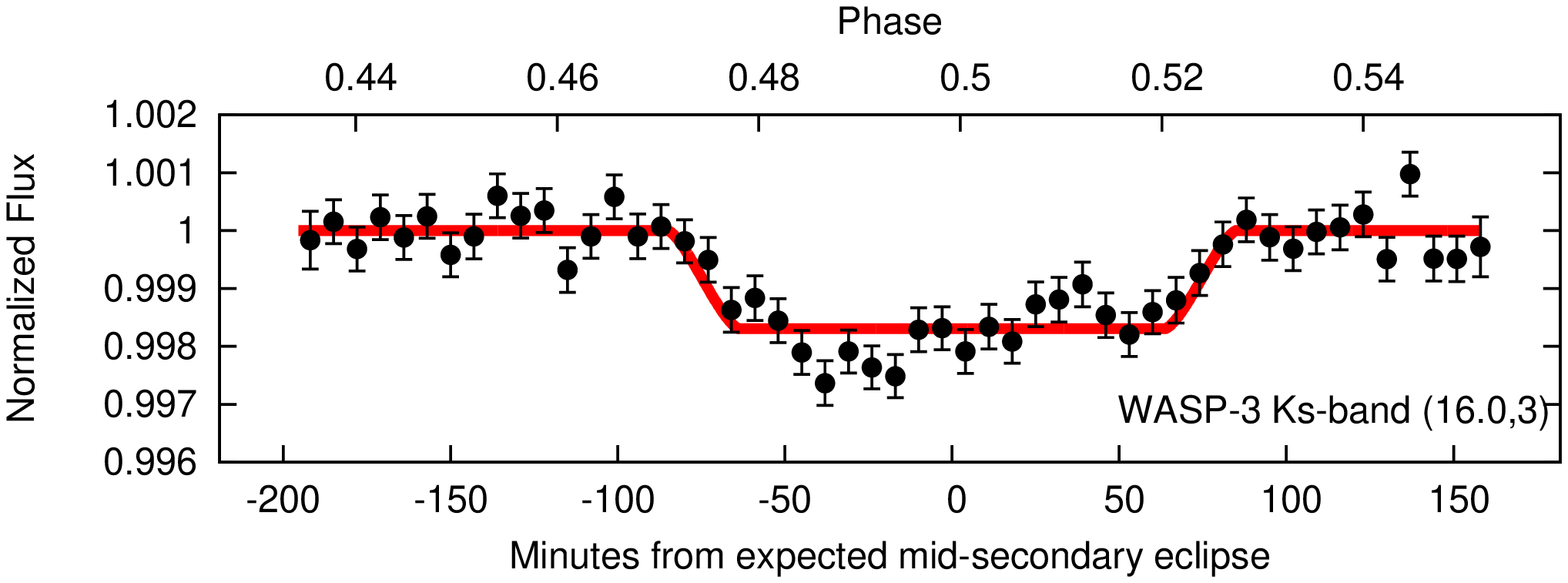}
\includegraphics[scale=0.27, angle = 270]{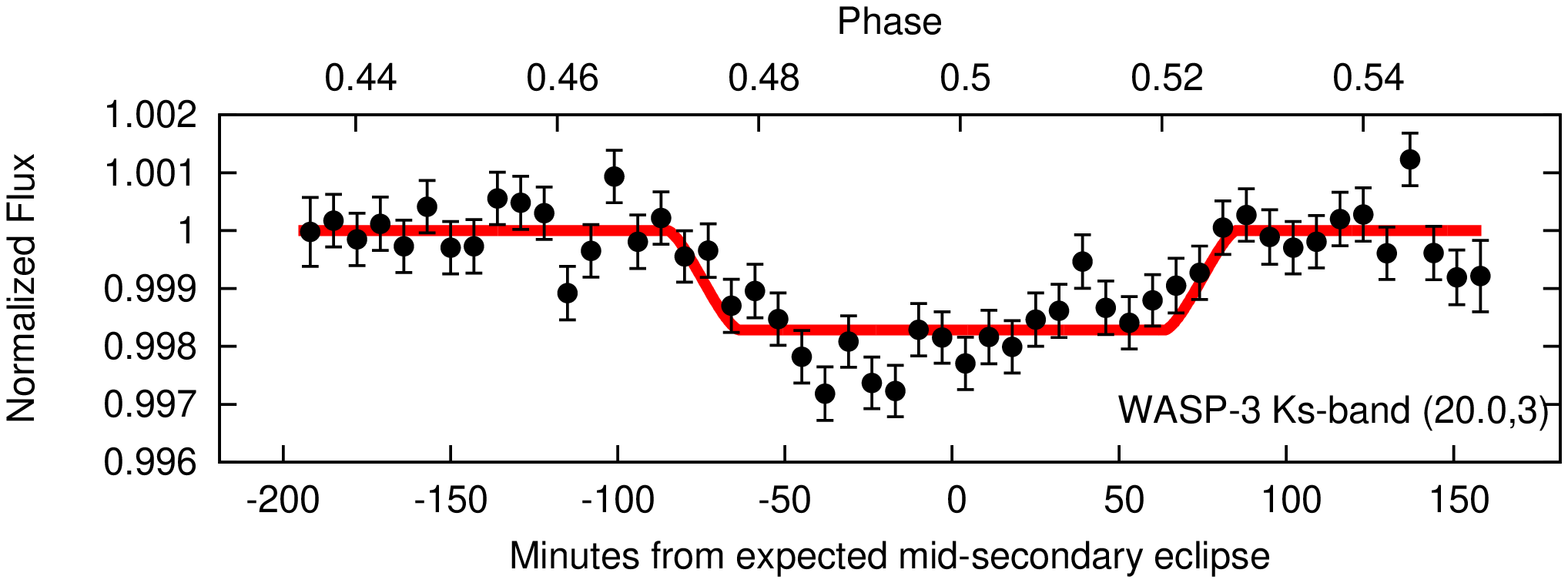}

\includegraphics[scale=0.27, angle = 270]{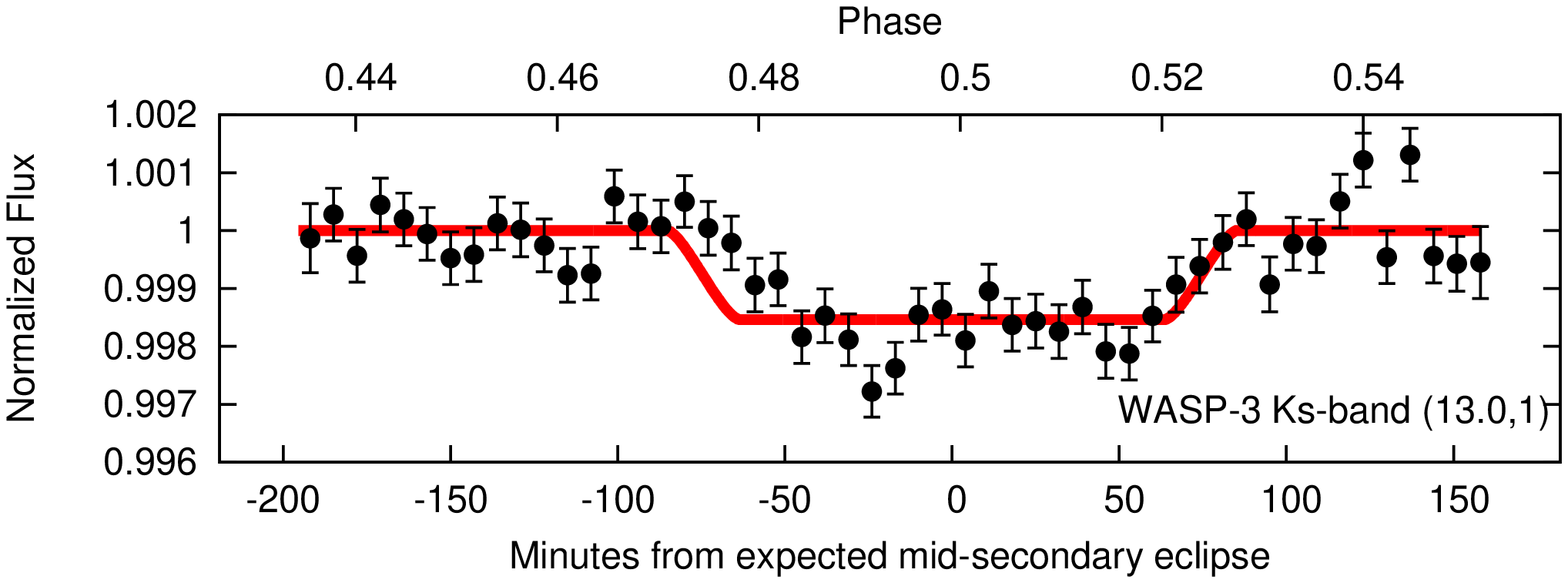}
\includegraphics[scale=0.27, angle = 270]{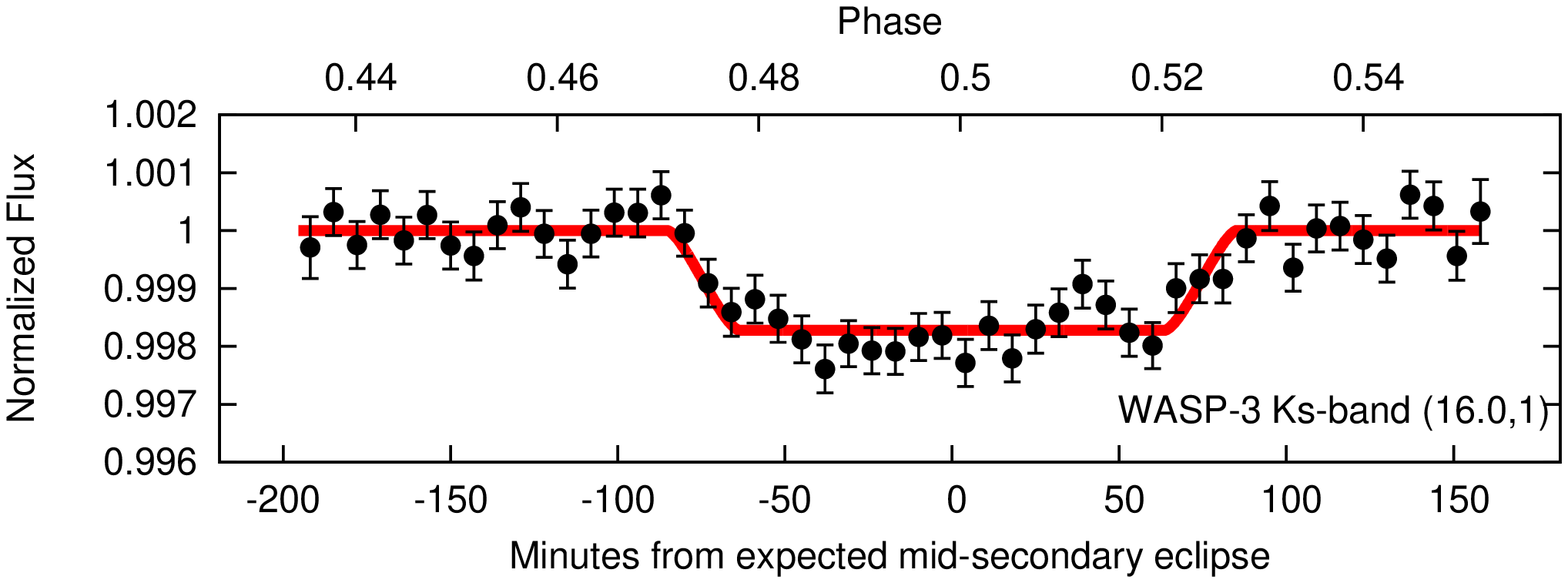}
\includegraphics[scale=0.27, angle = 270]{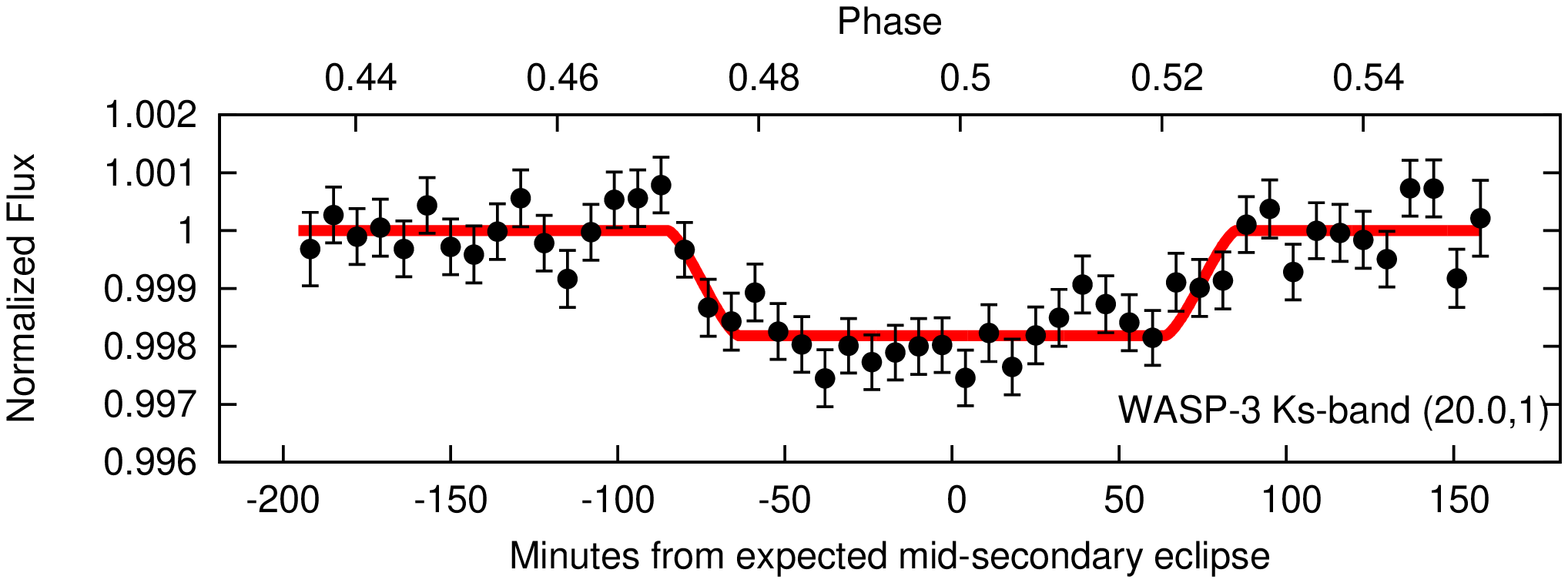}

\caption[BLAH]
	{	
		Same as Figure \ref{FigWASPTwelveKsbandFidelityManyOne} except for our second WASP-3 Ks-band secondary eclipse.	
		The scale of the bottom panels is identical to that of the other WASP-3 Ks-band
		eclipse (Figure \ref{FigWASPThreeKsbandFidelityManyOne}).
		For our second WASP-3 Ks-band eclipse, utilizing more than one reference star introduces correlated noise.
	}
\label{FigWASPThreeKsbandFidelityManyTwo}
\end{figure*}

\begin{figure*}
\centering
\includegraphics[scale=0.44, angle = 270]{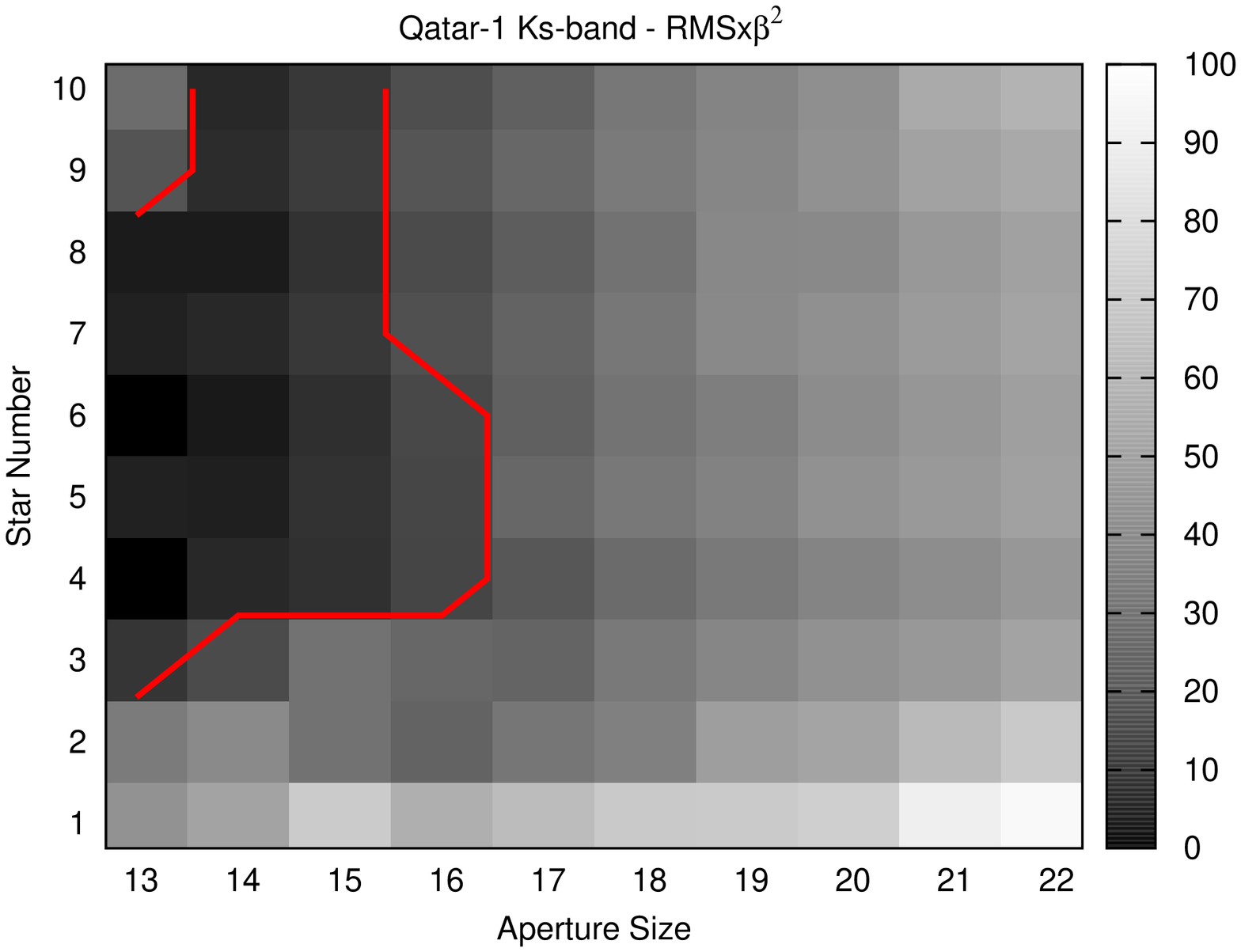}
\includegraphics[scale=0.44, angle = 270]{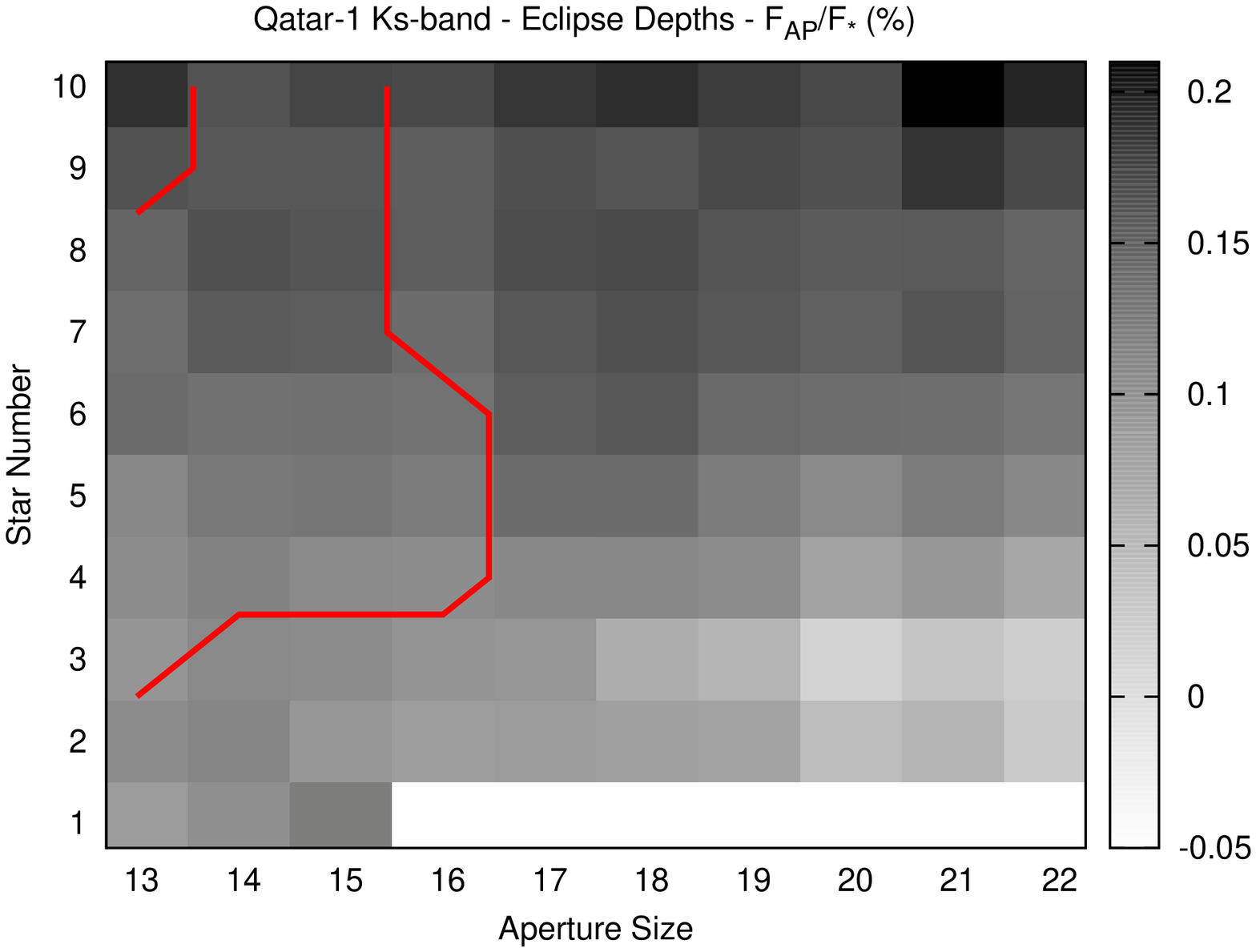}

\includegraphics[scale=0.27, angle = 270]{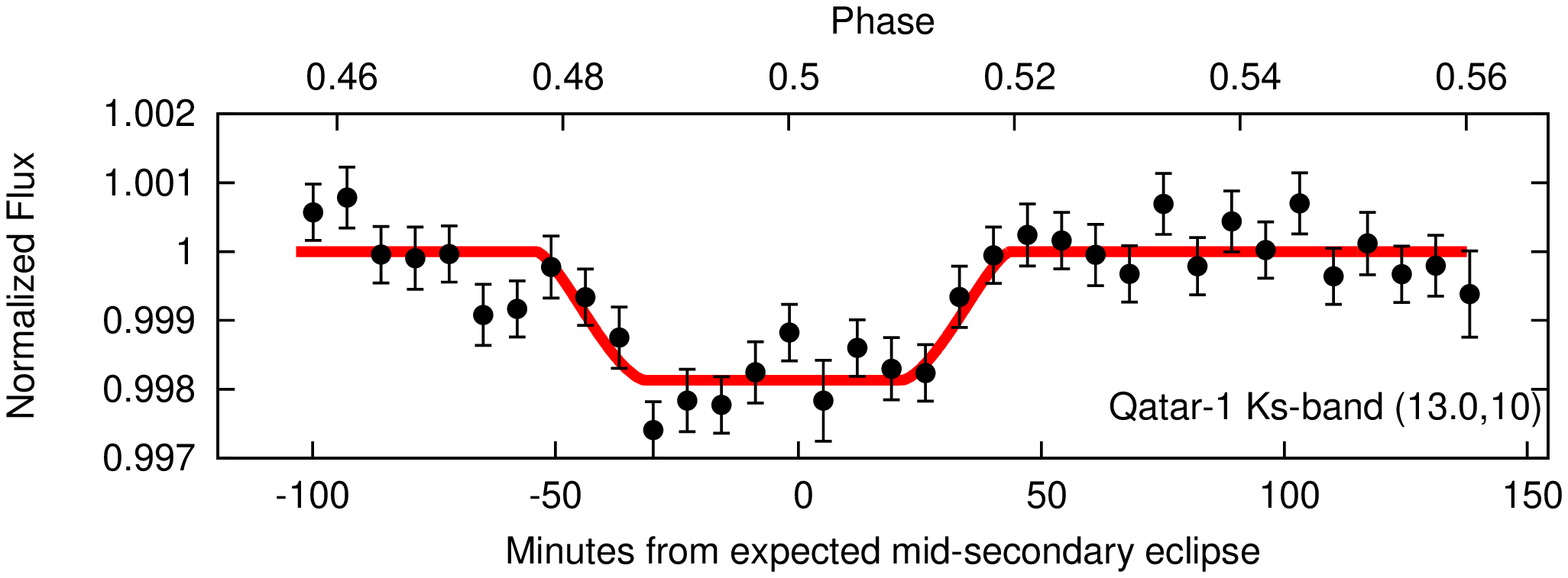}
\includegraphics[scale=0.27, angle = 270]{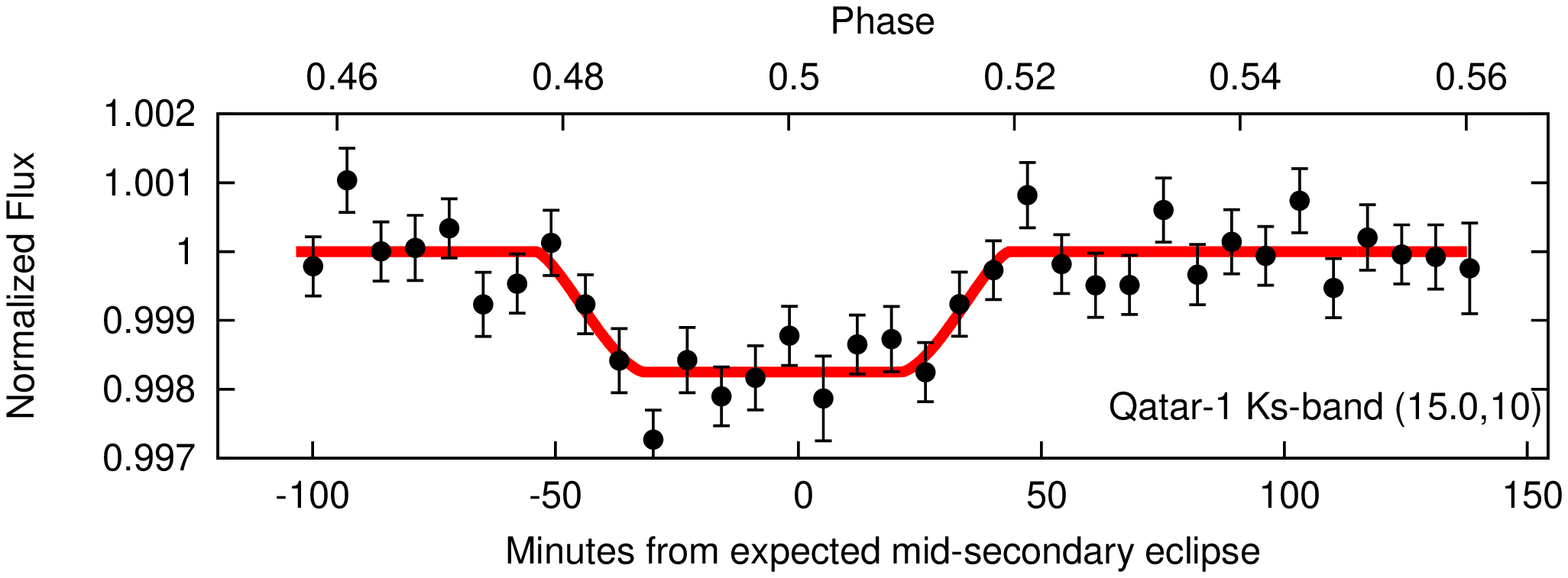}
\includegraphics[scale=0.27, angle = 270]{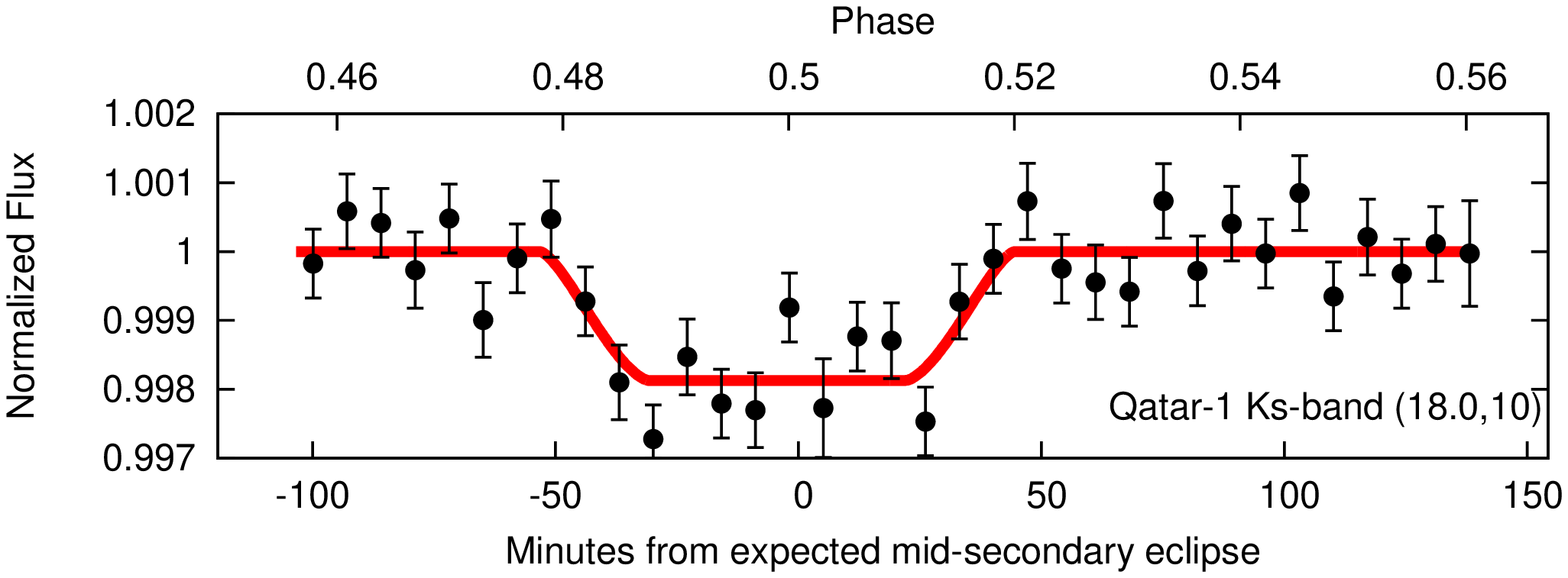}

\includegraphics[scale=0.27, angle = 270]{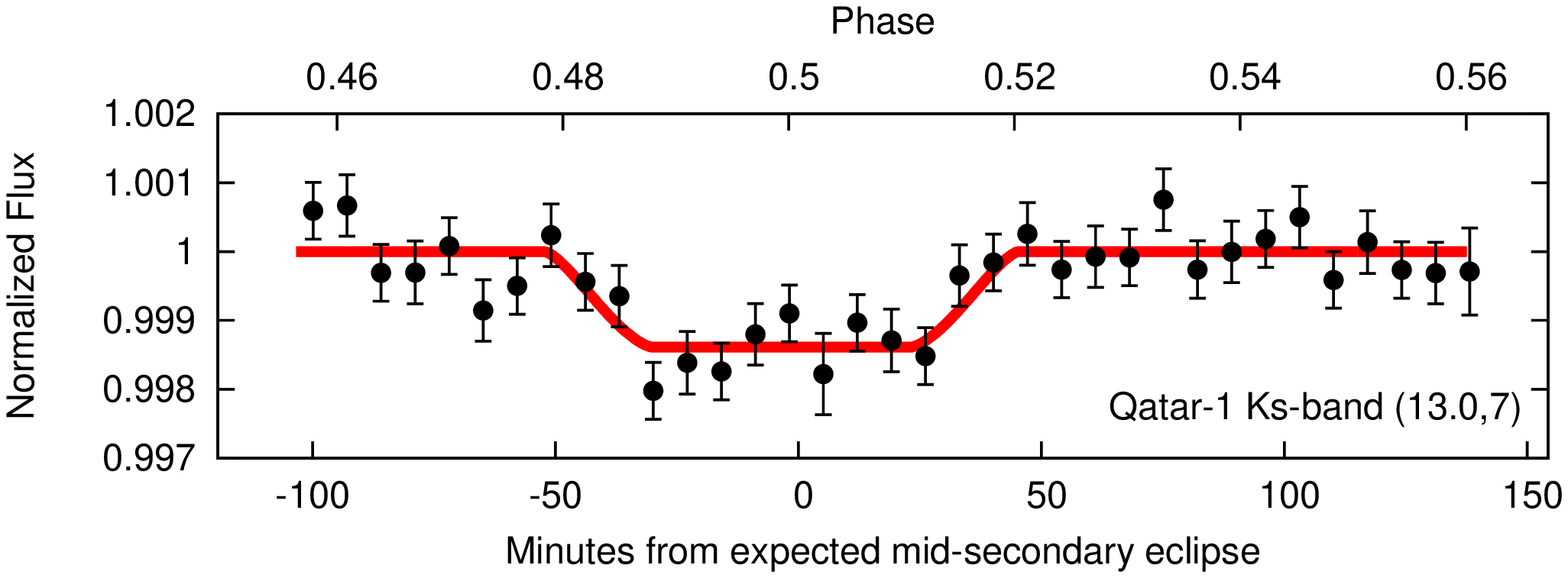}
\includegraphics[scale=0.27, angle = 270]{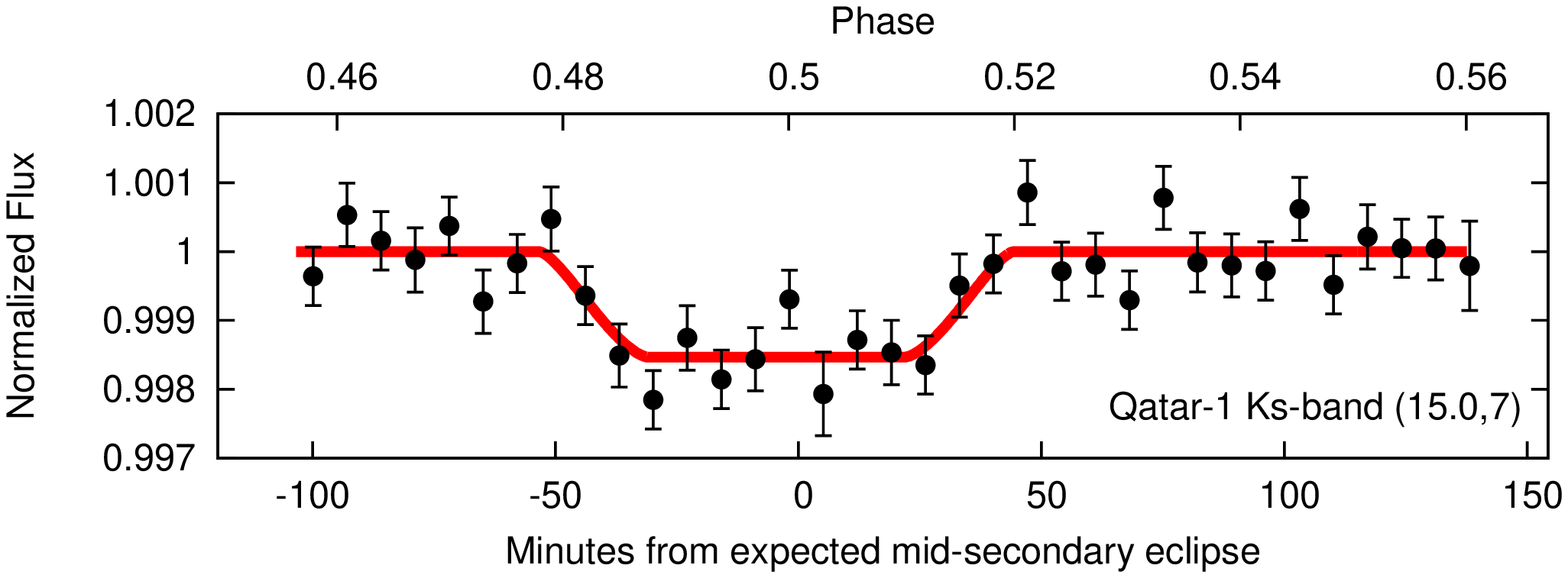}
\includegraphics[scale=0.27, angle = 270]{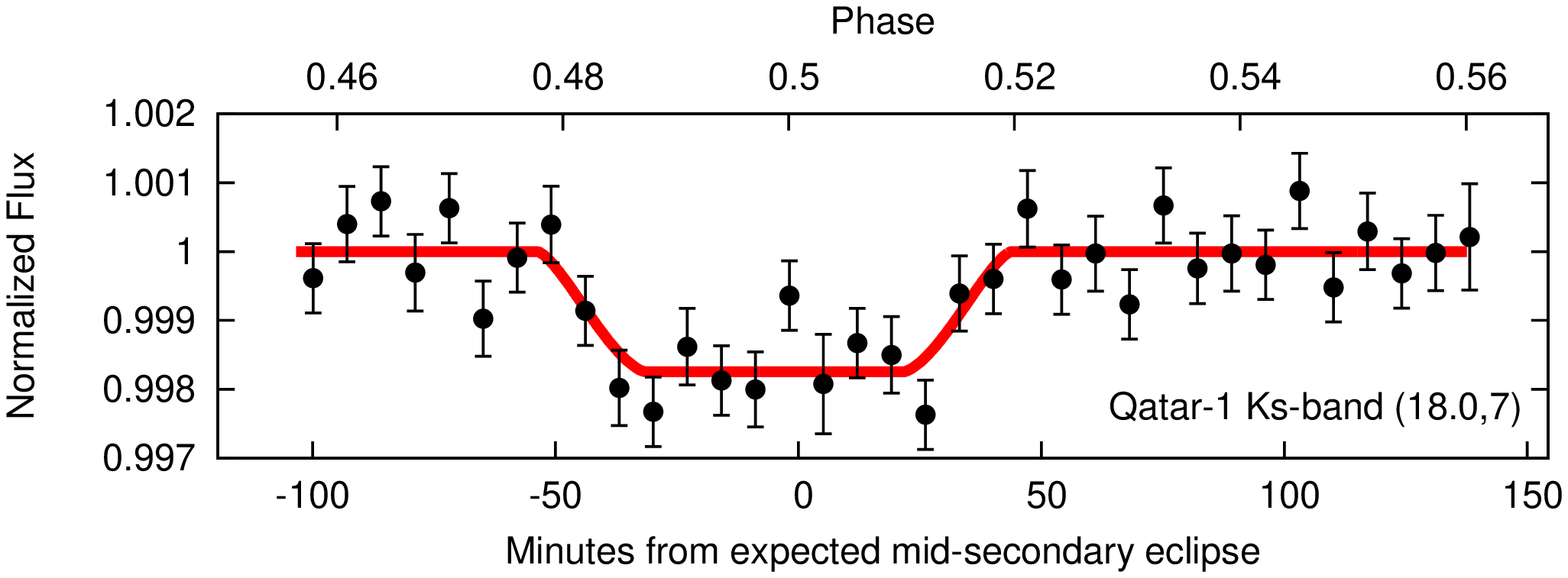}

\includegraphics[scale=0.27, angle = 270]{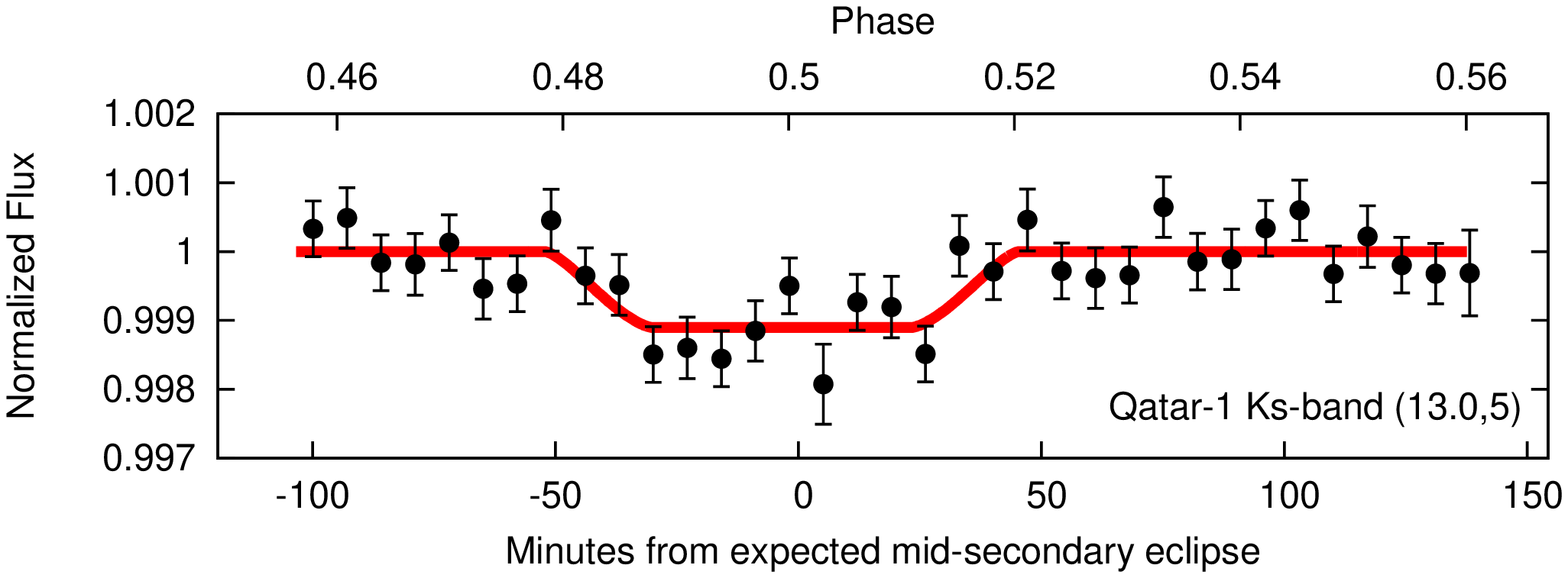}
\includegraphics[scale=0.27, angle = 270]{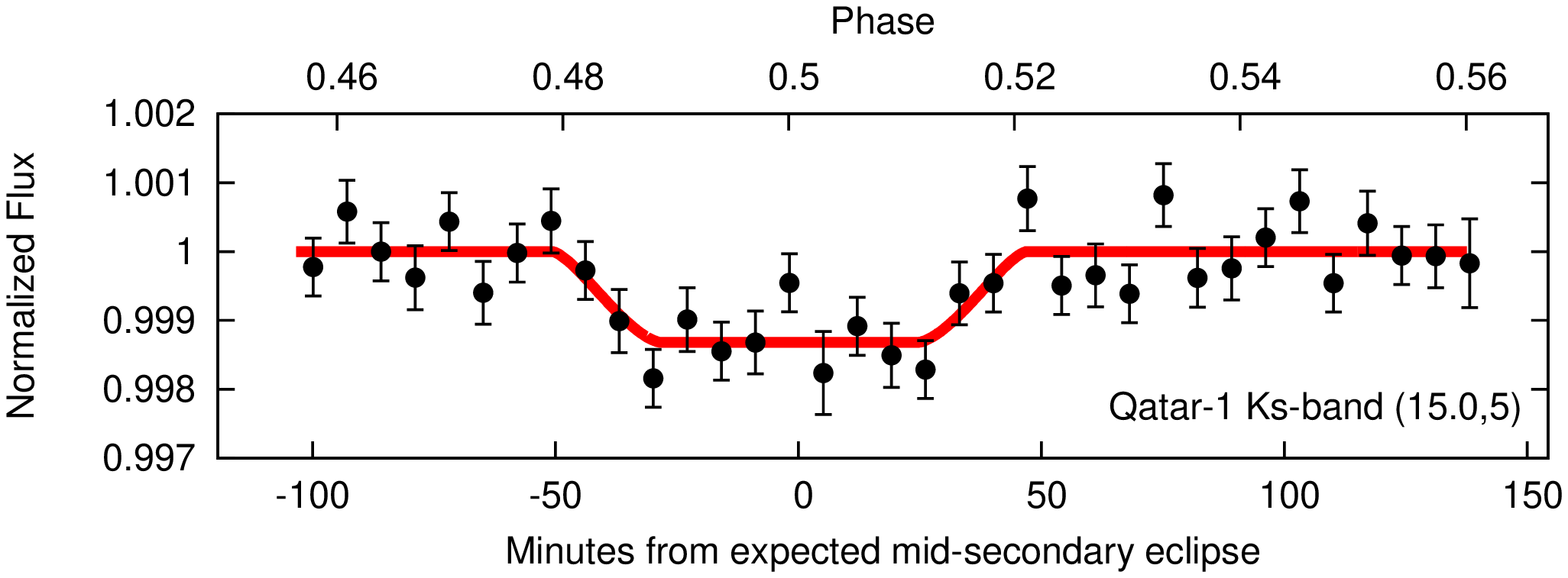}
\includegraphics[scale=0.27, angle = 270]{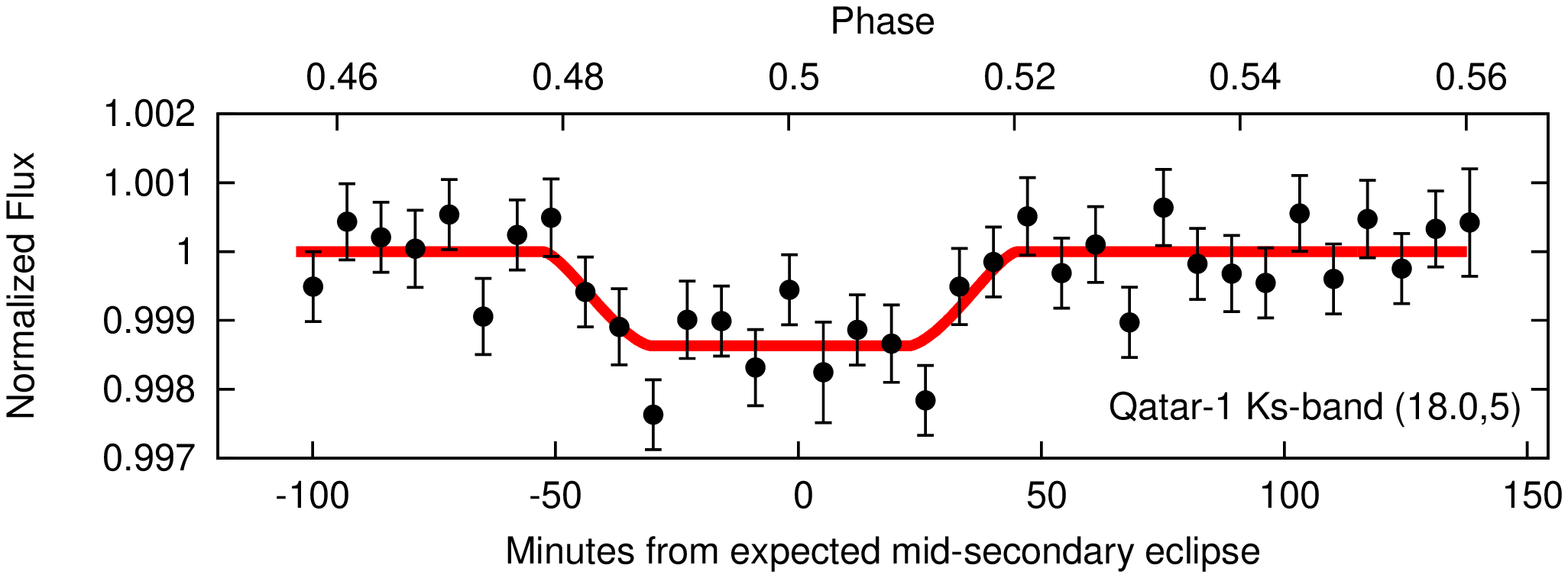}

\includegraphics[scale=0.27, angle = 270]{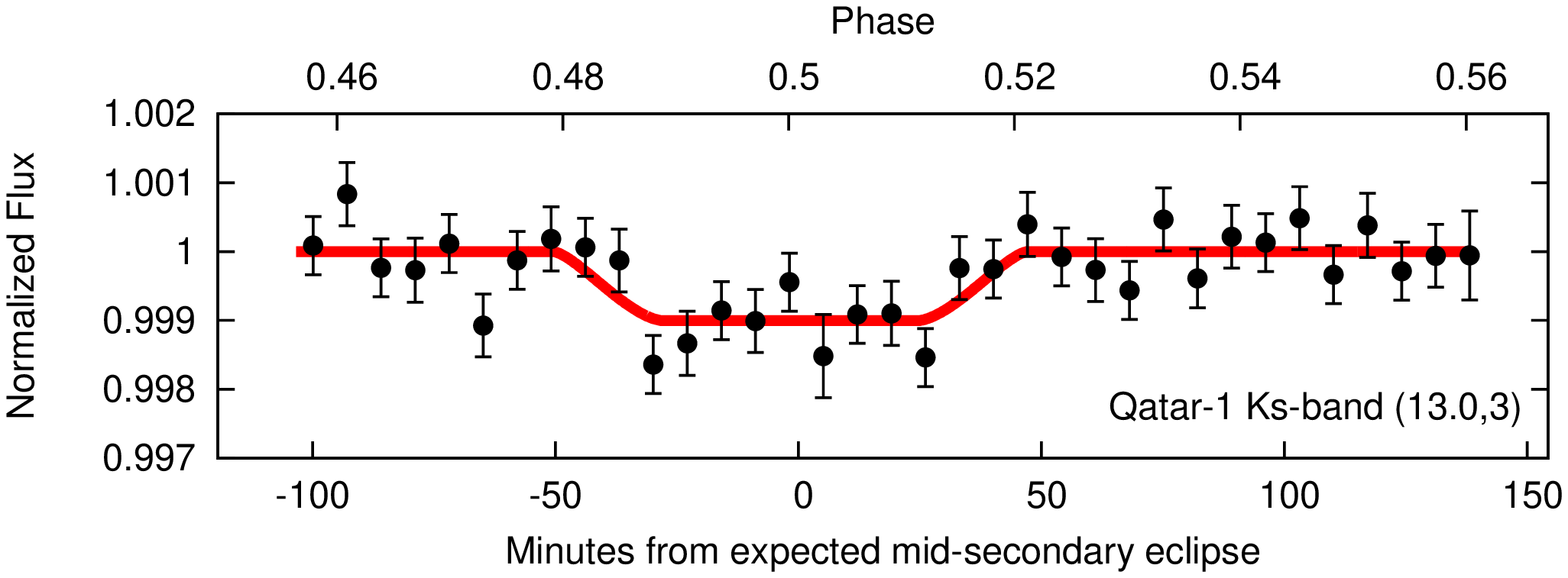}
\includegraphics[scale=0.27, angle = 270]{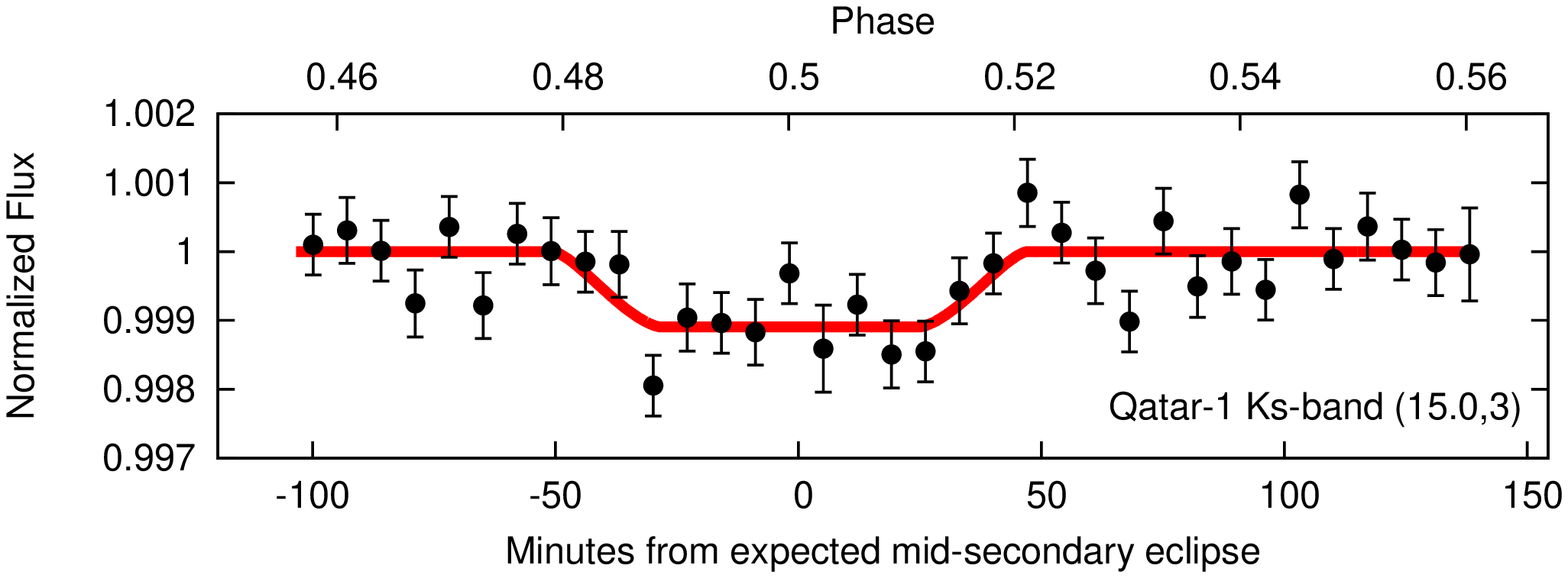}
\includegraphics[scale=0.27, angle = 270]{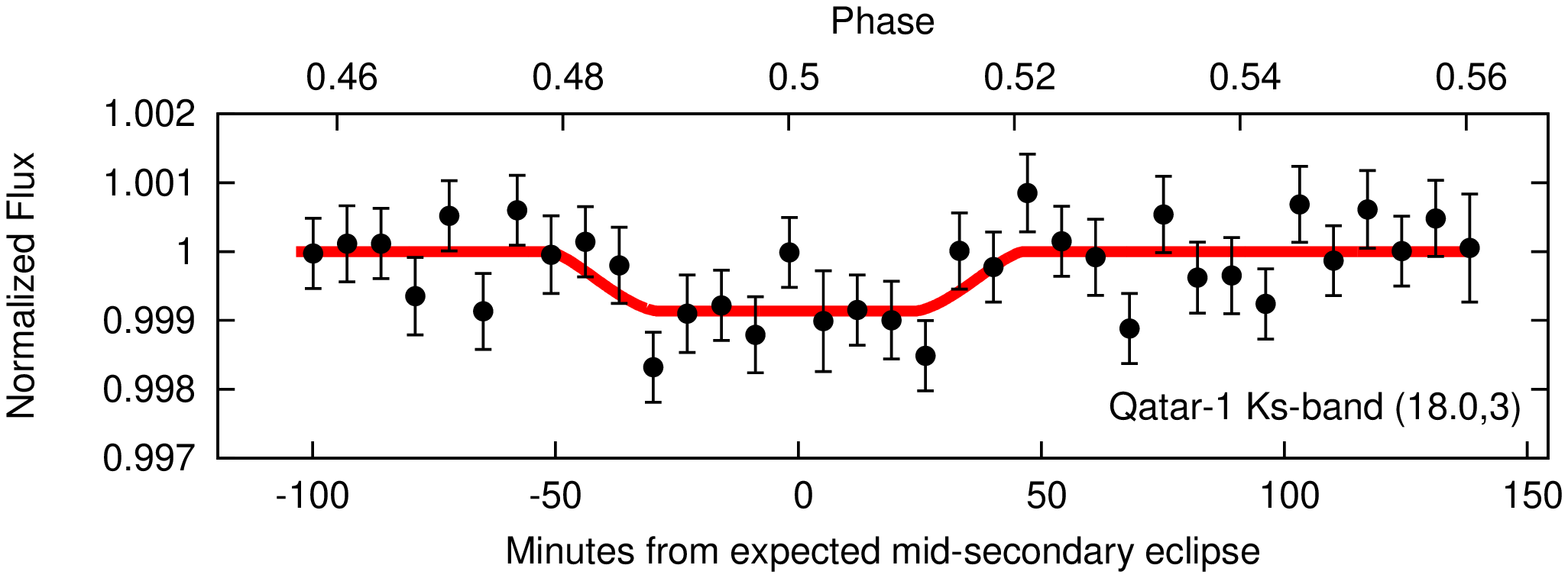}

\includegraphics[scale=0.27, angle = 270]{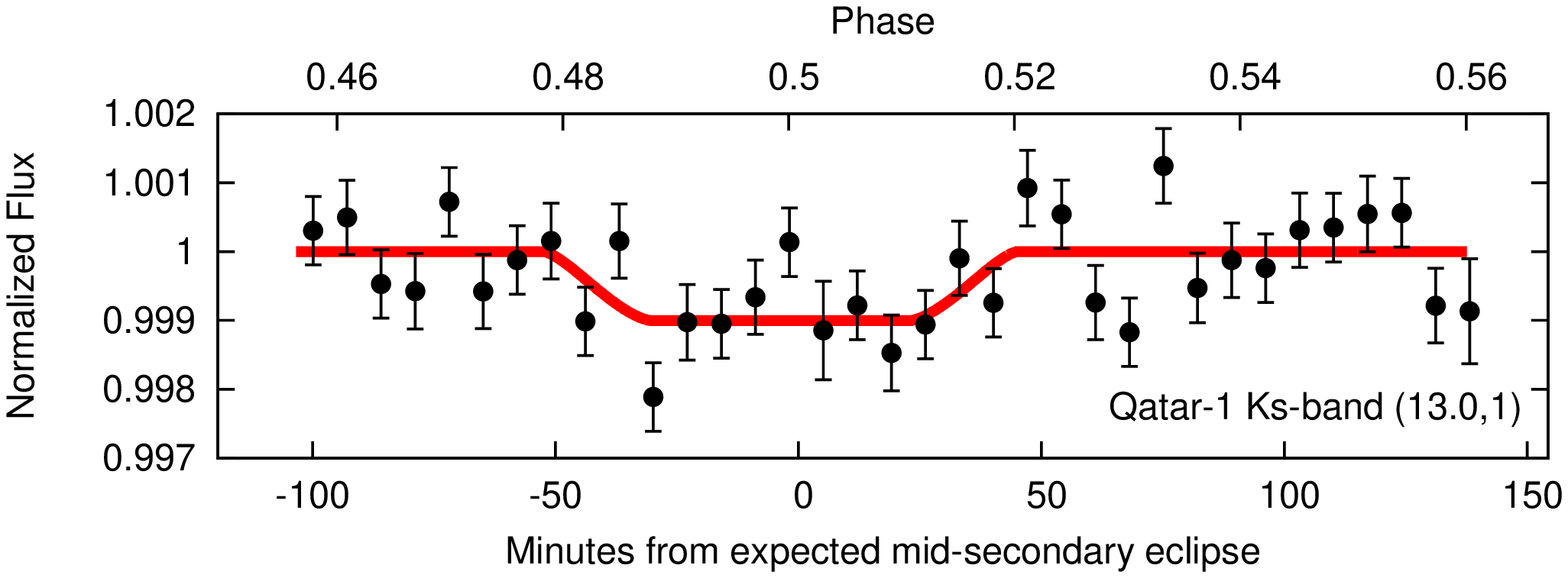}
\includegraphics[scale=0.27, angle = 270]{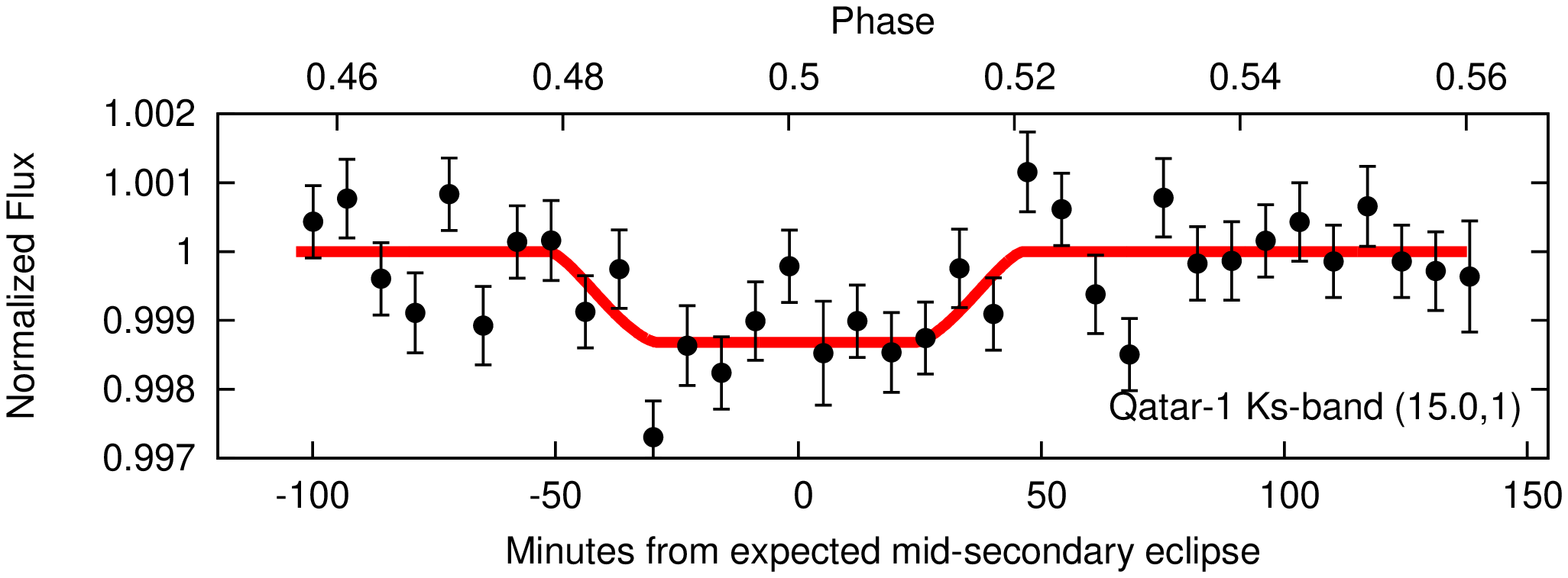}
\includegraphics[scale=0.27, angle = 270]{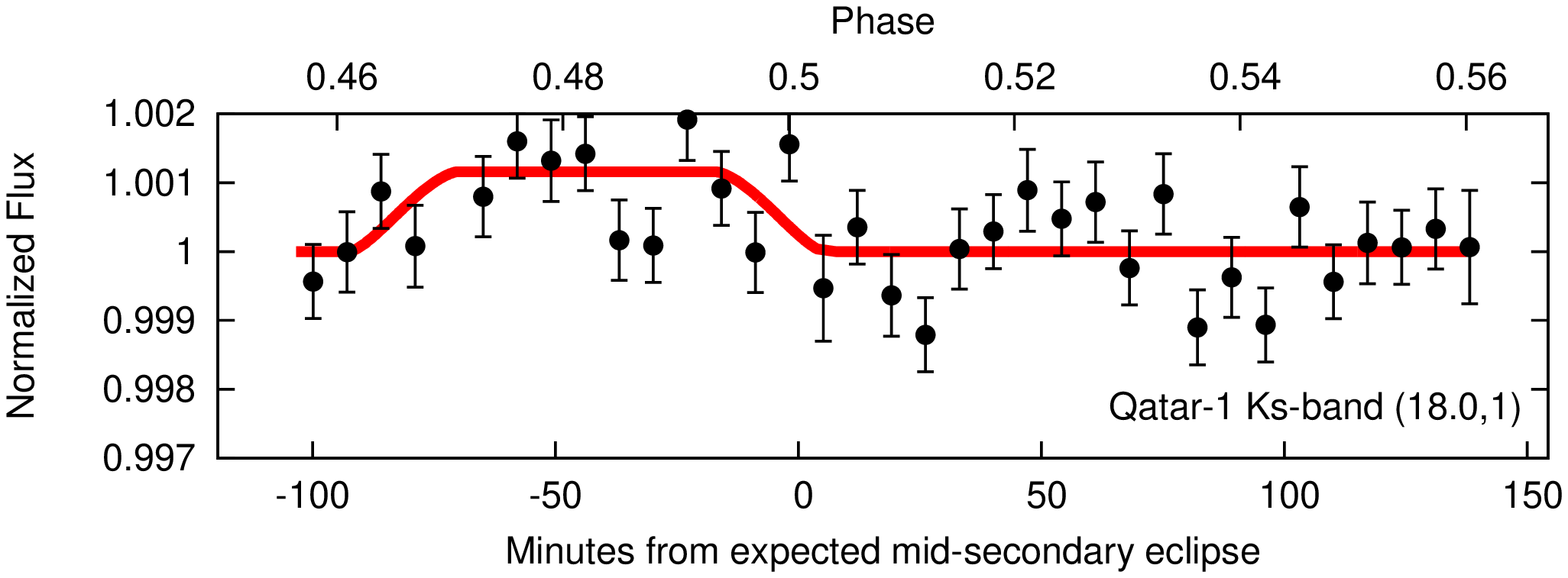}

\caption[BLAH]
	{	
		Same as Figure \ref{FigWASPTwelveKsbandFidelityManyOne} except for our Qatar-1 Ks-band secondary eclipse.
		For our Qatar-1 Ks-band eclipse, the eclipse depth varies significantly when different aperture sizes and reference star ensembles are chosen.
	}
\label{FigQatarOneKsbandFidelityMany}
\end{figure*}

\begin{figure*}
\centering

\includegraphics[scale=0.44, angle = 270]{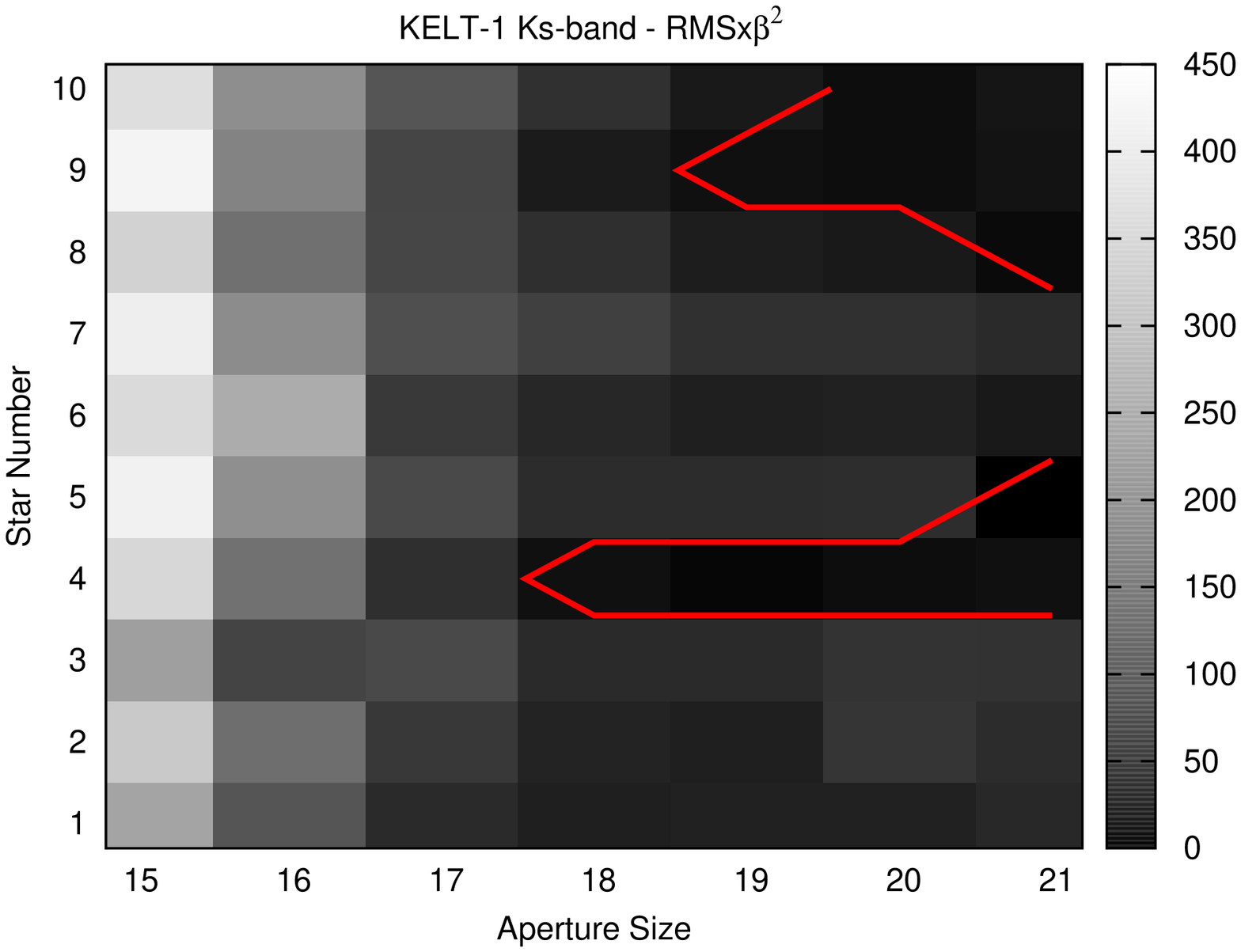}
\includegraphics[scale=0.44, angle = 270]{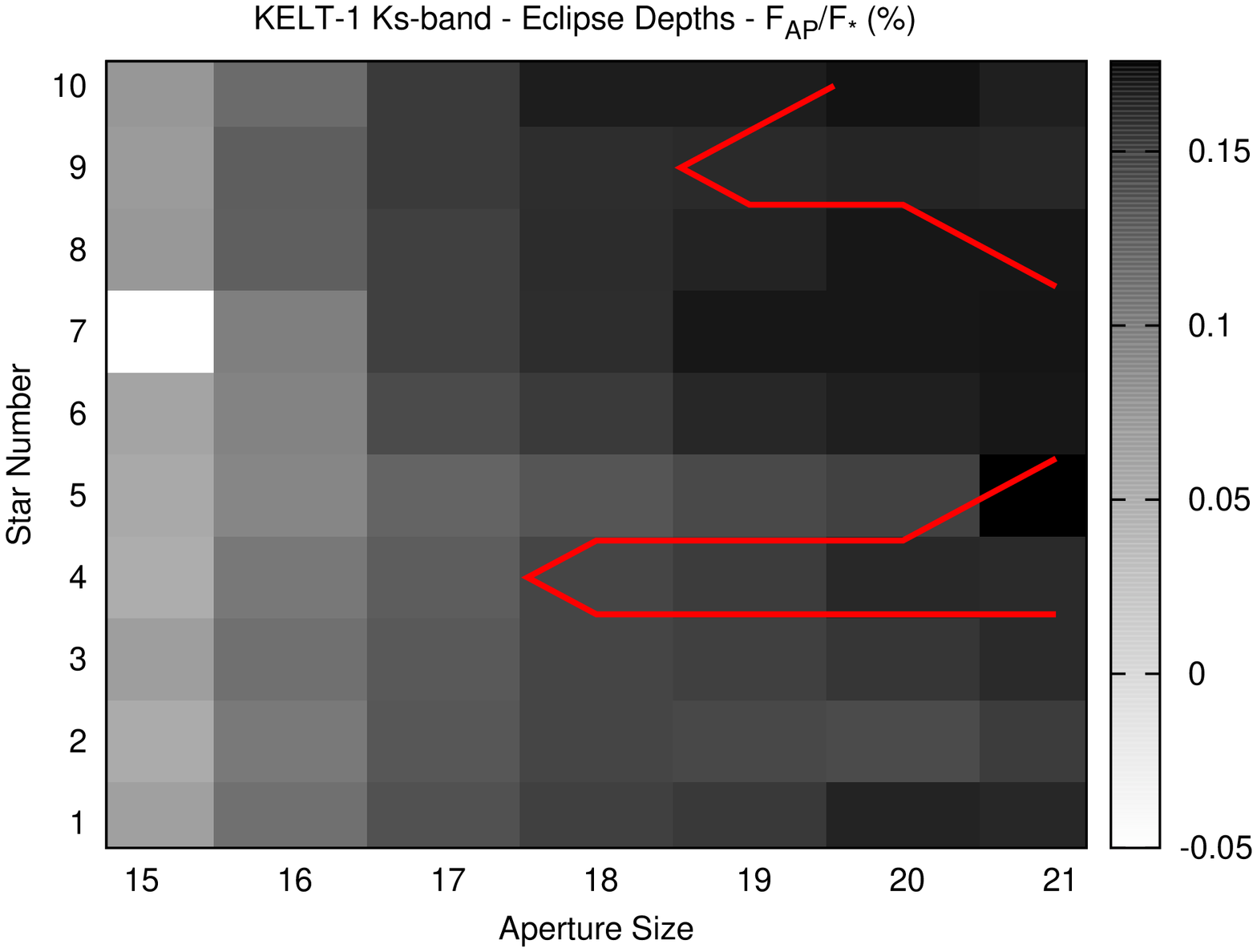}

\includegraphics[scale=0.26, angle = 270]{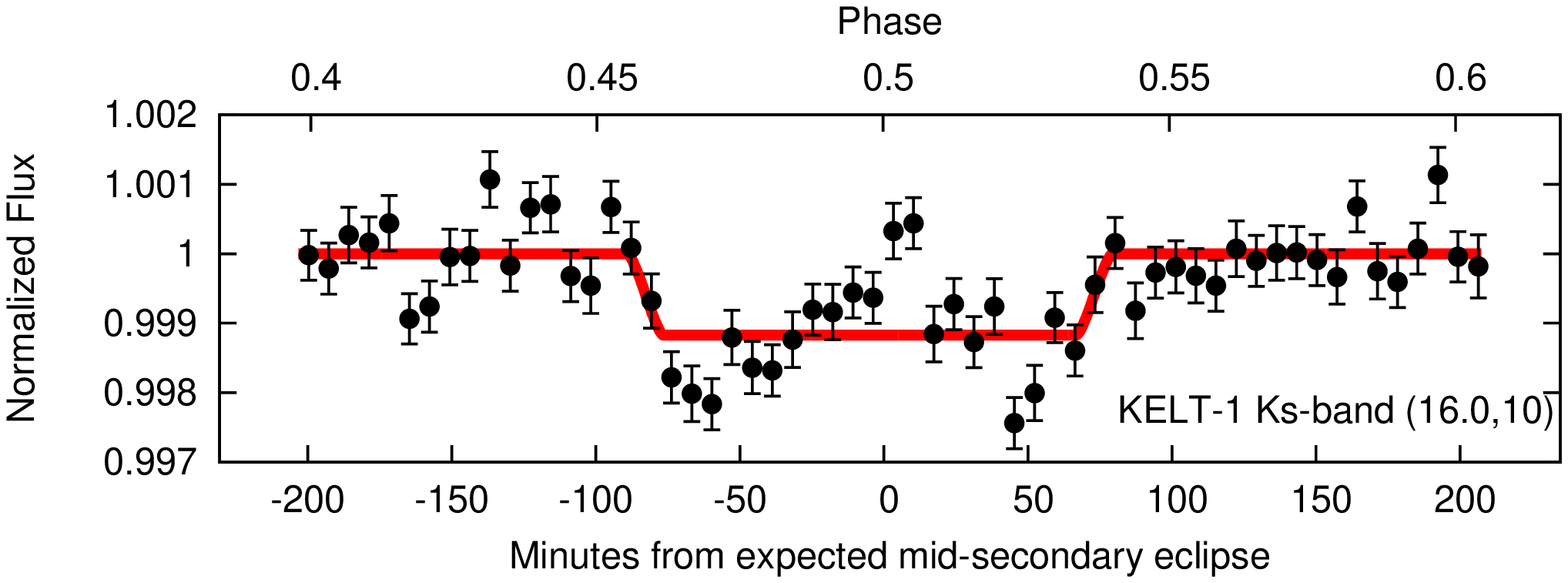}
\includegraphics[scale=0.26, angle = 270]{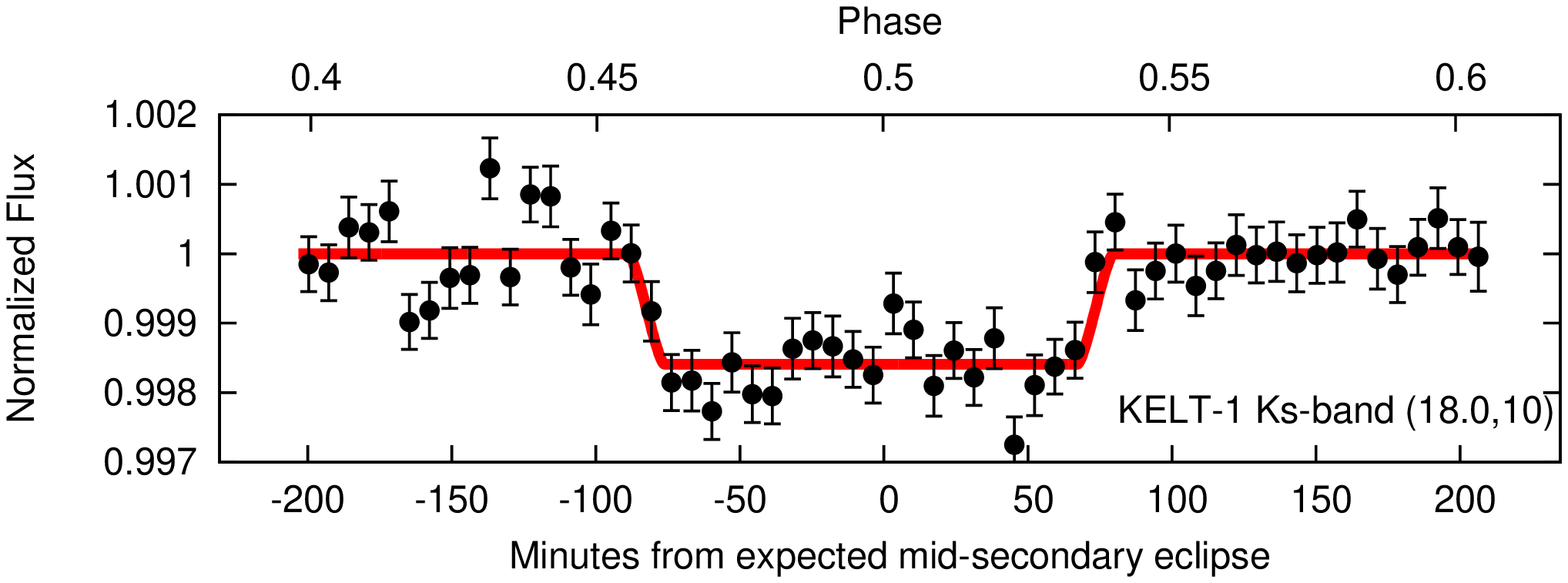}
\includegraphics[scale=0.26, angle = 270]{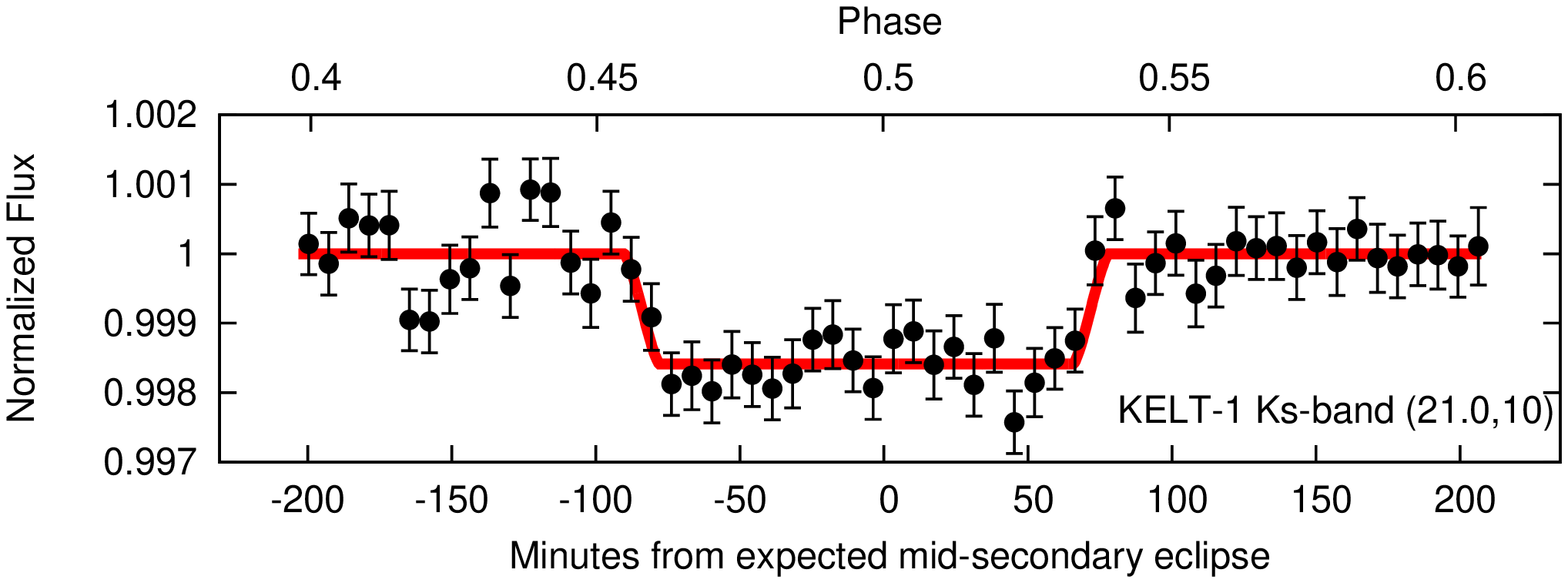}

\includegraphics[scale=0.26, angle = 270]{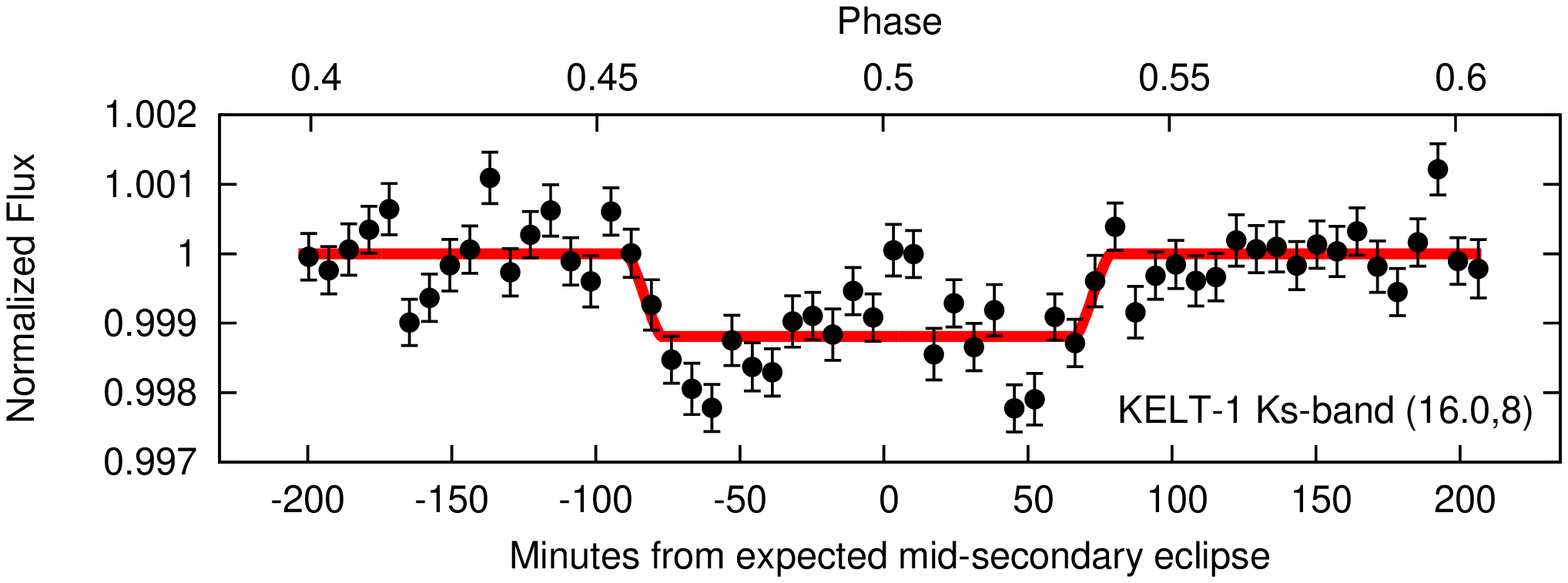}
\includegraphics[scale=0.26, angle = 270]{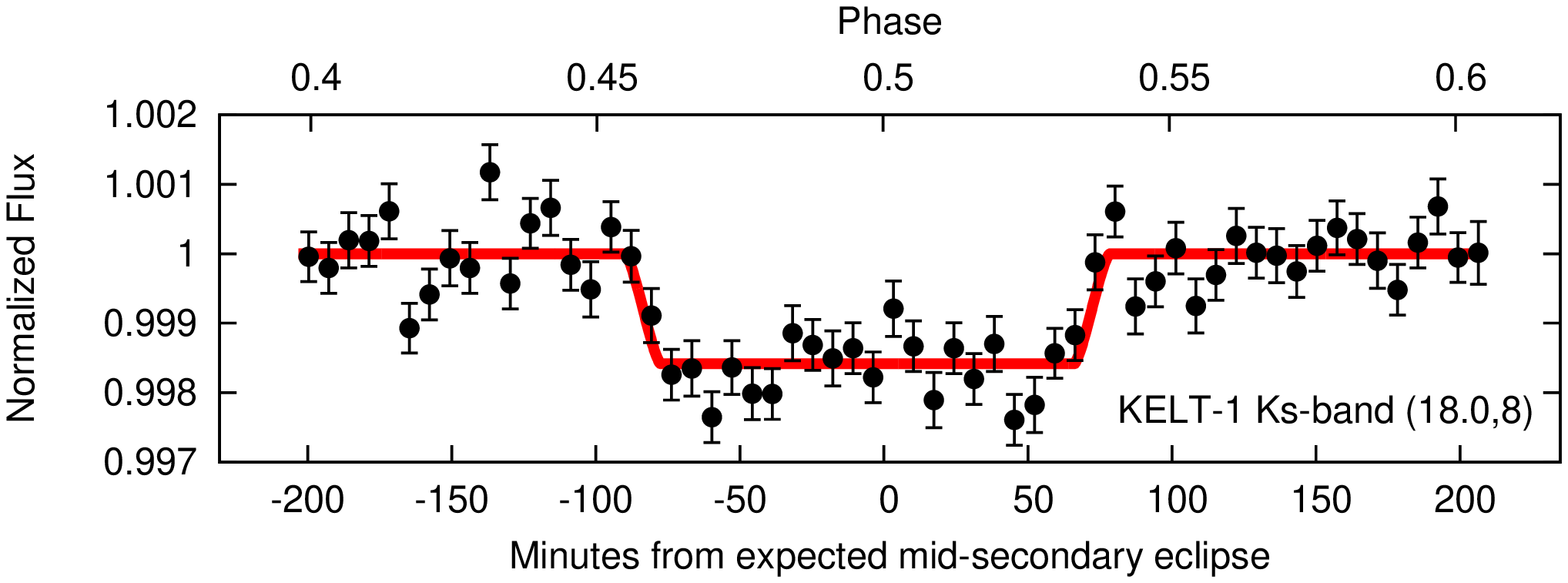}
\includegraphics[scale=0.26, angle = 270]{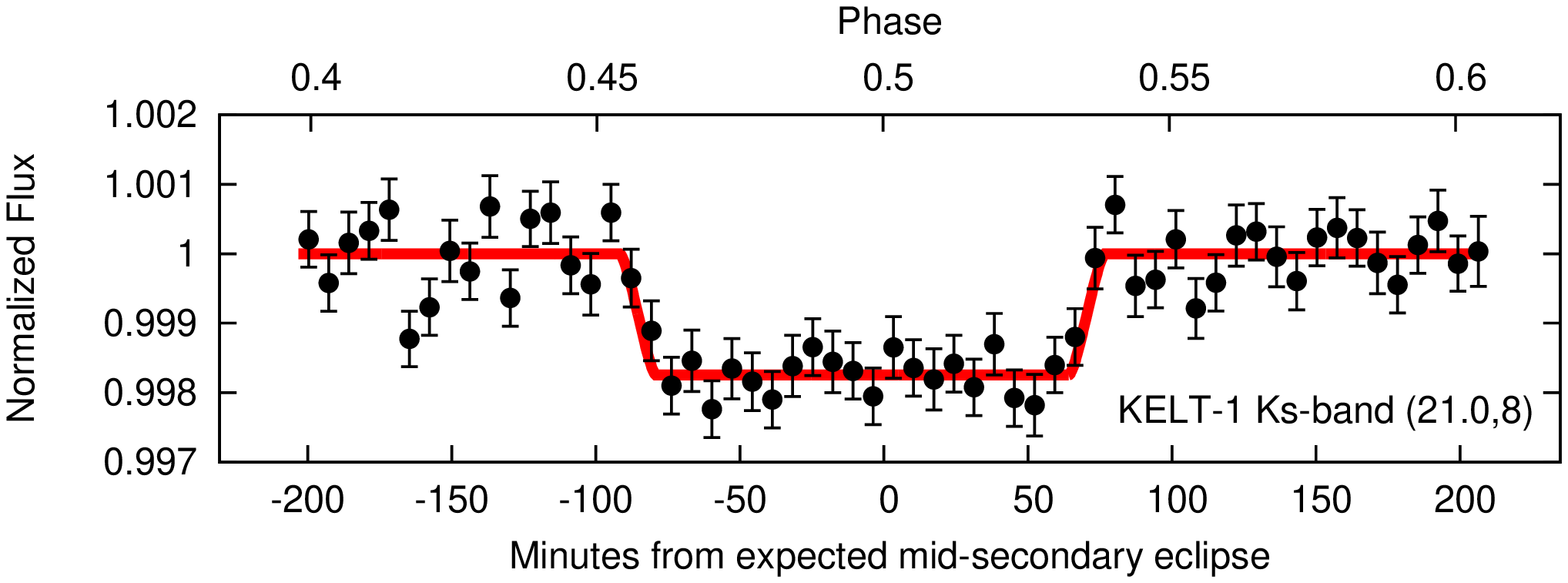}

\includegraphics[scale=0.26, angle = 270]{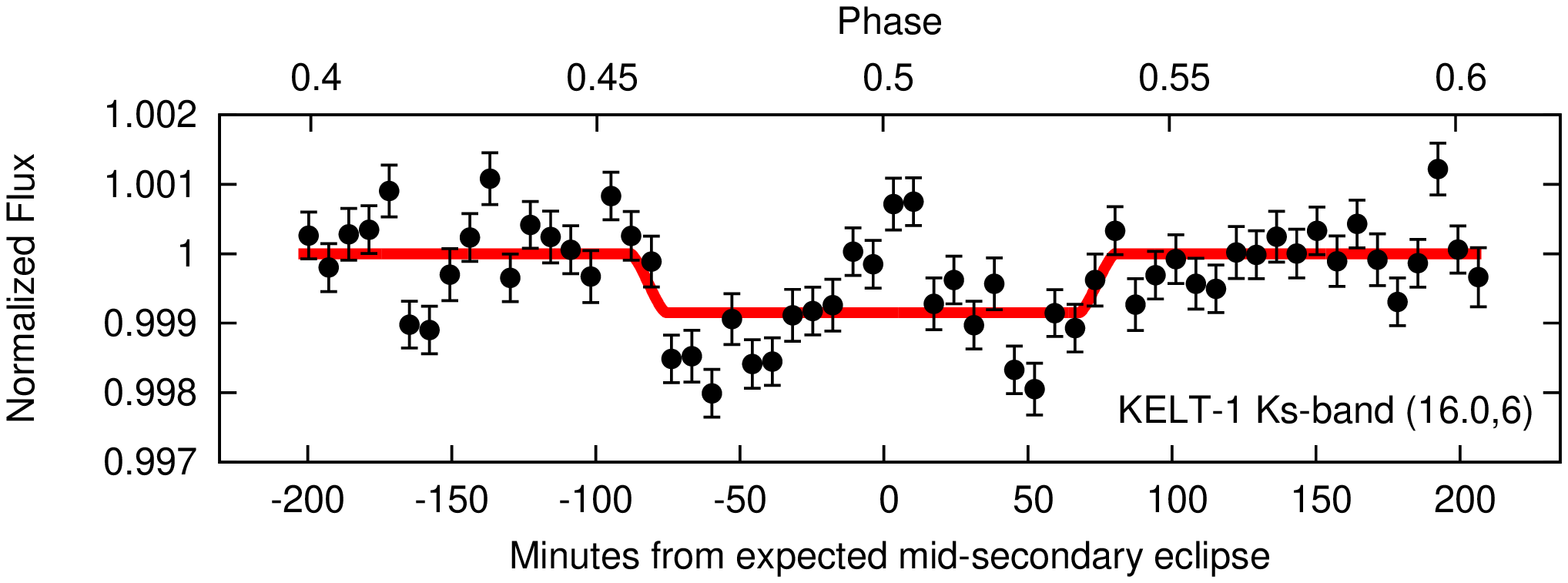}
\includegraphics[scale=0.26, angle = 270]{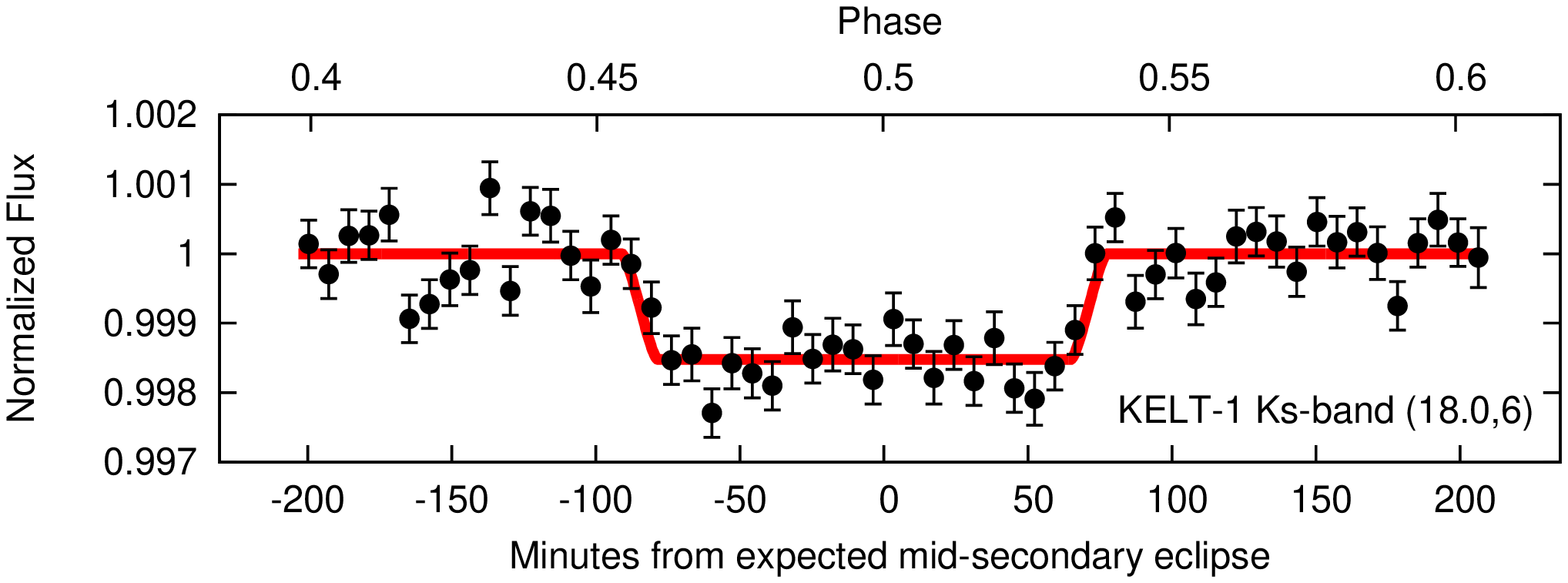}
\includegraphics[scale=0.26, angle = 270]{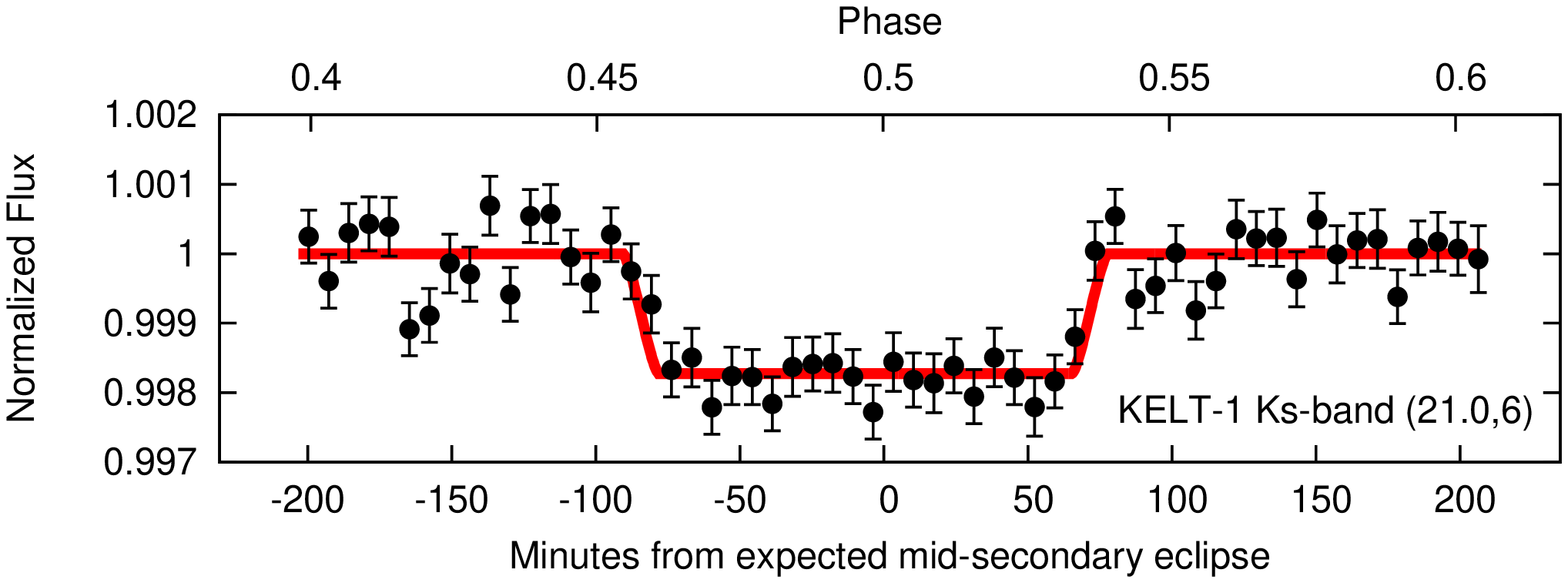}

\includegraphics[scale=0.26, angle = 270]{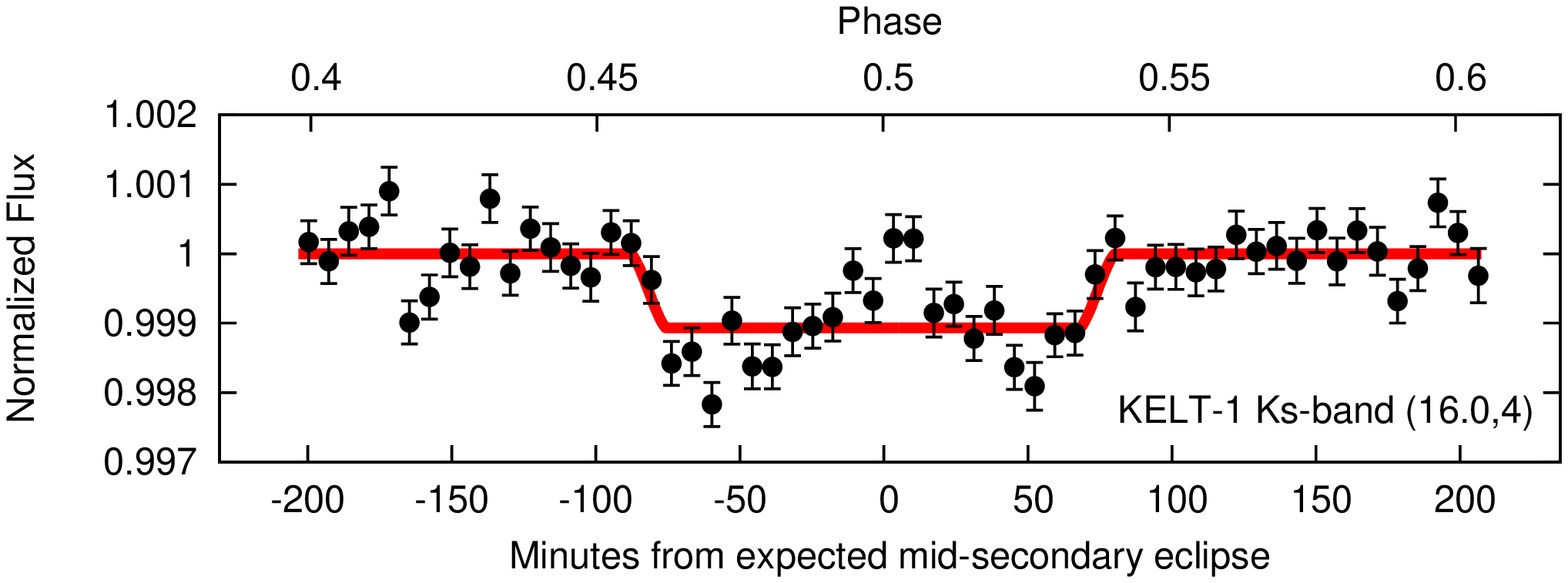}
\includegraphics[scale=0.26, angle = 270]{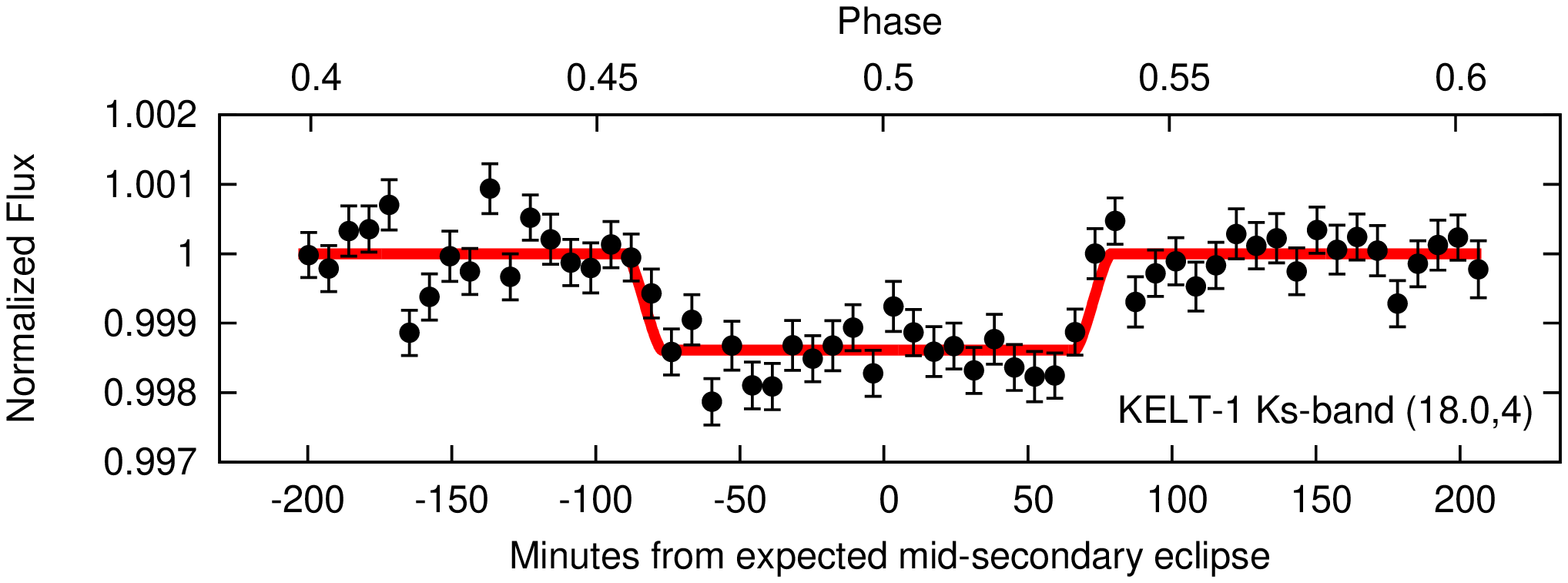}
\includegraphics[scale=0.26, angle = 270]{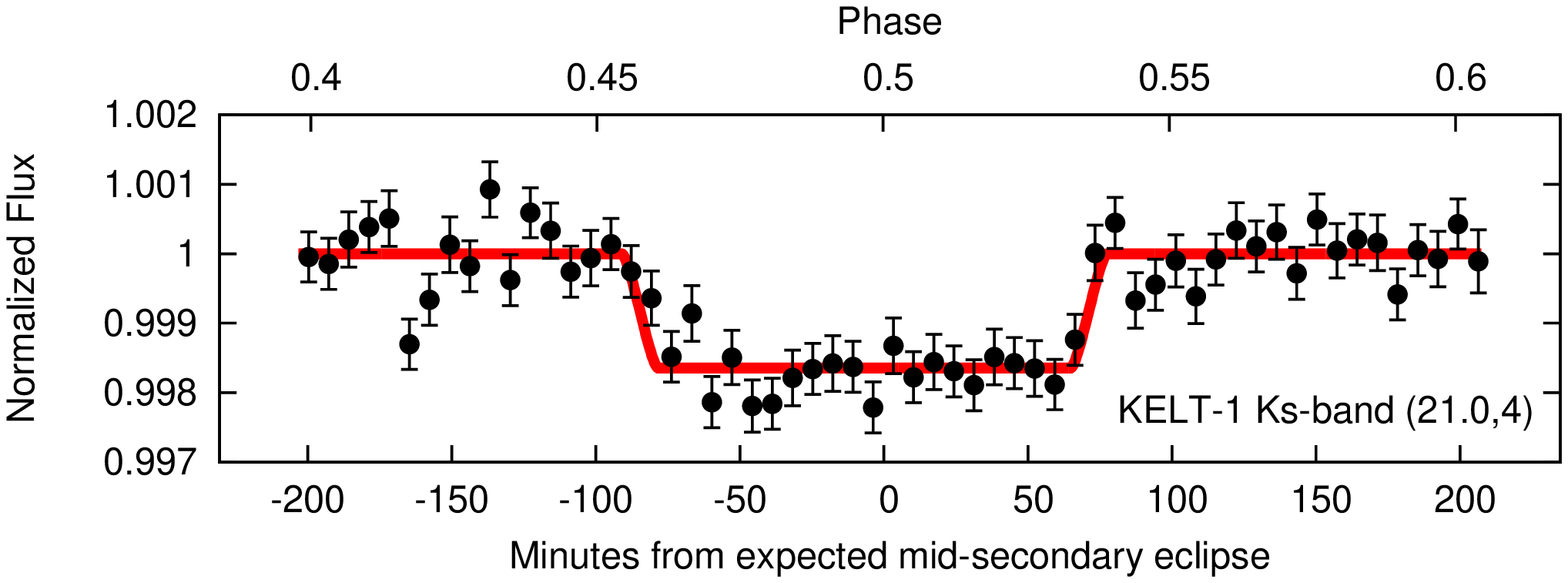}

\includegraphics[scale=0.26, angle = 270]{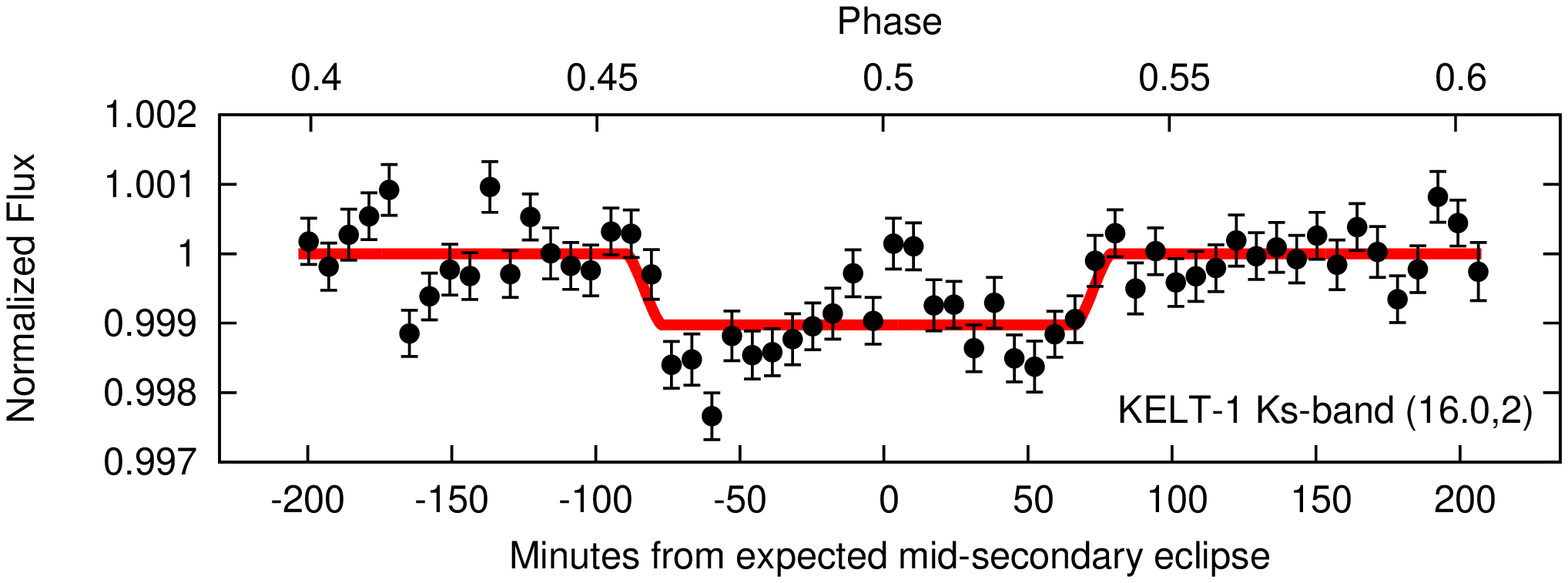}
\includegraphics[scale=0.26, angle = 270]{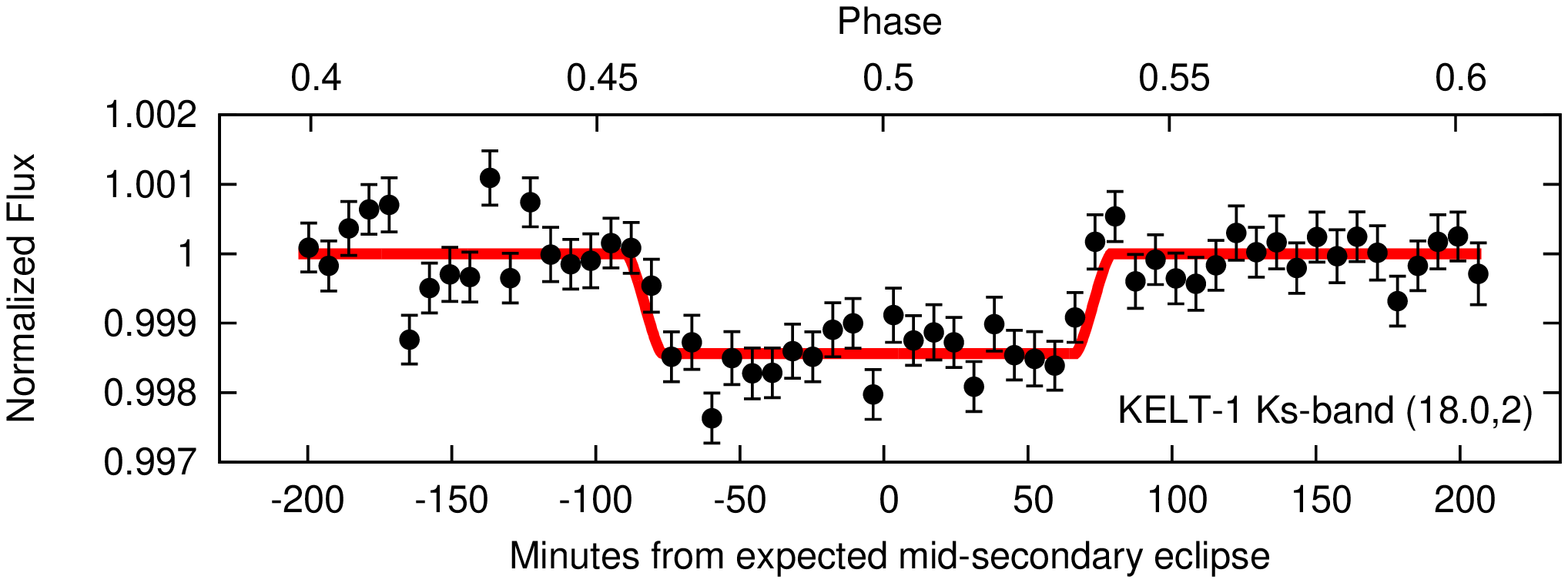}
\includegraphics[scale=0.26, angle = 270]{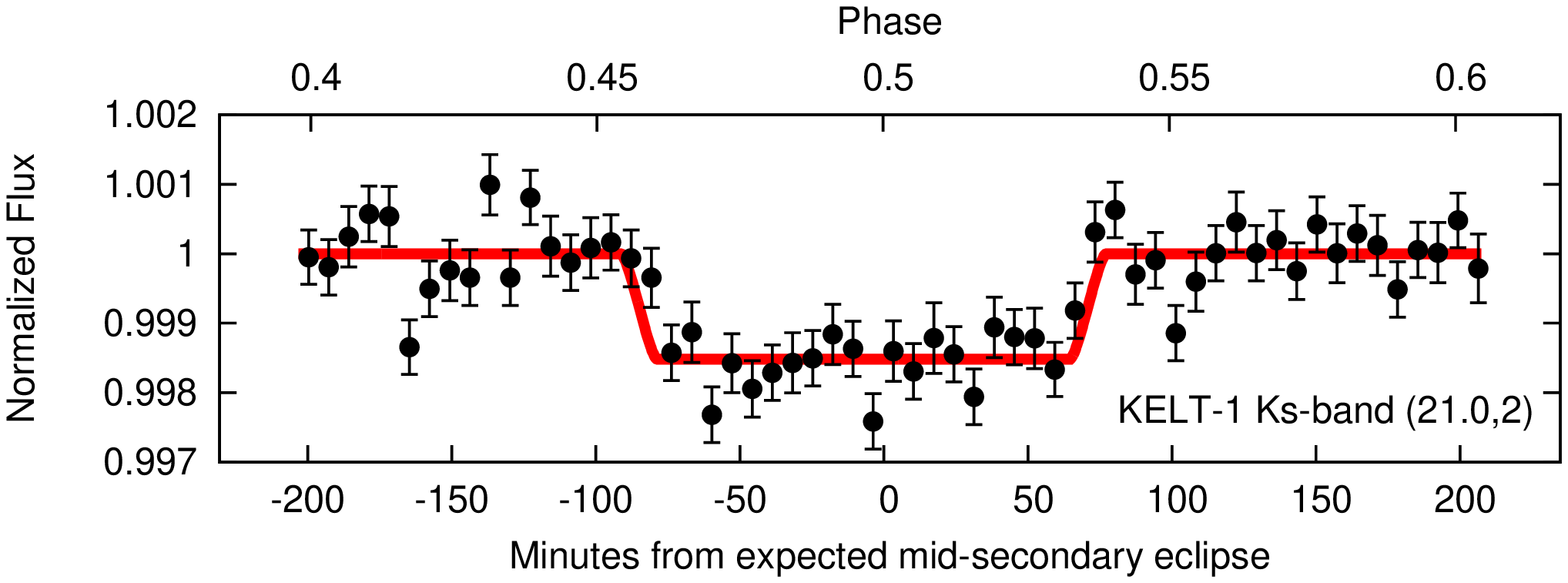}

\caption[BLAH]
	{
		Same as Figure \ref{FigWASPTwelveKsbandFidelityManyOne} except for our KELT-1 Ks-band secondary eclipse.
		For our KELT-1 Ks-band eclipse note the obvious correlated noise for the smallest apertures (the left set of panels for the bottom plots).	
	}
\label{FigKeltOneKsbandFidelityMany}
\end{figure*}

 Given these occasional correlations of the apparent eclipse depth, $F_{Ap}/F_{*}$, with the aperture size (Section \ref{SecAperture})
and the choice of
reference star ensemble (Section \ref{SecOptimalReferenceStars}), our
best-fit eclipse depth and associated errors need to take into account these correlations.
To do this we repeat our analysis for a variety of aperture size combinations and reference star ensembles and 
observe the variations in the precision (RMS$\times$$\beta^2$) and the eclipse depths;
Figure \ref{FigWASPTwelveKsbandFidelityManyOne} displays an example of these variations for our first WASP-12 Ks-band
eclipse. 
The same correlations with aperture size and the number of reference stars in the ensemble,
are displayed for our other data-sets in Figures \ref{FigWASPTwelveKsbandFidelityManyTwo}
- \ref{FigKeltOneKsbandFidelityMany}. 

We consider all aperture size and reference star ensemble combinations
that display nearly identical precision -- that is, we consider
all light curves within
\NUMBERHERERMSLIMIT\%\footnote{Our 				
choice of \NUMBERHERERMSLIMIT\% above the minimum RMS$\times$$\beta^2$ was an arbitrary choice,	
but experience has shown it to be a reasonable compromise
that captures a reasonable range of eclipse depth values for 
reasonably precise versions of our data-sets. For most of our data-sets
the 1$\sigma$ error on the eclipse depth is relatively insensitive to the precise value
we choose to accept above the minimum RMS$\times$$\beta^2$.
However, for data-sets
that show strong variations in the eclipse depth with aperture size and the choice
of reference star ensemble, such as Qatar-1 (Figure \ref{FigQatarOneKsbandFidelityMany}), the 
1$\sigma$ error-bar on the eclipse depth does depend on the precise percentage that we accept
above our minimum RMS$\times$$\beta^2$ value.}
of the minimum RMS$\times$$\beta^2$ observed in our grid of aperture sizes and reference
star ensembles. 
For our first WASP-12 Ks-band eclipse, the values within \NUMBERHERERMSLIMIT\% of the minimum 	
RMS$\times$$\beta^2$ observed are denoted by being enclosed in the red line in the top panels of Figure \ref{FigWASPTwelveKsbandFidelityManyOne}.

 To determine the accurate eclipse depth and error given these
correlations
with aperture size and the number of reference
stars in the ensemble, we marginalize over the variations in the eclipse depths for the best RMS$\times$$\beta^2$ values (all aperture size and reference star ensembles with RMS$\times$$\beta^2$
values within \NUMBERHERERMSLIMIT\% of the minimum RMS$\times$$\beta^2$), by combining the MCMC chains
of all these aperture size and reference star ensembles. We determine our best-fit parameters by 
simply applying our MCMC analysis to these combined Markov Chains. The uncertainties before and after we've corrected for these correlations for our various data-sets 
are given in Tables \ref{TableParams} and \ref{TableParamsOther},
and Figures \ref{FigWASPTwelveKsbandFidelityManyTwo} - \ref{FigKeltOneKsbandFidelityMany}.

 In most, but not all cases, by employing this technique the errors on our apparent
best-fit secondary eclipse depths, $F_{Ap}/F_{*}$,
marginally increase (Tables \ref{TableParams} and \ref{TableParamsOther}).
The necessity
of taking into account correlations of the eclipse depths with the aperture size and the number
of stars in the reference star ensemble is best demonstrated by 
our Qatar-1 Ks-band eclipse. In Figure \ref{FigQatarOneKsbandFidelityMany} there are a variety of 
reference star ensemble and aperture size combinations that fit the data with relatively similar goodness-of-fits 
(RMS$\times$$\beta^2$ values), that have significantly different eclipse depth values. Therefore, it is imperative
that our reported eclipse depth and the associated error, take into account these correlations.
The reported apparent secondary eclipse depth for our Qatar-1 Ks-band eclipse
changes from $F_{Ap}/F_{*}$ = \FpOverFStarPercentAbstractQatarOneKsBand$^{+\FpOverFStarPercentAbstractPlusQatarOneKsBand}_{-\FpOverFStarPercentAbstractMinusQatarOneKsBand} $\%
by solely considering the single best aperture and reference star combination,
to $F_{Ap}/F_{*}$  = \FpOverFStarPercentAbstractCorrQatarOneKsBandAll$^{+\FpOverFStarPercentAbstractCorrPlusQatarOneKsBandAll}_{-\FpOverFStarPercentAbstractCorrMinusQatarOneKsBandAll}$ \%
by marginalizing over the various selected aperture sizes and reference star ensembles.
For this reason our method should be
superior to a method that just scales up the errors on $F_{Ap}/F_{*}$ 
by an arbitrary factor to account for systematic errors, as our method
easily differentiates between those data-sets that display strong correlations with aperture
size and reference star ensembles, and those that do not.

 There appear to be no hard or fast rules in selecting aperture sizes and reference star combinations.
Although relatively small aperture sizes are occasionally favoured (Figure \ref{FigQatarOneKsbandFidelityMany}),
in other cases large apertures are favoured (Figure \ref{FigKeltOneKsbandFidelityMany}), and in others the
RMS$\times$$\beta^2$ is relatively insensitive to aperture size (Figure \ref{FigWASPThreeKsbandFidelityManyOne}).
Sometimes additional reference stars only contribute correlated noise and relatively few reference stars are
favoured (Figure \ref{FigWASPThreeKsbandFidelityManyTwo}), while in other cases a large number of reference 
stars are favoured (Figure \ref{FigWASPTwelveKContbandFidelityMany}).	


 We also repeat this analysis for the timing offset from the expected
mid-point of the eclipse, $t_{offset}$.
Although, we do not find strong correlations between the timing of the mid-point of our eclipses
with our data-sets analyzed using our various aperture size and reference star ensembles, the
different aperture size and 
reference star ensembles do appear to impart scatter in the timing offsets, that are greater than if they are
analyzed with a single aperture size or reference star ensemble. Therefore, the true uncertainty 
in the mid-point of the eclipse appears to be best returned once taking into account this scatter with the various 
aperture size and reference star ensembles.
We suspect that this method of correcting the mid-point of the eclipse
for correlations with aperture size and reference star ensemble will likely
limit the cases where spurious claims are made of eccentric close-in planets, due to a putative
measurement of a non-zero timing offset from the expected eclipse mid-point.

\section{Noise Budget of WIRCam ``Staring Mode'' Photometry}
\label{SecNoise}

\begin{figure}
\centering
\includegraphics[scale=0.45, angle = 90]{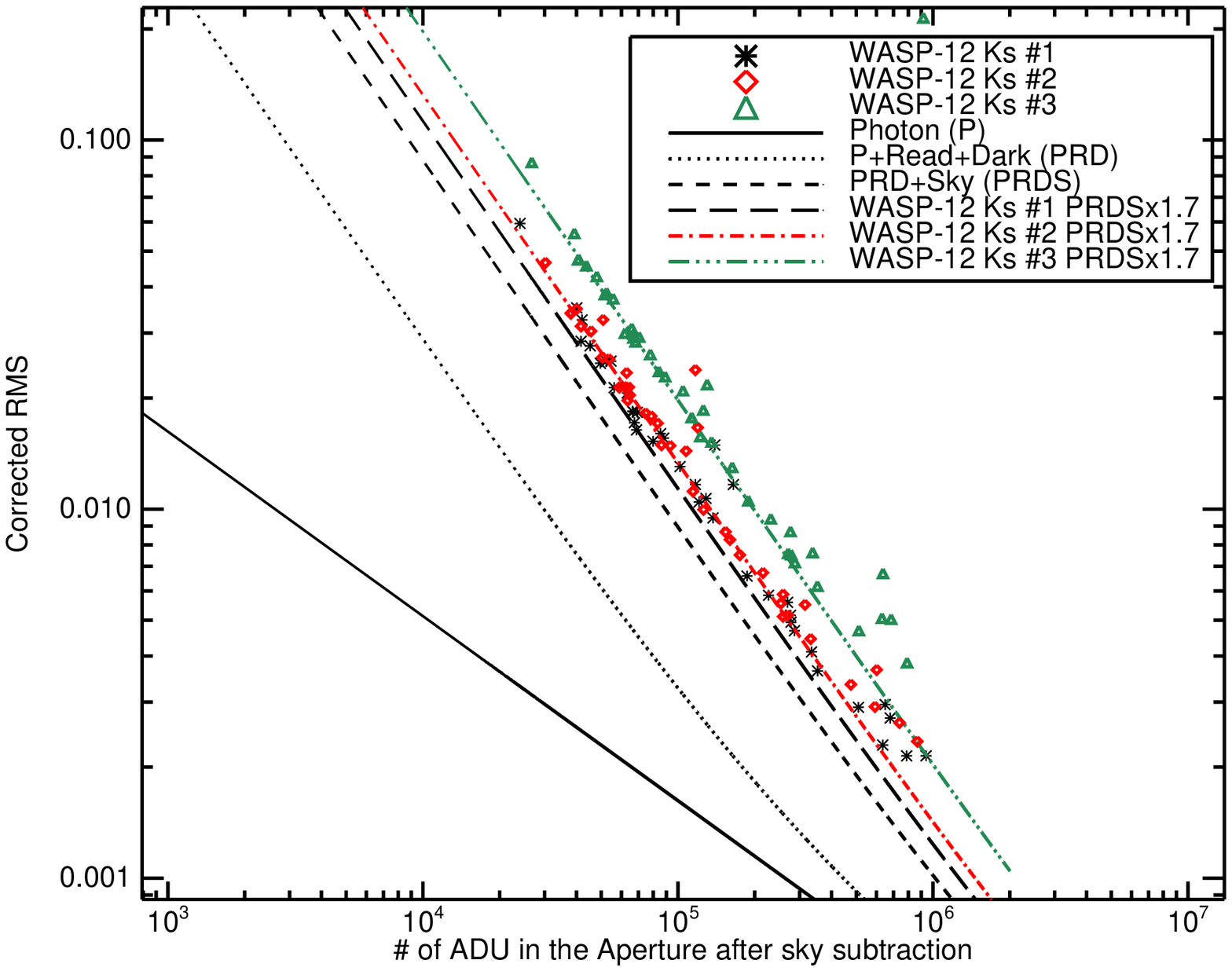}
\includegraphics[scale=0.45, angle = 90]{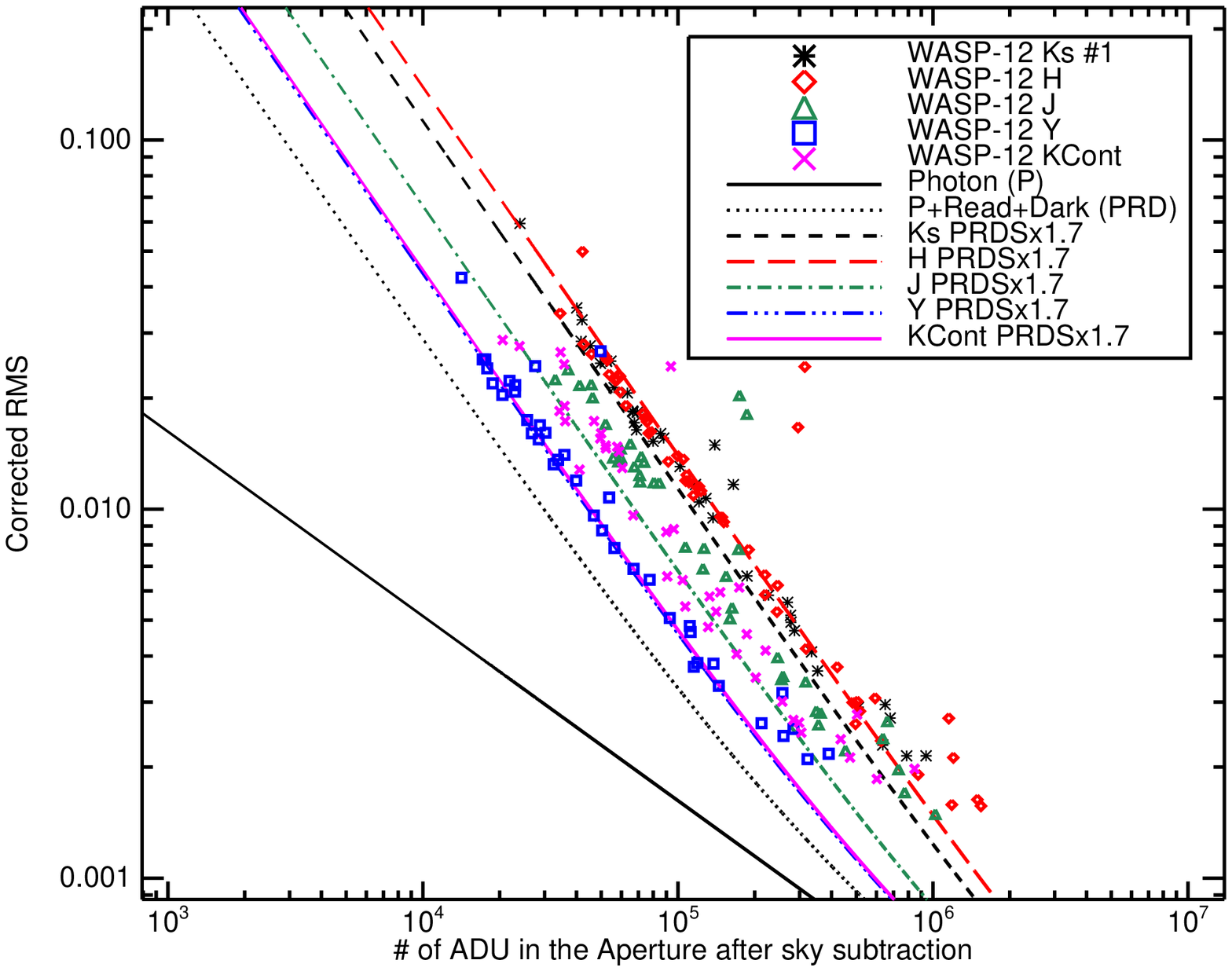}
\includegraphics[scale=0.45, angle = 90]{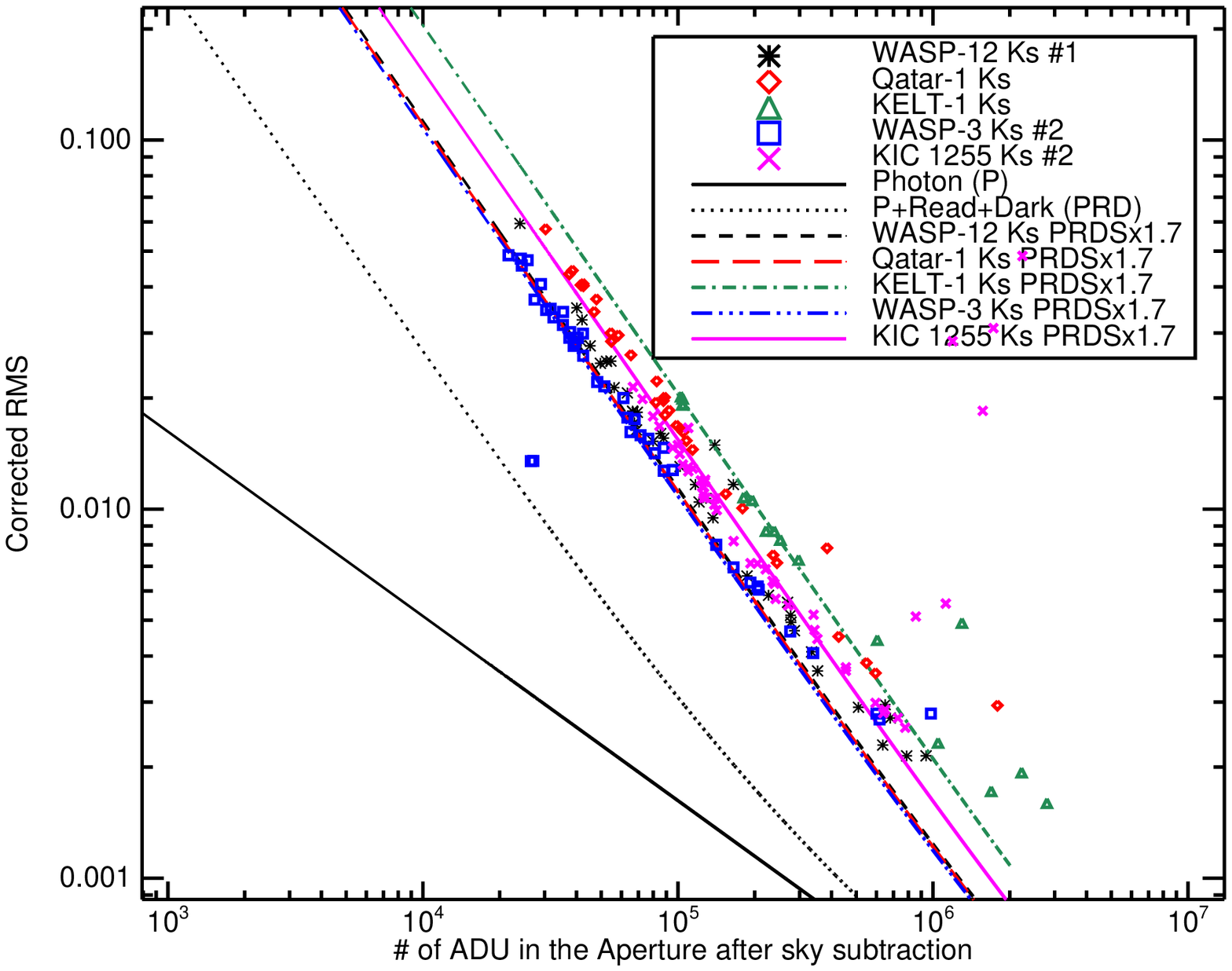}
\caption[Systematics All]
	{	\footnotesize
		The RMS of the reference stars for our various data-sets 
		(the various symbols as indicated in the legends)
		after correcting their flux with the 
		best ensemble of nearby reference stars.
		We subject each one of our reference stars to a near-identical
		differential photometric procedure that we normally apply to our target stars,
		as discussed in the text.
		For all panels the expected precision when accounting for simply photon noise (solid black line), 
		and photon, read-noise and dark current (indicated by ``PRD''; dotted black line), is given.
		The various other lines in these panels either indicate the precision when accounting
		for photon-noise, dark-current, read-noise and the sky (indicated by ``PRDS'' in the legend),
		or by all these factors and 1.7 the measured sky background (indicated by ``PRDS$\times$1.7''
		in the legend).
	}
\label{FigSystematicsAll}
\end{figure}

 We also explore the noise budget of our CFHT/WIRCam ``Staring Mode'' photometry; the goal
is to help identify the limiting systematic(s) in our ground-based, near-infrared photometry. 
To achieve this objective we utilize the fact that WIRCam's large field-of-view gives us access to a great
number of reference stars -- ranging from bright to faint -- that allow us 
to explore how the precision of our photometry scale with flux.
We therefore perform differential photometry on each one of our $\sim$15-50 or so reference stars
on the same WIRCam chip as our target star,
and correct their flux with the best ensemble of these nearby reference stars, using the exact same
method that we usually correct the stellar flux of our target star
(except for the fact that we do not
optimize our fits for each one of our reference stars for
the best reference star ensemble and 
aperture size combination as discussed in Section \ref{SecAperture} \& \ref{SecOptimalReferenceStars}).
We correct each one of our reference
stars with a four-star reference star ensemble\footnote{The four star
reference star ensemble for each ``chosen'' reference 
star are the four stars that minimize the RMS of the corrected
light curve flux for each ``chosen'' reference star.}.
We use aperture
sizes given in Table \ref{TableApertureNStar} for the WASP-3, Qatar-1, KELT-1, \& 
new WASP-12 light curves that we discuss here; for the other
light curves we use the best-fit apertures for those light curves as discussed in \citet{CrollKIC} for KIC 12557548,
and in \citet{CrollWASPTwelve} for the J, \& H-band light curves of WASP-12b.	
As our reference star light curves do not (presumably) display an obvious eclipse/transit, we
compute the RMS of the entire light curve; we subtract the best-fit quadratic
trend from the light curve (to correct for possible systematic
residual background trends as discussed in Section \ref{SecAnalysis}).

We display the resulting corrected RMS of all the reference stars 
of our various data-sets in Figure \ref{FigSystematicsAll}, compared
to the expectation with various noise sources taken into 
account\footnote{
We calculate the signal-to-noise, $SNR$, of our photometry using the ``CCD equation'' \citep{MerlineHowell1995}:
\begin{equation}
SNR = \frac{ N \times g }{ \sqrt{N \times g + n_{pix} (1 + \frac{n_{pix}}{n_{back}}) (N_{sky} \times g + N_D \times t + R_N ^2)} }
\label{EquationSignaltoNoise}
\end{equation}
where N is the number of ADUs from the star in the aperture,
and $g$ is the gain (3.8 $e-/ADU$ for WIRCam),
$n_{pix}$ is the number of pixels in the aperture,
$n_{back}$ is the number of pixels in the annulus used to estimate the sky background,
$t$ is the exposure time,
$N_{sky}$ is the sky noise per exposure in ADU/pixel,
$N_D$ is the dark current (which for the WIRCam array is only $\sim$0.05 $e-$/sec/pixel, and is therefore largely negligible),
$R_N$ is the RMS read-noise (for the WIRCam array, $R_N^2$=$30^2$ $e-$/pixel/read).
}.
As can be seen 
the read-noise, 
sky-background\footnote{We estimate the sky background for each individual exposure by taking 
the median of the annulus values for each target aperture; 
the sky background of the light curve is then the median of these values.}
and photon-noise all contribute appreciably to the expected noise budget.
There is an additional unknown systematic
that contributes noise at approximately the same level as the sky background; as can be seen, to first-order,
our data closely follows the noise limit if we multiply the sky-background by approximately $\sim$1.7 (the long-dashed
line in the top 							
panel of Figure \ref{FigSystematicsAll}, compared to 
the WASP-12 Ks first eclipse photometry (black stars)).
The various panels of Figure \ref{FigSystematicsAll} indicate
that the sky background is one of the dominant mechanisms in determining
the accuracy of our ``Staring Mode'' photometry; this is applicable for multiple observations of the same
star in the same band (multiple observations of the star WASP-12 in the Ks-band; the top panel of Figure \ref{FigSystematicsAll}), observations 
of the same star in
different bands (observations of WASP-12 in the YJHK and $K_{Cont}$-bands; the middle panel of Figure \ref{FigSystematicsAll}),
and observations of different targets in the same-band (the Ks-band; the bottom panel of Figure \ref{FigSystematicsAll}).
Unfortunately, the source of the systematic that causes our
light curve to scale at just less than twice the sky background level is unknown.



It is also evident that there is a subtle decrease in the expected
precision for the brightest stars (the highest illumination levels, and therefore the right side of the plots
in Figure \ref{FigSystematicsAll}); 
this is likely due to the fact
that for these stars there is a lack of equally bright reference stars to correct their
photometry, and that these stars often suffer from saturation or non-linearity effects.

\section{Discussion}
\label{SecDiscussion}

 We defer the constraints that our eclipse depths provide on the atmospheres of
hot Jupiters to our accompanying paper (Croll et al. in prep.).
Here we present both an investigation of the timing of the mid-point of our secondary eclipses in Section
\ref{SecTiming}, and a discussion of the repeatability of our ground-based, near-infrared
eclipse depths for the hot Jupiters WASP-12b and WASP-3b in Section \ref{SecWeather}.

\subsection{Phases of the mid-points of the transits}
\label{SecTiming}

 The mid-points of all the eclipses we present here are consistent with circular orbits,
as given in Tables \ref{TableParams} and \ref{TableParamsOther}.
We emphasize that one of the reasons that we find
a lack of offsets from the expected mid-point of the eclipse, $t_{offset}$, is due to the fact
that we have taken into account correlations of $t_{offset}$ with the aperture size and reference star ensemble,
as discussed in Section \ref{SecHonestEclipses}.

A secondary eclipse that occurs half-an-orbit after the transit, and therefore an eclipse 
detection that is consistent with a circular orbit,
agrees with previous 
secondary eclipse detections
for the hot Jupiter WASP-12b \citep{Campo11,CrollWASPTwelve,Cowan12},
and the brown-dwarf KELT-1 \citep{Beatty14}.		
For WASP-3b, \citet{Zhao12WASPThree} presented a previous detection 
of its thermal emission, and suggested the possibility 
that WASP-3b's orbit was mildly eccentric ($e \cos(\omega)$=0.0070$\pm$0.0032);
our finding of 
$e \cos(\omega)$ = \ECosOmegaCorrWASPThreeKsBandEclipseTwoAll$^{+\ECosOmegaCorrPlusWASPThreeKsBandEclipseTwoAll}_{-\ECosOmegaCorrMinusWASPThreeKsBandEclipseTwoAll}$
from our second Ks-band WASP-3 eclipse,
does not support their finding of an eccentric orbit.
Finally, this is the first thermal emission detection of Qatar-1b, and our secondary eclipse detection
provides no evidence in favour of an eccentric 
orbit\footnote{In the original Qatar-1 discovery paper \citep{Alsubai11} the formal radial-velocity
fit slightly favoured an eccentric orbit ($e$=0.24$^{+0.10}_{-0.12}$); however, the author's favoured a circular orbit,
and cautioned that the eccentric solution was likely spurious. A circular orbit was 
confirmed by \citet{Covino13} who performed radial velocity observations of Qatar-1 and were able
to place a limit on the eccentricity of the planet of: $e$=0.020$^{+0.011}_{-0.010}$.}.
Our findings are consistent with the notion that whatever primordial mechanism(s) dragged these hot Jupiters, and
the brown dwarf, to their present locations, if it imparted an initial
orbital eccentricity, then these eccentricities have been damped away
by tidal interactions with the host stars (e.g. \citealt{Lin96}).

\subsection{A limit on the temporal variability of two hot Jupiters}
\label{SecWeather}

\begin{figure}
\centering
\includegraphics[scale=0.60, angle = 270]{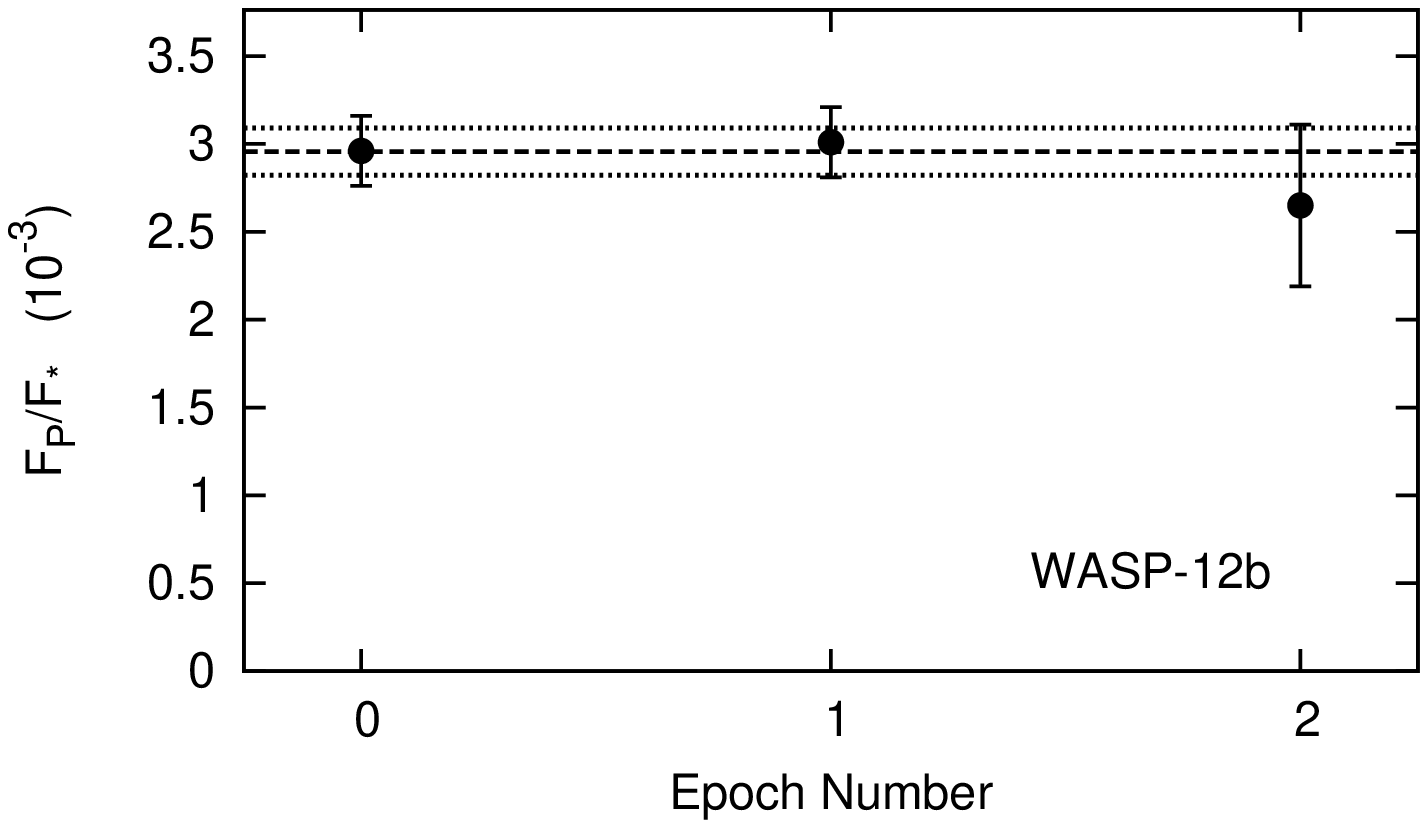}
\includegraphics[scale=0.60, angle = 270]{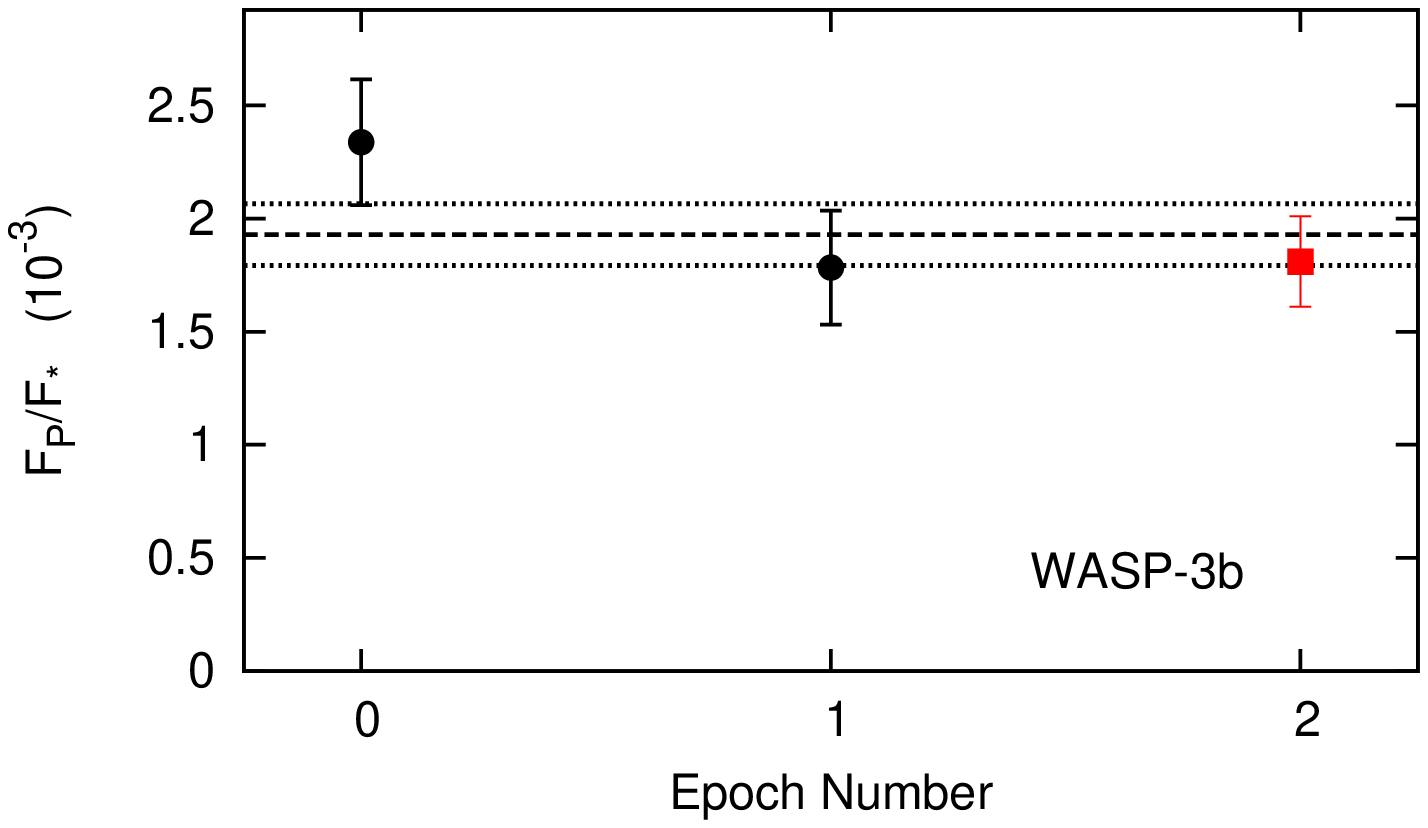}
\caption[Weather Plots for two hot Jupiters]
	{	The secondary eclipse depths of our three Ks-band eclipses of WASP-12 (top),
		and our two Ks-band eclipses of WASP-3 (black points), and another published by \citet{Zhao12} (red point).
		In both cases the dashed horizontal line indicates the weighted mean of the three observations,
		while the dotted horizontal line indicates the 1$\sigma$ upper and lower-limits
		on the weighted error of the three observations.
	}
\label{FigWeather}
\end{figure}

 A fundamental question that has to be asked of near-infrared detections of the thermal emission 
of hot Jupiters from the ground is whether these eclipses are repeatable within the sub-millimagnitude errors
that these depths are typically reported with
(e.g. \citealt{deMooijSnellen09,CrollTrESTwo,CrollTrESThree,CrollWASPTwelve,Bean13}).
Another question is whether the temperatures of the deep, high-pressure layers probed by near-infrared observations
\citep{Seager05,Fortney08,Burrows08} are stable, or whether they are variable due to violent storms that
have been predicted by some researchers \citep{Rauscher07,LangtonLaughlin08,MenouRauscher09}.
Here we demonstrate that multiple detections of the thermal emission in the Ks-band of WASP-3b and 
WASP-12b largely agree with one another, and we
are therefore able to place a limit on both
the impact of systematic effects on ground-based, near-infrared observations, and the presence of violent storms
in the deep, high pressure atmospheric layers of these two hot Jupiters.
The repeatability of these eclipse depths are presented in Figure \ref{FigWeather}, and summarized in Table \ref{TableCorr}.

The secondary eclipse depths from
our multiple detections of the thermal emission of WASP-12b in the Ks-band 
are displayed in the top panels of 
Figure \ref{FigWeather}.
The weighted mean and error of our WASP-12 Ks-band observations
are:
$F_{p}/F_{*}$ = \FpOverFStarPercentAbstractWeatherPlotWASPTwelveKsBandCombined \ $\pm$ \FpOverFStarPercentAbstractPlusMinusWeatherPlotWASPTwelveKsBandCombined\%
\ of the stellar flux after correcting for influence of the nearby M-dwarf binary companion \citep{Bergfors13,Crossfield12,Sing13}.
This corresponds to a limit on the brightness temperature in the Ks-band of 
$T_B$ = \TBrightWeatherPlotWASPTwelveKsBandCombined$^{+\TBrightPlusWeatherPlotWASPTwelveKsBandCombined}_{-\TBrightMinusWeatherPlotWASPTwelveKsBandCombined}$ $K$.
The reduced $\chi^2$ of our three WASP-12 eclipse depths are \ReducedChiSquaredWeatherPlotWASPTwelveKsBandCombined;
given the size of our 1$\sigma$ error bars on these points, it is likely only by
chance that these three eclipses are in such close agreement. 
Our third WASP-12 Ks-band eclipse is the most discrepant of our three eclipses, and is discrepant by less than 1$\sigma$. 
It is consistent to within $\sim$\FpOverFStarPercentDiffWeatherPlotWASPTwelveKsBandCombined\% of the stellar flux of our weighted mean,
or to within a temperature variation of $T_B$$\sim$\TBrightDiffHeatherFuhrWeatherPlotWASPTwelveKsBandCombined $K$.

 The Ks-band thermal emission of WASP-3 has already 
been presented in \citet{CrollThesis}, and \citet{Zhao12WASPThree}.
The analysis presented in \citet{CrollThesis} was a previous analysis of the data that we present here.
We combine the analysis that we present here of two eclipses of WASP-3b, and the eclipse depth reported
by \citet{Zhao12WASPThree} here.
The weighted mean and error of these three WASP-3 Ks-band detections 
are:
$F_{p}/F_{*}$ = \FpOverFStarPercentAbstractWeatherPlotWASPThreeKsBandCombined \ $\pm$ \FpOverFStarPercentAbstractPlusMinusWeatherPlotWASPThreeKsBandCombined\% of
the stellar flux, corresponding to a limit on the brightness temperature in the Ks-band of
$T_B$ = \TBrightWeatherPlotWASPThreeKsBandCombined$^{+\TBrightPlusWeatherPlotWASPThreeKsBandCombined}_{-\TBrightMinusWeatherPlotWASPThreeKsBandCombined}$ $K$.
The reduced $\chi^2$ of these three WASP-3 eclipse depths are \ReducedChiSquaredWeatherPlotWASPThreeKsBandCombined.
Our first WASP-3 Ks-band eclipse depth falls slightly outside the 1$\sigma$ 
error on our combined depth; skepticism may be warranted with our first Ks-band eclipse of WASP-3 due to the fact
that it features very little out-of-eclipse baseline just prior to the eclipse. 
This depth is consistent to within $\sim$\FpOverFStarPercentDiffWeatherPlotWASPThreeKsBandCombined\%
of the stellar flux of the weighted mean,
or to within a temperature variation of $\sim$\TBrightDiffHeatherFuhrWeatherPlotWASPThreeKsBandCombined $K$.

The consistent eclipse depths of our WASP-12, as well as our WASP-3 eclipses, 
allow us to place a strict limit on the systematics
inherent in ground-based near-infrared photometry.
For our WASP-12 eclipses, we emphasize that the close agreement in the eclipse depths
may be due to the fact that 
the eclipses were observed with the same
telescope/instrument configuration; for the WASP-3 Ks-band eclipses 
the \citet{Zhao12WASPThree}
eclipse depth provides an independent confirmation with another telescope/instrument - the Palomar 5-m telescope 
and the WIRC instrument. Further independent measurements of
the Ks-band thermal emission of WASP-3b and WASP-12b are encouraged. 
We note two such additional detections of WASP-12b's thermal emission in the Ks-band have been presented by
\citet{Zhao12}; these detections feature reduced precision compared to the results we presented 
here, but nonetheless the weighted mean of these observations agree with our own
results\footnote{We note the important caveat that 
the WASP-12b eclipse depths presented in \citet{Zhao12} were not corrected for the dilution due to the  
M-dwarf binary companion to WASP-12, as these binary companions were not known at the time.}.

In addition to placing a limit on systematics on ground-based near-infrared photometry, our
repeated eclipse depths also allow us to place a limit on epoch-to-epoch, temperature differences 
of the deep, high pressure region of these two hot Jupiters, due to violent storms.
\citet{MenouRauscher09} predicted temperature changes on the
order of $\sim$100 $K$ for a canonical HD 209458-like hot Jupiter from a three-dimensional numerical
model.
We are able to rule-out such large temperature variations at the epochs of
two of our observations for WASP-12b, and for two of the WASP-3b observations; our first WASP-3b observation
varies by $\sim$\FpOverFStarPercentDiffWeatherPlotWASPThreeKsBandCombined\%
of the stellar flux (or $\sim$\TBrightDiffHeatherFuhrWeatherPlotWASPThreeKsBandCombined $K$) from the mean eclipse level, but
the large uncertainty in the eclipse depth for our second WASP-3 Ks-band eclipse
($F_{p}/F_{*}$ = \FpOverFStarPercentAbstractCorrWASPThreeKsBandEclipseOneAll$^{+\FpOverFStarPercentAbstractCorrPlusWASPThreeKsBandEclipseOneAll}_{-\FpOverFStarPercentAbstractCorrMinusWASPThreeKsBandEclipseOneAll}$\%)
means that there is no compelling evidence for any sort of temperature fluctuations
in WASP-3b.
Our limits on eclipse depth variability
for the deep, high pressure regions probed by our Ks-band, near-infrared observations for the hot Jupiters
WASP-12b and WASP-3b, complement previous limits in the stratospheres
of the hot Jupiters using the {\it Spitzer}/IRAC instrument, including HD 189733 \citep{Agol10}, and
the hot Neptune GJ 436 \citep{Knutson11}. 

\acknowledgements

B.C.'s work was performed under a contract with the California Institute of
Technology funded by NASA through the Sagan Fellowship Program.
The Natural Sciences and Engineering Research Council of Canada supported the research of B.C. during much of the completion
of the work described here.
The authors especially appreciate the hard-work and diligence of the CFHT staff
for both scheduling the challenging CFHT observations described here and ensuring these ``Staring Mode'' observations were successful.
We thank Kevin Schlaufman and Gordon Walker for providing helpful comments that improved this manuscript.


\end{document}